\documentclass[12pt]{report}
\usepackage{fancyhdr}
\usepackage{tesisty}
\usepackage{mathtools}
\usepackage[T1]{fontenc}
\usepackage{babel}
\usepackage{a4, color}
\usepackage{amsmath, amsfonts, amssymb}
\usepackage{siunitx}
\usepackage{lipsum}
\usepackage{chngcntr}
\usepackage[hidelinks]{hyperref}
\usepackage{amsthm}
\usepackage[compat=1.1.0]{tikz-feynman}
\tikzfeynmanset{warn luatex=false}
\usepackage{braket}

\DeclareMathOperator{\Tr}{Tr}
\DeclareMathOperator{\sign}{sign}

\DeclareMathOperator{\sgn}{sgn}
\DeclareMathOperator{\curl}{curl}
\DeclareMathOperator{\sech}{sech}

\usepackage{braket}
\usepackage{comment}
\newcommand{\pad}{4pt}
\DeclarePairedDelimiter{\angi}{\langle}{\rangle}
\usepackage{slashed}
\usepackage{xcolor}
\definecolor{RED}{named}{red}
\usepackage{float}
\usepackage{siunitx}
\usepackage{cite}
\usepackage{amsmath,amssymb}
\usetikzlibrary{matrix,fit,positioning}

\definecolor{energy}{HTML}{D81B60}   % magenta-ish
\definecolor{eflux}{HTML}{1E88E5}    % blue
\definecolor{mdens}{HTML}{43A047}    % green
\definecolor{press}{HTML}{FB8C00}    % orange
\definecolor{shear}{HTML}{8E24AA}    % purple
\usepackage{bm}
\newcommand{\pref}[1]{(\ref{#1})}
\usepackage[a4paper,
left=4 cm,
right=3 cm,
top=3 cm,
bottom=3 cm,
]{geometry}
\usepackage{tabto}
\usepackage{systeme}
\usepackage{bookmark}
\numberwithin{equation}{section}
\usepackage{graphicx}
\usepackage{hyperref}
\usepackage{physics}
\usepackage[style=super]{glossaries}
\usetikzlibrary{shapes.geometric, arrows}

\tikzstyle{startstop} = [rectangle, rounded corners, 
minimum width=3cm, 
minimum height=1cm,
text centered, 
draw=black, 
fill=red!30]

\tikzstyle{io} = [trapezium, 
trapezium stretches=true, % A later addition
trapezium left angle=70, 
trapezium right angle=110, 
minimum width=3cm, 
minimum height=1cm, text centered, 
draw=black, fill=blue!30]

\tikzstyle{process} = [rectangle, 
minimum width=3cm, 
minimum height=1cm, 
text centered, 
text width=3cm, 
draw=black, 
fill=orange!30]

\tikzstyle{decision} = [diamond, 
minimum width=1cm, 
minimum height=1cm, 
text centered, 
draw=black, 
fill=green!30]
\tikzstyle{arrow} = [thick,->,>=stealth]

\newenvironment{acknowledgements}%
{\cleardoublepage\thispagestyle{empty}\null\vfill\begin{center}%
		\bfseries Acknowledgements\end{center}}%
{\vfill\null}
\newtheorem{obs}{Osservazione}[section]

\newtheorem{pro}{Problema}[chapter]

\newtheorem{teor}{Teorema}[section]

\newtheorem{defn}{Definizione}[section]

\fancypagestyle{plain}{
	\fancyhead{} } 

\numberwithin{equation}{section}

\newcounter{abstractpage}
\begin{document}
\cleardoublepage
\pdfbookmark[1]{Frontespizio}{frontespizio}
	\corso{Dottorato in Scienze Fisiche e Chimiche} \titoloTesi{Non-equilibrium Quantum Field Theory and Axion Electrodynamics in curved spacetimes} \ciclo{XXXVIII ciclo}
	\anno{2024-2025}
	\relatore{ Prof Roberto Passante}
	\autore{Amedeo Maria Favitta}
	\correlatore{Prof Lucia Rizzuto}
	\baselineskip=25pt
	\intestazione	
	\newpage
\thispagestyle{empty}
\newpage

\cleardoublepage
\pdfbookmark[1]{Dedica}{dedica}
\begin{dedication}
\vspace*{4cm} % Adjust vertical spacing from the top

To my grandpa Amedeo, who would have been very proud and happy of me.\\[0.5em]
To Sirio, Nicola, Giulia, my book friends and my family, I couldn't have done this without you. Thank you for all of your support along the way.\\[2em]

\begin{center}
\textit{``Touch the wooden gate in the wall you never saw before.\\
Say ‘please’ before you open the latch, go through, walk down the path.''}
\end{center}

\begin{flushright}
— Neil Gaiman, \textit{Instructions}
\end{flushright}
\end{dedication}
\cleardoublepage
\pdfbookmark[1]{Abstract}{abstract}	\begin{abstract}\label{abstract}
		\thispagestyle{plain}
		\setcounter{page}{\value{abstractpage}}
Axions are a class of hypothetical fundamental particles introduced formerly as a solution to the Strong CP problem of Quantum Chromodynamics (QCD), but also have been obtained in several low-energy compactification models of String Theory.\\
Various astronomical and experimental constraints imply that the axion is 'invisible' in the sense that its interactions with Standard Model (SM) particles are significantly weak, which is why the axion is regarded as a viable candidate for Dark Matter.

In this thesis, we discuss our new results on the topics that have been developed during the three PhD years, in particular on Axion Cosmology and Axion Electrodynamics, research areas of strong interest nowadays, where the production of axion particles with their topological defects, along with the interaction between SM particles and the axions themselves, are respectively studied and have been object of published papers and conference and workshop presentations. In particular, with a light on methods and results in non-equilibrium Quantum Field Theory and Quantum Field Theory in curved spacetimes.

Those topics are addressed in the present PhD thesis. We first review the basics of Quantum Field Theory in curved spacetimes and some elements of Cosmology. We will introduce the Strong CP problem in the Standard Model of particle physics, which heavily justifies the introduction of a new particle, the QCD axion. This includes the Peccei-Quinn solution for the Strong CP problem in QCD, the first Peccei-Quinn-Wilczek-Weinberg (PQWW) axion model, and the invisible QCD axion models. 
Furthermore, we analyze the two main classes of UV completions to QCD axion theory: the "field theory" completion models and the extradimensional ones, both related to the so-called "quality problem" for the axions and the last one justifying the possible existence of further axions, known as axion-like particles (ALPs).

All these introductory parts are relevant to understanding the current state and problems in the literature on axions and on Quantum Field Theory in curved spacetimes; they are also significant to understanding and putting in the correct perspective the relevance of our original results, new theoretical results, and new experimental methods.

We first investigate a practically interesting aspect of Axion physics, which is Axion Electrodynamics.
Axion Electrodynamics is the study of the modifications to Electrodynamics due to the presence of an interacting classical axion field.

It is deeply connected to applications for axion detection, since many of the most important experimental devices for detecting the axion utilize a strong magnetic field, and we expect a more significant classical behavior for the DM axion field, although some recent claims of stochasticity from surviving axion miniclusters.
However, it is also related to the theoretical aspects we deal with in the involved Axion Cosmology case, particularly energy-momentum conservation and the interplay between the condensate and particle kinetic regimes.

Furthermore, we study some aspects that are less investigated in the literature, such as the effects of the axion field on electromagnetic Casimir forces. This leads to modifications of dispersion relations and zero-point energies that could be detected by Casimir force experimental setups or by astronomical observations. 
We found the study of the Casimir force for a spatially dependent axion field to be of remarkable interest, as it is deeply connected to cosmological thermal friction. 
We will consider in particular the following aspects due to the axion-photon interactions:
\begin{itemize}
\item Basic aspects and modifications to Maxwell equations
\item Energy-momentum conservation in Axion Electrodynamics
    \item Cavity haloscope models
    \item Green's functions in Axion Electrodynamics
    \item Casimir physics in Axion Electrodynamics
    \item Thermal friction on the effective axion domain wall
    \item Optical properties of the axion medium
\end{itemize}

We then investigate the non-equilibrium quantum field theory dynamics of a self-interacting axion field interacting with a generic Standard Model sector or a Dark Sector, considering the path integral approach and the 2PI effective theory. 

We demonstrate that the $\rm nPI$ approach enables us to extend previous approaches and overcome their limitations in the approximations, in particular in the dynamical regimes where non-linear effects are relevant.  

We use the considerations above to start with dealing with two main cosmological scenarios: 
\begin{enumerate}
    \item Preinflationary scenario for high-mass photophilic ALPs: We discuss the cosmological constraints on the parameter space $(m_a,g_{a \gamma \gamma})$, in particular in the region of the high-mass axions with $m_a >10\, \mathrm{keV}$, from the contribution to $\Delta N_{eff}$ due to the irreducible axion freeze-in production.
    %The basic idea in the way of calculating this minimal possible production is the following: we consider inflationary scenarios with low reheating, since 
    %We will assume an initial axion field, after the inflation,
    \item Postinflationary scenario for the QCD axion and high-mass photophilic ALPs: We discuss the Domain Wall problem for such models and analyse the dynamics of the networks of these axion topological defects, from the current analytical models based on the Velocity-One scale framework and the extensions of it through a non-equilibrium QFT method. 
    This last method comes with adopting an extension of moduli space quantization.

    In both cases, we analyze the friction effects for the photophilic ALPs due to the interaction with the SM primordial plasma by taking care of plasma effects, which we claim to be mostly coming from the presence of electrons and muons, and show the validity of our model and the connection with former approaches and constraints, along with preliminary results.
\end{enumerate}

\end{abstract}
%\setcounter{page}{\value{abstractpage}}
%\stepcounter{page}
\newpage
\pagenumbering{Roman}
\cleardoublepage
	\tableofcontents
	\newpage
	\cleardoublepage
	\pagenumbering{arabic}
	
	\fancyhead[R]{Introduction} \fancyfoot[L]{Introduction}
	\fancyfoot[R]{\thepage}

	\part{Introduction}\label{intro}
\chapter{Overview}
Axions were initially proposed to solve the Strong CP problem in QCD \cite{bigi2021new,PhysRevLett.40.223,schwartz2014quantum}, but they also got a strong theoretical interest because they are one of the most quoted suggested components of Dark Matter, one of the theoretical main problems in Cosmology and Physics nowadays  \cite{marsh2016axion,PhysRevD.85.105020,PhysRevD.91.065014,RevModPhys.93.015004,marsh2017axionsalpsshortintroduction,Marsh:2024ury}.\\ 
However, the Axion is still an undetected particle \cite{axiondmchadhaday, ohare2024cosmology}, so numerous efforts have been made to detect it. The most promising methods utilize its interaction with electromagnetic fields. 
The research area of Axion Electrodynamics is then historically focused on studying the interaction between the two fields to get theoretical results that can be useful to propose experimental devices designed to detect the axions \cite{RevModPhys.93.015004,app12136492}. Indeed, a properly designed axion electromagnetic detection device needs to be sensitive because it deals with electric and magnetic fields that are typically much smaller than those externally applied by orders of magnitude. 

That last point suggests that, as it has been tried in the literature and as we try to do in this thesis with our original work, one needs to design them to boost the axion-generated fields as much as possible.
Another sensible and wider problem concerns the axion theoretical and cosmological models: what should we expect to observe?
Indeed, since the allowed parameter space for axions, e.g., mass $m_a$ and coupling constant $g_{a \gamma \gamma} $ with SM photons, is very huge, the possible expectations on the input axion signals are quite extensive. We list here some of them:
\begin{itemize}
    \item Depending on the mass $m_a$ and the decay constant $f_a$, an axion dark matter model can cover a lot of classes of dark matter models, from cold dark matter, self-interacting dark matter, to fuzzy dark matter. Furthermore, several models of ALPs have been proposed in the literature for candidates to forming Dark Energy or be inflatons, with interesting effects when considering the Schwinger effect; unfortunately, such proposals are less popular than those primarily involving Dark Matter \cite{Pajer_2013,Smith:2024ayu,PhysRevD.104.123504,Smith:2025grk,iarygina2025schwingereffectaxioninflation}.
    \item Independently of the physical parameter, there exist uncertainties related to the axion quality problem and consequently to the UV completion of the theory \cite{choi2024axiontheorymodelbuilding}.
\end{itemize}All these aspects and problems are indeed treated in the object of the following PhD Thesis. 

Our original results will focus on the non-equilibrium QFT and nPI formalisms, and in particular, on relevant theoretical aspects related to the condensed and kinetic regimes, as well as the classical and quantum regimes, and their applications to two cosmological scenarios for the axion. They will be also object of next works , such as Ref.~\cite{Favitta2025-AxionBounds-prep,Favitta2025-AxionPoS-prep}. \\
The main limits of our works are the energy scales and the leading hypotheses we will limit on, which are for our theoretical formalism to assume a fixed classical background metric tensor, so we do work far below the Planck energy $E_{\text{Pl}}$, neglecting the backreaction of our quantum fields on the metric.

\begin{figure}[h]
    \centering
    \includegraphics[width=0.5\linewidth]{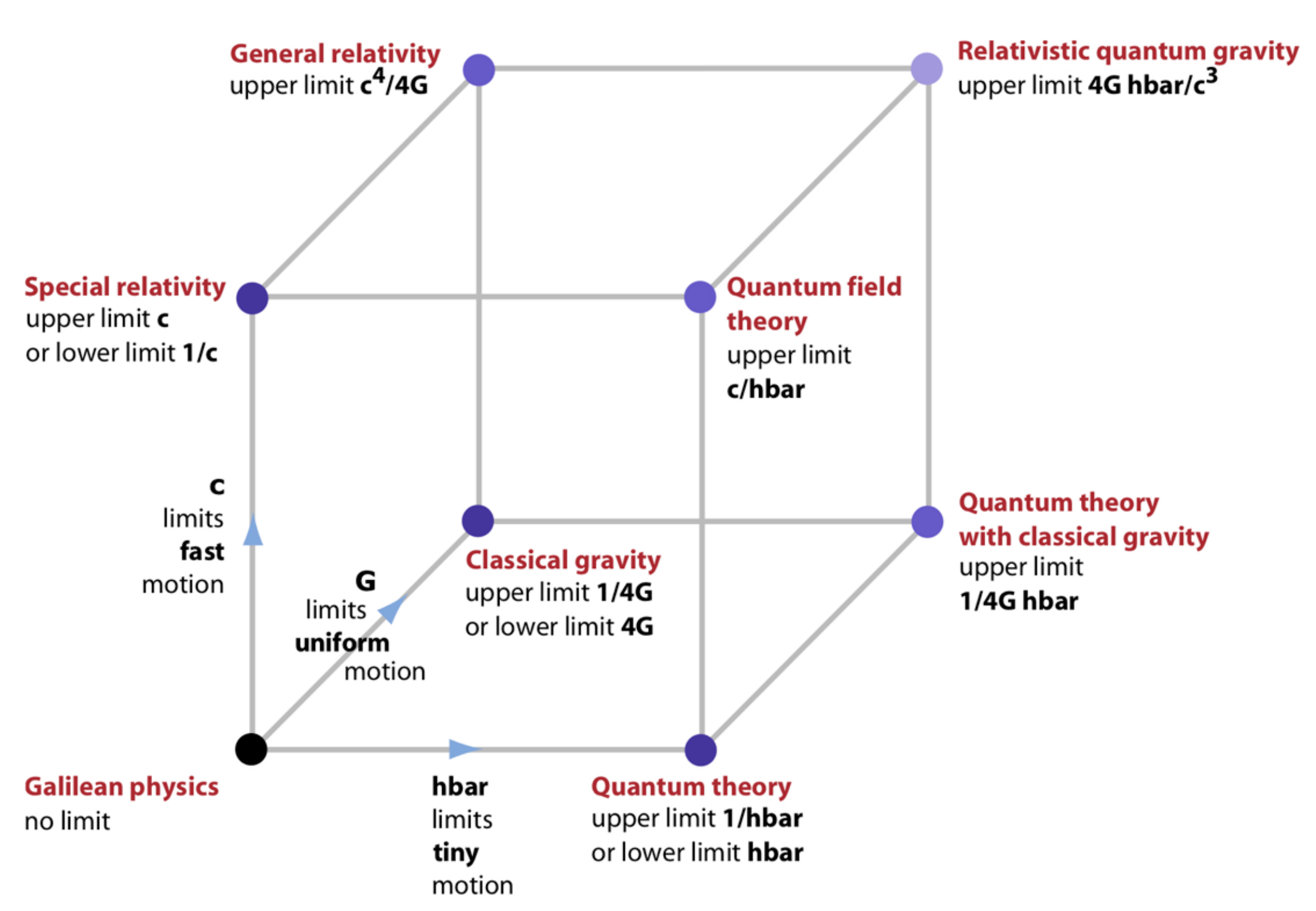}
    \caption{The Bronshtein's cube, showing the main current physics theories and how they are connected. They can be distinguished by which combination of the fundamental constants and the typical speed of the physical system of interest is contained more significantly. Our work is in the $(1,1,0)$ cube vertex. Image from Ref.~\cite{unknown1}.}
    \label{planckcube}
\end{figure}

This aspect is also to underline an essential element of the axion that is intrinsic in its name: the axion is a "detergent" in the sense that, if we consider all the axion models, they cover a range of energies and scales of the whole Planck cube we show in Figure \ref{planckcube}.
In this sense, the axion is a significant class of particles since it is a "reminder" of all the current problems in Fundamental Physics.

In this work, we first explore the fundamentals of Axion theory, beginning with the solution of Roberto Peccei and Helen Quinn to the Strong CP problem in QCD, starting from the model's Lagrangian and the resulting modified Maxwell equations and non-equilibrium dynamics.

We extend the theory about the wave equations and dispersion relations treated by Refs.~\cite{PhysRevD.101.123503,zhang2015time} for a dynamical axion field in Refs.~\cite{PhysRevD.107.043522,FAVITTA2023169396,doi:10.1142/S0217751X24500040}.\\
We also extend the theory of the axion-to-photon conversion process and some methods developed in the former literature for example the Haloscope model and the optical properties.

One of the most successful and established methods for detecting axions is the Haloscope model, which exploits the axion-to-photon processes resulting from the interaction between axions and an electromagnetic field in an externally applied static magnetic field. This process is enhanced by placing the system inside an electromagnetic cavity.\\
%We show that, according to our results, we get Boost factors of electric fields that are usually one order of magnitude lower than the ones claimed by \cite{Millar_2017} and \cite{Majorovits_2020}, so power boosts that are two orders of magnitude lower, in the regime of a big number N of plates and big refractive index (e.g. n=5). \\
%We find,
%For example, an electromagnetic planar wave with frequency $\omega_0$ interacting with the Axion field in LW approximation (or equivalently in the reference frame of the Axion fluid at rest) is characterized by a Raman effect-like emission
%at $\omega_a+\omega_0$ and $|\omega_0-\omega_a|$, where $\omega_a$ is the frequency of the axion field, and this is also gotten for the case of our 'Axion Echo Haloscope' system, where we have an electromagnetic cavity with an externally applied electromagnetic wave.
%One could also study the Raman-like emission by interferometric methods, and this is our new idea of 'Axion Echo interferometer'.
%They are methods that could be promising if built in practice because they could have boost factors bigger than Haloscope and Dielectric Haloscope.\\
%Furthermore, the two new experimental methods we are developing, which we named 'Axion echo Haloscope' and 'Axion echo interferometer', are two novelties for the literature.\\
The effects of the interaction between the axion field and the electromagnetic field on electromagnetic Casimir forces are less investigated in the literature. One mostly finds works on this topic for the effective 'Axion Electrodynamics' of topological insulators \cite{PhysRevD.100.045013} and they are the main object of our work~\cite{FAVITTA2023169396}.
We find that the interaction with the axion field leads to modifications of the dispersion relations and zero-point energy of the electromagnetic field, which could be detected experimentally (realistically not for standard Casimir force experimental setups, but by modifying them or more easily in cosmological scenarios) or through astronomical observations (modifications of emission spectra, e.g., the blackbody spectrum) or helpful for topological materials. All of these points can be of interest for Casimir-Polder interactions too as discusses in a next work \cite{Campello2025}. \\
These two last aspects are discussed along with the calculation of the Green functions in Axion Electrodynamics, where we find some new complementary results to the ones in \cite{PhysRevD.102.123011}.
This work also demonstrates that axion-modified Casimir forces may be relevant to Early Universe models.\\
The methods we develop in the following are also useful for treating the Green functions of the electromagnetic field interacting with the axion field. In this way, we investigate the modifications to zero-point energy of the electromagnetic field, the Axion-modified Casimir effect, and also calculate them for the case of an axion domain wall in the Early Universe. In such a case, we find that radiation pressure acts on the axion topological defects in the Early Universe, preventing them from stabilizing in some region of parameter space with $N_{DW}>1$. 

If those stable configurations survived until our time it would conflict with cosmological observations\cite{PhysRevD.85.105020}.
\section{Outline of the thesis}
The Thesis is divided into two parts, as follows.
\begin{itemize}
    \item 
The Part \ref{intro} is the introductory part of the thesis, where we review the current status of the research of Axion Physics and introduce relevant aspects of Cosmology and Quantum Field Theory in curved spacetimes.

In particular, we introduce relevant aspects of the cosmological metric, the history of the Universe and the quantization of quantum fields in a flat FLRW metric in Chapter ~\ref{curvedqft}, while we introduce the Strong CP problem in QCD and current main axion models, including both QCD axions and axion-like particles, in Section~\ref{strongCP}.

These are the first theoretical aspects which are treated in Chapter~\ref{curvedqft} useful to understand the next chapters of Part~\ref{intro}, which are dedicated to experimental and observational
constraints (Chapter~\ref{experiments}), Axion Cosmology (Chapter~\ref{axioncosmo}) and Axion Electrodynamics (Chapter~\ref{baseAED}).
\item The Part~\ref{II} develops our original results. It starts with the related ideas for Axion Electrodynamics (Chapter~\ref{DeepAED}). Subsection \ref{greenaqua} is dedicated to our general treatment of Green's function in Axion Electrodynamics, while the following sections are dedicated to toy models and applications of our formalism for Cosmology and topological materials, ranging from the case of a purely time-dependent axion field, to the optical properties of the axion medium   until the case of a purely space-varying axion field  , which has applications for topological materials and axion domain walls.

We will focus in particular to this last point, since the phenomena of thermal friction and planar compression on the domain wall are relevant aspects for the next chapter, where we will face them with non-equilibrium formalism.

Chapter~\ref{noneqQFT} is focused to our approach with non-equilibrium Quantum Field Theory in curved spacetimes.
We consider two toy models in Section~\ref{Berges},which are helpful to understand the results we will obtain in the successive sections of this work.

We then consider our approach to non-equilibrium quantum field theory in fixed curved spacetime from first principles, by introducing the full path integral and then obtaining the generating functional, from which 2PI approach can be dealt.

We then apply this formalism to two cases of interest.

First case is about the pre-inflationary scenario with high-mass axion-like particles with a significant freeze-in production and the case of post-inflationary scenarios of axion models with an instanton potential and $N_{\rm DW}>1$, which are characterized by the domain wall problem.

The second case is about the QCD axion wall networks and the photophilic ALPs wall networks where friction is significant, in relation with the problem of  explaining and correcting the differences in the theoretical predictions of the mass of a dark matter QCD axion in a postinflationary scenario. At the same time, the two cases show the importance of quantum effects for misalignment mechanism and networks of axion topological defects.
\end{itemize}

\chapter{Notation}
We adopt the $g_{\alpha \beta}=diag\{+1,-1,-1,-1\} $ signature for the metric tensor.\\3D vectors suffixes are denoted by Latin letters \textit{i, j, k}..., while 4-vectors ones by Greek letters $\alpha$, $\beta$, $\gamma$ ...
Furthermore, we also adopt the notation $g_{M N}$ with capital Latin letters to denote the indices for extradimensional scenarios.
We adopt the following abbreviations:
\begin{itemize}
\item LHS=Left-hand side
\item RHS=Right-hand side
	\item QED= Quantum Electrodynamics
    \item FLRW metric= Friedman-Lemaitre-Robertson-Walker metric\footnote{We will adopt the name used in Refs.~\cite{weinberg1972gravitation,misner2017gravitation,baumann2022cosmology} }
	\item AED= Axion Electrodynamics
    \item DM= Dark Matter
    \item SM=Standard Model
    \item BSM= Beyond the Standard Model
    \item QFT=Quantum Field Theory
	\item LW approximation=Long Wavelength approximation
	\item HF approximation= High frequency approximation
    \item EoM=Equation of Motion
    \item $\Lambda\text{CDM}$ model=Standard Model of Cosmology, name referring to the two main components which are the Cosmological Constant $\Lambda$ and the Cold Dark Matter 
    \item CMB=Cosmic Microwave Background
\end{itemize}
We use both $\mathcal{F}(g)$ and $\hat{g}$ to denote the Fourier Transform of the function g, and here the Fourier Transform is defined by the following relations:
\begin{equation}
	\begin{split}
		\mathcal{F}(g)(\omega)&=\int_{-\infty}^{+\infty} d^d x \; e^{-i\omega x} g(x) \\
		g(x)&=\frac{1}{2 \pi}\int_{-\infty}^{+\infty} d\omega \; e^{i\omega x} \hat{g}(\omega)\\
		\delta(\omega)&=\int_{-\infty}^{+\infty} d^d x \; e^{-i\omega x}
	\end{split}
\end{equation}
If not written differently, a function h=h(x) is to be understood in the following as a function of time \textit{t} and position $\vec{x}$.\\
We denote the axion field as $a(x)=a(\vec{x},t)$ as usual in the literature and we define $ \theta(x)=\frac{a(x)}{f_a}$ and $\Theta(x)=g_{a \gamma \gamma} a(x)$. \\
The measure system I am adopting is the natural units system, defined by: $\hbar = c = 4 \pi \epsilon_0 = 1$. The remaining unit is chosen to be Energy (measured in $\mathrm{eV}$, particularly in most cases in $\mathrm{MeV}$ or $\mathrm{GeV})$.\\
This is a list of useful conversions between natural units and CGS units:
\begin{itemize}
	\item $ eV \sim 1.602 \cdot 10^{-12}\, \mathrm{erg}$
	\item $ eV \sim 1.783 \cdot 10^{-33}\, \mathrm{g}$
	\item $ eV^{-1} \sim 1.973 \cdot 10^{-5}\, \mathrm{cm} $
	\item $ eV^{-1} \sim 6.582 \cdot 10^{-16}\, \mathrm{s} $
\end{itemize}
Furthermore, the gravitational constant is $G \simeq 6.70711 \times 10^{-38}\,\,\mathrm{GeV}^{-2} $. 
For describing electromagnetic fields (and writing Maxwell equations), we use the Heaviside-Lorentz system of units.

The metric signature we adopt for the Minkowski spacetime is
 \begin{equation}
     \eta_{\mu \nu}=(+1,-1,-1,-1),
 \end{equation}
%We express the Einstein equation as 
 %\begin{equation}
 %    G_{\mu \nu}=R_{\mu \nu} +\frac{1}{2} R g_{\mu \nu}=8 \pi G\, T_{\mu \nu}
 %\end{equation}
%where $R_{\mu \nu}$ is the Ricci tensor, $R$ is the Ricci scalar, $G_{\mu \nu}$ the Einstein tensor and $T_{\mu \nu}$ the matter stress-energy tensor\footnote{%add here comment from SET part
% }. 
 We will also adopt the convention of Ref.~\cite{misner2017gravitation} of the Levi-Civita tensor that we write in our coordinate basis
\begin{equation}
    \varepsilon_{\alpha \beta \gamma \delta}=\frac{1}{\sqrt{|g|}}[\alpha \beta \gamma \delta],
\end{equation}
where 
\begin{equation}
   [\alpha \beta \gamma \delta]=\begin{cases} 
+1 \quad&\text{if \,$\alpha \beta \gamma \delta $ is an even permutation of 0123}\\
-1 \quad&\text{if \,$\alpha \beta \gamma \delta $ is an odd permutation of 0123}\\
0 \quad&\text{if \,$\alpha \beta \gamma \delta $ are all different}
   \end{cases}
\end{equation}
is the totally antisymmetric pseudotensor and $g$ is the determinant of the metric tensor $||g_{\alpha \beta}||$.
\begin{equation}
    g \coloneqq \det{||g_{\alpha \beta}||}.
\end{equation}
This will be useful since we use them and it allows us to see explicitly that an interaction term of the form
\begin{equation}
    \sqrt{|g|} \tilde{F}_{\alpha \beta} F^{\alpha \beta}=[\alpha \beta \gamma \delta] F^{\gamma \delta} F^{\alpha \beta}
\end{equation}
does not depend explicitly on the metric, where $\tilde{F}_{\alpha \beta}=\varepsilon_{\alpha \beta \gamma \delta} \,F^{\gamma \delta}$.

	\fancyhf{} %elimina header/footer vecchi

	\fancyhead[R]{\rightmark} \fancyhead[L]{\leftmark}
	\fancyfoot[R]{\thepage}

%	\pagenumbering{roman}
%	\title{A review on Axion Electrodynamics}
%	\author{Amedeo Maria Favitta\thanks{Universit\`{a} degli Studi di Palermo}}
%	\date{Io}
%	\maketitle
%\begin{acknowledgements}
%	To Timothy, a great loving father  \#justicefortimothy
%\end{acknowledgements}

\thispagestyle{empty}

%\listoffigures
%\listoftables	
%\pagenumbering{arabic}
%#\include{./Assioni}
%#\part{Axion theory and searches}
%#\chapter{The Strong CP problem of QCD and the axion}\label{AED}
\chapter{Theoretical background}\label{theoback}

\section{Quantum Field Theory in a curved spacetime}\label{curvedqft}
In this chapter, we provide a brief introduction to Quantum Field Theory (QFT) in a curved spacetime. In particular, we present the fundamental aspects of the Friedmann-Lemaître-Robertson-Walker (FLRW) metric that will be adopted throughout this work. We then focus on several key elements of the theory that will serve as the foundation for the original results of this thesis, such as the canonical quantisation procedure for a scalar field in a time-dependent background, and its applications to spin-0 and spin-1 particles in an FLRW metric.

Finally, we address some advanced theoretical subtleties that highlight an important conceptual point of Quantum Field Theory, namely that the fundamental physical entities are the fields themselves, since the very notion of a “particle” depends on the reference frame and the observer.
\subsection{Some fundamental aspects of Gravity and Cosmology}
In this thesis, we will limit ourselves to one concrete example of a homogeneous and isotropic metric, which is the flat Friedman-Lemaitre-Robertson-Walker metric \cite{misner2017gravitation, weinberg2008cosmology, baumann2022cosmology}
\begin{equation}\label{metric0}
    ds^2=dt^2-R^2(t)\, \delta_{ij} \,dx^i dx^j,
\end{equation}
where $t$ is the cosmological time, $R(t)$ is the scale factor and $\delta_{ij}$ is the Kronecker delta.
%We will limit ourselves, since %add

We will also adopt the conformal time $\eta$ defined as $R \, d\eta=dt$, the Hubble parameter $H=\frac{\dot{R}}{R}$ and the conformal Hubble parameter $\mathcal{H}=\frac{R'}{R}$, where the dot denotes the derivative with respect to the cosmological time $t$, while the prime the derivative respect to the conformal time $\eta$.

We will consider very minimal extensions of the $\Lambda \text{CDM}$ model with just QCD axions or Axion-Like particles.
The "Standard Model of Cosmology" predicts our current universe to be homogeneous and isotropic at large scales today\footnote{We mean by "large scales" length scales much bigger than roughly $L \sim 10 \,\mathrm{Mpc}$   \cite{baumann2022cosmology}} and behave like a perfect fluid with three main components.
In particular, we assume the stress-energy tensor of each component $i$ to be of the simple form
\begin{equation}
    T^i_{\mu \nu}=(\rho_i+p_i)\, u_{\mu}^a \,u_{\nu}^a+p_i \,g_{\mu \nu}
\end{equation}
and the total stress-energy tensor to be the sum of them\footnote{Interaction terms are neglected in the $\Lambda\text{CDM}$ cosmology.}. A simple pic of a matrix representation of a stress-energy tensor is shown in Fig.~\ref{stressenergy}.
 \begin{figure}\label{stressenergy}
	\centering
	\begin{tikzpicture}[every node/.style={font=\Large}]
		%======== Matrix (parentheses, no mu/nu labels) ========%
		\matrix (m) [
		matrix of math nodes,
		nodes={anchor=center, inner sep=3pt},
		column sep=10pt, row sep=8pt,
		left delimiter={(}, right delimiter={)}
		] {
			|(n00)| T^{00} & |(n01)| T^{01} & |(n02)| T^{02} & |(n03)| T^{03} \\
			|(n10)| T^{10} & |(n11)| T^{11} & |(n12)| T^{12} & |(n13)| T^{13} \\
			|(n20)| T^{20} & |(n21)| T^{21} & |(n22)| T^{22} & |(n23)| T^{23} \\
			|(n30)| T^{30} & |(n31)| T^{31} & |(n32)| T^{32} & |(n33)| T^{33} \\
		};
		
		%======== Rectangular outlines ========%
		% T^{00} alone
		\node[draw=energy, thick, rounded corners, fit=(n00), inner sep=\pad] {};
		
		% Energy flux T^{01}-T^{03} (unified)
		\node[draw=eflux,  thick, rounded corners, fit=(n01)(n03), inner sep=\pad] {};
		
		% Momentum density T^{10}-T^{30} (unified)
		\node[draw=mdens,  thick, rounded corners, fit=(n10)(n30), inner sep=\pad] {};
		
		% Pressures (normal stresses): three separate rectangles
		\foreach \x in {11,22,33} {
			\node[draw=press, thick, rounded corners, fit=(n\x), inner sep=\pad] {};
		}
		
		% Shear stresses: six separate rectangles
		\foreach \x in {12,13,21,23,31,32} {
			\node[draw=shear, thick, rounded corners, fit=(n\x), inner sep=\pad] {};
		}
		
	\end{tikzpicture}
	\caption{Matrix representation of a generic stress-energy tensor, which will be useful for the following of the thesis. Image inspired by a similar one from Elizabeth Winstanley.}
\end{figure}
The three components differ in the equation of state, which is the perfect fluid relation between the pressure $p_i$ and the energy density $\rho_i$, which is assumed to be linear
\begin{equation}
    p_i=w_i \rho_i
\end{equation}
\begin{itemize}
    \item Radiation: Component of ultrarelativistic and massless particles, with $w=\tfrac13$.
    It can be intuited from the Perfect Gas Law and Planck's law $p=\tfrac13\rho$
    %talk about argument from introduction to cosmology or just do my way?
     \item Matter: Component of non-relativistic particles, with $w=0$. It can be intuited analogously, since we have differently $m \gg |\vec{p}|$, where $m$ is the mass of the particle and $\vec{p}$ the 3-momentum.
     \\ We furthermore divide the matter component into two subcomponents:
     \begin{itemize}
         \item Baryonic Matter, the ordinary matter we normally see. %add more on bM
         \item Dark Matter, a suggested component that explains several gravitationally anomalous phenomena \cite{baumann2022cosmology}, which is a motivation for the axions. %add more on DM
     \end{itemize}
     \item Cosmological constant: A mysterious component that allows us to explain the current acceleration of the expansion of the Universe, and with $w=-1$. It is the Dark Energy\footnote{We mean by "Dark Energy" any component with $w<-\tfrac13$.} component assumed in the $\Lambda \text{CDM}$ model.
\end{itemize}
It is worth mentioning that the recent Dark Energy Spectroscopic 
 Instrument(DESI) results are putting in serious discussion the $w=-1$ paradigm, with a possibility for $-1<w<-1/3$ \cite{AbdulKarim2025_DR2I,AbdulKarim2025_DR2II,Brodzeller2025_DampedLyA,Elbers2025_NeutrinoConstraints,Lodha2025_ExtendedDE,Andrade2025_Validation}, which was already slightly challenged by the Hubble problem, for which the estimations of the Hubble constant $H_0$ from CMB and supernovae are found different.%add reference

It is customary in the Cosmology of the Early Universe to work out its time evolution in terms of the primordial plasma temperature $T$ instead of the cosmological time $t$, since it helps a lot to connect with the observations, e.g. one can directly measure the temperature of the CMB and the cosmological redshift $z$ while the time $t$ is affected from the model dependence of $R(t)$. Furthermore, it helps connect with the microphysics, and the Boltzmann equations are easier to handle (we will introduce them in Section \ref{freezein}). 

Furthermore, the scattering rates $\Gamma$ of the SM particles in the plasma are typically much bigger than the Hubble rate during most of its history, so we expect for the plasma, with an excellent approximation, a local thermal equilibrium (LTE) at each instant of time \cite{kolb1991early}.

The "dynamics" of the FLRW metric are set up from the scale factor $R(t)$, which tells us about the geometric expansion of the Universe.

We talk about the expansion of the Universe, since the typical scale factors we will deal with, also for our BSM models,  are characterized by $\dot{R}>0$, and we will assume a standard history of the Universe, where a period of inflation is present.\footnote{ Cosmological bounce models try to explain the problems of the homogenenity, isotropy   assuming an epoch before the Big Bang singularity where $\dot{R}<0$. They are very common in String Cosmology \cite{CICOLI20241}.%add ref here
}.

As Wheeler once wrote, "matter tells spacetime how to curve," \cite{Wheeler1990} and we need the Einstein equation to relate the scale factor and our components of the Universe.
In particular, this can be done with the Friedmann equations %\footnote{These are the versions with $k=0$ \cite{baumann2022cosmology}}
\begin{equation}
\begin{split}
    &\text{First Friedmann equation:}\quad H^2(t)=\frac{8 \pi G}{3} \rho+\frac{\Lambda}{3},\\
    &\text{Second Friedmann equation:}\quad \frac{\ddot{R}}{R}=-\frac{4 \pi G}{3} (\rho+3p)+\frac{\Lambda}{3}.
    \end{split}
\end{equation}
Combining the two, we obtain the energy conservation law
\begin{equation}
    \dot{\rho}+3H(\rho+p)=0.
\end{equation}

A rough estimate of the connection between cosmological time $t$ and plasma temperature $T$ is the following
\begin{equation}\label{TMeVvst}
    \frac{T}{1 \, \mathrm{MeV}} \sim \Bigg(\frac{t}{1 \mathrm{s}}\Bigg)^{-\tfrac12}
\end{equation}
during the radiation-dominated era.
This result comes from just inserting in the RHS of the first Friedmann equation a dominant component  with $\rho(t) \propto a^{-3(1+w)}$ with $w >-1$:
\begin{equation}
    H=\frac{2}{3(1+w)t}\,,
\end{equation}
take then just $\rho_{\text{rad}} \propto T^4$ and $w=\tfrac13$ .
\footnote{For $w=-1$ which is the case of a Cosmological Constant, the Hubble parameter is a constant $H_0$ in time
%, which is what DESI results seem to spoil.
}
Such expressions are just good for rough estimates, since they do not consider the dependence of radiation energy density from the relativistic degrees of freedom $g_{*}(T)$, as we will show in more detail in Section  .
%add section in the text on numerical boltzmann

However, they can usually be adopted having a specific value of $g_{*}$ in mind for a specific temperature range.
Anyway, a more precise expression is the following \cite{kolb1991early,baumann2022cosmology}
\begin{equation}\label{HRD}
    H\simeq 0.66\times  \sqrt{g_*(T)}\, \frac{T^2}{M_{\rm Pl}}
\end{equation}%adjust
%We will show a more complete approach within the $\Lambda \text{CDM}$ model %more uniform writing for \LambdaCDM
%in Section %add section
Some useful physical quantities we will adopt in the following are
\begin{itemize}
\item The Planck mass $M_{\text{Pl}}$ and the reduced one $\bar{M}_{\text{Pl}}$ we mentioned in the notation, with $\bar{M}_{\text{Pl}}=\sqrt{\frac{1}{8 \pi G}}$ in our units. In this way, we can write Eq.~(\ref{TMeVvst})  as $H \sim \frac{T^2}{\bar{M}_{\text{Pl}}}$.
    \item $H(T=T_{QCD})\simeq 1.972 \times 10^{-11} \, \mathrm{eV}$ with a reference value of the temperature\footnote{The QCD transition is a smooth crossover, since there is no discontinuity in thermodynamical quantities \cite{Borsanyi:2020fev}.This is a relevant point for comparisons with other works, since the number of relativistic degrees of freedom varies of roughly a multiplicative factor $3$ between $100 \, \mathrm{MeV}$ and $200 \, \mathrm{MeV}$ \cite{baumann2022cosmology} } at the QCD crossover transition $T_{QCD}\simeq 150 \, \mathrm{MeV}$.
\end{itemize}

\begin{figure}
    \centering \includegraphics[width=0.75\linewidth]{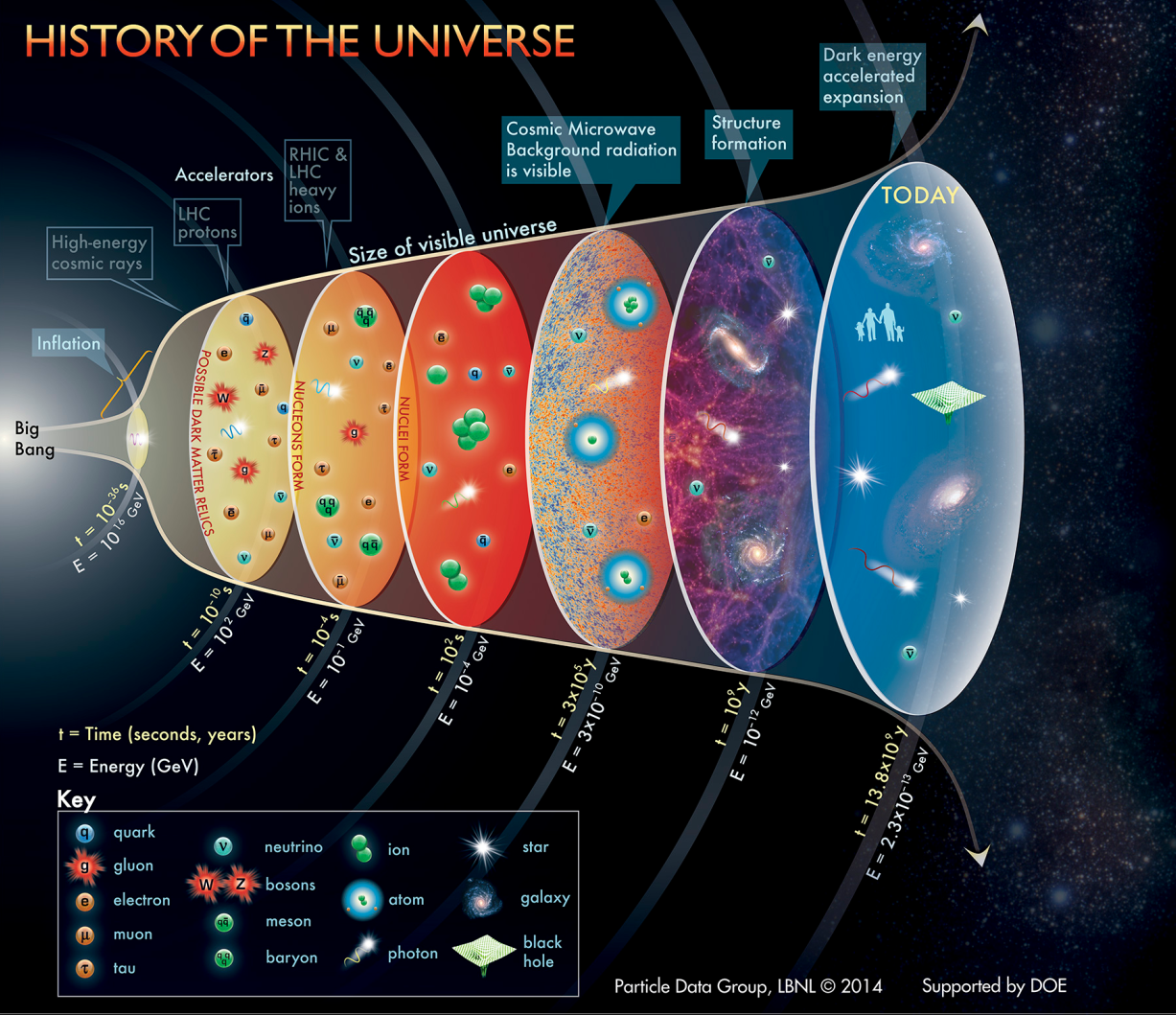}
    \caption{Brief history of the Universe by Particle Data group\cite{olive2014review}. 
.}
    \label{fig:placeholder}
\end{figure}

We will focus in particular on the following Axion Dark Matter scenarios:
    \begin{itemize}
\item Preinflationary scenarios for high-mass ALPs.

        \item Postinflationary scenario for a high-mass ($m_a > 10 \,\mathrm{keV}$ axion-like particle) and in particular, to the scenario between the moment the axion domain walls are created (when $m_a \sim H$ for which we can define a time $t_1$ \cite{Sikivie2008}), then from the relation~(\ref{HRD})
        we consider a scenario where they are formed at $T> 10^{5} \,\mathrm{GeV}$.

        \item Postinflationary scenario for the QCD axion: In such a case we have more limited constraints and less model-dependence.
        Here the time $t_1$ is
        with a good approximation inside or near to the early QCD era.
    \end{itemize}

\subsection{Particle production by a time-dependent potential}\label{UtQFT}
In the Early Universe, we will work with quantum fields living in a FLRW background, so we need to know how to handle quantum fields in time-dependent backgrounds.
Let us start by considering the simplest example of a massive quantum scalar field $\hat{\phi}$ in Minkowski spacetime, which obeys the usual Klein-Gordon equation, but is subjected to an external potential $U(t)$
\begin{equation}
    [\Box+m^2+U(t)]\hat{\phi}(x)=0,
\end{equation}
Since the potential is only time-dependent and without space dependence, the equation is linear in the spatial coordinates and a convenient form of the orthonormal basis is $u_{\vec{k}}=N_{\vec{k}}\, \hat{f}_{\vec{k}}(t) \,e^{i \vec{k} \cdot {\vec{x}}}$
with 
\begin{equation}
    \hat{\phi}(x) =\int_{\Omega_{\vec{k}}} \frac{d^3 \vec{k}}{(2 \pi)^3} N_{\vec{k}}\, \hat{f}_{\vec{k}}(t) \,e^{i \vec{k} \cdot {\vec{x}}},
\end{equation}
where $N_{\vec{k}} $ is a normalization factor, $\Omega_{\vec{k}}$ is the phase space with coordinates $\vec{k}=(k_1,k_2,k_3)$ associated to the spatial coordinates of theMinkowski spacetime $\vec{x}=(x_1,x_2,x_3)$ %\footnote{For rigorous mathematicians: we can write }%add and adjust this part

The Equations of Motion (EoMs) for the mode functions are 
\begin{equation}\label{fkhoeom}
\ddot{f}_{\vec{k}}(t)+\omega^2_{\vec{k}}(t) f_{\vec{k}}(t)=0
\end{equation}
where we have defined the time-dependent frequency 
\begin{equation}
    \omega_{\vec{k}}^2(t)=|\vec{k}|^2+m^2+U(t).
\end{equation}
If $U(t)$ is periodic in time, the equation (\ref{fkhoeom}) has the form of the Flouquet equation which has a characteristic band-like structure and regions of instabilities with $f_{\vec{k}} \sim e^{\mu_{\vec{k}} t}$ for its solutions.
These parametric amplifications in the region of instabilities are relevant for inflationary models \cite{CICOLI20241}.

We will deal with an unusual scenario in the literature where we consider
\begin{equation}
\begin{split}
    \lim_{t \rightarrow +\infty}\omega_{\vec{k}}(t)&=\omega_{\text{obs}},\\
    \lim_{t \rightarrow +\infty} \dot{\omega}_{\vec{k}}(t)&=0.
    \end{split}
\end{equation}
In such a way, we are assuming we have an asymptotic observer at $t \rightarrow +\infty$ which does not have a time-dependent frequency, since $\dot{U}(t)$ goes to zero and then it can define its vacuum state $\ket{0}_{\text{obs}}$
from
\begin{equation}
    \hat{a}_{\vec{k}}\ket{0}_{\text{obs}}=0,
\end{equation}
where
\begin{equation}
    \hat{\phi}(x) =\int_{\Omega_{\vec{k}}} \frac{d^3 \vec{k}}{(2 \pi)^3} N_{\vec{k}}\, \hat{f}_{\vec{k}}(t) \,e^{i \vec{k} \cdot {\vec{x}}}
\end{equation}
If we are however interested in a field at $t=t_0$, it has a mode structure which is different from the asymptotical observer and, since the definition of particles comes from the creation and annihilation operators along with the vacuum state, they will observe different particle numbers (i.e. the concept of Bogoliubov coefficients, see Refs.~\cite{Mukhanov:2007zz,parker2009quantum}) in general and this is much clearer thinking of it in terms of the Heisenberg picture.

\begin{figure}
    \centering
    \includegraphics[width=0.45 \paperwidth]{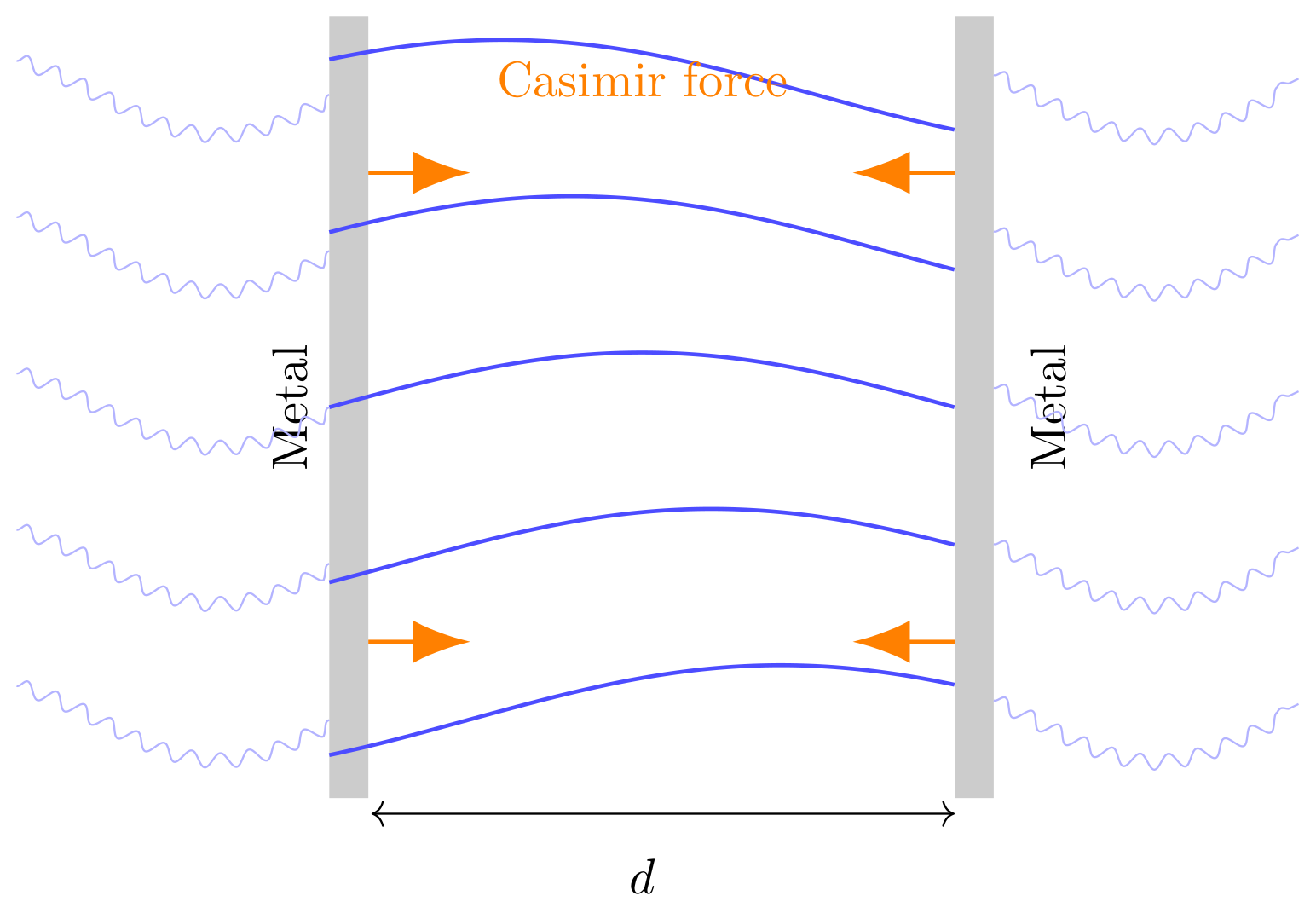}
    \caption{A schematic representation of the static Casimir effect between two parallel metallic plates at distance $d$. The effect provides a purely macroscopic manifestation of quantum mechanics, reflecting the nontrivial structure of the quantum vacuum and its fluctuations. We illustrate how vacuum modes are continuous in the exterior regions, while the inner ones are discrete due to boundary conditions. This mismatch leads to a discontinuity in the vacuum expectation value of the normal component of the stress tensor $T_{zz}$, resulting in a net pressure.}
    \label{casimir_metal}
\end{figure}

We would like to mention two other interesting quantum phenomena that furthermore show the relevance of the field as the basic physical object in contrast to particles. The first one is the Casimir effect \cite{Casimir1948,CasimirPolder1948,Lamoreaux:2011,Bimonte:2016}, showing that the quantum vacuum is not "void" as it is thought in classical physics, but can determine observable interactions between macroscopic objects. An example we will develop in the following is the Casimir force between two static parallel metallic plates as shown in Figure~\ref{casimir_metal}. 

A second one is the Unruh-Hawking effect, where a uniformly accelerated observer or near a black hole perceives the inertial vacuum state as a thermal state with a temperature $T$ \cite{Fulling:1973,Davies:1975,Unruh:1976,Hawking:1974,GibbonsHawking:1977,ChenTajima:1999,PhysRevD.94.105025}.

Furthermore, the related question about thermodynamics of black holes is still a big open question \cite{witten2025introductionblackholethermodynamics}.

\subsection{Quantization in FLRW metric and relevant aspects}% on the Heisenberg equation and the stochastic quantization}
\label{quantize}
Here, we briefly show a canonical quantization procedure to quantize the axion and SM fields in the background metric tensor ~(\ref{metric0}), which is analogous to what is done in references ~\cite{BERTONI1998331,Finelli1999,Mukhanov:2007zz, parker2009quantum,Cao2024vfu,PhysRevD.111.016028}. We also show some relevant aspects, in particular about the mode functions and the time-dependent frequencies, which will be helpful in the following, particularly Section~(\ref{Schwinger}).

% and the oscillator frequencies are $\omega_p =\sqrt{p^2/a^2+m^2}$.
We consider the action of a pseudoscalar field, write its Euler-Lagrange equation, and adopt mode expansion for the field, analogously to the Klein-Gordon field in the external potential $U(t)$, discussed in the previous paragraph.
The mode functions are the solutions of the classical equations of motion with proper normalization conditions and we set commutation rules\footnote{In the cases we show, they are bosonic commutation rules, but they can be extended to the fermionic anticommutation\cite{parker2009quantum} } and properly define annihilation and creation operators.
We then have the free Lagrangian density of the axion:
\begin{equation}
\begin{aligned}
    \mathcal{L}_{a}=\sqrt{|g|} \left[ \frac{1}{2} \partial_{\mu} a \, \partial^{\mu} a-\frac{1}{2} m^2_{\Phi} a^2  \right]=\frac{R^3}{2} \left[ \dot{a}^2- \frac{1}{R^{2}} (\nabla a)^2-  m^2_{\phi} a^2    \right].
\end{aligned}
\end{equation}
and we expand the axion field $a$ as:
\begin{equation}\label{aI}
    a(x)=\frac{1}{\sqrt{V}} \sum_{\vec{k}} \left[  a_{\vec{k}} f_{\vec{k}}(t) e^{i \vec{k} \cdot \vec{x}}  + a^{\dagger}_{\vec{k}}f_{\vec{k}}^{*}(t) e^{-i \vec{k} \cdot \vec{x}}    \right].
\end{equation}
The classical equation of motion for the scalar field mode function is then
\begin{equation}\label{modfunceoms}
\ddot{f}_{\vec{k}}(x)+3\frac{\dot{R}}{R} \dot{f}_k(x)+ \left[m_\phi^2+\frac{k^2}{R^2} \right] f_{\vec{k}}(x)=0,
\end{equation}
from which we can define $\omega_{k}^2(t)=m_a^2+\frac{k^2}{R^2(t)}$, and obtain the mode functions, taking into account of the required initial or final conditions. As in the former paragraph we will consider, after a time limit $t^*$,  the scale factor $R(t) \rightarrow R_*$ to be asymptotically flat, and then the mode functions will need to correspond to Minkowski mode functions with radiation boundary conditions
\begin{equation}
    \psi_k(x) \sim (V R_*^3)^{-1/2} (2 \omega_k(t^*))^{-1/2} \exp{ \left\{i (\vec{k} \cdot \vec{x}-\omega_k(t^*)\, t )  \right\}},\label{asymptmodefunk}
\end{equation}
where we introduce the function $\psi_k(x)=V^{-1/2} e^{i \vec{k} \cdot \vec{x} } f_{\vec{k}}(t)$.
From the condition (\ref{asymptmodefunk}) the following normalization condition arises
\begin{equation}\label{consmodfunc}
    R^3(t) \left[f_{\vec{k}}(t) \,\partial_t f^*_k(t)-f^*_k(t)\, \partial_t f_{\vec{k}}(t)        \right]=i,
\end{equation}
coming from the conservation law of the quantity on the left side of Eq.~(\ref{consmodfunc}) due to the EoM (\ref{modfunceoms}) \cite{parker2009quantum}.
We also observe that if we define the function $h_{\vec{k}}(t)$ as
\begin{equation}
   f_{\vec{k}}(t)= R(t)^{-3/2} h_{\vec{k}}(t), 
\end{equation}
we obtain the following equation of motion
\begin{equation}\label{modfuncond}
\ddot{h}_{\vec{k}}+\left[\omega^2_{k}(t)+\sigma(t)\right] h_{\vec{k}}=0,
\end{equation}
where $\sigma(t)=-\frac{3}{4}H^2-\frac{3}{2} \frac{\ddot{R}}{R}$.
In this form, it is visible that the dynamics of the mode function are based on the hierarchy between the comoving wavenumber $k/R=|\vec{k}|/R$, the mass $m_a$, and the Hubble parameter $H(t)$. Furthermore, we also have a contribution from the deceleration parameter $q \coloneqq -\Big( \frac{\ddot{R}(t)}{R(t) H^2} \Big)$ \cite{hogg2000distancemeasurescosmology,baumann2022cosmology}.

A possible way of dealing with this is the following: a mode function of the form
\begin{equation}
   f_{\vec{k}}(t)= (2 \Omega_k(t))^{-1/2} \exp{ \left\{- i\int_{t_0}^t dt' \, \Omega_k(t')  \right\}}, 
\end{equation}
can be a solution of the EoMs (\ref{modfunceoms}) with the proper normalisation conditions if $\Omega_k(t)$ solves the following differential equation

\begin{equation}\label{eomcomplicata}
    \Omega_k^2 =\frac{3 \dot{\Omega}_k^2}{4  \Omega_k^2}-\frac{\ddot{\Omega}_k}{2  \Omega_k}  +   \frac{k^2}{R^2}+m_{\phi}^2+\sigma(t).
\end{equation}
%where we have defined $\sigma(t)$ formerly, and, 
Being a second order differential equation in the cosmological time $t$, we need two conditions which we take accordingly to the asymptotical Minkowski limit we impose, $\Omega_k \rightarrow \omega_k(t^*)$ for $t \rightarrow t^*$ and accordingly $ \dot{\Omega}_k\rightarrow 0$.
The equation (\ref{eomcomplicata}) is challenging to solve analytically. Still, in the case of a slowly varying scale factor, it can be expanded in the so-called adiabatic expansion, which is an asymptotical, not convergent, series \cite{parker2009quantum}.
However, the advantage of considering the full $\Omega_k$ is that we can write our field in the form %add discuss choice of vacuum here
\begin{equation}
     \Phi=\frac{1}{\sqrt{V}} \sum_{\vec{k}} \frac{1}{\sqrt{\Omega_k(t)}}\left[  a_{\vec{k}}  e^{i \int_{t_0}^{t} \Omega_k(t') dt'- i \vec{k} \cdot \vec{x}}  + a^{\dagger}_{\vec{k}} e^{-i \int_{t_0}^{t} \Omega_k(t') dt'+ i\vec{k} \cdot \vec{x}} \right].
\end{equation}%check

and we will avoid the usual problem with the vacuum choice at each instant of time \cite{parker2009quantum} in such scenarios by just adopting a Heisenberg representation. 

Following the same references \cite{BERTONI1998331,Finelli1999, PhysRevLett.21.562, PhysRev.183.1057}, we get analogously the free quantized Hamiltonian for the axion field.
\begin{equation}
    \hat{H}_a= \sum_{\vec{k}} \Omega_{k}\left(\hat{a}_{\vec{k}}^{\dagger} \hat{a}_{\vec{k}}+\frac{1}{2}\right),
\end{equation}
where $\left[\hat{a}_{\vec{k}}, \hat{a}_{\vec{k}^{\prime}}^{\dagger}\right]=\delta_{\vec{k} \vec{k}^{\prime}}$.
We have discussed in more details about quantizing the scalar field and its mode function. Quantization is in fact more involved when a mass is present since there is no conformal invariance, differently from the case of massless gauge fields, such as a $U(1)$ field.

Indeed, starting from the free Lagrangian density for the electromagnetic potential $A_{\mu}$ given by \cite{Finelli1999}
\begin{equation}
\begin{aligned}
\mathcal{L}_{E M} =\sqrt{-g}\left[-\frac{1}{4} F_{\mu \nu} F^{\mu \nu}\right]
 =-\frac{R^{3}}{4}\left[+\frac{2}{R^{2}} F_{0 j}^{2}-\frac{1}{R^{4}} F_{i j}^{2}\right],
\end{aligned}
\end{equation}
where $F_{\mu \nu} \equiv \partial_{\mu} A_{\nu}-\partial_{\nu} A_{\mu}$.We choose the generalized Lorentz gauge $\nabla_{\mu} A^{\mu}=0$ (with $\nabla_{\mu}$ the covariant derivative) and take $A_{0}=0$ (possible in the source-free case), from which we obtain the radiation gauge $(\vec{\nabla} \cdot \vec{A}=0)$.
Using this gauge, one gets
\begin{equation}
\begin{aligned}
F_{0 j} & =-\dot{A}_{j},\\
F_{i j} & =(\vec{\nabla} \times \vec{A})_{i j}.
\end{aligned}
\end{equation}
 Let us note that the vector potential $\vec{A}$ has been chosen to be the covariant $A_{i}$, analogously to Ref.~\cite{Finelli1999}.
 We rewrite the Lagrangian density in the form
\begin{equation}
\mathcal{L}_{E M}=-\frac{1}{2}\left\{a \dot{\vec{A}}^{2}+\frac{1}{a} \vec{A} \cdot \nabla^{2} \vec{A}-\frac{1}{a} \vec{\nabla} \cdot\left[A_{i} \vec{\nabla} A_{i}-(\vec{A} \cdot \vec{\nabla}) \vec{A}\right]\right\}.
\end{equation}

We expand $\vec{A}$ as

\begin{equation}\label{AI}
\vec{A}=\frac{1}{\sqrt{V}} \sum_{\vec{k}, \lambda}\left[c_{k}^{(\lambda)}(t) \vec{\varepsilon}^{(\lambda)} e^{i \vec{k} \cdot \vec{x}}+c_{k}^{(\lambda) *}(t) \vec{\varepsilon}^{(\lambda)} e^{-i \vec{k} \cdot \vec{x}}\right],
\end{equation}
where $\lambda$ runs over the two polarization states and $\vec{\varepsilon}^{(\lambda)}$ is a unit vector that satisfies $\vec{k} \cdot \vec{\varepsilon}^{(\lambda)}=0$ and $\vec{\varepsilon}^{(\lambda)} \cdot \vec{\varepsilon}^{\left(\lambda^{\prime}\right)}=\delta_{\lambda \lambda^{\prime}}$.

Thus one obtains that the different modes $k, i,(\lambda)$ decouple with the following total Hamiltonian:
\begin{equation}
\begin{aligned}
    H_{\gamma}=\frac{1}{2} \sum_{k, i, \lambda}\left(\frac{\pi_{i, k}^{(\lambda) 2}}{a}+a \omega_{k}^{2} c_{k i}^{(\lambda) 2}\right),
\end{aligned}
\end{equation}
where $\pi_{i, k}^{(\lambda)}=a \dot{c}_{k i}^{(\lambda)}$ (with $i=1,2)$, ${c}_{k i}$ are respectively the real part ($i=1$) and imaginary part ($i=2$) of $\sqrt{2} \,c_{k}^{(\lambda)}$ and finally $\omega_{k}^{2}=k^{2} / a^{2}$. 
The classical equation of motion is

\begin{equation}\label{ckjl}
\ddot{c}_{k i}^{(\lambda)}+\frac{\dot{R}}{R} \dot{c}_{k i}^{(\lambda)}+\omega_{k}^{2} c_{k i}^{(\lambda)}=0.
\end{equation}
On canonically quantizing, the Hamiltonian  can be factorized as (henceforth we shall denote collectively $\mathbf{k}, \mathbf{i},(\lambda)$ by $\sigma$):

\begin{equation}
\hat{H}= \sum_{\sigma} \omega_{k}\left(\hat{b}_{\sigma}^{\dagger} \hat{b}_{\sigma}+\frac{1}{2}\right),
\end{equation}

with

\begin{equation}
\begin{aligned}
&\hat{b}_{\sigma}=\left(\frac{R(t) \,\omega_{k}}{2}\right)^{\frac{1}{2}}\left(\hat{c}_{\sigma}+i \frac{\hat{\pi}_{\sigma}}{R(t)\, \omega_{k}}\right) \\&\hat{b}_{\sigma}^{\dagger}=\left(\frac{R(t)\, \omega_{k}}{2 }\right)^{\frac{1}{2}}\left(\hat{c}_{\sigma}-i \frac{\hat{\pi}_{\sigma}}{R(t)\, \omega_{k}}\right),
\end{aligned}
\end{equation}

with $\left[\hat{b}_{\sigma}, \hat{b}_{\sigma}^{\dagger}\right]=\delta_{\sigma \sigma^{\prime}}$.

From classical equations of motion, as before, we obtain the general solution 
\begin{equation}\label{confinvmodfunc}
    c_{\sigma}=\frac{1}{\sqrt{\omega_k}} \exp{\left[\,i \,\omega_k \, (\eta-\eta_0) \right]},
\end{equation}
where we used that they can be reduced to a Sturm-Liouville equation form.

%have used the fact that, in such a case, the equation (\ref{ckjl}) can be reduced to a Sturm-Liouville equation.

Such results can also been obtained using only the conformal invariance of the Lagrangian, as discussed in Ref.~\cite{parker2009quantum}.

The quantized fields at the time $t=t^*$ have the trivial form:
\begin{equation}
    \Phi(\vec{x},t=t^*)=\frac{1}{\sqrt{V}} \sum_{\vec{k}} \left[a_{\vec{k}} e^{i \vec{k} \cdot \vec{x}} +a_{\vec{k}}^{\dagger}e^{-i \vec{k} \cdot \vec{x}}     \right],
\end{equation}
and
\begin{equation}
    \vec{A}(\vec{x},t=t^*)=\frac{1}{\sqrt{V}} \sum_{\vec{k}, \lambda}\left[c_{k}^{(\lambda)}\vec{\varepsilon}^{(\lambda)} e^{i \vec{k} \cdot \vec{x}}+c_{k}^{(\lambda) \dagger} \vec{\varepsilon}^{(\lambda)} e^{-i \vec{k} \cdot \vec{x}}\right],
\end{equation}
employing the conditions we impose on mode functions and the mode function basis is the one of the asymptotic observer.
As demonstrated in Refs.~\cite{PhysRevLett.21.562,PhysRev.183.1057,parker2009quantum,Mukhanov:2007zz}, the Heisenberg equation for an operator $A(t)$ in the Heisenberg representation
\begin{equation}\label{Heisenberg}
    \frac{d\hat{A}}{dt}= \frac{\partial \hat{A}}{ \partial t}+i\left[\hat{H}(t),\hat{A}(t)    \right]
\end{equation}
is valid for a general quantum field theory in an FLRW metric or a general time-dependent background. This is a relevant point for the following. 
We will further consider, for simplicity, as usually done \cite{PhysRevD.107.063518,Cao2024vfu}, an initial factorized density matrix
\begin{equation}
    \rho(t_0)=\rho_{\phi}(t_0) \otimes \rho_{\chi}(t_0)
\end{equation}
where $\chi$ indicates a general SM or Dark sector, as in the following of the thesis.
We assume accordingly that the $\chi$ fields expand adiabatically at all times with plasma temperature $T(t)=1/\beta(t)$ and, if we neglect plasma effects and variations in the effective relativistic degrees of freedom $g_{*}$, it gets the simplifying form 
\begin{equation}
    \rho_{\chi}(t_0)=\frac{e^{-\beta H_{\chi}}}{\Tr e^{-\beta H_{\chi}}}.
\end{equation}
Interestingly, this density matrix is stationary if $\beta(t) \propto a(t)$, as expected (in this case, the exponent is an adiabatic invariant \cite{BERTONI1998331,Finelli1999}). However, as discussed in the following, various factors are involved when dealing with the primordial plasma and all the plasma interactions that can occur.

We will consider the axion field in an initial coherent state:
\begin{equation}
    \ket{\Delta}=\Pi_{\vec{k}} e^{-\frac{1}{2} |\Delta_{\vec{k}}|^2} e^{- \Delta_k a^{\dagger}_{\vec{k}}} \ket{0},
\end{equation}
that is an eigenstate of the annihilation operator
\begin{equation}
    \hat{a}_{\vec{k}}  \ket{\Delta}= \Delta_{\vec{k}} \ket{\Delta}.
\end{equation}

The expectation values of the axion field $\Phi$ and its canonical momentum $\pi$ in such a state are
\begin{equation}
    \angi{\Delta|\Phi(\vec{x},t=t^*)|\Delta}=\frac{1}{\sqrt{V}} \sum_{\vec{k}} \frac{1}{\sqrt{2 \omega_k}}
    \left[\Delta_k e^{i \vec{k} \cdot \vec{x}}+\Delta_k^*e^{-i \vec{k} \cdot \vec{x}}\right],
\end{equation}
and 
\begin{equation}
    \angi{\Delta|\pi(\vec{x},t=t^*)|\Delta}=\frac{i}{\sqrt{V}} \sum_{\vec{k}} \sqrt{\frac{\omega_k}{2}}
    \left[\Delta_k e^{i \vec{k} \cdot \vec{x}}-\Delta_k^*e^{-i \vec{k} \cdot \vec{x}}\right].
\end{equation}

It is then worth noting that if we use $\Phi=\varphi+\phi$, the annihilation operators $\hat{b}_k$ for the field $\phi$ are $\hat{b}_k= \hat{a}_k-\Delta_k$, and so the state $\ket{\Delta}$ is the vacuum state of the field $\phi$.

Then, accordingly, the initial density matrix for the axion field is
\begin{equation}
    \rho_{\phi}(t_0)=\ket{\Delta}\bra{\Delta}.
\end{equation}
This makes sense in the above mentioned scenarios  for the axions we deal with, such as freeze-in scenarios or significant non-thermal production, where we can neglect or assume neglibible the initial thermal and squeezed population.

\section{The Strong CP problem in QCD}\label{strongCP}
The strong CP problem in QCD \cite{PhysRevLett.38.1440,Peccei:1977ur,PhysRevLett.40.223,PhysRevLett.40.279} is a problem of the Standard Model and is related to the non-trivial structure of the vacuum state of Quantum ChromoDynamics (QCD). Indeed, if we consider the Lagrangian density of gluons, without interactions with other particles,

\[ L=-\frac{1}{4} G_{\mu \nu}^b G^{\mu \nu}_b\]
where $G_{\mu \nu }^b $ is the gluonic field and $b$ the color index. This Lagrangian density is invariant under $SU(3)$ local gauge transformations. It is also invariant under \textbf{CP}, since \textbf{P} and \textbf{C} transformations act on the color electric and magnetic field, $\vec{E}_b$ and $\vec{B}_b$, as follows \cite{bigi2021new}:
\begin{align*}
	\vec{E}_b \xrightarrow{\textbf{P}} -\vec{E}_b & \qquad \vec{E}_b \xrightarrow{\textbf{T}} +\vec{E}_b \\
	\vec{B}_b \xrightarrow{\textbf{P}} +\vec{B}_b & \qquad \vec{B}_b \xrightarrow{\textbf{T}} -\vec{B}_b
\end{align*} 
It is thus consequential that
\begin{equation}
	G \cdot G =G_{\mu \nu}^a G^{\mu \nu}_a \propto \sum_a \Big( |\vec{E}_a|^2+|\vec{B}_a|^2\Big)\xrightarrow{\textbf{P},\textbf{T}} \sum_a \Big( |\vec{E}_a|^2+|\vec{B}_a|^2\Big) .
\end{equation}
There is also another gauge-invariant four-dimensional scalar operator that is 
\begin{equation}
	G \cdot \tilde{G}= G_{\mu \nu}^a \tilde{G}^{\mu \nu}_a , 
\end{equation}
where $\tilde{G}^{\mu \nu}_a=\varepsilon^{\mu \nu \alpha \beta } G_{\alpha \beta}^a$.
This term violates $P$ and $T$, but not $C$ transformations, and is a total divergence
\begin{equation}
    G_{\mu \nu}^a \tilde{G}^{\mu \nu}_a=\partial_{\mu} K^{\mu},
\end{equation}
where 
\begin{equation}
    K^{\mu}=\varepsilon^{\mu \alpha \beta \gamma} A_{ \alpha}^a \Big(G_{a \beta \gamma}- \frac{g_s}{3} f_{abc} A_{b \beta} A_{c \gamma} \Big).
\end{equation}
$g_s$ is the strong coupling constant and $f_{abc}$ denote the structure constants of the color $\mathfrak{su}(3)$ Lie algebra \cite{schwartz2014quantum}.

We would then not expect any CP violation in the Strong Interaction sector, if we add
to the effective Lagrangian density, coming from Feynman's path integral,  a term of the form
\begin{equation}
\mathcal{L}_{\theta}=\theta_{\text{QCD}} \,\tilde{G}_{\mu \nu}^a\, G^{\mu \nu}_a\, ,
\end{equation}
as it happens in QED \cite{schwartz2014quantum}.

However, this simple scenario was first doubted by Gerard t'Hooft \cite{PhysRevLett.37.8}, who introduced the instanton solutions to solve the so-called $U(1)$ problem of QCD.
Instantons are field configurations keeping the action finite in any 4-dimensional non-Abelian gauge theory and they are localized.
Due to the
existence of such configurations, we must consider the vacuum structure of a quantum
field theory in an unusual way to a $U(1)$ field, since they do correspond to solutions for which the vector potential $A_{\mu}^a$ does not approach zero in the limit of $\vec{r} \rightarrow +\infty$.

The theta vacuum comes from a topologically interesting property of $ SU(3) $ group, which is
\begin{equation}
    \pi_3{(SU(3))}=\mathbb{Z}.
\end{equation}
where $\pi_3$ denotes the third homotopy group associated with continuous mappings from the three-sphere $S^3$ into the group manifold, expressed inside the brackets \cite{chaichian2012introduction}. 
This mathematical property has a significant impact on physics, relying on the properties of vacuum states, and tells us that an integer number, a topological invariant named the Pryviakov number \footnote{Also known as the winding number.}, labels many distinct vacuum kets and the stability of instantons is connected with the topological distinguishability of the vacuum manifold \cite{vilenkin1994cosmic,Polchinski1998,Peccei2008,chaichian2012introduction}.

\begin{figure}
    \centering
    \includegraphics[width=0.7\linewidth]{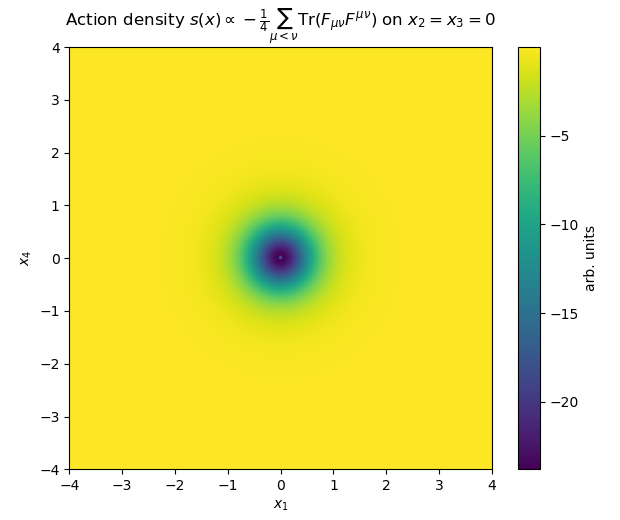}
    \caption{Plot of the action density of the instanton configuration of Eq.~(\ref{instanto}) in arbitrary units, with a slice at $x_2=x_3=0$. More details are shown in Ref.~\cite{FavittaDWAnimation}}.
    \label{fig:instanton}
\end{figure}

For each winding number $n$, one can associate an instanton configuration whose behaviour at spatial infinity is not required to satisfy $A_{\mu}^a \rightarrow 0$, but rather approaches a field configuration that is gauge-equivalent to zero. 
Such a configuration corresponds to a ket denoted by $\ket{n}$.
An analytical example of a $SU(2)$ instanton \cite{chaichian2012introduction,Saikawa:2013thesis} is the configuration shown in Fig.~\ref{fig:instanton} , which is associated to the configuration
\begin{equation}\label{instanto}
    A_{\mu}(x)=\frac{i}{g} \Bigg( \frac{r^2}{r^2+\rho^2}\Bigg) U_1^{-1}(\hat{x}) (\partial_{\mu} U_1(\hat{x})),
\end{equation}
with 
\begin{equation}
    U_1(\hat{x})=\frac{x_1 \sigma_1+x_2 \sigma_2+x_3 \sigma_3}{r},
\end{equation}
and the matrices
\begin{equation}
    \sigma_1=\begin{pmatrix}
    0 &1\\
    1 &0
    \end{pmatrix}, \quad
     \sigma_2=\begin{pmatrix}
    0 &-i\\
    i &0
    \end{pmatrix},\quad
     \sigma_3=\begin{pmatrix}
    1 &0\\
    0 &-1
    \end{pmatrix},
    \end{equation}
are the Pauli matrices.

The kets $\ket{n}$ are not, however, physical states, since winding numbers are not invariant under gauge transformations. If we calculate the transition amplitude between the two of them, we get the following
\begin{equation}
	{}_{+}\angi{n|e^{-Ht}|m}_{-}= \int DA_{n-m} \exp{\left[-\int d^4 x \,\mathcal{L}\right]}
\end{equation}
where the subscripts $+$ and $-$ are referred respectively to limits $t \rightarrow +\infty$ and
$t \rightarrow -\infty$.
Although the $\ket{n}$ kets are not gauge invariant, we can build physical vacuum states, the $\theta$ vacua, as linear combinations of all of them
\[  \ket{\theta}=\sum_{n} e^{-i n \theta} \ket{n}\],
where $\theta$ is a real number. Such states are analogous to Bloch states in Bravais crystals \cite{ashcroftmermin} and have fascinating properties that lead to CP violations in the theory.
If we calculate the transition amplitude between two theta vacua, we get
\begin{equation}
 {}_{+}\angi{\theta'|\theta}_{-}=\sum_{m,n} e^{-i(n-m)\,\theta} {}_{+}\angi{n|m}_{-}=\sum_{\nu} \sum_n  e^{-i \,\theta \,\nu}  {}_{+}\angi{n|n+\nu}_{-}\, .
 \end{equation}

It is then easy to see that the difference in the winding numbers $\Delta \nu$ is given by 
\begin{equation}
    \Delta \nu =\frac{g^2_s}{32 \pi^2} \int d \sigma_{\mu} \Big. K^{\mu} \Big|_{t_0 \rightarrow -\infty}^{t_f \rightarrow+\infty}=\frac{g^2_s}{32 \pi^2}  \int d^4 x \,G^{\mu \nu}_a \tilde{G}_{a \mu \nu}.
\end{equation}

Using the usual path integral representation for the vacuum-to-vacuum amplitude 
${}_{+}\angi{\theta'|\theta}_{-}$ we have
\begin{equation}
     {}_{+}\angi{\theta'|\theta}_{-}= \sum_{\Delta \nu} \int \delta A e^{i S_{\text{eff}}[A]} \delta{\Big(\Delta \nu- \frac{g^2_s}{32 \pi^2}  \int d^4 x \,G^{\mu \nu}_a \tilde{G}_{a \mu \nu} \Big )}
\end{equation}
where 
\begin{equation}
    S_{\text{eff}}[A]=S_{\text{QCD}}[A]+ \theta \frac{g^2_s}{32 \pi^2}  \int d^4 x \,G^{\mu \nu}_a \tilde{G}_{a \mu \nu}.
\end{equation}

Solving the $U(1)_A$ problem, by recognizing the complicated nature of the QCD vacuum, effectively adds an extra term to the QCD lagrangian density
\begin{equation}
    \mathcal{L}_{\theta}=\theta \frac{g^2_s}{32 \pi^2}  \,G^{\mu \nu}_a \tilde{G}_{a \mu \nu}.
\end{equation}
 This term violates parity $P$ and time reversal invariance $T$, but conserves charge conjugation $C$, so it violates $CP$. 

 It induces a neutron electric dipole moment, which can be estimated within effective chiral theory by upgrading isospin to SU(3) and using the baryon mass \cite{CREWTHER1979123,schwartz2014quantum} 
 \begin{equation}
    d_N=\frac{m_N}{4 \pi^2} g_{\pi N N} \,\bar{g}_{\pi NN} \ln{\frac{m_N}{m_\pi}}=(5.2 \times 10^{-16} e \cdot \textrm{cm}) \,\bar{\theta}
\end{equation}
where $g_{\pi N N} \simeq 13.4$ is the known standard pion-neutron coupling and $\bar{g}_{\pi NN} $ the CP-violating one.
 
\begin{figure}[h]
\centering
\includegraphics[width=0.2\linewidth]{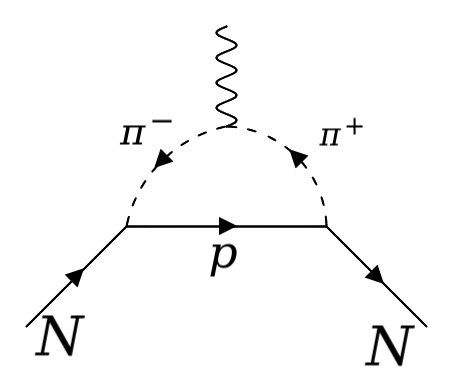}
    \caption{Example of diagram of a pion loop generating a neutron dipole moment}
    \label{neutrondipole}
\end{figure}

Loops of pions, such as the one in Figure~\ref{neutrondipole}, have the CP violation coming from the $\bar{g}_{\pi NN}$ vertex and generate the neutron dipole moment.

% In your document body

\begin{comment}
\begin{equation}
    d_n \simeq \frac{e \theta m_q}{m^2_N}
\end{equation}
\end{comment}
The current experimental bound on the neutron EDM is $|d_N|<1.8 \times 10^{-26}\,e \cdot \textrm{cm}$ from PSI \cite{PhysRevLett.124.081803}, from which it comes
$  \bar{\theta} \lesssim 10^{-11}$.
This is then a problem of explaining why the angle $\bar{\theta}$ is so small, if not equal to zero, since it holds two contributions, $\theta_{\text{QCD}}$ and $\theta_F$, coming from two apparently different and distinct sectors. Since $\bar{\theta}$ is a dimensionless parameter of the theory, we would naively expect
to be $\mathcal{O}(1)$. Hence, it would be important to explore a natural way to explain why $\bar{\theta}$ is so
small. In this sense, the strong CP problem is a fine-tuning problem.

 The resolution of the old $U(1)$ problem then creates  another issue, which is the Strong CP problem and it is closely tied to the existence of the QCD theta vacuum.

\subsection{A possible solution for the Strong CP problem: the Peccei-Quinn theory}
Roberto Peccei and Helen Quinn proposed the most attractive solution to the strong CP problem by introducing a new pseudoparticle \cite{PhysRevLett.38.1440,Peccei:1977ur}. The crucial point is to introduce a dynamical quantity which mimics the $\bar{\theta}$
parameter and takes a zero value in the low-energy Lagrangian. Soon after the proposal, Wilczek pointed out that this dynamical variable should be identified as a light spin-zero particle called the axion.%check references

Further aspects are the introduction of a global $U(1)_{\text{PQ}}$ axial symmetry, which we call the Peccei-Quinn (PQ) symmetry, and impose appropriate PQ charges into quarks so that there exists $ U (1)_{\text{PQ}} -SU (3)_c -SU (3)_c $ anomaly. 

This $U(1)_{\text{PQ}}$ symmetry is spontaneously broken at some energy scale higher than the QCD scale $\Lambda_{\text{QCD}}$, which is at the order of $ 150 \,\mathrm{MeV}$ \cite{Borsanyi:2020fev,ohare2024cosmology}.%put other reference here

The dynamical degree of freedom $a$ can be identified as a Goldstone boson associated
with the spontaneous breaking of the $ U (1)_{PQ}$ symmetry, which acts as a shift in the axion field:

\begin{equation}
	U(1)_{PQ} : a \rightarrow a+ \epsilon \eta 
\end{equation}
where $\epsilon$ is a real parameter and $\eta$ is the energy scale of the spontaneous symmetry breaking of the $U(1)_{PQ}$.

In the following subsection, we provide a brief review of some explicit models of the QCD axion. As discussed here, the presence of the QCD anomaly is necessary to induce the axion potential whose minimum is located at $\bar{\theta}=0$. This requires some extensions of the Standard Model and an appropriate arrangement of the $U(1)_{PQ}$ multiplet.

\subsection{The original PQWW model}\label{PQWW}
Weinberg and Wilczek proposed the original model of the QCD axion \cite{PhysRevLett.40.223,PhysRevLett.40.279}, based on the idea of Peccei and Quinn, and it is commonly known as the Peccei-Quinn-Weinberg-Wilczek (PQWW) model, or the "visible axion" model. 

In this model, the axion field is identified
as a phase direction of the Standard Model Higgs field, but it is necessary to introduce two or more Higgs doublets, since an axion degree of freedom is not present in the theory with a single Higgs doublet.

Furthermore, more doublets are necessary to make the Standard Model (SM) invariant under a $U(1)_{PQ}$ transformation to absorb independent chiral transformations of the quarks and leptons. 
If we denote two Higgs doublets as $\Phi_1$ and $\Phi_2$, we can assign the $U(1)_{PQ}$ charges $q_1$ and $q_2$ to Higgs doublets and quarks such that
\begin{equation}
  U(1)_{\text{PQ}} : \begin{cases}
      a \rightarrow a+\alpha v_F, \\ u_{R_j} \rightarrow e^{-i \alpha x} u_{R_j},\\d_{R_j} \rightarrow e^{-i \alpha/x} d_{R_j}, \\ \ell_{R_j} \rightarrow e^{-i \alpha x} \ell_{R_j},
  \end{cases} 
\end{equation}
where $v_1$ and $v_2$ are the Vacuum Expectation Values (VEVs) respectively of $\Phi_1$ and $\Phi_2$, $x=v_2/v_1$ and $v_F=\sqrt{v_1^2+v_2^2}$. The axion is the common phase field of $\Phi_1$ and $\Phi_2$, which is orthogonal to the weak hypercharge:
\begin{equation}
    \Phi_1=\frac{v_1}{\sqrt{2}} e^{iax/v_F} \begin{pmatrix}
        1\\0
    \end{pmatrix} \quad   \text{and}\quad  \Phi_2=\frac{v_1}{\sqrt{2}} e^{ia/(x\,v_F)} \begin{pmatrix}
        0\\1
    \end{pmatrix}
\end{equation}

The symmetry current for the $U(1)_{\text{PQ}}$, focusing on just the quark pieces, is 
\begin{equation}
    J^{\mu}_{\text{PQ}}= -v_f \partial^{\mu} a + x \sum_i  \bar{u}_{iR} \gamma^{\mu} u_{iR}+ \frac{1}{x} \sum_i \bar{d}_{iR} \gamma^{\mu} d_{iR},
\end{equation}
from which we can identify for the PQWW model the value of the anomaly coefficient $\xi$ as
equal to
\begin{equation}
    \xi=\frac{N}{2} \Big(x+\frac{1}{x}\Big)=N\Big(x+\frac{1}{x}\Big)
\end{equation}
where $N$ is the color anomaly.
We will also name it in the following as the domain wall number $N_{\text{DW}}=N$ and it will be the number of physically distinct vacua of the axion potential generated by QCD instantons.

%If the scalar field breaking PQ has charge 

In some models (e.g. DFSZ), the normalization leads also to
$N_{DW}=2N$, depending on how the quark and Higgs PQ charges are defined.
The factor 2 (or other integer) is not universal — it depends on whether the definition of 
$N$ includes the PQ charge of the Higgs or the scalar responsible for the axion.%adjust

The relevant Yukawa interactions involving these Higgs fields in the SM are

\begin{equation}
    \mathcal{L}_{\text{Y}} = \Gamma^u_{ij} \bar{Q}_{L_i} \Phi_1 u_{R_j} +\Gamma^d_{ij} \bar{Q}_{L_i} \Phi_2 d_{R_j}+ \Gamma^{\ell}_{ij} \bar{L}_{L_i} \Phi_2 {\ell}_{R_j}+h.c.
\end{equation}

To compute the axion mass in this model, it is convenient to distinguish the interaction terms between the axion and the light quarks from the rest, which can be obtained from the starting theory by constructing an effective chiral Lagrangian.
The effects of heavy quarks can be accounted for with their contribution to the chiral anomaly of $ J^{\mu}_{\text{PQ}}$.

 For the two light up and down quarks we introduce a $2 \times 2$ matrix of Nambu-Goldstone fields

\begin{equation}
\Sigma=\exp \left(\mathrm{i} \frac{\tau \cdot \pi+\eta}{f_{\pi}}\right),
\end{equation}
where $f_{\pi}$ is the pion decay constant. The meson sector of the light-quark theory, neglecting the effect of the Yukawa interactions, is then embodied in the $U(2)_{V} \times U(2)_{A}$ invariant effective Lagrangian

\begin{equation}
\mathcal{L}_{\text {chiral }}=-\frac{f_{\pi}^{2}}{4} \operatorname{Tr}\left(\partial_{\mu} \Sigma \,\partial^{\mu}\Sigma^{\dagger}\right) .
\end{equation}

We must add $U(2)_{V} \times U(2)_{A}$ breaking terms which mimic the $U(1)_{\mathrm{PQ}}$ invariant Yukawa interactions of the up and down quarks to the $L_{\text {chiral }}$. This can be obtained introducing the Lagrangian
\begin{equation}
    \mathcal{L}_{\text {mass }}=\frac{1}{2}\left(f_{\pi} m_{\pi}^{0}\right)^{2} \operatorname{Tr}\left[\Sigma A M+(\Sigma A M)^{\dagger}\right],
\end{equation}
where we have defined
$$
A=\left(\begin{array}{cc}
\mathrm{e}^{-\mathrm{i} a x / v_{\mathrm{F}}} & 0 \\
0 & \mathrm{e}^{-\mathrm{i} a / (x \,v_{\mathrm{F}})}
\end{array}\right) \quad \text { and the quark mass matrix } \quad M=\left(\begin{array}{cc}
\frac{m_{u}}{m_{u}+m_{d}} & 0 \\
0 & \frac{m_{d}}{m_{u}+m_{d}}
\end{array}\right) .
$$
Now, $x=\frac{m_u}{m_d}$, where $m_u$ is the up quark mass and $m_d$ is the down quark mass. 
We observe that the invariance of $\mathcal{L}_{\text {mass }}$ under $U(1)_{\mathrm{PQ}}$ implies the transformation

\begin{equation}
\Sigma \rightarrow \Sigma\,\left(\begin{array}{cc}
\mathrm{e}^{\mathrm{i} \alpha x} & 0 \\
0 & \mathrm{e}^{\mathrm{i} \alpha / x}
\end{array}\right) .
\end{equation}
$\mathcal{L}_{\text {mass }}$, however, only gives part of the physics associated with the symmetry breakdown of $U(2)_{A}$. The quadratic terms of $\mathcal{L}_{\text {mass }}$ involving neutral fields are indeed

$$
\begin{aligned}
\mathcal{L}_{\text {mass }}^{(2)}=-\frac{\left(m_{\pi}^{0}\right)^{2}}{2}  {\left[\frac{m_{u}}{m_{u}+m_{d}}\left(\pi^{0}+\eta-\frac{x f_{\pi}}{v_{\mathrm{F}}} a\right)^{2}\right.}
 \left.+\frac{m_{d}}{m_{u}+m_{d}}\left(\eta-\pi^{0}-\frac{f_{\pi}}{x v_{\mathrm{F}}} a\right)^{2}\right],
\end{aligned}
$$
where the terms in round brackets  are respectively the physical pion $\pi^0_{\text{phys}}$ and $\eta_{\text{phys}}$.
Consequently, we have the following relation between the mass of $\eta$ particle and the pion mass 

$$
\frac{m_{\eta}^{2}}{m_{\pi}^{2}}=\frac{1}{x} \simeq 1.6,
$$
in contradiction with the experimental results \cite{ParticleDataGroup:2024cfk}. Indeed, if it were just for $L_{\text{mass}}$, we would have again the $U(1)_{A}$ problem in the effective Lagrangian language with a massless axion.

The resolution of the $U(1)_{A}$ problem within the effective Chiral Lagrangian theory is achieved by adding a mass term that takes account of the anomaly in both $U(1)_{A}$ and $U(1)_{\mathrm{PQ}}$. This mass term gives the $\eta$ particle the right mass and a mass to the axion.

It is then easy to notice that such a term needs to be of the form
\begin{equation}
\mathcal{L}_{\text {anomaly }}=-\frac{\left(m_{\eta}^{0}\right)^{2}}{2}\left[\eta+\frac{f_{\pi}}{v_{\mathrm{F}}} \frac{\left(N_{g}-1\right)(x+1 / x)}{2} a\right]^{2},
\end{equation}
where $\left(m_{\eta}^{0}\right)^{2} \simeq m_{\eta}^{2} \gg m_{\pi}^{2}$. The coefficient in front of the axion field in $\mathcal{L}_{\text {anomaly }}$ reflects the relative strength of the couplings of the axion and the $\eta$ to $G \tilde{G}$ as the result of the anomalies in $U(1)_{\mathrm{PQ}}$ and $U(1)_{A}$. Naively, the ratio of these couplings is just $(f_{\pi} / 2 v_{\mathrm{F}})\, \xi$. However, the reason that $N_{g}-1$ appears above rather than $N_{g}$, is that $\mathcal{L}_{\text {mass }}$ already includes the light quark interactions of axions, so only the contribution of heavy quarks to the PQ anomaly should be taken into account in $\mathcal{L}_{\text {anomaly }}$.

Diagonalization of the quadratic terms in $L_{\text{mass}}$ and $L_{\text{anomaly}}$ yields both the axion mass and the parameters for axion-pion and axion-eta mixing in the PQ model. It is convenient to define

$$
\bar{m}_{a}=m_{\pi} \frac{f_{\pi}}{v_{\mathrm{F}}} \frac{\sqrt{m_{u} m_{d}}}{m_{u}+m_{d}} \simeq 25 \,\rm{keV} .
$$

We then find and can write

$$
m_{a}=\lambda_{m} \bar{m}_{a}, \quad \xi_{a \pi}=\lambda_{3} \frac{f_{\pi}}{v_{\mathrm{F}}}, \quad \xi_{a \eta}=\lambda_{0} \frac{f_{\pi}}{v_{\mathrm{F}}},
$$

where

$$
\begin{aligned}
\lambda_{m} & =N_{g}\left(x+\frac{1}{x}\right),\quad
\lambda_{3} =\frac{1}{2}\left[\left(x-\frac{1}{x}\right)-N_{g}\left(x+\frac{1}{x}\right) \frac{m_{d}-m_{u}}{m_{u}+m_{d}}\right], \\
\lambda_{0} & =\frac{1}{2}\left(1-N_{g}\right)\left(x+\frac{1}{x}\right) .
\end{aligned}
$$

In addition to the three parameters above, every axion model is also characterized by the coupling of one axion to two photons. The interaction Lagrangian describing this coupling is of the form
\begin{equation}
\mathcal{L}_{a \gamma \gamma}=-\frac{\alpha}{8 \pi} C_{a \gamma \gamma} \frac{a}{f_{a}} F^{\mu \nu} \tilde{F}_{\mu \nu},
\end{equation}

To find the coupling $C_{a \gamma \gamma}$ for the PQWW model, we use the electromagnetic anomaly of the PQ current

\begin{equation}
\partial_{\mu} J_{\mathrm{PQ}}^{\mu}=\frac{\alpha}{4 \pi} \xi_{\gamma} F_{\mu \nu} \tilde{F}^{\mu \nu},
\end{equation}

where $\xi_{\gamma}$ gets contributions from quarks and leptons, and it is

\begin{equation}
\begin{aligned}
\xi_{\gamma} &=-\frac{8}{3} N_{g}\left(x+\frac{1}{x}\right).
\end{aligned}
\end{equation}

As before, in computing $C_{a \gamma \gamma}$, one must separate the light quark contribution of the axion in the anomaly, so that $\xi_{\gamma}^{\text {eff }}=\frac{8}{3} x+ \frac{2}{3x}-\frac{8}{3} N_{g}(x+\tfrac1x)$, and add back the axion to two-photon contribution that arises from the coupling of the $\pi^{0}$ and $\eta$ to two photons, via the axion-pion and axion-eta mixing
%: $\lambda_{3}+\frac{5}{3} \lambda_{0}$. This then gives

\begin{equation}
C_{a \gamma \gamma}=-2N_{g}\left(x+\frac{1}{x}\right) \frac{m_{u}}{m_{u}+m_{d}}.
\end{equation}

\subsection*{Invisible Axion Models}\label{invisibleaxion}
The problem with the original model of the QCD axion is that it was ruled out by experiments data and observations. For the example, the bounds on the branching ratio of the decay of one positive kaon to a pion and an axion, which it is estimated within the PQWW model as \cite{BARDEEN1987401, Sikivie2008}
$$
\mathrm{BR}\left(K^{+} \rightarrow \pi^{+}+a\right) \simeq 3 \times 10^{-5} \lambda_{0}^{2}=3 \times 10^{-5}(x+1 / x)^{2}
$$
is well above the bound obtained at the KEK in 1981 of $\text{BR} <3.8 \times 10^{-8}$ \cite{Asano:1981nh}.

However, the so called "invisible axion" models, with $f_{a} \gg v_{\mathrm{F}}$, are still viable.

Invisible axion models introduce scalar fields that carry PQ charge but are $S U(2) \times U(1)$ singlets.

This allows to obtain VEVs of these fields with scales much larger than the ones set by the weak interactions. Two classes of models have been proposed.

The first, due to Kim \cite{PhysRevLett.43.103} and Shifman, Vainshtein and Zakharov \cite{SHIFMAN1980493} is the so-called KSVZ model and introduces a scalar field $\sigma$ with $f_{a}=\langle\sigma\rangle \gg v_{\mathrm{F}}$ and a very heavy quark $Q$ with $M_{Q} \sim f_{a}$ as the only fields carrying PQ charge. 

The second class, due to Dine, Fischler and Srednicki \cite{DINE1981199} and Zhitnisky \cite{Zhitnitsky:1980tq} for the first model, is the so-called, DFSZ models which add to the original PQ model one or more scalar fields $\phi$ that carry PQ charge and $\langle\phi\rangle \sim f_{a}  \gg v_{\mathrm{F}}$.
In the literature \cite{choi2024axiontheorymodelbuilding,ohare2024cosmology}, these models are labeled with DFSZ II, DFSZ III, and the original model being the DFSZ I, which just introduces one scalar field.

We will not repeat here the same kind of calculations as above to get the axion mass and couplings, but we will do some of them for the KSVZ model. The reason is for simplicity, and to highlight the most basic features and challenges of QCD invisible axion models. We will adopt an approach similar to the ones of Refs.~\cite{Sikivie2008,choi2024axiontheorymodelbuilding}.

By assumption, the KSVZ axion only interacts with light quarks as a result of the strong and electromagnetic anomalies

$$
L_{\mathrm{axion}}^{\mathrm{KSVZ}}=\frac{a}{f_{a}}\left(\frac{g_{\mathrm{s}}^{2}}{32 \pi^{2}} G_{b}^{\mu \nu} \tilde{G}_{b \mu \nu}-3 e_{Q}^{2} \frac{\alpha}{8 \pi} F^{\mu \nu} \tilde{F}_{\mu \nu}\right)
$$
where $e_{Q}$ is the electromagnetic charge of the super-heavy quark $Q$.

As in the KSVZ model, the ordinary Higgs particle does not carry PQ charge; the only interactions of the axion with the light-quark sector come from the effective anomaly mass term, which is given here by

$$
\mathcal{L}_{\text {anomaly }}=-\frac{\left(m_{\eta}^{0}\right)^{2}}{2}\left(\eta+\frac{f_{\pi}}{2 f_{a}} a\right)^{2} .
$$

To the above, we must add the standard quadratic term coming from the light quarks

$$
\mathcal{L}_{\mathrm{mass}}^{(2)}=-\frac{\left(m_{\pi}^{0}\right)^{2}}{2}\left[\frac{m_{u}}{m_{u}+m_{d}}\left(\pi^{0}+\eta\right)^{2}+\frac{m_{d}}{m_{u}+m_{d}}\left(\eta-\pi^{0}\right)^{2}\right] .
$$

Diagonalizing $\mathcal{L}_{\text {anomaly }}$ and $\mathcal{L}_{\text {mass }}^{(2)}$ gives

$$
m_{a}=\frac{v_{\mathrm{F}}}{f_{a}} \bar{m}_{a}, \quad \xi_{a \pi}=-\frac{m_{d}-m_{u}}{2\left(m_{u}+m_{d}\right)} \frac{f_{\pi}}{f_{a}}, \quad \xi_{a \eta}=-\frac{1}{2} \frac{f_{\pi}}{f_{a}},
$$
so we obtain

$$
\lambda_{m}=1, \quad \lambda_{3}=-\frac{m_{d}-m_{u}}{2\left(m_{u}+m_{d}\right)}, \quad \lambda_{0}=-\frac{1}{2} .
$$

Note that in the KSVZ model the axion mass is given by the formula
\cite{PhysRevLett.40.223}
\begin{equation}
m_{a}=\frac{v_{\mathrm{F}}}{f_{a}} \bar{m}_{a} \simeq \frac{\sqrt{m_u m_d}}{m_u+m_d} \frac{f_{\pi} m_{\pi}}{f_a} \simeq 6.3 \times 10^{-6} \; \SI{}{eV} \left(\frac{10^{12} \; \SI{}{GeV}}{f_a} \right),
\end{equation}
where $m_{\pi} \simeq \SI{140}{MeV}$ is the pion mass and $f_{\pi} \simeq \SI{93}{MeV} $ is the pion decay constant.

The calculation of $C_{a \gamma \gamma}$ in this model is analogously easy. We must add the contribution from the mixing of the axion with the $\pi^{0}$ and the $\eta$, (giving the term $\lambda_{3}+\frac{5}{3} \lambda_{0}$), and the one from the super-heavy quark in the electromagnetic anomaly, $3 e_{Q}^{2}$. This gives us, finally,
$$
C_{a \gamma \gamma}=6 e_{Q}^{2}-2\frac{4 m_{d}+m_{u}}{3\left(m_{u}+m_{d}\right)} .
$$

 For the original DFSZ model, it is convenient to define the quantities

$$
X_{1}=\frac{2 v_{2}^{2}}{v_{\mathrm{F}}^{2}} \quad \text { and } \quad X_{2}=\frac{2 v_{1}^{2}}{v_{\mathrm{F}}^{2}},
$$
where again $v_{\mathrm{F}}=\sqrt{v_{1}^{2}+v_{2}^{2}}$ and $v_{1}$ and $v_{2}$ are the two Higgs VEVs. Furthermore, if one rescales $f_{a} \rightarrow f_{a} / N_{\text{DW}}$, the axion mass in the DFSZ model is given the same equation as for the KSVZ model, corresponding to $\lambda_{m}=1$. When understanding the idea of rescaling, it becomes easy to obtain

$$
\lambda_{3}=\frac{1}{2}\left(\frac{X_{1}-X_{2}}{2 N_{g}}-\frac{m_{d}-m_{u}}{m_{d}+m_{u}}\right), \quad \lambda_{0}=\frac{1-N_{g}}{2 N_{g}},
$$
and

$$
C_{a \gamma \gamma}=\frac{4}{3}-\frac{4 m_{d}+m_{u}}{3\left(m_{d}+m_{u}\right)}
$$

Although the KSVZ and DFSZ axions are very light, very weakly coupled, and very long-lived, they are not totally invisible. 

As it can be seen from the expressions of $C_{a \gamma \gamma}$ and since it is also convenient for the DFSZ II model, we can write \cite{ohare2024cosmology}

\begin{equation}
    C_{a \gamma \gamma}=\frac{E}{N}-1.92(4) \simeq\begin{cases}
    -1.92, \quad&\text{PQWW}\\
        -1.92, \quad& \text{KSVZ}\\
        +0.75, \quad& \text{DFSZ I}
        \\
        -1.25, \quad& \text{DFSZ II}
    \end{cases}
\end{equation}
We will mostly adopt in the following the coupling constant $g_{a \gamma \gamma}$ which is related as
\begin{equation}
    g_{a\gamma\gamma} = \frac{\alpha_{\rm em}}{2\pi f_a}\, C_{a\gamma\gamma},
    \label{eq:gagg_definition}
\end{equation}
where $\alpha_{\rm em}$ is the fine-structure constant \cite{JEGERLEHNER2008135,schwartz2014quantum} and the axion decay constant $f_a$ is related to the magnitude of the vacuum expectation value that breaks the $U(1)_{PQ}$ symmetry by $f_a = \eta/N_{DW}$. $N_{DW}$ is an integer characterizing the color anomaly of $U(1)_{PQ}$.
We observe that all axion couplings are
inversely proportional to $f_a$.
 
The simple relation between the QCD axion mass and its decay constant, which in general does not hold for generic axion-like particles (ALPs), greatly simplifies the treatment of experimental constraints. 
In the parameter space $(m_a, g_{a\gamma\gamma})$, a generic axion has two independent parameters, whereas the QCD axion effectively depends on only one. 
Nevertheless, some model dependence remains in the invisible axion models, and thus one usually refers to the "QCD axion band" rather than to a "QCD axion line".

\subsection{The axion effective potential}\label{axiopotential}
We have considered previously several aspects of the quantum field theory of an axion coming from the chiral effective field theory at temperature $T=0$ .
However, if we are interested in its dynamics in the Early Universe, we need to know how the theory behaves at temperature $T$, in particular, its effective potential.

An effective potential for the axion field needs to satisfy some characteristics stemming from the symmetries of the theory.

It needs to be invariant under $a(x)\rightarrow a(x)+2 \pi \eta  $ from which
$V(a)=V(a+2 \pi \eta)$ and, if it is an analytical function, it can be written as
\begin{equation}
    V(a)=\frac{1}{2}m_a^2 a^2+ (m_a f_a)^2 \sum_{n=2}^{+\infty} \frac{\lambda_{2n}}{(2n)!} \Bigg(\frac{a}{f_a} \Bigg)^{2n}.
\end{equation}

The $\mathcal{Z}_2$ symmetry implies that $V(a)$ is an even function of $a$, then the number of axions is conserved modulo 2 \cite{marsh2016axion}. %and we can also write 
%\begin{equation}
    %a(\vec{x},t)=\sum^{+\infty}_{n=0} a_{2n+1}(r) \cos{[(2n+1) \omega t]}
%\end{equation}

We can already see from the low-energy two-flavor chiral lagrangians that we can obtain an effective theory at temperatures below $T_{\text{QCD}}$.

We start from the potential coming from the axion-pion-eta interaction
 and take the minimum towards the $\eta$ field, namely fixing $a$ and $\pi$, and we find the following axion-pion potential
  \begin{align}
    \mathcal{L}_\text{eff}
    &= \frac{f_\pi^2}{4}\,\mathrm{Tr}\!\left[\partial_\mu U^\dagger \partial^\mu U \right]
       + \frac{f_\pi^2 B}{2}\,\mathrm{Tr}\!\left[M_a U + M_a^\dagger U^\dagger \right]
       + \frac{1}{2} (\partial_\mu a)(\partial^\mu a),
       \end{align}
       where
       \begin{align}
    M_a &=
    \begin{pmatrix}
      m_u\, e^{i a/f_a} & 0 \\[0.3ex]
      0 & m_d\, e^{i a/f_a}
    \end{pmatrix}, \qquad
    U = \exp\!\left(i \frac{\vec{\pi}\cdot\vec{\tau}}{f_\pi}\right).
  \end{align}

  If we take the minimum again, but now over the pion fields instead of the $\eta$ field, we obtain the following potential \cite{DILUZIO202010}
  \begin{equation}
    V(a) = -\,m_\pi^2 f_\pi^2
    \sqrt{
      1 -
      \frac{4 m_u m_d}{(m_u + m_d)^2}
      \sin^2\!\left(\frac{a}{2f_a}\right)
    },
  \end{equation}
  This potential has a minimum at $a=0$, solving the Strong CP problem, and has all the properties required for an effective potential. Furthermore, if we expand it for small $a/f_a$ near to the minimum
  \begin{equation}
    V(a) \simeq  -\,m_\pi^2 f_\pi^2+\frac{1}{2} m_a^2 a^2 + \mathcal{O}\Bigg(\frac{a^4}{f_a^4}\Bigg),
    \qquad
    m_a^2 = 
    \frac{m_u m_d}{(m_u + m_d)^2}
    \frac{m_\pi^2 f_\pi^2}{f_a^2}.
  \end{equation}

  We observe that the product $f_a^2 \,m_a^2$ is independent of the axion parameters and is simply equal to $\Lambda_{\text{QCD}}^4$ \cite{di2016qcd}.

However, things become very complicated at temperatures comparable with $T_{\text{QCD}}$, since the effective axion mass depend on the fluctuations of $\tilde{G}\, G$, in particular from the topological susceptibility 
\begin{equation}
    \chi_{\text{QCD}}(T)=\int d^4 x  \angi{q(x) q(0)},
\end{equation}
where $q(x)=\frac{g_s}{8 \pi} \tilde{G}_{\mu \nu}^a\, G^{\mu \nu}_a$ and several difficulties come from the non-perturbative nature of QCD.

At sufficiently high temperatures one expects that its behaviour
is described by the so-called
\emph{dilute instanton gas approximation} (DIGA)~\cite{Gross1981}.

At asymptotically high temperatures, the dilute instanton gas approximation
predicts that the susceptibility is governed by the instanton density
$n(T)$, which yields an explicit analytic form~\cite{Gross1981}:
\begin{equation}
  \chi_{\mathrm{DIGA}}(T)
  = 2 \int d\rho \, n(\rho, T)
  \simeq
  2 \, C_{N_c} \,
  \left(\frac{4\pi^2}{g^2(T)}\right)^{2N_c}
  \exp\!\left[-\frac{8\pi^2}{g^2(T)}\right]
  \prod_{f=1}^{N_f} (m_f \rho),
  \label{eq:DIGA_explicit}
\end{equation}
where $\rho$ is the instanton size, $g(T)$ is the running coupling at
the scale $\mu \simeq \pi T$, and $C_{N_c}$ is a numerical group factor, which is $C_{3}\!\approx\! 0.0015$ for SU(3).
Integrating over $\rho$ and using the one-loop running of the coupling,
$g^{-2}(T) = \frac{\beta_1}{8\pi^2} \ln(T/\Lambda_{\mathrm{QCD}})$,
one obtains a power-law behaviour
\[
\chi_{\mathrm{DIGA}}(T)
  \propto T^{-\beta_1}
  \left[\ln\!\left(\frac{T}{\Lambda_{\mathrm{QCD}}}\right)\right]^{2N_c}.
\]
This scaling defines the asymptotic high-temperature behaviour of the QCD topological susceptibility, with $N_c$ and $N_f$ denoting the number of colours and light flavours,
respectively.
Theoretically, $\beta_1=\frac{11}{3} N_c-\frac{2}{3} N_f-1$, which is equal to 8 for QCD ($N_f=N_c=3$), while the fitted values from lattice methods disagree, with a value varying between 7 and 8.

Instead, in the deep confined phase it should approach a constant non-zero value, reproducing the QCD axion mass at low energies, which is $\chi_{\mathrm{QCD}}^{1/4}(0) = 75.5(5)\,\mathrm{MeV}$~\cite{di2016qcd}.

Due to the non-perturbative nature of QCD, the most
reliable approach to determine $\chi_{\mathrm{QCD}}(T)$ would be through
lattice simulations.
However, results from different lattice groups and methodologies
(see e.g.~\cite{Chen:2022fid,Athenodorou:2022aay,Petreczky:2016vrs,
Borsanyi:2016ksw,Kotov:2021rah})
still show quantitative discrepancies and also in the QCD axion mass, as we show in Fig.~\ref{QCDaxionband} of Chapter \ref{experiments}, calling for further detailed studies.

We have shown part of this theoretical structure to highlight how it just depends on having a confined $SU(3)$ field and it is also valid with axion-like particles coupled to a dark $SU(3)$ sector.

Such models of axion-like particles have been treated formerly in Refs.~\cite{Rubakov:1997vp,Berezhiani:2000gh,Hook:2016mqo,Arias:2012az}   , where they showed how we can have interesting axion models at various masses from them, in particular also in the region of our interest of masses above $1\,\mathrm{keV}$, where there are not significant constraints from structure formation.

They are of particular interest for us since the high-mass ALPs cosmological models can significantly affected from the thermal friction acting on topological defects.
 A general axion-like particle (ALP) need not couple to QCD and may instead couple to another confining gauge sector, obtaining a similar temperature-dependent
  mass with $\Lambda_d$ replacing $\Lambda_{\mathrm{QCD}}$ \cite{Rubakov:1997vp}, or simply possess an explicit mass term that is approximately
  temperature-independent \cite{Arias:2012az}.
We will be interested to the case of Ref.~\cite{Rubakov:1997vp}.

%ROBERTO

%low-energy chiral lagrangian approach loses its validity and the  

\begin{figure}
    \centering
    \includegraphics[width=0.9\linewidth]{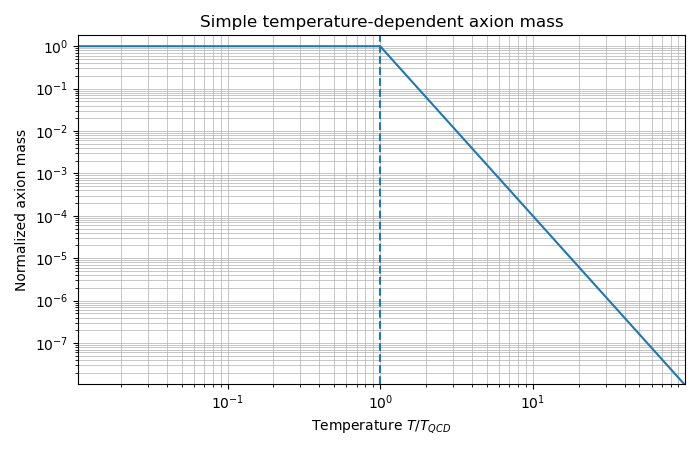}
    \caption{Log-log plot of the temperature-dependent axion mass with normalized units for the plasma temperature $T$ and the axion mass $m_a(T)$, accordingly to our adopted model in Eq.~(\ref{axionmassa}).}
    \label{fig:axionmassa}
\end{figure}

In the following we adopt the expression of the QCD axion temperature-dependent mass adopted in the usual axion literature to fit the results of Ref.~\cite{Borsanyi:2016ksw} %bring refs by Sikivie eBorsany and ones by O'Hare

\begin{equation}\label{axionmassa}
    m_a(T) \simeq \begin{cases}
    m_{a0}  \qquad &T \lesssim T_{QCD}\\
        4\times 10^9 \,\mathrm{eV}\, \Big(\frac{10^{12} \, \mathrm{GeV}}{f_a}   \Big) \,\Big( \frac{\mathrm{GeV}}{T}\Big)^{n/2}\quad &T \gtrsim T_{QCD}
    \end{cases}
\end{equation}
where $n=8$, corresponding to the choice of $\beta_1=8$ and $m_{a0}$ is such that the value of the mass is continous at $T=T_{\rm QCD}$ .
A visual log-log plot of this behaviour of the axion mass is shown in Fig.~\ref{fig:axionmassa}.
%for $T \simeq 150 \,\mathrm{MeV}$ and below.%put here relation T vs t

\subsection{The Axion quality problem}\label{axionquality}
\vspace{0.5em}
In the former subsection, we have celebrated the very nice  "quality" of the global symmetry $U(1)_{\text{PQ}}$, under which the theory is invariant under shift symmetry.
However, quantum gravity is expected to violate global symmetries~\cite{Kallosh:1995hi,Banks:2010zn} and 
Planck-suppressed higher-dimensional operators such as
\begin{equation}\label{highdop}
\delta V \supset \frac{c_n}{M_{\rm Pl}^{n-4}} \phi^n + \text{h.c.},
\end{equation}
where $\phi$ is the PQ-breaking field, should explicitly break $U(1)_{\rm PQ}$ and shift the axion potential.
These terms reintroduce an effective $\bar{\theta}_{\rm eff}$, spoiling the PQ solution unless they are
extremely suppressed:
\begin{equation}\label{requerim}
\left(\frac{f_a}{M_{\rm Pl}}\right)^{n-4} \lesssim \frac{1}{c_n}10^{-10} \Bigg(\frac{\Lambda_{\rm QCD}}{f_a} \Bigg)^4.
\end{equation}
This requirement corresponds typically to $n \gtrsim 10$ for $f_a \sim 10^{12}\,\text{GeV}$ and can be understood as follows.
The additional higher-dimensional operator (\ref{highdop}) can be parametrized as 
\begin{equation}
    \delta V \sim \frac{c_n}{M_{\rm Pl}^{n-4}} f_a^n \cos{\Bigg(\frac{a}{f_a}+\delta\Bigg)}
\end{equation}
In the case $\Lambda_{\rm UV}^4 \ll \Lambda_{\rm QCD}^4$, where we have defined
\begin{equation}
\Lambda_{\rm UV}^4
= \frac{c_n f_a^n}{M_{\rm Pl}^{\,n-4}}
= c_n\, f_a^4 \left(\frac{f_a}{M_{\rm Pl}}\right)^{n-4},
\end{equation}

we can expand perturbatively in the factor $(\Lambda_{\rm UV}/\Lambda_{\rm QCD})^4$ and obtain an estimation of the shift of the theta value
\begin{equation}
    \bar{\theta}_{\text{eff}}  \simeq \bar{\theta}+
\frac{\Lambda_{\rm UV}^4}{\Lambda_{\rm QCD}^4}\,\sin\delta.
\end{equation}
Requiring the experimental bound $\bar{\theta}_{\text{eff}}\lesssim 10^{-10}$, we obtain Eq.~(\ref{requerim}).

More precise and general details on this requirement, in particular for more general axion models and with $N_{\rm DW}>1$, are treated in Ref.~\cite{ardu2020axion}.

The need for such an extraordinarily "high-quality" PQ symmetry leads to the
axion quality problem~\cite{Barr:1992qq,Holman:1992us,choi2024axiontheorymodelbuilding}.

This problem cannot be solved within the usual "Field Theory" UV completion, which we describe in the following subsection and is useful to introduce the axion cosmic strings.

Possible approaches to the Axion quality problem include realizing PQ as an accidental or gauge-protected
symmetry, embedding it in discrete gauge symmetries, or deriving it from UV-complete constructions
such as string theory where axions emerge as pseudo-Nambu–Goldstone bosons of higher-dimensional gauge fields~\cite{Banks:2003sx,Svrcek:2006yi}, which we will discuss in the other following subsection.
%\subsection{Axion field theory UV completion}
 \subsection{Field Theory UV completion}\label{ftUV}
 \begin{figure}[h]
    \centering
    \includegraphics[width=0.8\textwidth]{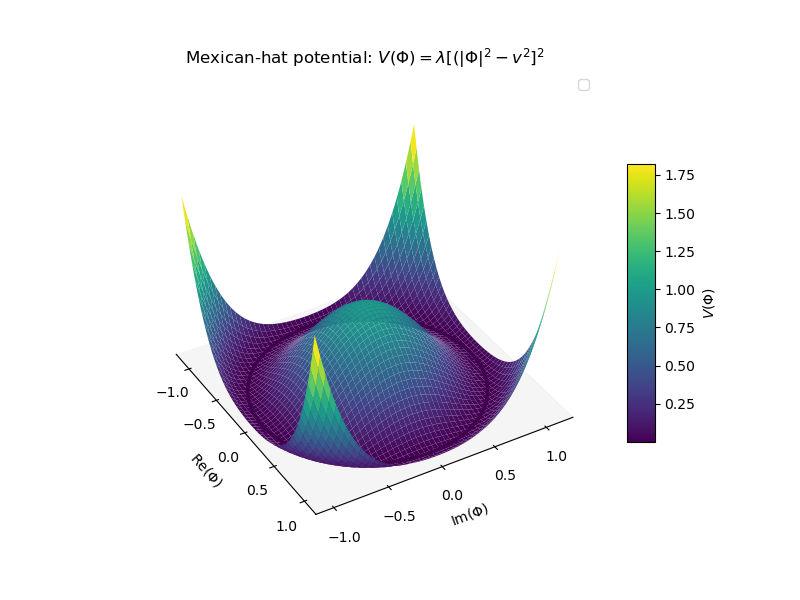} 
    \caption{Plot of the Mexican hat potential of the Peccei-Quinn field after the $U_{\rm PQ}(1)$ phase transition in normalized units.}
    \label{mexicanhat}
\end{figure}
 We define our axion theory with a minimal scalar sector, which we refer to as the Peccei-Quinn field $\Phi$, whose phase is the axion.
 
 We assume a Lagrangian density with a Higgs-like potential \cite{Sikivie2008, Benabou2023npn}

\begin{equation}
    \mathcal{L}=|\partial_{\mu} \Phi|^2-\lambda \Big(|\Phi|^2-\frac{\eta^2}{2}\Big)^2
\end{equation}
where we can write the Peccei-Quinn field $\Phi$ as
\begin{equation}
    \Phi(x)=\Bigg(\frac{\eta+s(x)}{\sqrt{2}}\Bigg) e^{\frac{ia(x)}{\eta}}.
\end{equation}

This theory holds the axion as a Goldstone boson and adds a radial mode $s(x)$ with a mass $m_s=\sqrt{2\lambda}\,\eta$.
The classical theory admits static infinitely straight string solutions
\begin{equation}
    \Phi(r, \theta,z)=\frac{\eta}{\sqrt{2}} g(m_s r) e^{in \theta}
\end{equation}
which are associated to the spontaneous breaking of the $U(1)_{\rm PQ}$ symmetry
and we call them "cosmic strings". $n$ is the topological winding number of the string.
.
Such a string has an effective energy per unit length, then a tension $\mu_{\rm eff}$ which is
    \begin{equation}
    \mu_{\rm eff}=\int^{2 \pi}_0 d\theta \int_0^{r_{\text{IR}}} dr \,r\, \rho_{s}=\pi \eta^2 \ln{\Big(\gamma m_s r_{\text{IR}}}\Big)
\end{equation}
where $r_{{\text{IR}}}$ is the infrared (IR) cutoff we impose to regulate the logarithmically divergent contribution to the tension arising from gradient energy, and it is typically limited in Cosmology from the Hubble radius, while $\gamma$ is a numerical factor of order unity.

\subsection{String axion UV completion}\label{saUV}
Previously, we have discussed several QCD axion models with a linearly realized $U_{PQ}(1)$\cite{choi2024axiontheorymodelbuilding}.
However, another intriguing possibility is to consider models in which the 4D axion arises from a higher-dimensional gauge field. 
Such models do not obtain a linear $U_{PQ}(1)$ in the low-energy limit, but a nonlinear one which is intrinsically related to the higher-dimensional gauge symmetry of the multidimensional model.
We discuss one example of a simple 5D toy model from Ref.~\cite{Benabou2023npn}, whose axion shows the fundamental features of axions from p-form gauge fields in string theory.
%\subsubsection{Example of axion from a 5D gauge field : Flat extra dimension with a radion as the modulus regulating the string core}

We consider a 5D $U(1)$ gauge field $A_M$ and a radion $\rho$. 
The latter field appears in the parametrization of the 5D metric that is 

\begin{equation}
    ds^2= \tilde{g}_{MN} \, dx^{M} dx^{N}=\frac{b}{\rho(x)} g_{\mu \nu}(x)\, dx^{\mu}dx^{\nu}+ \rho(x)^2 d\phi^2,
\end{equation}
and the radion potential can stabilize the size of the fifth dimension to the VEV $\angi{\rho}=b$. The 5D spacetime is a manifold of the form $\mathcal{M}_4 \times S^1/\mathbb{Z}_2$, where $\mathcal{M}_4$ is a four-dimensional Minkowski spacetime, $S^1$ the 1-sphere, i.e. the circle and $\mathbb{Z}_2$ the two-element cyclic group.

The 5D action is then of the form
\begin{equation}
    S=S^{(5)}_{EH}+ S^{(5)}_{U(1)}+S_{\rho}
\end{equation}

where it includes the 5D Einstein-Hilbert action $S^{(5)}_{EH}$
\begin{equation}
\begin{aligned}
    S^{(5)}_{EH}=\int d^4x \, d\phi \, \sqrt{|\tilde{g}|} \Big( M^3_5 \, R^{(5)}(\tilde{g}) \Big)=\frac{1}{2} M^2_{pl} \int d^4 x \sqrt{|g|}  \Big( \, R^{(4)}(g) - \frac{3}{2 \rho^2} \partial_{\mu} \rho \,\partial^{\mu} \rho  \Big)
    \end{aligned}
\end{equation}
, the action $S^{(5)}_{U(1)}$of a five-dimensional $U(1)$ gauge field 
\begin{equation}
   S^{(5)}_{U(1)}=  - \int d^4 x \frac{1}{4 g_{5A}^2} A^{MN} A_{MN}  -\frac{1}{4 g_{5S}^2} G^{aMN} G^a_{MN}
\end{equation}

and a term $S_{\rho}=-\int d^4 x \, d\phi \, \sqrt{|g|} \, V(\rho)$ including the radion potential and whose form depends on the details of the stabilization mechanism.

We then consider the field $\theta(x)$ defined as the gauge-invariant Wilson loop 
\begin{equation}
    \theta(x) \coloneqq 2 \int^\pi_0 d\phi \,A_5.
\end{equation}
We observe that $\theta(x)$ has periodicity $2\pi$ due to the large $U(1)$ gauge transformations and with a spectrum of intenger charges.

The $\phi-$dependent modes acquires mass through Kaluza-Klein mechanism, while the $\phi$ independent modes leas to the action for the $\theta(x)$
\begin{equation}
    S_{\theta}=-\int d^4 x \sqrt{|g|} \frac{1}{8 \pi^2 g_4^2 \rho^2} g^{\mu \nu} \partial_{\mu} \theta \,\partial_{\nu} \theta
\end{equation}
where we have defined the convenient quantity 
\begin{equation}
    \frac{1}{g_4^2} \coloneqq \frac{2 \pi b}{g^2_5}.
\end{equation}
It is then now convenient to define $a(x)=f_a \theta$, which we can interpret as a massless axion, with $f_a \coloneqq \frac{1}{2 \pi b g_4}$.

This example highlights some important features of the axion decay constant for a stringy axion, which are the dependence on the dimension scale $b$ and the 4-dimensional coupling $g_4$.
This simple picture shows us that, in a more complicated Calabi-Yao manifold with more loops, we can expect to have multiple axions. Furthermore, as e.g. shown in Ref.~\cite{Benabou2023npn}, we can think of models where the QCD axion arises as a linear combination of these stringy axions and could also admit the presence of other ALPs, including the high-mass axions of our interest.
These works can then be a theoretical justification for being interested to post-inflationary scenarios for both a QCD axion and a high-mass ALP.

%%%adjust from Safdi string axion cosmic strings and add suggestion from Choi 2025 and more recent ones 

Other interesting axion low-energy theories can be found between the variety of low-energy string compactification theories and the possible topologies of the various Calabi-Yau manifolds, and a good selection is Refs.~\cite{Acharya:2010zx,Cicoli:2012sz,Reig:2019vgh,Agrawal:2022lsp,Gorghetto:2020qws,Cicoli:2022uqa,Agrawal:2024heterotic,Petrossian-Byrne:2025jhf,Agrawal:2025baryon}

\begin{comment}
\subsubsection{Axion from a 5D gauge field in toy model 2:  }

This is a simple example of the model proposed in   and discussed 

The 5D action of the model is
\begin{equation}
\begin{aligned}
	S_{5D}= \int d^5 x \sqrt{\tilde{g}} \left[\frac{1}{2} M^3_5 \mathcal{R}_5(\tilde{g})-\frac{1}{4 g_{5A}^2} A^{MN} A_{MN} -\frac{1}{4 g_{5S}^2} G^{aMN} G^a_{MN} + \right.\\\left. i \sum_{I} \bar{Q}_I\,(\gamma^M D_M+\mu_I\, A_{MN} \gamma^{MN})\, Q_I+ \frac{k_{CS}}{32 \pi^2} \frac{\epsilon^{MSPQR}}{\sqrt{-\tilde{g}}} A_M G^a_{SP} G^a_{QR}+...   \right]
    \end{aligned}
\end{equation}

%\subsubsection{Axion from a 5D gauge field in toy model 2:  }

%\subsubsection{Axion from a 5D gauge field in toy model 3: Higher axions  }

%\subsubsection{Axions from p-form gauge fields in string theory}
%\subsubsection{Axions in type IIB and IIA string theories}

%\subsubsection{Axions in heterotic string theory}
\end{comment}

%\subsection{Phenomenology of QCD axion and Axion-like particles}

%\subsection{Other possible solutions}

%\subsection{About static and dynamical solutions of the Strong CP problem of QCD}

%put discussion by Vilenkin
%\subsection{Codes and numerical implementations}

%\subsection{Original results}

%\chapter{Axion inflation with or without gauge fields}

\chapter{Relevant experimental aspects and constraints on the axion}\label{experiments}

 In this chapter, we will consider general aspects of experimental searches and constraints for axions relevant to this thesis.

 To fix the ideas, we will consider a Lagrangian density of the form
\begin{equation}
    \mathcal{L} =- \frac{1}{4} G^b_{\mu \nu} G_{b \mu \nu}  +\frac{1}{2} \partial_{\mu} a \, \partial^{\mu} a
 +\sum \bar{q} (i\gamma_{\mu} \partial^{\mu} - m_q ) q+ \frac{g_s^2}{32 \pi} a\,G_{b \mu \nu} \tilde{G}_b^{\mu \nu}-\frac{1}{4}g_{a \gamma \gamma} a F_{\mu \nu} \tilde{F}^{\mu \nu}
\end{equation}
 and we will then focus on laboratory constraints on the parameter space $(m_a,g_{a \gamma \gamma})$.

%Equation uses standard notation for the chromomagnetic field strength
%tensor $G$, the strong coupling constant $g_s$, and the quark fields $q$. The axion
%mass, after mixing with the $\eta$ and $\pi_0$ mesons, is given in terms of $f_a$ by

There are three main ways in which the invisible QCD axion, along with axion-like particles, has been searched for in recent years: 
\begin{itemize}
	\item The first possibility is to detect it directly
	by means of laboratory experiments.
	
	\item  The second way is to observe it indirectly in astronomical objects.
	
	\item The third way is to constrain its properties and models from cosmology.
\end{itemize}  
Indeed, the original work presented in this thesis is useful for all three approaches, with a particular emphasis on the first and third.
%These ways of constraining the axion parameters meet with three different classes
We briefly summarize the constraints obtained in other research activities \cite{RevModPhys.93.015004,ohare2024cosmology,aybas2025exploringdarkuniverseeuropean}.

\section{Astrophysical cooling and superradiance constraints}
Astrophysical systems provide powerful, largely model-independent, bonds on axion properties, in particular through axion emission. Relevant processes in the stars are the Compton production $\gamma+e^- \rightarrow a+e^-$ and the Primakoff process $\gamma+ Ze \rightarrow a+Ze$, which cause energy loss in stars, along with neutron bremsstrahlung $N\, N \rightarrow N \, N\, a$.

%We highlight the difference between the former two processes.
The energy loss processes mentioned above are usually inversely proportional to $f_{a}^{2}$ and hence proportional to $m_{a}^{2}$ for QCD axions. Consequently, QCD axions must be light enough so as not to affect stellar evolution. 

Main constraints are the following:
%This is not the only way in which one obtains an upper bound on $m_{a}$. 
\begin{itemize}
    \item Horizontal-branch stars, red giants, white dwarfs: Excess energy loss via axion emission would alter stellar lifetimes and luminosity functions, bounding $g_{a\gamma\gamma}$ and $g_{ae}$ \cite{Raffelt2008,Giannotti2017}.
    \item SN1987A: Excess axion emission from the supernova core, through the process $N N \rightarrow N N a$,  would have shortened its observed neutrino burst. This sets limits on axion-nucleon couplings in hadronic axion models \cite{RAFFELT19901,Payez2015}.
    \item Solar axions: CAST's non-observation of solar axion conversion to keV X-rays yields $g_{a\gamma\gamma} \lesssim 6\times10^{-11}$~GeV$^{-1}$ for $m_a \lesssim 0.02$~eV \cite{CAST2017}.
\end{itemize}
Together, these arguments disfavour QCD axions heavier than $\sim 0.1$~eV, because such axions would couple strongly enough to drain energy from stars and supernovae in conflict with observations \cite{Raffelt2008,Giannotti2017,Payez2015,CAST2017}.

Other very interesting astrophysical bounds come from the Black-hole superradiance.
Rotating Kerr black holes can undergo \emph{superradiance} if a light boson of mass $m_a$ exists with a Compton wavelength comparable to the black hole's gravitational radius, then bound "gravitational atom" levels can form and exponentially grow by extracting spin from the black hole \cite{Arvanitaki2010,Brito2015,PhysRevLett.128.221102}.  
This process would spin down black holes in specific mass ranges over astrophysically short timescales and then observed black holes with high spins exclude bosons whose masses would have triggered spin-down.
Observationally:
\begin{itemize}
    \item Stellar-mass black holes ($\sim 5$-50 $M_\odot$) observed with high spin exclude axion-like particle masses of order $m_a \sim 10^{-13}$--$10^{-11}$~eV \cite{Arvanitaki2010,Stott2021,Abbott2021,Brito2015}.
    \item Supermassive black holes ($\sim 10^{6}$-$10^{9} M_\odot$) with large spins rule out $m_a \sim 10^{-18}$-$10^{-16}$~eV \cite{Stott2021,Brito2015}.
\end{itemize}
Importantly, they arise from purely gravitational dynamics, so they are independent from other coupling constant, e.g. $g_{a \gamma \gamma}$, and they could be a QCD axion or an ALP.

Recent work is pushing these constraints further, notably around $m_a \sim 10^{-13}$~eV from high-spin stellar-mass black holes observed in gravitational-wave events \cite{CaputoRaffelt2024,Caputo2025,WitteMummery2025}.

\section{Cosmological bounds}
Cosmology constrains both very light and relatively heavy QCD axions:
\begin{itemize}
\item Standard misalignment mechanism:  
If axions are produced by the standard misalignment mechanism with a natural initial angle ($\theta_i \sim \mathcal{O}(1)$), requiring $\Omega_a \le \Omega_{\rm DM}$ sets a \emph{lower} bound on the mass in conventional cosmology.  
Typical estimates give $m_a \gtrsim 25~\mu$eV  if axions are to make up all of dark matter under standard assumptions \cite{Arias2012,DiLuzio2020,PDG2024}. We will discuss this point in more detail in Section~\ref{misa}.
\item Hot relic limits and structure formation:  
Axions that thermalize in the early Universe behave like hot relics. For $m_a \gtrsim 0.1$~eV, they suppress small-scale structure in ways disfavored by CMB + BAO + Lyman-$\alpha$ data \cite{Arias2012,PDG2024}.  
    This again disfavors very heavy QCD axions ($\gtrsim 0.1$~eV).
\end{itemize}

Thus cosmology points to a cold dark matter window for QCD axions roughly in the $10^{-6}$-$10^{-2}$~eV band, modulo assumptions about the post-inflation vs pre-inflation Peccei-Quinn breaking history \cite{Arias2012,DiLuzio2020,PDG2024}.

A critical aspect is the estimation of the QCD axion mass from numerical simulation of axion cosmic string networks \cite{Benabou2023ghl,Benabou:2024msj,Saikawa2024}.

The basic idea of such approaches is that cosmic strings decay to axions since they are unstable due to Derrick's theorem, and they emit significantly roughly up to a reference time before the QCD phase transition, which can be taken to be $t_1$ or $t^*$ defined by $m_a(t^*)=3H(t^*)$.

In particular, the numerical analysis is focused on the evolution of the string length per Hubble volume \begin{equation}
    \xi \coloneqq \frac{l t^2}{V},
\end{equation}
where $l$ is the extracted total string length and $V$ is the simulation volume, and the istantaneous emission spectrum
\begin{equation}
    F\Bigg(\frac{k}{H}\Bigg) \coloneqq \frac{1}{R^3(t)}\partial_t [R^3(t) \partial_k \rho_a ]
    \end{equation}
    where $\rho_a(k)$ is the axion energy density in momentum space \cite{Saikawa2024, Benabou:2024msj}.
If the axion field is in the scaling regime, analogously to what we mention for axion domain wall networks in the following, the axion DM abundance can be calculated from the number density at $t=t^*$:

\begin{equation}
    \Omega_a^{\rm str} h^2 \approx 0.12 \times \Bigg(\frac{f_a}{1.4 \times 10^{11} \rm GeV}    \Bigg)\,\frac{318}{\delta} \,\sqrt{\frac{\xi^*}{13}}\, \frac{Log*}{70}
\end{equation}
where $g_*(T) \approx g_*^0 \Big(\frac{T}{1\,\rm MeV}   \Big)^\gamma$ assumed for the temperature range $800 \,\,\rm MeV<T<1800 \,\,\rm MeV$ with $g_*^0 \approx 50.8$ and $\gamma \approx 0.053$.

The parameter $\delta$ is related to the average $\angi{...}$ of the ratio $H/k$, calculated through the istantaneous emission spectrum,  with the expression
\begin{equation}
   \angi{H/k} =\delta \sqrt{\xi},
\end{equation}
while the quantity $\rm Log*=\ln{\Big(\frac{m_s}{H(t^*)}\Big)}$ is commonly adopted in the literature \cite{Saikawa2024,Benabou:2024msj,Correia:2024cpk}.

This highlights the importance of better valuating the spectral index $q$ of the instantaneous emission spectrum, since it is relevant for the contribution to the axion density parameter coming from cosmic strings.

\begin{figure}
    \centering
    \includegraphics[width=0.8\linewidth]{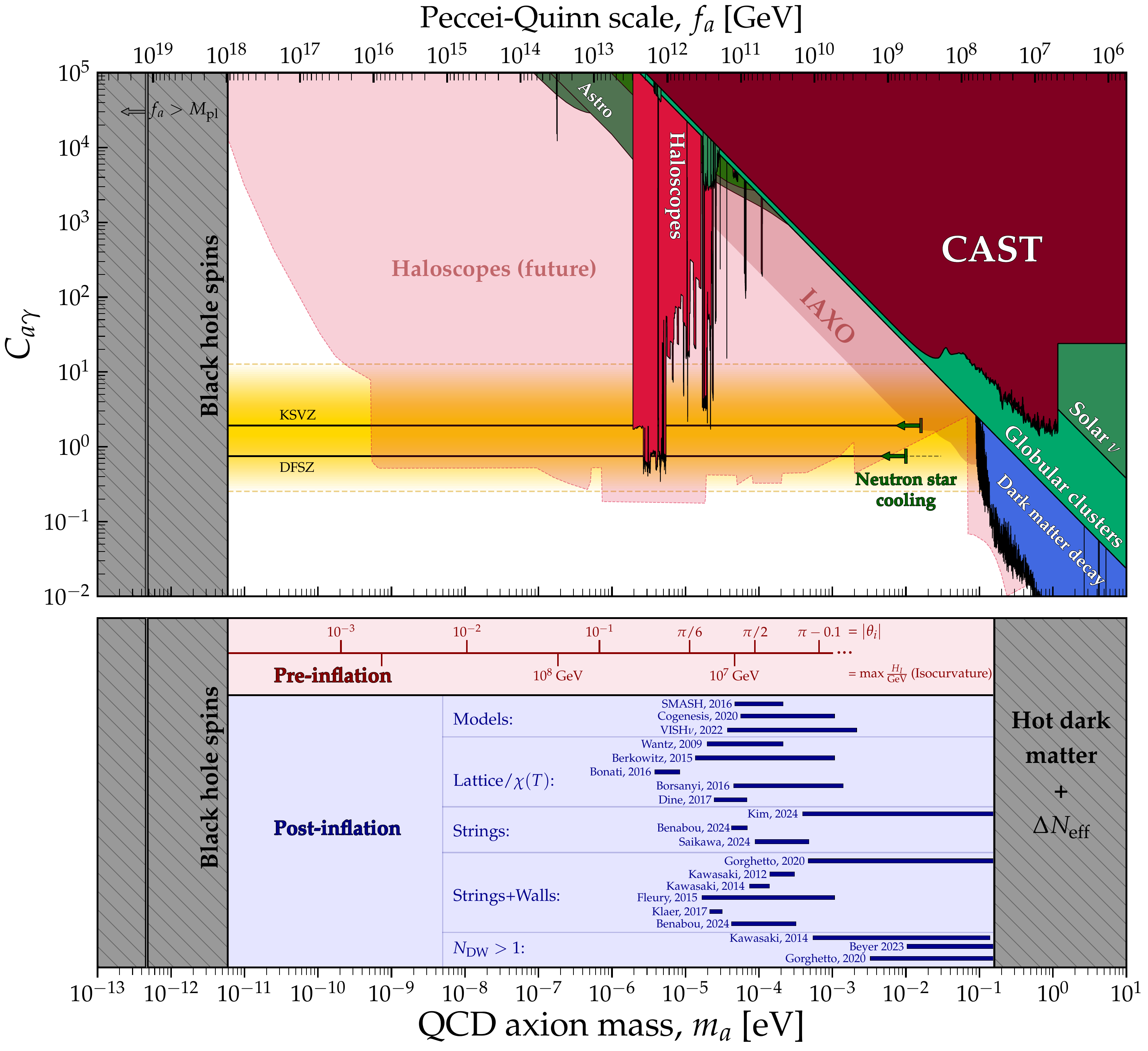}
    \caption{Closeup of the parameter space $(m_a, C_{a \gamma \gamma})$ in the radio band from the Github Axion Limits by Ciaran O'Hare \cite{axionLimits}, showing clearly the QCD axion band, the current constraints, and the projected areas of next haloscopes experiments or of the new runs of older experiments. The lower image shows the different, and contradicting, estimations for the QCD axion from several methods and groups, in particular the ones from lattice methods we discussed in Subsection~(\ref{axiopotential})  and the one from numerical simulations of the topological defect network we have introduced here. Update (10 December 2025).}
    \label{QCDaxionband}
\end{figure}

This also implies that it affects the cosmological bounds on the QCD axion in the post-inflationary scenario, as visible in Fig~(\ref{QCDaxionband}), where it is shown how different groups obtain different estimations, which is problematic for theoretical prediction.

Furthermore, axion domain walls contribute more significantly, and it is more reasonably relevant to understand the problems related to the numerical methods.

The approach we are starting in Section~\ref{Schwinger} is very straightforward: we aim to improve the analytical models.

\section{Laboratory bounds}
We then find a "sweet spot" where QCD axions can be all or a large fraction of dark matter, evade stellar cooling bounds, and evade superradiance, and this is the canonical QCD axion window, which lies approximately at
\begin{equation}
m_a \sim 10^{-6}-10^{-2}~\mathrm{eV},
\end{equation}
i.e.\ $\mathcal{O}(\mu{\rm eV}-10~{\rm meV})$,  
corresponding to Peccei-Quinn scales $f_a \sim 10^{9}-10^{13}$~GeV and photon couplings
$g_{a\gamma\gamma} \sim 10^{-16}$-$10^{-10}$~GeV$^{-1}$ \cite{di2016qcd,DILUZIO202010,DiLuzio20201,CaputoRaffelt2024}.

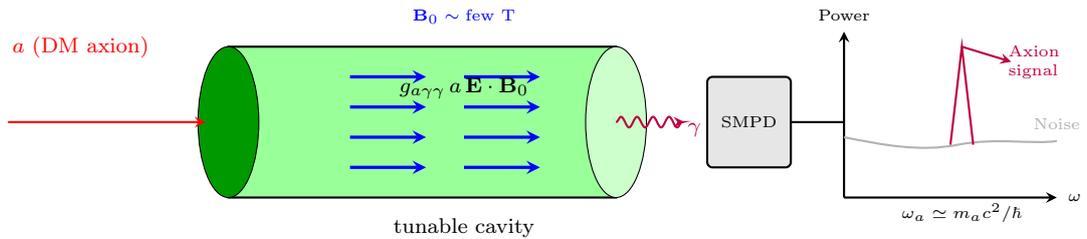
\begin{figure}
\begin{center}
\begin{tikzpicture}[scale=1.0,>=stealth]
% Cavity
\shade[ball color=green!50!black,opacity=0.3] (-3.1,-1) rectangle (2,1);
\draw[fill=green!40,draw=black,thick] (-3.1,-1) rectangle (2,1);
\draw[fill=green!60!black,draw=black] (-3.1,0) ellipse (0.4 and 1);
\draw[fill=green!20,draw=black] (2,0) ellipse (0.4 and 1);
\node[font=\scriptsize,align=center] at (0,-1.4)
{tunable cavity};

% B field
\foreach \y in {-0.6,-0.2,0.2,0.6} {
  \draw[very thick,blue,->] (-1.5,\y) -- (-0.5,\y);
  \draw[very thick,blue,->] (0.0,\y) -- (1.0,\y);
}
\node[blue,font=\tiny] at (0,1.4) {$\mathbf{B}_0 \sim \mathrm{few\ T}$};

% Axion
\draw[->,thick,red] (-6,0) -- (-3.4,0);
\node[red,font=\scriptsize,anchor=east] at (-4,1) {$a$ (DM axion)};

\node[font=\scriptsize,align=center] at (-0.0,0.45)
{$g_{a\gamma\gamma}\,a\,\mathbf{E}\cdot\mathbf{B}_0$};

Photon out
\draw[decorate,decoration={snake,amplitude=2pt,segment length=6pt},
      thick,purple,->]
(2,0) -- (2.9,0);
\node[purple,font=\scriptsize,anchor=west] at (2.8,-0.1)
{$\gamma$};

% Readout chain box
\draw[fill=gray!20,draw=black,rounded corners=2pt,thick]
(3.2,-0.6) rectangle (4.3,0.6);
\node[font=\tiny,align=center] at (3.75,0.0)
{SMPD};
\draw[thick] (4.3,0) -- (5.0,0);

% Spectrum
\draw[->,thick] (5.0,-1.0) -- (7.8,-1.0)
 node[anchor=west,font=\tiny] {$\omega$};
\draw[->,thick] (5.0,-1.0) -- (5.0,1.2)
 node[anchor=south,font=\tiny] {Power};

\draw[thick,gray!60]
(5.0,-0.2) .. controls (5.5,-0.3) and (6.0,-0.4) ..
(6.5,-0.3) .. controls (7.0,-0.2) and (7.5,-0.3) .. (7.8,-0.25)node[anchor=south,font=\tiny] {Noise};

\draw[thick,purple]
(6.4,-0.3) -- (6.55,1.0) -- (6.7,-0.3);

\draw[->,purple,thick] (6.55,1.0) -- (7.2,0.8);
\node[purple,font=\tiny,align=left] at (7.5,0.8)
{Axion\\signal};

\node[font=\tiny] at (6.55,-1.2)
{$\omega_a \simeq m_a c^2/\hbar$};

\end{tikzpicture}
\end{center}
\caption{Basic scheme of a Sikivie's haloscope, similarly as in ADMX \cite{ADMX2018,admxcollaboration2025searchaxiondarkmatter} and QUAX experiments \cite{Braggio2025, SardoInfirri2025}.}
\label{sikiviehaloscope}
\end{figure}

\begin{figure}
    \centering
    \includegraphics[width=1\linewidth]{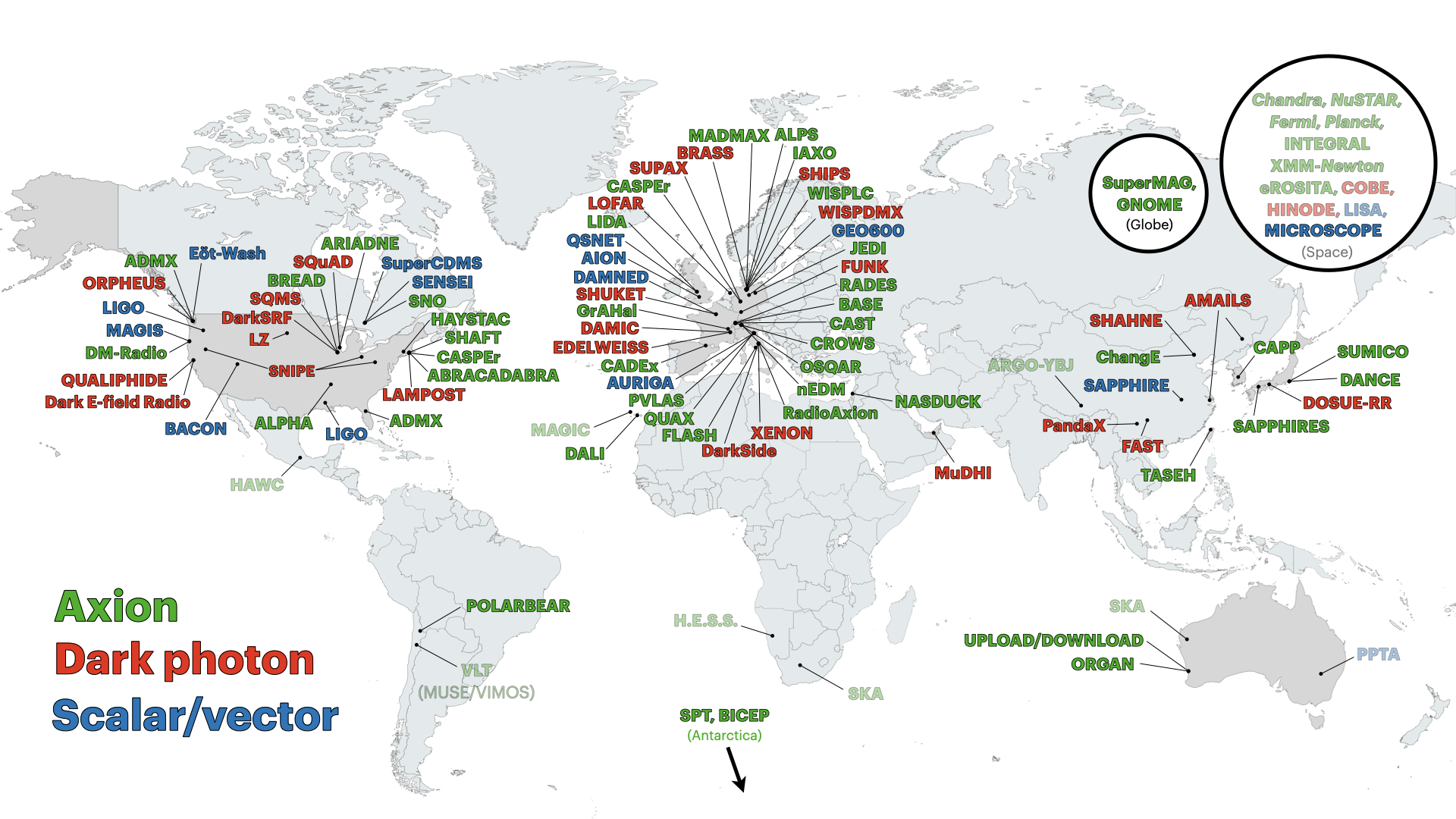}
    \caption{A summary of current experiments searching for axions and WISP particles. Image from the Github Axion Limits by Ciaran O'Hare \cite{axionLimits}. Update (10 December 2025). }
    \label{fig:axionworld}
\end{figure}

The main haloscopes are targeting exactly this band from the $\mu$eV scale upward \cite{ADMX2018,Millar2022,Braggio2025,SardoInfirri2025}, 
while helioscopes, e.g. CAST, IAXO, and optical/ALPS-type experiments,  analyze complementary portions of $(m_a,g_{a\gamma\gamma})$ space without necessarily assuming it to be dark matter \cite{CAST2017,IAXO2020,ALPSII2024,PVLAS2008,BMV2015}.
Several other experiments searches for the axions and other BSM particles are developed around the world and a current summary is shown in Fig.~\ref{fig:axionworld}.

\section{Axion-like particles}
\begin{figure}
    \centering
    \includegraphics[width=0.9\linewidth]{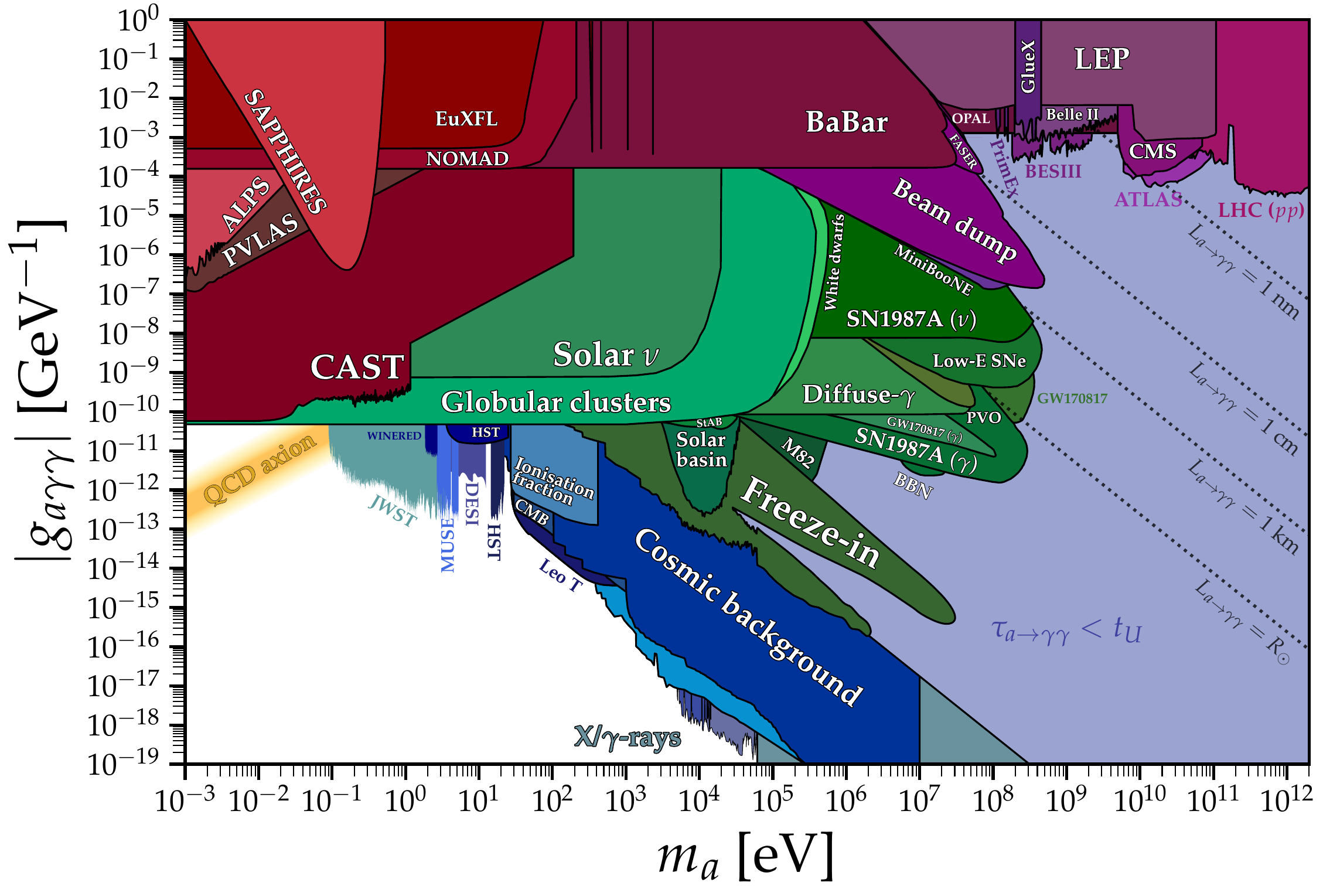}
    \caption{Plot of the parameter space $(m_a, g_{a \gamma \gamma})$, showing the region of the high-mass axions. Image from the Github Axion Limits by Ciaran O'Hare.\cite{axionLimits}.Update (10 December 2025). }
    \label{highmass}
\end{figure}

Experimental aspects and constraints are more involved for ALPs, since they do not follow the nice direct relation between the mass and the decay constant, which is held from the QCD axion, and they are shown briefly in Fig~\ref{highmass}. The main interesting constraints to us are the ones from freeze-in, BBN, and the decay time, which we discuss better in Section~\ref{axionfreezein}. 

\chapter{Axion Cosmology}\label{axioncosmo}

\section{Basic aspects on the cosmology of the axions}\label{axioncosmointro}
The object of this chapter is to discuss and review cosmological properties of the axions, useful for the thesis.
We introduce the Boltzmann kinetic equation, which we will adopt for the rest of the chapter and we will generalize it in Chapter~\ref{Schwinger} with an application in Section~\ref{axionfreezein}.

We discuss briefly the thermal QCD axions produced in the early universe, and will focus more on non-thermal production processes: the evolution of the 
average axion field between the Peccei-Quinn and QCD crossover, axion domain walls with the domain wall problem ,and its possible resolutions.

%, string decay and domain wall decay, and the population of cold axions produced by vacuum realignment, and finally, axion
%miniclusters and axion isocurvature perturbations.
The cosmological scenarios for the QCD axion depend deeply on when the PQ phase transition happens before or after the end of the inflation\footnote{Another interesting scenario involves considering the PQ phase transition happening during the inflation. Unfortunately, we will not discuss it here.} and lead to two main scenarios:
\begin{itemize}
\item Pre-Inflationary scenario: The PQ phase transition happens before the inflation. Cosmic strings are created, but they are homogenized from the inflation. The standard scenario for the QCD axion is then that the main non-thermal production mechanism is the misalignment mechanism, while the main thermal production come from the interactions with gluons and quarks. 
    \item Post-inflationary scenario: The PQ phase transition happens after the inflation and topological defects are produced, leading to inhomogeneitis. If the scenario is less problematic for cosmic strings, it can be problematic for axion domain walls with instanton potential for $N_{\rm DW}>1$ ,leading to the so-called domain wall problem.

\end{itemize}
The scenario with ALPs is still more variable than the QCD axion case, but it still deals with these two scenarios. We will only consider the photophilic model, for which analogous aspects are present. More details on uch scenarios are shown in Ref.~\cite{FavittaDWAnimation}.

\section{The Boltzmann kinetic equation}
The Boltzmann equation describes the collisional dynamics of a generic physical system. It plays a crucial role in calculating particle abundances in the Early Universe, since it was filled with a hot primordial plasma and the Hubble radius was smaller, favouring collisions between particles. 

We express all these aspects quantitatively by introducing the phase-space distribution function $f(x,p)$  which describes our system, where $x$ is 4-position and $p$ the 4-momentum,. We will be interested to its evolution in a curved Riemannian spacetime with metric $g_{\mu\nu}$.

We expect its dynamics to be influenced by the collisional forces, the local interactions among particles, which we characterize with the collisional operator $\mathcal{C}$ acting on the distribution function as the collisional term $\mathcal{C}[f]$. Furthermore, we need a free-streaming inertial term, which we take to be the Liouville operator $\mathcal{L}$ acting on $f$ as \begin{equation}
    \mathcal{L}=p^{\alpha} \frac{\partial}{\partial x^{\alpha}}- \Gamma^{\alpha \mu}_{\nu}\, p_{\mu}\, p_{\alpha}\, \frac{\partial}{\partial p_{\nu}}
\end{equation}
which is linear in the first derivatives $\frac{\partial}{\partial x^{\alpha}}$ and $\frac{\partial}{\partial p^{\alpha}}$.

The Liouville term can be derived from the total derivative of the phase-space distribution along the particle worldline, with $f(x(\tau), p(\tau))$, writing $\frac{d}{d \tau}f(x(\tau), p(\tau))$ and expressing it in terms of partial derivatives. Finally, one invokes the geodesic equation from General Relativity \cite{misner2017gravitation}.

The dynamics is then obtained equating the two terms
\begin{equation}\label{basicboltz}
    \mathcal{L}[f]=\mathcal{C}[f].
\end{equation}
Assuming spatial homogeneity and a FLRW background metric, along with $p_\mu=(E,-\vec p)$, we can write the LHS of Eq.~ (\ref{basicboltz}) as \footnote{To not be confused with the Lagrangian density $\mathcal{L}$.}
\begin{equation}
    \mathcal{L}=E \frac{\partial}{\partial t}-H(t) \,|\vec{p}|^2 \frac{\partial}{\partial E}.
\end{equation}
Integrating over momenta, it gives the following equations
	\begin{align}
		\dot n + 3Hn &= \int \frac{d^3p}{(2\pi)^3}\,\mathcal{C}[f],\\
		\dot \rho + 3H(\rho+p) &= \int \frac{d^3p}{(2\pi)^3}\,E\,\mathcal{C}[f],
	\end{align}
	with
	\begin{align}
		n &= \int \frac{d^3p}{(2\pi)^3}\,f(t,\vec p),\qquad
		\rho = \int \frac{d^3p}{(2\pi)^3}\,E \, f(t,\vec p),\qquad
		p = \int \frac{d^3p}{(2\pi)^3}\,\frac{|\vec p|^2}{3E}\,f(t,\vec p).
	\end{align}
where $n$ is the average number density, $\rho$ and $p$ are respectively the average energy density and pressure.

	For processes of the form $1+2\leftrightarrow 3+4$,
	\begin{equation}
		\begin{aligned}
			\mathcal{C}[f_1] = -\frac{1}{2E_1}\!\int d\Pi_2 d\Pi_3 d\Pi_4 (2\pi)^4 \delta^{(4)}(p_1{+}p_2{-}p_3{-}p_4)\, \Big[ &|\mathcal{M}_{12\to34}|^2 f_1 f_2(1\pm f_3)(1\pm f_4)\\
			&-|\mathcal{M}_{34\to12}|^2 f_3 f_4(1\pm f_1)(1\pm f_2)\Big],
		\end{aligned}
	\end{equation}
	with $d\Pi_i=d^3p_i/[(2\pi)^3 2E_i]$ is the Lorentz-invariant phase space element \cite{schwartz2014quantum}.
    
We observe that, if we assume 
 our particle species to follow a Maxwell-Boltzmann(MB) statistics at all times,  the integrated number-density equation becomes \cite{baumann2022cosmology}
	\begin{equation}
		\frac{dn_1}{dt}+3Hn_1 = -\langle\sigma v\rangle\, n_1^{(\text{eq})} n_2^{(\text{eq})} \!\left[\frac{n_1 n_2}{n_1^{(\text{eq})}n_2^{(\text{eq})}} - \frac{n_3 n_4}{n_3^{(\text{eq})}n_4^{(\text{eq})}}\right].
	\end{equation}
    where $\langle\sigma v\rangle\,$ is the thermally averaged cross section and $n_i^{(\text{eq})}$ are the equilibrium number density in the MB approximation.
	It is convenient to use $x\equiv m/T$.

\section{Thermal production of axions}\label{axionthermo}
\subsection{Massive axions case}
Axions can be created or annihilated during interactions among particles in the
primordial plasma and a part of them can also thermalize with the plasma.
This population is what we will call the “thermal axions”, to distinguish them from the population of
“non-thermal axions” which we shall discuss in the next section.
The scenario of thermal axions is very wide (theoretically and in the parameter space) and is not the main focus of our thesis. We will limit ourselves in discussing the thermal QCD axions and their interactions with gluons and pions, which are significant for the cosmological hot relics constraints we have mentioned in Chapter~\ref{experiments}.
\subsection{Thermal QCD axions}

The most significant process for such axions is the process $g\,g\rightarrow g\, a$ coming from the coupling with gluons, but other relevant processes are the scatterings $
q\,g \leftrightarrow q\,a$ and $
q\,\bar q \leftrightarrow g\,a$.

\begin{figure}[h]
\centering
\begin{tikzpicture}
    % Diagram (b): q g -> q a (thermal, Compton-like)
    \begin{feynman}
        % Define the vertices
        %\vertex (i1) at (0, 2) {$g$};  % Incoming scalar particle
       % \vertex (f1) at (0, 3) {$g$};  % Outgoing photon 1
       % \vertex (a) at (2, 0) {$\gamma$};  % Outgoing photon 2
         %\vertex (f2) at (2, -2) {$a$}; 

        % Define the interaction vertex
        \vertex (f2) at (2, -1) {$a$}; % Decay vertex

        % Draw the diagram
        %\diagram*{
           % (a1) -- [scalar] (c),       % Scalar particle decays
            %(c) -- [photon] (b1),       % First photon
            %(c) -- [photon] (b2),       % Second photon
        %};
   
    \diagram[small, horizontal=a to b] {
      i1 [particle=\(g\)] -- [gluon] a -- [gluon] f1 [particle=\(g\)],
      i2 [particle=\(g\)] -- [gluon] b,
      a -- [gluon] b,
      b -- [scalar] (f2) ,
    };
    %\node at (0,-1.9) { Feynman diagram of the production process \(g\,g \to g\,a\). };
  \end{feynman}
  
\end{tikzpicture}
\label{ggga}
\caption{Feynman diagram of the production process \(g\,g \to g\,a\).}
\end{figure}
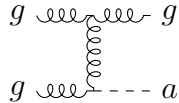

We can write the Boltzmann equation

\begin{equation}
    \frac{d Y_p}{dt}=-\angi{\sigma v} s(y_a y_g-y_a^{(\text{eq})} y_g^{(\text{eq})})
\end{equation}
which can be rewritten explicitly in terms of the plasma temperature $T$ as 

\begin{equation}
    \frac{dY_a}{dT}=\frac{\Gamma}{HT} (y_a-y_a^{(0)}),
\end{equation}
where roughly $\Gamma \sim g_s^3\,T^3/f_a^2
$.
In general, we can infer we have for all three processes a factor of the kind $\kappa{g_s}$, which is visibly of the order $g_s^3$ from Fig~\ref{ggga}, since we have one 3-gluon vertex and one 2-gluon-axion vertex. 
This leads to a rough estimation of the freeze-out temperature as 
\begin{equation}
    T_{\rm freeze-out}\sim \frac{g_s^3}{1.66\,\sqrt{g_*}}\,\frac{M_{\rm Pl}}{f_a^2}
\end{equation}
This is not far from a more precise calculation giving \cite{ohare2024cosmology} 
\begin{equation}
   T_{\rm freeze-out} \simeq C\,\frac{\alpha_s^3(T)}{\sqrt{g_*}}\,\frac{M_{\rm Pl}}{f_a^2}
\end{equation}
which involves in any case to introduce the numerical factor $C \sim 0.1-1$.

\begin{comment}
The number density nth
a (t) of thermal
axions solves the Boltzmann equation [1]
where

is the rate at which axions are created and annihilated. $H(t)$ is the Hubble
expansion rate and

is the number density of axions at thermal equilibrium, where $\zeta(3) = 1.202$
is the Riemann zeta function of argument 3. In (2.4), the sum is over processes
of the type $a + I \leftrightarrow  1 + 2$, where 1 and 2 are other particle species, ni is the
number density of particle species i, $\sigma_i$ is the corresponding cross section, and .... indicates averaging over the momentum distributions of the particles
involved.
Unless unusual events are taking place, $T \propto R^{-1}$ where $R(t)$ is the scale
factor, and (2.5) implies, therefore
%\end{equation}
\subsection{Ultralight axions}
%\section{Connections to the bounds to axions produced by supernovae}
\end{comment}
\section{Non-thermal production of axions}\label{axionothermo}
In this chapter, we will focus on the class of axion non-thermal production mechanisms, which are of primary interest to us. 

In particular, we will introduce the freeze-in production, the standard misalignment mechanisms
and topological defects.

\subsection{Freeze-in }\label{freezein}
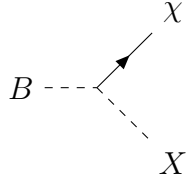
\begin{figure}[h]
\centering
\begin{tikzpicture}
  \begin{feynman}
    % Diagram (c): B -> chi + X  (freeze-in decay)
    \vertex (f1) at (2, +1){$\chi$};
     \vertex (f2) at (2, -1){$X$};
    \diagram[small] {
      i1 [particle=\(B\)] -- [scalar] a,
      a -- [fermion] (f1) [particle=\(\chi\)],
      a -- [scalar] (f2),
    };
   % \node at (0,-1.2) { \(B \to \chi + X\)};
  \end{feynman}
\end{tikzpicture}
 \label{BxX}
 \caption{Freeze-in production of dark matter particle $\chi'$through the channel \(B \to \chi' + X\)}
\end{figure}
In freeze-in, the dark sector $\chi'$ remains out of thermal equilibrium ($\Gamma\ll H$) while production accumulates slowly, until it "cools down" and behaves as a cold dark matter component. For example, if we consider a simple decay $B\to \chi'+X$, where $B$ is a primordial bath particle and $X$ is a SM particle, it leads to the Boltzmann equation for our $\chi $ particle
	\begin{equation}
		\frac{dY_\chi}{dx}\simeq \frac{n_B^{\rm eq}\,\Gamma_{B\to\chi}}{s\,H\,x},
		\qquad
		n_B^{\rm (eq)}=\frac{g_B m_B^2 T}{2\pi^2}K_2(m_B/T).
	\end{equation}
	The saturated yield is approximately
	\begin{equation}
		Y_{\chi'}(\infty)\simeq C_{\rm dec}\,
		\frac{g_B\,\mathrm{Br}(B\to\chi'+X)}{g_{*s}\sqrt{g_*}}\,
		\frac{M_{\rm Pl}\Gamma_B}{m_B^2},
	\end{equation}
with $C_{\rm dec}\approx 0.5$.
	The relic density is then $\Omega_\chi h^2 \simeq 2.75\times10^8\,\mathrm{GeV}^{-1}\,m_{\chi'}\,Y_{\chi'}(\infty)$.
    Further graphical details are shown in Ref.~\cite{FavittaDWAnimation}.

%In the case of a usual freeze-in production mechanism, we always work in the collisional regime, and then adopt Boltzmann equations, but the DM particles never reach the thermal equilibrium, so it is not a thermal process. 

\begin{comment}
Indeed, and this defines the freeze-in in contrast to freeze-out, $\Gamma \ll H$

Let us consider for simplicity a $\text{B} \rightarrow \text{SM}+\text{DM}$
The moment it becomes
\begin{equation}
    Y_{\chi} \sim \angi{\Gamma_B} t_H \sim \Gamma_B \,x_B \Bigg(\frac{M_{\text{PL}}} 
 {m_B^2} x^2  \Bigg)=\Gamma_B
\end{equation}

with \begin{equation}
    t_H=H^{-1} \sim \frac{M_{\text{PL}}{T^2}=\frac{M_{\text{PL}{m^2_B}x^2
\end{equation}
\end{comment}

\subsection{Standard Misalignment angle mechanism}\label{misa}
%\subsection{

The Standard misalignment mechanism is present quite generally for any particle in the early Universe with a significant non-relativistic population. It is also the simplest model to obtain a cold dark matter component, since it obtain straightforwardly oscillating field $\phi \sim \sin{(\omega t +\phi_0)}$ in the late-time limit (if we consider a scalar field), which clearly is a cold dark matter component.

Consider $a$ with action $\mathcal{S}=\int d^4x\sqrt{|g|}\,[\frac12(\partial a)^2 - V_{\rm eff}(a,T)]$. The homogeneous EoM is
	\begin{equation}
		\ddot a + 3H\dot a + V_{\rm eff}'(a,T)=0.
	\end{equation}

    We will consider the standard scenario where we have the initial misaligned-angle is $\theta_0=\frac{a(t_0)}{f_a}$ and $\dot{\theta}=0$ at $t=t_0$\footnote{The scenario with a "non-zero velocity" is currently treated in the literature and leads to interesting scenarios such as the kinetic misalignment.We will not treat it in this thesis\cite{PhysRevLett.124.251802,PhysRevD.102.015003,Eroncel2024rpe,Eroncel2025bcb,Eroncel2025qlk} }.

    For a usual axion-like particle with constant mass $m_a$, the natural comparison of the relevant quantities for the dynamics comes between $m_a$ and $H$, while for an instanton potential is between $H$ and $m_a(T)$.
	For $H\gg m_a(T)$ Hubble friction dominates and the field is frozen at the initial value $a_0=f_a\theta_0$. When $3H(T_{\rm osc})\simeq m_a(T_{\rm osc})$ oscillations start to dominate, yielding to the cold dark matter behavior.

\begin{figure}
    \centering
    \includegraphics[scale=0.7]{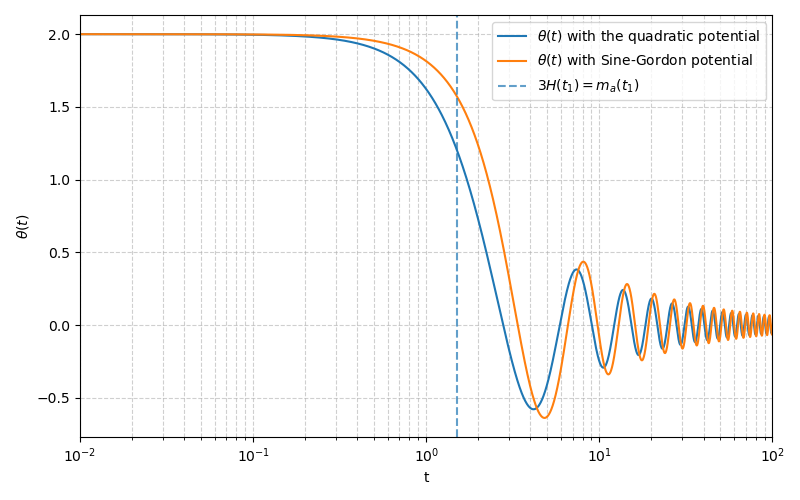}
    \includegraphics[scale=0.8]{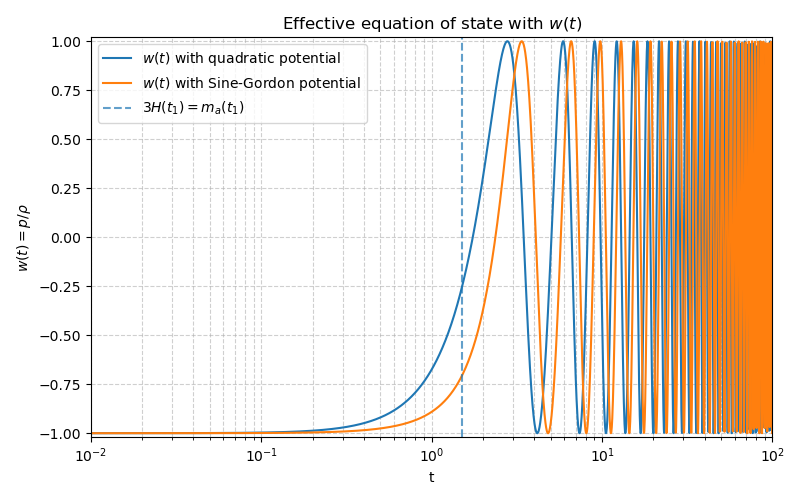}
    \caption{Plots showing the evolution of misalignment angle (above) and the effective equation of state parameter $w(t)=\frac{p}{\rho}$ (below) with the cosmological time in units of $m_a^{-1}$. The misalignment angle shows the expected behaviour and $w(t)$ grows from $-1$ to a fast oscillatory regime where it averages to zero. }
    \label{thetamisa}
\end{figure}

    We show in the plots~(\ref{thetamisa}) a visual representation of this.
    
    For a constant $m_a$ in Radiation-Dominated era,
	\begin{equation}
		T_{\rm osc}\simeq \left(\frac{m_a M_{\rm Pl}}{3\times1.66\sqrt{g_*}}\right)^{1/2}.
	\end{equation}
	For the QCD axion with $V(a)=\chi(T)[1-\cos(a/f_a)]$ and $m_a^2(T)=\chi(T)/f_a^2$,
	\begin{equation}
		\Omega_a h^2 \simeq 0.12\,\theta_i^2\,\mathcal{F}(\theta_i)\,
		\left(\frac{f_a}{5\times 10^{11}\,\mathrm{GeV}}\right)^{\alpha},\end{equation}
		with $\alpha\simeq 1.16$\cite{Sikivie2008,ohare2024cosmology}. 

        Having this result, it is then straightforward to obtain the bounds on QCD axions mentioned in Chapter~\ref{experiments}.

The physical reason connected with the bounds themselves and the difference between the bound with constant mass and variable mass can be understood as follows. When the Universe goes through the PQ phase transition at $T \sim \eta \gg \Lambda_{\mathrm{QCD}}$, the QCD anomaly is ineffective, then $\left\langle a\right\rangle$ is arbitrary. Eventually, when the Universe cools down to temperatures $T \sim \Lambda_{\mathrm{QCD}}$, the axion acquires a mass and $\left\langle a\right\rangle \rightarrow 0$. This is not an instantaneous process, and $\left\langle a\right\rangle$ oscillates to its final value. These coherent oscillations contribute to the Universe's energy density as a cold dark matter component. The energy density is directly proportional to $f_{a}$ and thus bounds on the energy density of cold dark matter in the Universe provide an upper bound on $f_{a}$ for any axion and a lower bound on $m_{a}$ for QCD axions.

\subsection{Topological defects}\label{topo}

Topological defects are irregularities in the ordering of a physical system, present in various theoretical models ranging from Cosmology to condensed matter physics, and they are created due to the Symmetry Spontaneous Breaking.

\begin{figure}
    \centering
    \includegraphics[width=0.39\linewidth]{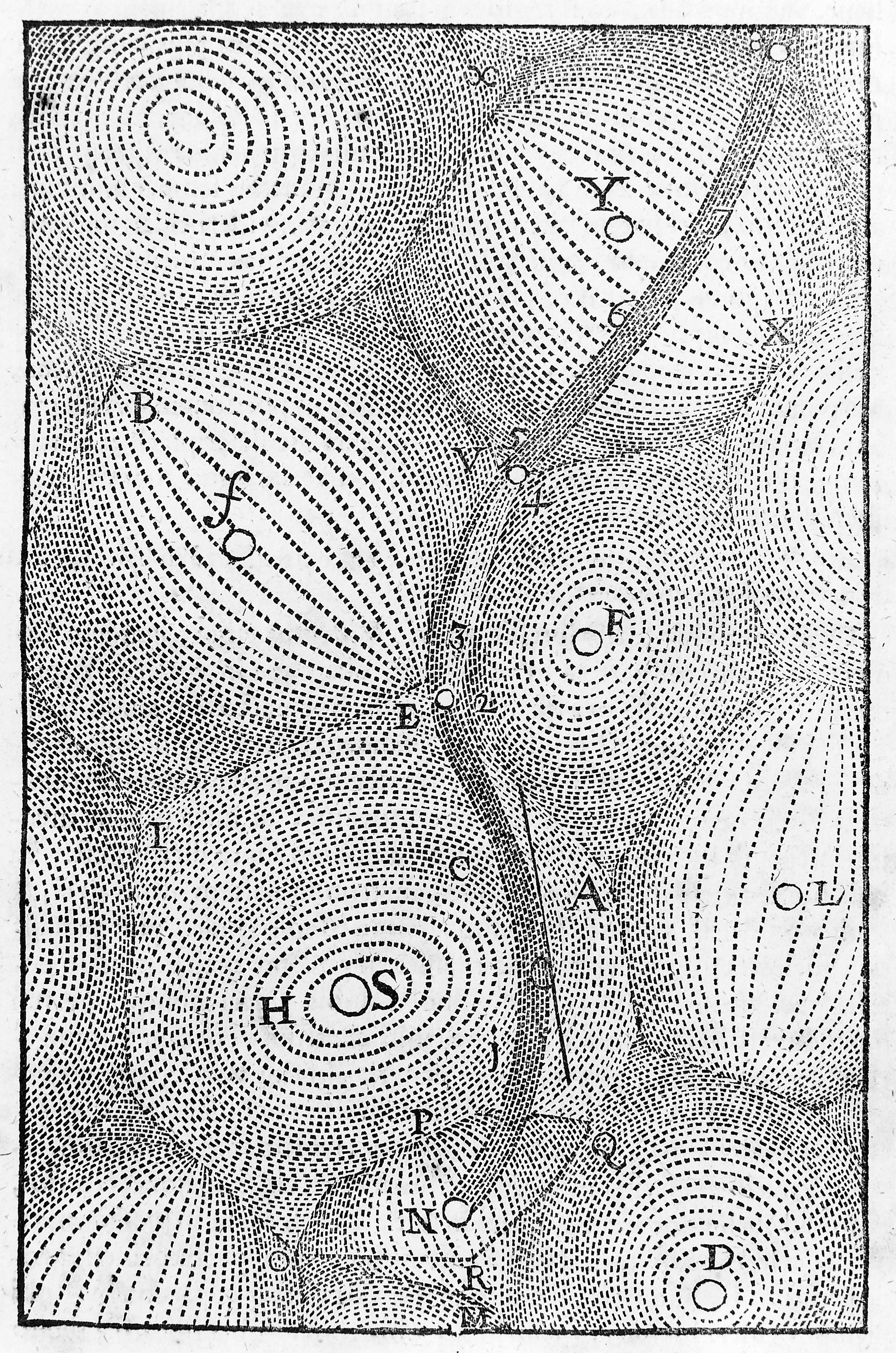}
    \caption{Example of a vortex from "Principia philosophiae" by Reneé Descartes, a first example of a "cosmic string" \cite{Descartes1644}.}
    \label{cartesio}
\end{figure}

They can be more rigorously described as stable and localised defects of an order parameter in the system, arising from the topology of the vacuum manifold of the theory \cite{vilenkin1994cosmic,chaichian2012introduction}.

The classical examples in $1{+}1$ dimensions are the $\phi^4$ kink and sine–Gordon soliton, which come from scalar field theories with topological charge $N=\phi(+\infty)-\phi(-\infty)$, to which is connected their classical stability \cite{vilenkin1994cosmic,chaichian2012introduction}.

Another example is the cosmic strings, which we already introduced in Sections~\ref{ftUV} and~\ref{experiments}.
They are, in general, associated with the spontaneous breaking of an axial symmetry, while domain walls are associated with a discrete symmetry.
We will concentrate on the axion domain walls, since they are our main focus of the thesis and since we already introduced cosmic strings.

\subsubsection{Kinks and axion domain walls}\label{axioncosmodw}
	A $\mathbb{Z}_2$ Goldstone model with $V=\frac{\lambda}{4}(\phi^2-\eta^2)^2$ admits kink stable solutions with
	\begin{equation}
		\phi_{\rm kink}(z)=\eta\,\tanh\!\Big(\frac{z}{\delta_{\rm kink}}\Big),
        \end{equation}
        where
        \begin{equation}
		\delta_{\rm kink}=\sqrt{\frac{2}{ \lambda \eta^2}},
	\end{equation}
    which is localized and centered at $x=0$, taking values $\phi=-\eta$ for $x \rightarrow -\infty$ and  $\phi=+\eta$ for $x \rightarrow +\infty$.

\begin{figure}
    \centering
    \includegraphics[scale=0.35]{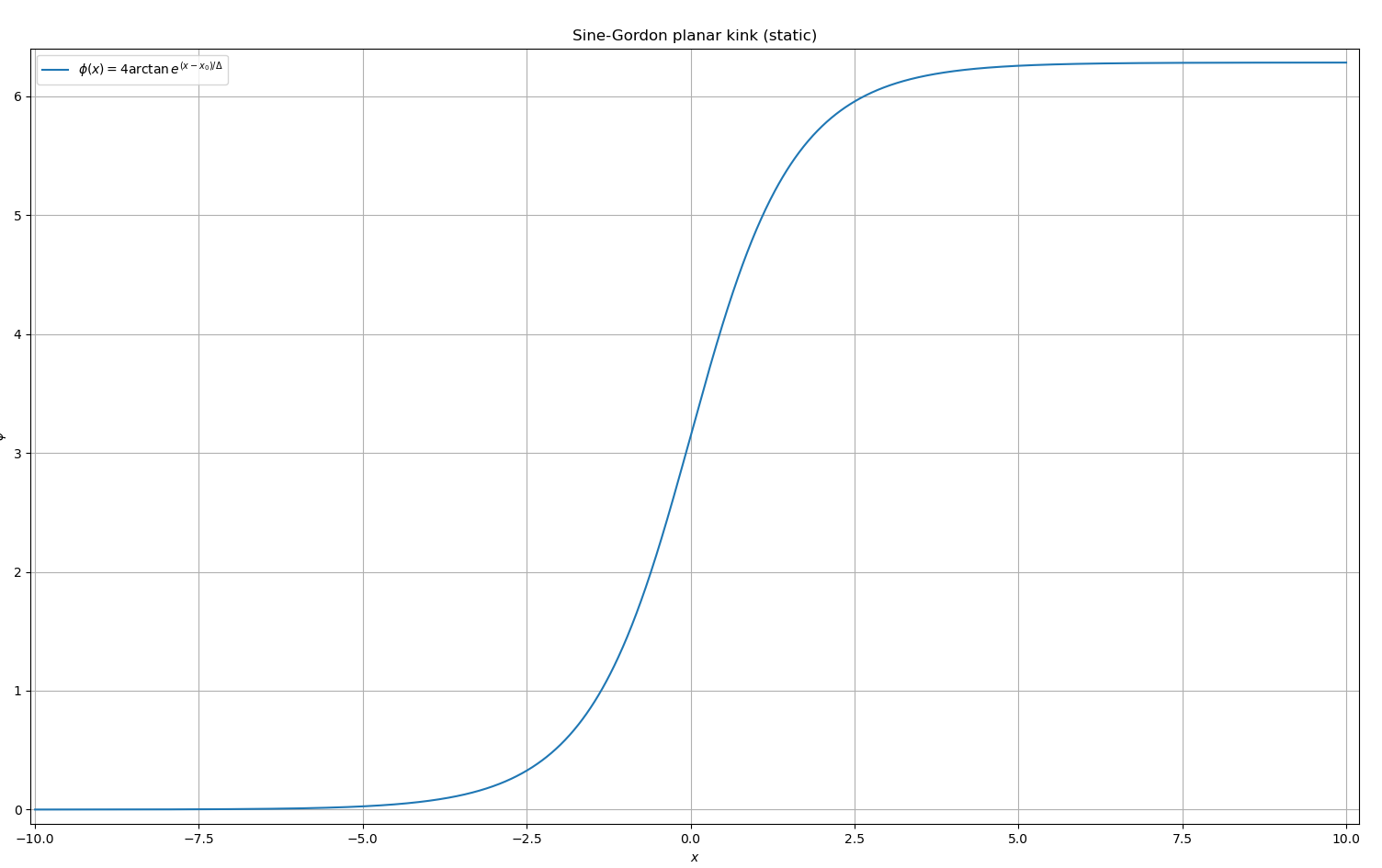}
    \caption{Plot of the configuration of equation~(\ref{kink}) in normalized units, interpolating between the two domains with $\phi=0$ and $\phi=2 \pi$. }
    \label{fig:kink}
\end{figure}

Similarly, a sine-Gordon soliton comes from an analogous theory with a periodic potential of the form
\begin{equation}
     V(\phi)=2\lambda \eta^4 \Big[1- \cos{\Big(\phi/\eta\Big)}   \Big],
\end{equation}
which is the same form of istanton potential.
The analytic soliton solution is
\begin{equation}\label{kink}
    \phi(x)=4 \eta \arctan{\exp{\Big( \sqrt{\lambda} \eta x \Big)}},
\end{equation}
for which we display a plot in Fig.~\ref{fig:kink}.
All these solutions are remarkably non-dissipative, time-independent, localized and have finite energy. 
Their stability is a consequence of a topological conservation law, with each defect state having a conserved quantum number.
For the one-dimensional case, the topological current is 
\begin{equation}
    j^{\mu}=\epsilon^{\mu \nu} \partial_{\nu} \phi.
\end{equation}
Indeed, it is trivially conserved, since it is the divergence of an antisymmetric tensor. However, it does not arise from a continuous symmetry, then it is not conserved by Noether's theorem.
It has an associated conserved charge $N$
\begin{equation}
    N=\int dx\, j^0=\phi(x)|_{x\rightarrow +\infty}-\phi(x)|_{x\rightarrow -\infty}.
\end{equation}
showing that it is the  presence of a soliton with $\phi$ in different vacua at $x\rightarrow \pm \infty $ to give rise to a non-zero charge $N$ which gives it the classical stability.

Furthermore, they can be boosted up to arbitrary velocities thanks to the Lorentz invariance of the theory.

For example, in the instanton case of our interest, we have a boosted profile along the wall normal
	\begin{equation}\label{kink1}
		a(\vec x,t)=f_a\Big[2\pi k + 4\arctan e^{\gamma m_a \hat n\cdot(\vec x - \vec v t)}\Big],
	\end{equation}
    with a thickness of the domain wall
    \begin{equation}
     \delta_{\rm DW} \sim (\gamma m_a)^{-1}.
     \end{equation}
	The rest wall tension is
	\begin{equation}
		\sigma_{\rm DW}=8\,m_a f_a^2.
	\end{equation}
We show more on this configuration in Ref.~\cite{FavittaDWAnimation}.
    
The stress–energy for a static planar wall is similarly to the case treated by Ref.~\cite{kolb1991early}
	\begin{equation}
		T^\mu{}_\nu(z)=\rho(z)\,\mathrm{diag}(1,-1,-1,0),\qquad
		\rho_{\rm kink}(z)=4 m_a^2 f_a^2\,\sech^2\!\Bigg(\frac{z}{\delta_{\rm DW}}\Bigg),
	\end{equation}
	
    The Newtonian limit of relativistic Poisson's equation \cite{kolb1991early,misner2017gravitation}, coming from a stress-energy which is of the form of a perfect fluid, apart from a factor $\sech^2\!\Bigg(\frac{z}{\delta_{\rm kink}}\Bigg) $ which localizes the wall distribution, is of the form 
\begin{equation}
    \nabla^2 \Phi_{\text{grav}}=4 \pi G (\rho+p_1+p_2+p_3)=-4 \pi G \rho .
\end{equation}
This means that we have a counterintuitive result, where an infinite flat domain wall repels a gravitational test particle rather than attracting it, and two infinite flat domain walls repel each other gravitationally.
This strange gravitational behaviour is only related to assuming an infinite planar wall, and the spatial dimensions are limited by the Hubble radius in Cosmology.
%, and at large distances both a planar wall and a spherical one behave gravitationally as a particle with mass $m_{DW} \sim \sigma L^2 $.
	
	If we work again in the wall rest frame,
	\begin{equation}
		T^\mu{}_\nu(z')=\rho(z')\,\mathrm{diag}(1,-1,-1,0),\qquad z'=\gamma(z-vt),\ \gamma=(1-v^2)^{-1/2}.
	\end{equation}
	Boosting along $z$, the direction normal to the wall, gives
	\begin{equation}
		T'^{00}=\gamma^2\rho(z'),\quad T'^{0z}=\gamma^2 v\rho(z'),\quad
		T'^{zz}=\gamma^2 v^2\rho(z'),\quad T'^{xx}=T'^{yy}=-\rho(z').
	\end{equation}
	The wall is Lorentz contracted and $\sigma'=\gamma\sigma$.
These last points will be relevant when we discuss the thermal friction in Section~\ref{dio}.

%The difficulty with scalar models has been faced by several means in the literature, from extending the models, as in the Skyrme model where higher derivatives are added, to allowing time-dependence, as it happens for spheraleons, or relaxing the localization assumption \cite{vilenkin1994cosmic}.
\begin{comment}
\begin{itemize}
    \item Extending with more complicated models: this is the approach, for example, of the Skyrme model, where higher derivative terms are added, or abelian-Higgs and Yang-Mills-Higgs models, where gauge fields are added.
    \item Allowing time-dependence: 
    \item Relaxing the localization assumption: this allows for several admissible defects from scalar theories. For example, a $U(1)$ Goldstone model with a complex scalar field has then physical vortex solutions in two dimensions, and isolated global strings are  %adjust here
\end{itemize}
\end{comment}

%\subsubsection{Axion domain walls}\label{axioncosmodw}

\subsubsection{Further aspects on axion domain walls}

In this subsection, we treat in more detail how we precisely connect the theory about domain walls with the cosmological axion domain walls.

We begin by considering a non-Abelian $SU(3)$ group, which can be related to the SM gluon field for the QCD axion or a dark gluon sector for ALPs. %adjust for didactic and add references

We have a $U_{\text{PQ}}(1)$ symmetry anomalous under $SU(3)$, whose pseudo-Nambu-Goldstone boson is the axion of our interest.
Then, the Lagrangian of the axion contains the following interaction term
\begin{equation}
    \mathcal{L}_a= \frac{g_s}{8 \pi} \frac{N_{\text{DW}}}{\eta} a\, G_{\mu \nu}^b \,\tilde{G}^{\mu \nu}_b
\end{equation}
$g_s$ is, as before, the gauge coupling constant, $G_{\mu \nu}^b$ the gauge boson field strength, and $\eta$ is the $U_{\text{PQ}}(1)$ symmetry-breaking VEV, related to the axion decay constant by 
\begin{equation}
    f_a=\frac{\eta}{N_{\text{DW}}},
\end{equation}

\begin{figure}[h]
    \centering
    \includegraphics[width=0.99\linewidth]{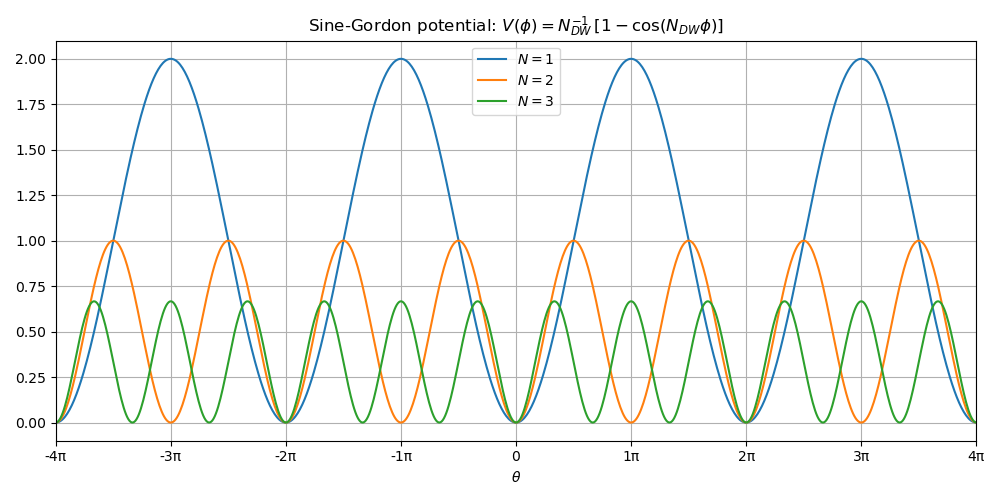}
    \caption{Plot of the instanton potential $V(a)=\frac{\eta^2 m_a^2}{N_{\rm DW}^2} \Big[   1-\cos{\Big(N_{\rm DW}\frac{a}{\eta}\Big)}\Big]$ in normalized units and with various values of the domain wall number $N_{\rm DW}$.}
    \label{fig:dwnumber}
\end{figure}

For $N_{DW}=1$, the vacuum manifold for the ALP effective potential induced by $SU(3)$ gauge interaction is trivial since it contains only identified minima as $a=0$ and $a= 2\pi f_a= 2 \pi \eta$, and then it is not topologically stable for Derrick's theorem \cite{10.1063/1.1704233}. 
However, for $N_{DW}>1$, the vacuum is composed of disconnected points corresponding to the discrete $Z_{N_{DW}}$ symmetry, and the theory admits domain wall solutions that interpolate between neighbouring minima, as we show with an intuitive picture in Fig.~\ref{fig:dwnumber}.

As done in the former literature \cite{PhysRevLett.48.1156,PhysRevD.30.712,Hindmarsh:1996xv,Garagounis:2002kt,TakashiHiramatsu2013,Vaquero2019,Gorghetto2021,Pierobon2023ozb}, and since we are interested in the dynamics after the creation of domain walls, at temperatures below the QCD phase transition $T_{QCD}$, we first capture the fundamental features of the axion potential by adopting the instanton potential, as before.

The simple instanton form of the potential is helpful since it allows us to obtain simple analytical solutions of domain walls with uniform velocity $\vec{v}$
\begin{equation}
    a(z)= f_a \Big[2 \pi k + 4 \arctan{\Big[e^{\gamma m_a \hat{n} \cdot (\vec{x}+ \vec{v} t)}\Big]\Big\}}         ,  \qquad k=0,...,N_{\text{DW}}-1
\end{equation}
where $\gamma=\frac{1}{\sqrt{1-v^2}}$ is the corresponding Lorentz factor of the velocity $v$ and $\hat{n}$ is the direction of motion normal to the wall worldsheet.
The typical width of a DW moving with velocity $\vec{v}$ is then $\delta_{\rm DW} \sim (\gamma m_a)^{-1}$
with tension at rest
\begin{equation}
    \sigma_{\text{DW}}=8 m^2_a f_a.
\end{equation}
\begin{figure}
   \centering
\includegraphics[width=0.3\textwidth]{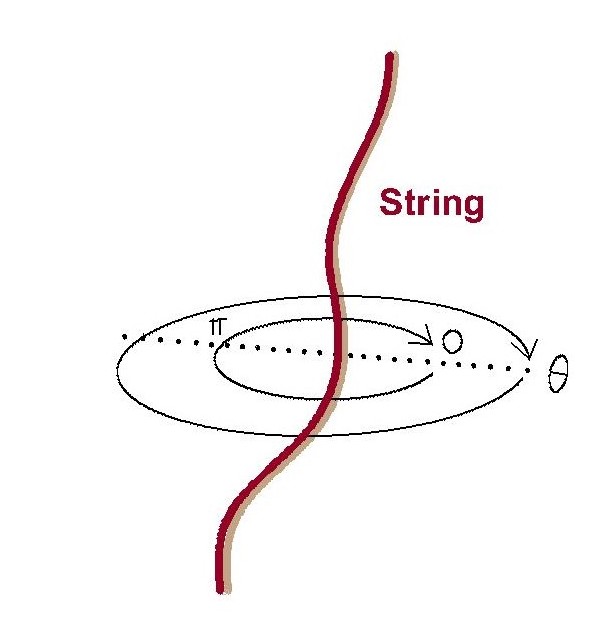}
\includegraphics[width=0.3\textwidth]{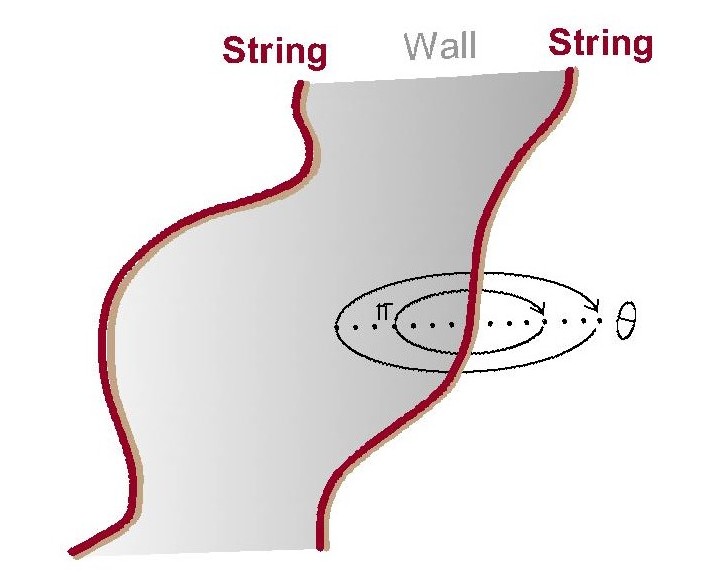}
\caption{
Schematic illustration of the fundamental features of axion cosmic strings and domain walls.
The cosmic string (left) is characterized by the axion field $\theta = a / \eta$, which undergoes an excursion of $2\pi$ when rotating around the string.
The domain wall (right), attached to two strings with $N_{\rm DW}=2$, is associated with the axion field $\theta = a / f_a$, which also varies by $2\pi$ across the wall.
Figure adapted from the analogous illustration in Ref.~\cite{ohare2024cosmology}.
}
\end{figure}
After the formation of domain walls and according to numerical simulations, the energy density of the resulting string-wall network is soon dominated by the walls and, if $N_{\text{DW}}>1$, the DW network is stable and reach the scaling regime where the energy density of the DW network redshifts as $\rho_{DW} \sim \sigma_{DW} H$. This regime corresponds to having an order of one domain wall per Hubble sphere and a mildly relativistic\footnote{Tipical values of rms velocities $v$ can be 0.5 or higher\cite{PhysRevD.93.043534}. } regime for DWs \cite{Hindmarsh:1996xv,Garagounis:2002kt,Oliveira:2004he,Avelino:2005pe,PhysRevD.84.103523}. Differently, for $N_{\text{DW}}=1$ the DW network is unstable and decays soon \cite{Sikivie2008}.

A DW network in a scaling regime would eventually dominate the energy density of the Universe, in contrast to cosmological observations.
Since the Universe is radiation-dominated in the epoch of interest, the temperature of radiation-DW equality can be roughly obtained by equating $\rho_{DW}$ with $\rho_{\rm rad} \sim 3H^2 M^2_{\text{Pl}}$. 

Adopting the Hubble parameter $H$ expression in terms of plasma temperature $T$, the following dominance temperature $T_{\text{dom}} $ is obtained \cite{Blasi:2023sej}
\begin{equation}
    T_{\text{dom}} \simeq 14 \, \mathrm{MeV} \Bigg(\frac{\sigma_{\text{DW}}}{100 \, \mathrm{TeV}}        \Bigg)^{3/2}\,\Big(\frac{g_{*}}{10}   \Big)^{-1/4}, 
\end{equation}
where it is used $\rho_{DW}= 2 \sigma_{DW} \mathcal{A}\, H $ with $\mathcal{A}=0.8$ from numerical DWs simulations   .
To collapse the DW network before dominating, we need a mechanism that annihilates the network at a temperature $T_{*}>T_{\text{dom}}$. 

This is a requirement to determine whether the thermal friction, or a bias potential, is sufficient to solve the DW problem for a specific point in the axion parameter space.
Furthermore, a convenient  quantity for the observations is $\alpha_{*}$, the DW energy normalised to the total energy 
\begin{equation}
    \alpha_*=\frac{\rho_{DW}}{\rho_{tot}} \simeq \frac{\rho_{DW}}{\rho_{r}} \simeq 0.02\,  \Big(\frac{\sigma_{\text{DW}}}{100  \mathrm{TeV}}        \Big)^{3}  \Big(\frac{T_*}{100  \mathrm{MeV}}        \Big)^{-2}   \Big(\frac{g_{*}}{10}   \Big)^{-1/2}
\end{equation}

 Another aspect for which the annihilation temperature $T_*$ can also be relevant for cosmological observation is that, as the numerical simulations have proven \cite{Gorghetto2021,Benabou:2024msj}, the DW network in the scaling regime generates a large Stochastic Gravitational Wave Background (SGWB) \cite{husa2009michele}.
 
 The expected density parameter $\Omega_{\text{gw}}(f)$ is with a broken power law in frequency $f$ and the signal is dominated by the last moment of emission, so it depends explicitly on $T_{*}$.
The density parameter redshifted to today is \cite{Benabou2023npn}
\begin{equation}
    \Omega_{\text{gw}}(f, T_*) =\Omega_{\text{peak}} \times \begin{cases}
        \Big(\frac{f}{f_{\text{peak}}} \Big)^3 \qquad &\text{if}\quad f \leq f_{\text{peak}}\\  \Big(\frac{f}{f_{\text{peak}}} \Big)^{-1} \qquad &\text{if}\quad f > f_{\text{peak}}
    \end{cases}
\end{equation}
with 
\begin{equation}
\begin{aligned}
 &\Omega_{\text{peak}} \simeq 1.64 \times 10^{-6} \times\Bigg(\frac{\tilde{\epsilon}}{0.7}   \Bigg) \Bigg(\frac{\mathcal{A}}{0.8}   \Bigg)^2 \Bigg(\frac{g_{*}}{10}   \Bigg)    \Bigg(\frac{g_{s}}{10}   \Bigg)^{-4/3} \times \Bigg(\frac{T_{\text{dom}}}{T_*}   \Bigg)^4    
 \end{aligned}
\end{equation}
and \begin{equation}
     f_{\text{peak}} \simeq 1.15 \times 10^{-9}\, \mathrm{Hz} \times \Bigg(\frac{g_{*}}{10}   \Bigg)^{1/2}    \Bigg(\frac{g_{s}}{10}   \Bigg)^{-1/3}\Bigg(\frac{T_*}{10 \,\mathrm{MeV}}   \Bigg)  
\end{equation}
Another interesting example of SGWB from axions comes from the Axion U(1) inflation \cite{Cook_2012},which can lead to an amplification of gravitational waves
that could be detectable by Advanced LIGO or Advanced Virgo in the next few years.
\subsubsection{Velocity-One scale model for domain walls}\label{firstVOS}

The idea of analytical models for topological defects networks stands on the idea of avoiding the difficulties coming from numerical simulations by evaluating the evolution dynamics of average properties of the network. In particular, in Cosmology one can be interested in
the mean energy density $\rho$ and the root-mean-squared (rms) velocity $v$.

We outline in this subsection the approach by Ref.~\cite{PhysRevD.93.043534} for the Velocity-One Scale (VOS) model, since it will be helpful in order to clarify the approach we will adopt in Section \ref{Schwinger}.
\begin{figure}
    \centering
    \includegraphics[width=0.5\linewidth]{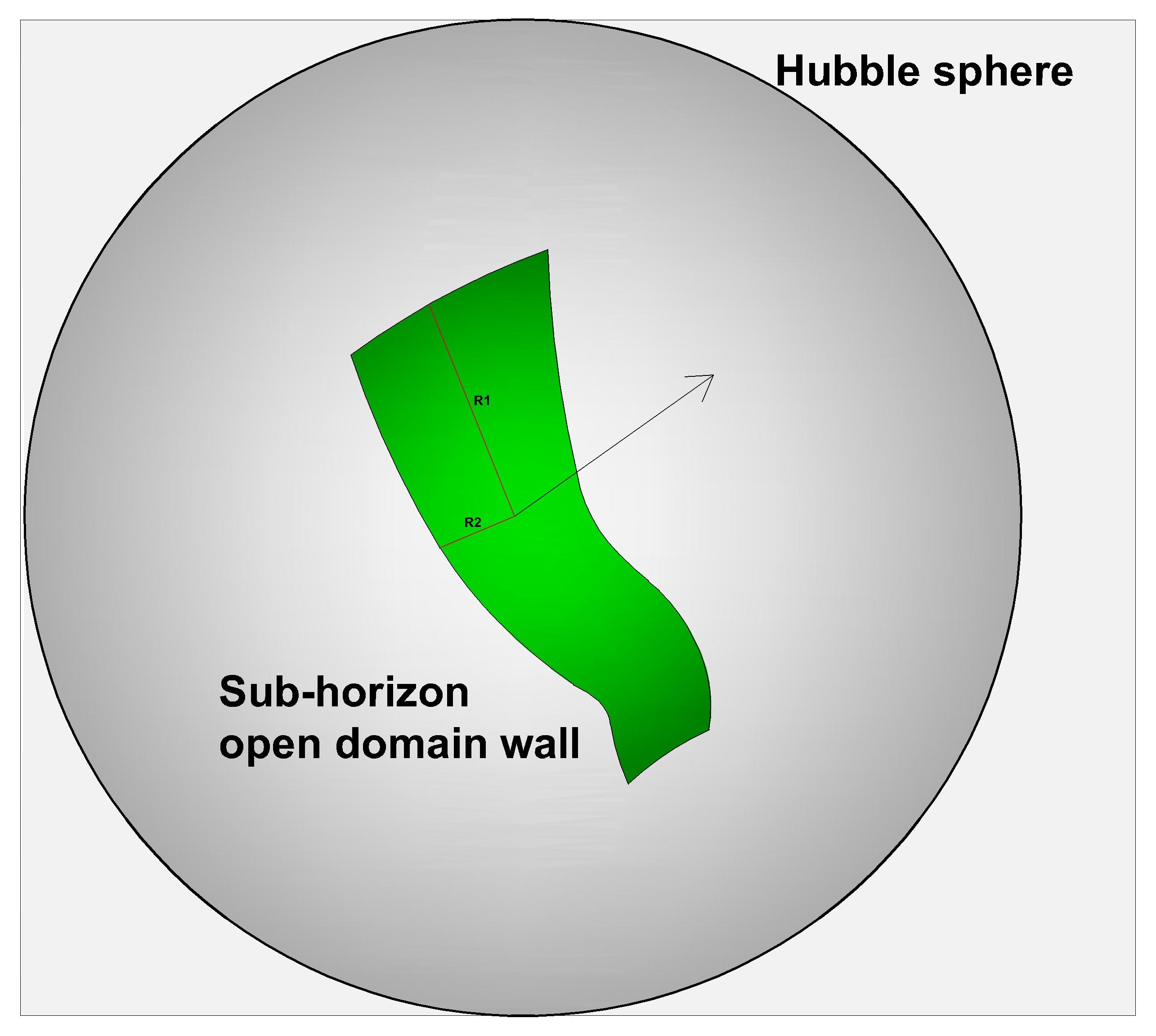}
    \caption{Pic of a subhorizon domain wall inside the Hubble sphere centered in its center of mass.}
    \label{subhorizondw}
\end{figure}
Let us consider a subhorizon open domain wall as in Fig~\ref{subhorizondw}, inside the Hubble horizon.

If we parametrise the wall surface $\mathcal{M}$ by $x^\mu(\sigma_1,\sigma_2)$, inside a classical 4-dimensional spacetime with a metric $g_{\mu \nu}$ and fix it to be the FLRW metric, the wall evolution is described by the 4-vector $x^{\mu}(\sigma_0,\sigma_1,\sigma_2)=x^{\mu}(\sigma_1,\sigma_2,\tau)$ where we have identified $\sigma_0=\tau$, where $\tau$ is a real number parameter.

If the function $x^{\mu}$ is smooth, we can parameterise the wall surface such that two tangential vectors will be orthogonal 
\begin{equation}
    \partial_{\sigma_1}x^{\mu} \, \partial_{\sigma_2} x_{\mu}=0
\end{equation}
and the velocity of the wall $\dot{x}^{\mu}$ to be only normal to the tangent surface $\mathcal{T}_{\mathcal{M}}$.
We use the minimal worldvolume Dirac action for the domain wall
\begin{equation}
    \mathcal{S}=-\sigma_{DW} \int \sqrt{\gamma} \,\,d^3\sigma,
\end{equation}
where $\sigma_{DW}$ is a constant surface energy density, the induced metric is $\gamma_{ab}=g_{\mu \nu}\, x^{\mu}_{,a}\, x^{\nu}_{b}$ with the determinant $\gamma=\frac{1}{6} \epsilon^{ab} \epsilon^{cd} \gamma_{ac}\gamma_{bd}$ and where $\epsilon^{ab}$ is the 2D Levi-Civita symbol.

We can easily obtain the EoMs for a domain wall by differentiating the Lagrangian density $\mathcal{L}=\sigma_{DW}  \sqrt{\gamma}$
\begin{equation}
    d\mathcal{L}=\frac{1}{2} \sqrt{\gamma} \gamma^{ab} d\gamma_{ab},
\end{equation}
and then obtain the Euler-Lagrange equations
\begin{equation}\label{singleDWFLRW}
    \frac{\dot{a}}{a} \delta_{0 \lambda} \sqrt{\gamma} \gamma^{ab}  \gamma_{ab} -\partial_c \left(\sqrt{\gamma} \gamma^{ab} g_{\mu \lambda}\, x^{\mu}_{,a}\, \delta^{c}_{b}     \right)=0.
\end{equation}

Redefining the coordinates $(\sigma_1,\sigma_2) \rightarrow (s_1,s_2)$ such that       $ |\frac{\partial x^i}{\partial s_{\alpha}}|^2=1$, with $\alpha=1,2$, we can introduce an orthonormal basis with $\zeta^i_{\alpha}=\frac{\partial x^i}{\partial s_{\alpha}}$ and $n^i=\frac{\dot{x}^i}{|\dot{x}^i|}$.

This allows us to simplify Eq.~(\ref{singleDWFLRW}) and write its zeroth component $(\lambda=0)$ as

\begin{equation}\label{singleDWFLRW0comp}
    \dot{\varepsilon}+3 \frac{\dot{a}}{a} \varepsilon \dot{x}^i \dot{x}_i=0,
\end{equation}
 and the spatial part $(\lambda=i)$ contracted with the normal vector $n_i$
 \begin{equation}\label{singleDWFLRWspatcomp}
     \ddot{x}^i n_i+3 \frac{\dot{a}}{a} \dot{x}^i n_i (1-\dot{x}^i \dot{x}_i)=(1-\dot{x}^i \dot{x}_i) k^i_1 n_i+(1-\dot{x}^i \dot{x}_i)k^i_2 n_i,
 \end{equation}
 with $k_{\alpha}^i=\frac{\partial \zeta_{\alpha}^i}{\partial s_{\alpha}}$.

We note that the operations of the scalar products $k_{\alpha}^i n_i$ are projections of the curvatures corresponding to $\sigma_1$ and $\sigma_2$ along the normal vector $n^i$. Consequently, we can write $k_{\alpha}^i=\frac{a}{R_{\alpha}} u_{\alpha}^i$ where $u_{\alpha}^i$ are unit vectors and $R_{\alpha}$ are the relative curvature radii.

Averaging equations~\pref{singleDWFLRW0comp} and \pref{singleDWFLRWspatcomp} over a network of subhorizon walls yields the velocity-one-scale equations for the mean energy density $\rho$ and the rms velocity $v$:
	\begin{align}
		\dot\rho &= -H\rho(1+3v^2) - \frac{c_w\rho}{L}\,v ,\\
		\dot v &= (1-v^2)\left[\frac{K}{L}-3Hv\right],
	\end{align}
where we have introduced three macroscopic quantities: the averaged energy density $\rho$
\begin{equation}
    \rho=\frac{E}{V}=\frac{\sigma_{DW} a^2}{V} \int \varepsilon \,\,d^2 \sigma,
\end{equation}
the root-mean-squared (rms) velocity $v$
\begin{equation}
    v^2=\frac{\dot{x}^2 \int \varepsilon \,\,d^2 \sigma }{\int \varepsilon \,\,d^2 \sigma},
\end{equation} 
and the correlation length $L=\sigma_{\rm DW}/\rho$, which gives a characteristic length scale of the network.

The equations can be rewritten in simpler terms for the correlation length and $v$ as
\begin{equation}
    \begin{aligned}
    \dot{L}&=H \rho\, (1+3v^2)-c_w  v,\\
    \dot{v}&=(1-v^2) \Bigg[\frac{K}{L}-3Hv    \Bigg],
    \end{aligned}
\end{equation}
displaying the scaling fixed point with $L\propto t$ and constant $v$, both in a radiation-dominated and matter-dominated universe.

We finally observe that the same Ref.~\cite{PhysRevD.93.043534} observes that such equations cannot fit properly the numerical simulations and we need to consider two energy-loss mechanisms which we insert phenomenologically and are inspired from analogous terms from string networks %cosmic string dyn with friction 
\cite{PhysRevD.43.1060, PhysRevD.48.2502}:
%adjust checking the terms
\begin{itemize}
    \item Chopping mechanism with a term $\frac{c_w\rho}{L}\,v $, it takes care of the energy loss due to possible intersections of domain walls and creation of sphere-like objects that collapse
    \item Scalar radiation term $d\,[k_0-k(v)]^r$ produced by zero-mode perturbations on the wall surface. %adjust check the terms
\end{itemize}

Then it can be obtained from these additional energy-loss mechanisms, although not directly justified from a microscopic theory, the following equations 
\begin{align}
		\dot L &= H L(1+3v^2) + c_w v + d\,[k_0-k(v)]^r,\\
		\dot v &= (1-v^2)\left[\frac{k(v)}{L}-3Hv\right].
	\end{align}

The dimensionless momentum parameter $k(v)$ was obtained to be
\begin{equation}
    k(v)=k_0 \, \frac{1-(qv^2)^{\beta}}{1+(qv^2)^{\beta}} \, ,
\end{equation}
where $\beta$ and $q$ are other unknown parameters obtained by fitting the VOS model with the numerical simulations and its analytical form is suggested from the expected behaviour in the non-relativistic and ultrarelativistic limit.
The fitted values from numerical simulations for the DW network without friction effects are, in the treatment of Ref.~\cite{PhysRevD.93.043534}:

\begin{subequations}
		\begin{align}
			d&=0.28 \pm 0.01,\qquad r=1.30\pm 0.02,\\
			\beta&=1.69\pm 0.08,\qquad k_0=1.73\pm 0.01,\\
			q&=4.27\pm 0.10,\qquad c_w=0.00\pm 0.01.
		\end{align}
	\end{subequations}
Ref.~\cite{Blasi2023} added another possible aspect, more relevant for high-mass ALPS, which is thermal friction, a subject of Sections~\ref{axioncosmodw}
%adjustaaa
 and Chapter~\ref{noneqQFT}.
The basic idea, following the analogous one for cosmic strings in Ref.~\cite{PhysRevD.43.1060}, is to consider the possible relativistic expression for the 4-force acting on a domain wall from the primordial plasma and they obtain an effective contribution to the Hubble friction terms which is of the form $H+\frac{1}{l_f}$, where $l_f$ is the friction length and can be interpreted as a mean free path.

In the last section of the thesis, we will indeed start to extend the results from the VOS model with a new idea of analytical model.%check sections

%At the same time, the terms with the adimensional parameters $c_w$, $d$ (related to $d'$), $k_0$ , and 

\chapter{Axion Electrodynamics }\label{baseAED}

In this chapter, we outline in Subsection \ref{introAED} the basic aspects of Axion Electrodynamics, in particular the Lagrangian of the theory, the resulting axion-modified Maxwell equations and we discuss their properties, in particular the interpretation of the axion 4-current as polarization and magnetization currents and, also, the connection of this formalism to Condensed Matter Physics.

%\begin{enumerate}
 %\item We review the theory of Axion Electrodynamics and particularly the energy-momentum conservation in a linear dielectric and magnetic material. We treat this last aspect by extending the results of Brevik and Chaichian (2022) and Patkos (2022).
 %\end{enumerate}
\section{Basic elements of Axion Electrodynamics}\label{introAED}

\subsection{Lagrangian density of Axion Electrodynamics}
We consider a pseudoscalar axion field present in our laboratory apparatus, which has a two-photon interaction with the electromagnetic field, as previously discussed in Section \ref{invisibleaxion}.
The total Lagrangian density $\mathcal{L}$ describing the interaction between the electromagnetic field and the axion field in a Minkowski space inside a linear dielectric and magnetic material with dielectric permittivity $\varepsilon$ and magnetic permeability $\mu$ is \cite{doi:10.1142/S0217751X22501512, RevModPhys.93.015004, Sikivie2008, PhysRevD.85.105020,Millar_2017}:
\begin{equation} \label{eq0}
	\mathcal{L}= -\frac{1}{4} F^{\alpha \beta}H_{\alpha \beta}+\mathcal{L}_a-J^{\mu} A_{\mu}+ \frac{1}{4}g_{a \gamma \gamma} a(x) F_{\mu \nu} \tilde{F}^{\mu \nu},
\end{equation}
where $\tilde{F}^{\alpha \beta}=\frac{1}{2} \epsilon^{\alpha \beta \gamma \delta} F_{\gamma \delta} $ is the dual  Faraday tensor , $\epsilon^{\alpha \beta \gamma \delta}$ is the totally antisymmetric symbol with $\epsilon^{0123}=1$, and $m_a$ is the mass of the axion. The tensor $H$ is:
\begin{equation}\label{diofa}
	H_{\alpha \beta}=
	\begin{pmatrix}
		0 & -D_x & -D_y & -D_z \\
		D_x & 0 & H_z & -H_y \\
		D_y &-H_z & 0 & H_x \\
		D_z & H_y& -H_x & 0\\
	\end{pmatrix},
\end{equation}
where $\vec{D}$ and $\vec{H}$ are the usual displacement electric and magnetic fields. They are related for a linear dielectric and magnetic material in its rest frame to electric $\vec{E}$ and induction $\vec{B}$ fields by $\vec{D}=\varepsilon \vec{E}$ and $ \vec{B}=\mu \vec{H}$ . $J^{\mu}$ is an external classical electrical 4-current. 
%Finally, $g_{a \gamma \gamma}=g_{\gamma} \frac{\alpha}{\pi} \frac{1}{f_a}  $ where $\alpha$  is the usual fine
%structure constant and $f_{a}$ is the axion decay constant whose value is only insufficiently
%known.  
%Astrophysical and experimental bounds indicate that $f_{a} \sim 10^{9}-10^{12} \,\SI{}{GeV}$ \cite{RevModPhys.93.015004}.

We also define the dimensionless fields $\Theta(x) = g_{a \gamma \gamma} a(x)$ and $\theta(x) = a(x)/f_a$.

\subsection{Axion-Modified Maxwell equations}\label{AEDMaxwell}
The interaction term in Eq.~(\ref{eq0}) can be written as an interaction Lagrangian density of the form $ -J^{\nu} A_{\nu} $ by taking
\begin{equation}
	J^{\nu}=g_{a \gamma \gamma} \tilde{F}^{\mu \nu} \partial_{\mu}a =\frac{1}{2} g_{a \gamma \gamma} \partial_{\mu} a \; \epsilon^{\mu \nu \rho \sigma} \partial_{\rho} A_{\sigma}.
\end{equation}
The axions then generate an effective electromagnetic 4-current, which is thus $J_{a}=(\rho_{a}, \vec{J}_{a}) $, where
\begin{equation}\label{axionem4current}
	\begin{split}
	\rho_{a}&=g_{a \gamma \gamma} \vec{B} \cdot \nabla a,\\
	\vec{J}_{a}&= g_{a \gamma \gamma} \nabla a \wedge \vec{E}- g_{a \gamma \gamma} \dot{a} \vec{B},
	\end{split}
\end{equation}
and the continuity equation $\partial_{\mu} j_a^{\mu}=0$ is simply a topological conservation law $  \dot{\rho}_{a}+\nabla \cdot \vec{J}_{a}=0 $.

Another equivalent way to see these axion dielectric properties is by observing from the structure of Eq.~(\ref{eq0}) that it is possible to redefine the tensor $H$ as:
\begin{equation}
	H^{a}_{\mu \nu}=H_{\mu \nu}-g_{a \gamma \gamma}\, a(x)\,\tilde{F}_{\mu \nu} ,
\end{equation}
in order to rewrite Eq.~(\ref{eq0}) with the interaction term $\frac{1}{4} g_{a \gamma \gamma} a \tilde{F}_{\mu \nu} F^{\mu \nu}$ included inside the 'free term' $-\frac{1}{4} F^{\mu \nu} H_{\mu \nu}$.
We then find that we can use different constitutive relations for $\vec{D}$ and $\vec{H}$ in Axion Electrodynamics:
\begin{subequations}
	\begin{align}\label{HD}
		\vec{D}_a=\varepsilon \vec{E}+\vec{P}_a,\\
		\vec{B}=\mu \vec{H}_a+\vec{M}_a,
	\end{align}
\end{subequations}
with $\vec{P}_a=-g_{a \gamma \gamma} a \vec{B}$ the axion-induced polarization vector and  $\vec{M}_a=-\mu\,g_{a \gamma \gamma} a \vec{E}$ its magnetization.

We can then describe the charge and current densities (\ref{axionem4current}) as polarization and magnetization currents $\rho_a=-\nabla \cdot \vec{P}_a $ and $\vec{J}_a=\nabla \wedge \vec{M}_a+\dot{P}_a$, analogously to electrodynamics in gravitational backgrounds \cite{landau1951classical}. %cite

The Euler-Lagrange equations associated with the Lagrangian density (\ref{eq0}) are
\begin{subequations}
	\begin{align} \label{eq01}
		\ddot{a}-\nabla^2 a+V'(a)+g_{a \gamma \gamma} \vec{E} \cdot \vec{B} &=0,
	\end{align}
	\begin{align}\label{eq02}
		\nabla \cdot \vec{E}= \rho+g_{a \gamma \gamma} \vec{B} \cdot \nabla a,
	\end{align}
	\begin{align}\label{eq03}
		\nabla \wedge\vec{B}=\vec{J}+\dot{\vec{E}}-g_{a \gamma \gamma} \dot{a} \vec{B}+g_{a \gamma \gamma}  \nabla a \wedge \vec{E},
	\end{align}
	\begin{align}\label{eq04}
		\nabla \cdot \vec{B}=0,
	\end{align}
	\begin{align}\label{eq05}
		\nabla \wedge \vec{E}=-\dot{\vec{B}}.
	\end{align}
\end{subequations}
Equation (\ref{eq01}) is the classical equation of motion for the axion field interacting with the electromagnetic field. In contrast, equations (\ref{eq02}) and (\ref{eq03}) are the modified Maxwell equations with sources and the equations (\ref{eq04}) and (\ref{eq05}) are the "constraint Maxwell equations", that are unchanged since we assume the Bianchi identities $\partial_{\mu} \tilde{F}^{\mu \nu}=0$ to be valid.

The fields in (\ref{HD}) satisfy the following forms of dielectric Maxwell equations:
\begin{subequations}\label{Mxax}
	\begin{align}
		\nabla \cdot \vec{D}_a&= \rho,\\
		\nabla \wedge\vec{H}_a&=\vec{J}+\dot{\vec{D}}_a,\\
		\nabla \cdot \vec{B}&=0,\\
		\nabla \wedge \vec{E}&=-\dot{\vec{B}},
	\end{align}
\end{subequations}
in accordance with our interpretation of the axion currents.
\subsection{Electromagnetic 4-potential}
As in classical electrodynamics, the differential equations for the 4-potential $A_{\mu}(x)=(\phi,\vec{A})$ depend on the choice of gauge.
We can, in general, write the electric and magnetic fields as 
%\footnote{It is worth mentioning %add
%}
\begin{equation}
		\vec{E}=-\nabla \phi-\frac{\partial \vec{A}}{\partial t}, \qquad\vec{B}=\nabla \wedge \vec{A}.
\end{equation}
Consequently, we have in the vacuum:
\begin{align}\label{equA}
	\nabla^2 \Phi+\partial_t \left(\nabla \cdot \vec{A}\right)&=-\rho_a,\\
	( \partial_t^2-\nabla^2) \vec{A}+ \nabla \left( \nabla \cdot \vec{A}+ \partial_t \Phi\right)&=\vec{J}_a.
\end{align}
%In the following of this thesis we will use several gauge choices, particularly the Lorentz gauge:\\
We will adopt the following gauge choice, namely
%\begin{equation}\label{Lorentzg}
%	\epsilon \mu \partial_t \Phi + \nabla \cdot \vec{A}=0
%\end{equation}
the Coulomb gauge $\nabla \cdot \vec{A}=0$ (radiation gauge if $\rho_a=0$)
and the temporal gauge
	$A_0=\Phi=0.$
%If we use the Lorentz gauge (\ref{Lorentzg}), the inhomogeneous modified equations for the 4-potential are:\\
%\begin{subequations}\label{equLo}
	%\begin{align}
	%	(\epsilon \mu \partial_t^2-\nabla^2) \Phi=\frac{\rho+\rho_a}{\epsilon}\\
	%	(\epsilon \mu \partial_t^2-\nabla^2) \vec{A}=\mu (\vec{J}+\vec{J}_a)
%	\end{align}
%\end{subequations}
In the case of the temporal gauge, we have the following wave equation:
\begin{equation}\label{temp}
	\Box \vec{A}+\nabla(\nabla \cdot \vec{A})=-\dot{\Theta} \; \nabla \wedge \vec{A}-\nabla \Theta \wedge \frac{\partial \vec{A}}{\partial t}.
\end{equation}
%We observe that, accordingly to Refs.~\cite{RevModPhys.93.015004,doi:10.1142/S0217751X22501512}, if we have a medium with electrical permittivity $\varepsilon$ and $\mu$  then

In the case of the radiation gauge with a time-dependent axion field, we obtain the equations
	\begin{equation}\label{radiation}
		( \varepsilon \mu \partial_t^2-\nabla^2) \vec{A}= - \mu\dot{\Theta}\;  \nabla \wedge \vec{A}.
	\end{equation}

We can also derive the wave equations for the electric and magnetic fields
\begin{equation}
\nabla^2 {\bf E}-\varepsilon\mu \ddot{\bf E}=    {\bf \nabla (\nabla \cdot E)}
 +\mu \dot{\bf J}+ \mu \frac{\partial}{\partial t}\left[\dot{\Theta }{\bf B}+ {\bf \nabla}\Theta{\bf \times E}\right], \label{10}
\end{equation}
\begin{equation}
\nabla^2 {\bf H}-\varepsilon\mu \ddot{\bf H}= -{\bf \nabla \times J}-{\bf \nabla \times }[\dot{\Theta}{\bf B}+{\bf \nabla}\Theta{\bf \times E}]. \label{11}
\end{equation}

The wave equations above are complicated in the sense that they contain the second-order derivatives of $\theta$. These may be conveniently removed if we consider the approximations with which we work out in Chapter \ref{DeepAED} , by assuming constant axion derivatives. We will discuss the motivations of the utility of such an approximation in Subsections~\ref{time-dependent} and \ref{wavedomain}.
    \subsection{Energy-momentum balance with a classical axion background}\label{tensor}It is then easy to notice from the Lagrangian (\ref{eq0}) that the total energy-momentum balance reads (with $J^{\mu}=0$):
\begin{equation}\label{cons0}
	\partial_{\mu} T^{\mu \nu}=0,
\end{equation}
where $T^{ \mu \nu}$ is simply the sum of the free Minkowski stress-energy tensor $T^{ M\mu \nu}=-\frac{1}{4} \eta^{\mu \nu} F_{\alpha \beta} H^{\alpha \beta}+F^{\mu}_{\rho} H^{\nu \rho}$ and the free axion one  $T^{\mu \nu}_{a}=\partial^{\mu} a \,\partial^{\nu} a- \eta^{\mu \nu} (\frac{1}{2} \partial_{\rho} a \,\partial^{\rho} a-\frac{1}{2} m_a^2 a^2)   $, since the coupling between the axion and photons is topological, as already discussed in Section \ref{strongCP}. %add chapter on axion theory.
%This simple result is a consequence of the fact that, in this case, the physical system made up by the electromagnetic field and the axion field is treated as a closed system %{\footnote This is not rigorously exact since the potential $V(a)$ comes from the effective interaction with the gluonic fields, we will always deal with cases where backreactions from the gluons. This is pretty reasonable since }%ADD}
%However, if we consider the two fields alone, a relation such as Eq.(\ref{cons0}) is not valid, as they are open systems interacting with each other.
If we consider the electromagnetic field alone, one gets, according to Ref.~\cite{doi:10.1142/S0217751X22501512}:
\begin{equation}
	\partial_{\nu}T^{M \nu}_{\mu}=-f^M_{\mu},
\end{equation}
where $f^M_{\mu}=(f^M_0,\vec{f}^M)$ whose spatial components are the components of Abraham's force density $
	\vec{f}^A= (\varepsilon \mu-1) \frac{\partial}{\partial t} \Big(\vec{E} \wedge \vec{H}\Big)-(\vec{E} \cdot \vec{B}) \nabla \Theta
$
and $f^M_0=-g_{a \gamma \gamma } \dot{a} \vec{E} \cdot \vec{B}$.
%The 4-vector on the RHS can be interpreted as 
 \subsection{Axion Electrodynamics in Condensed Matter}
 A Weyl semimetal constitutes an intriguing phase of topological quantum matter, with fascinating physical properties, including protected surface states and a unique electromagnetic response, which is useful for practical applications, thus giving our theoretical results a broader interest, since our formalism can be adopted for the electromagnetic properties of such materials.
It is already subject of extensive literature (see e.g. Refs.\cite{nenno2020axion,PhysRevB.86.115133}), the electromagnetic properties of a Weyl semimetal can be described by a Chern-Simons theory with Lagrangian (\ref{eq0}), where the effective axion field is
\begin{equation}
	\Theta(x)=b_0 t-\vec{b} \cdot \vec{r}=b_{\mu} x^{\mu}.
\end{equation}
The quantities $b_{\mu} $ are of relevant physical meaning for the material, since they are related to the energy shift $b_0$ and the momentum shift $\vec{b}$ of the specific Weyl point of the material, whose Hamiltonian is
\begin{equation}
	h_W(\vec{k})=b_0+ v \vec{\sigma} \cdot (\vec{k}-\vec{b}).
\end{equation}

In the following, we will consider applications where the derivatives of the axion field are constant and uniform, which can be applied for the physical case of $b_0$ and $\vec{b}$, where they are really constant.
It is then interesting to know the expected orders of magnitude for Weyl semimetals.

Typical orders for the 4-momentum shifts of the Weyl point are $b_0 \sim 10^{-1} \, \SI{}{eV}$ and $|\vec{b}| \sim 10 \, \SI{}{eV}$ as mentioned in Ref.~\cite{PhysRevB.86.115133}.
These values are typically higher than the high-energy physics case, helping significantly on the possible applications which we develop in the following sections, in particular for the Casimir effect between two plates and the system we treat in Section~\ref{application}.
%add references in case

%In the first case, the time derivative of the axion field is constant, so it is easier to deal with since, e.g., the Green's function is dependent on $t-t'$, and can be regarded as an approximation of the second case when we consider times $0<t \ll \omega_a^{-1}$. The second case is a coherently oscillating axion field, %%ADJUST
%which can appropriately approximate a current laboratory axion field when the typical spatial dimension $L$ of the experimental system is much smaller than the de Broglie wavelength of the axion field (see Refs.~\cite{Millar_2017, RevModPhys.93.015004}).
%However, a possible extension would be to consider a velocity dispersion $f(\vec{v})$ for the dark matter axions, as treated in recent works about the postinflationary scenario %cite
. %Other authors, as in Ref.%add cite caustic
%, they have claimed that even with the creation of axion miniclusters in the postinflationary scenario, we could expect to have a further creation of the caustic rings, which would spoil the stochasticity of the axion spectrum emitted by miniclusters and obtain just a one-stream axion spectrum, which would lead to the same results as former works on oscillating axions.%adjust

%\subsubsection[The Axion]{A partial overview on Axion Cosmology}

%\begin{frame}{Axion Electrodynamics (AED)}
%\begin{block}{Axion-modified Maxwell equations}
%\begin{align*}
%\nabla \cdot \vec{D}= 0
%\end{align*}
%\begin{align*}
%\nabla \wedge\vec{H}=\vec{J}+\dot{\vec{D}} 
%\end{align*}
%\end{block}
%\begin{exampleblock}{Constitutive relations}
%\begin{align*}
%\vec{D}=\epsilon \vec{E}-g_{a \gamma \gamma} a \vec{B}\\
%\vec{B}=\mu \vec{H}+g_{a \gamma \gamma} a \vec{E}
%\end{align*}
%\end{exampleblock}
%\end{frame}

\part{Results and applications}\label{II}
\chapter{Deeper aspects of Axion Electrodynamics}\label{DeepAED}%

\section{Introduction}\label{secintro}Starting from the theory of Axion Electrodynamics that we have outlined in Chapter~\ref{baseAED}, we calculate the axionic modifications to the electromagnetic Casimir energy using the Green's function method. This will be done in both cases of the axion field initially assumed to be purely time-dependent and when the axion field configuration is a static domain wall, which we take to be strictly space-dependent.

For the first case, it means that the oscillating axion background is assumed to resemble a dark matter axion fluid at rest, approximated as cold, and neglecting the velocity dispersion of the input dark matter axions, in a conventional ideal Casimir setup with two infinite parallel conducting plates. 

In contrast, in the second case, we evaluate the radiation pressure acting on an axion domain wall. We extend previous theories to include finite temperatures. 
This work is also related to earlier investigations in Refs.~ \cite{Sikivie2008, PhysRevD.32.1560, Blasi2023,PhysRevD.102.123011,PhysRevD.41.1231,RevModPhys.93.015004, hassan2025chern}.

%add canfora2022casimir and oosthuyse2023interplay

%We present the basics of Axion Electrodynamics and introduce our notation in Section~\ref{intro}.
%Energy-momentum balance is considered in Section~\ref{tensor}.
This is an overview of this Chapter:
\begin{itemize} 
\item In Section~\ref{vacgreen} we develop the basics of the Green's function method in Axion Electrodynamics and use it in practice for time-dependent axion backgrounds, namely with constant time derivative and oscillating behaviour in Section~\ref{time-dependent}. We calculate the Casimir force between two perfectly conducting parallel plates for the first case and provide estimates for the second one, as we did in Ref.~\cite{FAVITTA2023169396}.
\item Section~\ref{optical} compares our results with formerly known phenomena in Classical Electrodynamics, such as the Faraday effect and optical activity in chiral media, along with other works on the optical activity of Axion Electrodynamics. We discuss an application in Section~\ref{application}, which was the main object of our Ref.~\cite{doi:10.1142/S0217751X24500040}.
\item In Sections~\ref{wavedomain} and \ref{spazio} we treat space-dependent axion backgrounds. We discuss their optical properties and dispersion relations.
%\item In Section~\ref{wavedomain}, we develop the optical properties of a toy model of the axion domain wall. The electromagnetic dispersion relations in such an axion background are given in Subsection~\ref{Casimirdisperdo}.
%\item General features of the Green's function method for domain walls are introduced in Section~\ref{spazio}.

\item Section~\ref{spazio}, in particular in Subsection~\ref{dio},  deals with the calculation of the electromagnetic radiation pressure acting on a planar axion domain wall. We use Casimir methods and connect them with former methods. 
We discuss our results for applications in Axion Cosmology and Condensed Matter Physics, as we did in Ref.~\cite{FAVITTA2023169396}.
% Subsection~\ref{exactcal-toy} treats the general features of Green's function method for such backgrounds, while the features of a toy model of domain wall are treated from Subsection~\ref{} to \ref{}. Namely, Subsection~\ref{} introduces general features of domain walls that are useful for what done in the following, Subsection~\ref{} develops the optical properties of a toy model of axion domain wall, while Subsection~\ref{Casimirdisperdo} deals with the electromagnetic dispersion relations in such an axion background. Subsection~\ref{dio} deals with the calculation of the temperature-dependent electromagnetic radiation pressure acting on axion domain walls and we exploit our results for applications in Axion Cosmology and Condensed Matter Physics.
\end{itemize}
The two classes of axion time-dependent backgrounds considered are also of experimental interest, for example, for observing the virialized axion dark matter of the Big Flow, the branch of cold dark matter halo of the Milky Way where the Earth is supposed to stay \cite{RevModPhys.93.015004}.

This is seen in the long-wavelength approximation, i.e., when we assume the experimental setup to have a typical spatial dimension $L$ much smaller than the de Broglie wavelength of the local axion field, so that we take $a(t) \simeq a_0 \sin(\omega_a t)$. 
For a dark matter QCD axion if we take $v \sim 10^{-3}$, which is the typical value of rms velocity in the Big Flow, and $\omega_a \sim m_a$, we have a typical value of 
\begin{equation}
    \lambda_{DB} \simeq 10^{-3} m_a^{-1}
\end{equation}

%\cite{Millar_2017}).%add details here

We can take a further approximation of taking the time derivative of the axion field to be constant, which makes sense when the argument of the sine of the oscillatory axion can be taken much smaller than 1.

From a physical point of view, it can be done when the spatial dimensions of the system are much smaller than the Compton wavelength, and can be just a pedagogical first approximation for the axion. In that case, it is exact for the effective axion field in Weyl semimetals.

The axion field is treated here as a fixed background field, since this is true at first order in the perturbation theory where the back reaction of the electromagnetic ﬁeld onto axions is neglected, partly analogous to what happens in the linear approximation for weak gravitational perturbations in General Relativity \cite{Kiefer2009}. We have already discussed this last point in Section~\ref{tensor}.
%One important point in order that this model can be useful to treat axion domain wall in Early Universe is if this model gives rise to a dynamics of the parameter L (given that, without taking care of electromagnetism, it is a parameter that depends on axion mass).
%One point is that according to our model L is dynamic.
%Let us consider the Euler-Lagrange equations for axion field (let us consider the axion field as a classical field).

\section{Green's functions}\label{vacgreen}
We first develop the Green's function approach in Axion Electrodynamics, using the temporal gauge.
We start from Eq.~(\ref{temp}) with $\varepsilon=\mu=1$ and define the kernel $ \vec{G}_{ij}(x,x')$, such that:
\begin{equation}\label{dyad}
	A_i(x)=\int d^4 x' \; G_{ij}(x,x') \, J^{j}(x').
\end{equation}
Due to causality, the variable $t'$ is only integrated over $t' \leq t$. From Eq.~(\ref{temp}) we find
%\begin{equation}
%	\Box \vec{G}+\nabla(\nabla \cdot \vec{G})=-\dot{\Theta} \; \nabla \wedge \vec{G}-\nabla \Theta \wedge \frac{\partial \vec{G}}{\partial t}+ \mathbb{I} \delta^{(4)}(x^{\mu}-y^{\mu}),
%\end{equation}
%which can be written in component form,
\begin{equation}\label{greenaqua}
	\begin{split}
		\Box G_{ij}(x,x')+\partial_i(\partial_k G_{kj}(x,x'))-\dot{\Theta}(x) \epsilon_{ikl} \partial_k G_{lj}(x,x')-\epsilon_{ikl} \partial_k \Theta(x) \; \partial_t G_{lj}(x,x')=\\\delta_{ij} \delta^{(4)}(x^{\mu}-x'^{\mu}).
	\end{split}
\end{equation}

The $G$ kernel (\ref{dyad}) is equal to $ iD_{i j}(x^{\mu} ,y^{\mu})$ , where $D$ is the retarded Green's function of the vector potential $\vec{A}$ defined as
\begin{equation}\label{GR}
	i D^{R}_{i j}(x^{\mu}, y^{\mu})=\begin{cases}
		\langle{A_{i}(x) A_{j}(y)-A_{j}(x) A_{i}(y)}\rangle \; \;  \; \textit{if} \;  \;  \;  \;x^{0} -y^{0} > 0,\\ 0 \;  \;  \;  \; \rm{otherwise}.
	\end{cases}
\end{equation}
It can be derived by the Schwinger-Dyson equations \cite{schwartz2014quantum} for Axion Electrodynamics,  which is exactly Eq.~(\ref{greenaqua}), when we take the temporal gauge. 

As shown in Refs.~\cite{landau2013electrodynamics,https://doi.org/10.48550/arxiv.hep-th/9901011} and \cite{birkeland2007feigel}, the calculation of the retarded Green's function  (\ref{GR})  is useful to get the significant two-point physical averages between electric and magnetic fields.
One can easily find that with the above choice of gauge:
\begin{subequations}\label{two}
\begin{align}
	\langle{E_{i}(x) E_{j}(x') }\rangle &=\partial_t \partial_{t'}\langle{A_{i}(x) A_{j}(x') }\rangle, \\
	\langle{B_{i}(x) B_{j}(x') }\rangle &=\curl_{il}  \curl'_{jm} \langle{A_{l}(x) A_{m}(x') }\rangle,  \\
		\langle{B_{i}(x) E_{j}(x') }\rangle &=-\curl_{il} \partial_{t'} \langle{A_{l}(x) A_{j}(x') }\rangle,
\end{align}
\end{subequations}
Similarly, this applies also to the other two-point functions. We will also use the Fourier transforms of the two-point functions, e.g.
\begin{equation}
\langle{A_{\alpha}(x) A_{\beta}(y) }\rangle_{\omega} =\int_{-\infty}^{+\infty} dt \, e^{i \omega t} \langle{A_{\alpha}(t,\vec{x}) A_{\beta}(0,\vec{y}) }\rangle,
\end{equation}
from which the relations (\ref{two}) can be used to find the corresponding ones for the two-point functions involving electric and magnetic fields.\\
%%ADD
%The axion field can be thought of as a fixed classical background $\Theta(\vec{x},t)$ in perturbative theory, and with the assumption that the  axion is the main dark matter component with   high occupation numbers.%%WHERE?\\
If the Green's function is a function of just the variables $t-t', \,x-x' \,\, \text{and} \,\, y-y'$, we can define the
reduced Green’s function $g_{ij}(z,z',\omega,\tilde{k})$:
\begin{equation}
	G_{i j}(x,x')=\int_{-\infty}^{\infty} \frac{d \omega}{2 \pi} \int_{-\infty}^{\infty} \frac{d^2 \tilde{k}}{(2 \pi)^2} e^{-i \omega (t-t')} e^{i \tilde{k} \cdot (\vec{r}-\vec{r}')} g_{ij}(z,z',\omega, \tilde{k}),
\end{equation}
with $\tilde{k}=(k_x,k_y)$. We also adopt the useful quantity $\kappa^2=\omega^2-k_x^2-k_y^2.$

The utility of the reduced Green's function is evident from the premise: it is advantageous in systems with translation symmetry along the $x-y$ plane.

In the following, we will indeed focus on the Green's function in the vacuum to calculate Casimir's forces in such a symmetric system.
We can find straightforwardly from the reduced Green's function, with $\varepsilon=\mu=1$:
\begin{equation}\label{Tzz}
	\begin{split}
		\langle{T^{M}_{zz}}\rangle_{\omega,\tilde{k}}=\frac{1}{2i} \left[-\kappa^2 g_{zz}+(\omega^2-k_y^2) g_{xx}+(\omega^2-k_x^2) g_{yy}+ik_{y} \left(\partial_z g_{yz}-\partial_z' g_{zy}\right) \right.\\ \left.  +ik_{x} \left(\partial_z g_{xz}-\partial_z' g_{zx}\right)
		+k_x k_y (g_{xy}+g_{yx}) + \partial_z \partial_z' (g_{xx}+g_{yy})         \right].
	\end{split}
\end{equation}
For our purpose of evaluating zero-point energies, we will adopt the retarded Green's function with $\mu \neq 1$ to find the expression in the vacuum by using the fluctuation-dissipation theorem \cite{landau1987statistical}:
\begin{equation}
	\langle{A_{\alpha}(x) A_{\beta}(y) }\rangle_{\omega}=\frac{i}{2} \left[ D^{R}_{\alpha \beta}(\omega,\vec{x}, \vec{y})-D^{R*}_{ \beta \alpha}(\omega, \vec{y},\vec{x})  \right],
\end{equation}
and then take the limit $\mu \rightarrow 1+i0$.

Notice that in the case of temporal gauge, the Green's function can be treated as a $3\times 3$ tensor, while this is not the case for other gauge choices, where we need to treat it fully as a $4\times 4$ tensor.

If the axion field is only a function of time $t$, it is convenient to adopt the radiation gauge, and we will do it in the following section. We will instead adopt the temporal gauge for space-dependent axion fields from Section~(\ref{spazio}) on.
\section{Purely time-dependent axion field}\label{time-dependent}
Assuming $\Theta(x)=\Theta(t)$ and a homogenous medium with electrical permittivity $\epsilon=1$ and magnetic permeability $\mu $, we obtain the following equation for the Green's function $G_{kj}(x,y)$ using Eq.~(\ref{radiation})%which ,:
\begin{equation}\label{Green1time}
	\left[   \delta_{ik} \Box(\mu) -\mu \,\dot{\Theta}(t)\, \epsilon_{lik} \nabla_l \right] G_{kj}(x,y)=\mu \delta_{ij}  \delta^{(4)}(x^{\mu} -y^{\mu}),
\end{equation}
where we have defined the operator $\Box(\mu)= \mu\, \partial_t^2- \nabla^2$.

%Explicitly, the coupled differential equations system is the following:
%\begin{equation}\label{dotgreen1}
%	\begin{cases}
	%	\Box G_{xx}+\dot{\Theta}  \partial_y G_{yx}-\dot{\Theta} \partial_z G_{zx}=\delta^{(4)}(x^{\mu} -y^{\mu})\\
	%	\Box G_{yy}-\dot{\Theta}  \partial_z G_{xy}+\dot{\Theta} \partial_x G_{zy}=\delta^{(4)}(x^{\mu} -y^{\mu})\\
	%	\Box G_{zz}+\dot{\Theta}  \partial_y G_{xz}-\dot{\Theta} \partial_x G_{yz}=\delta^{(4)}(x^{\mu} -y^{\mu})
%	\end{cases}
%\end{equation}
%with
%\begin{equation}\label{dotgreen2}
%	\begin{cases}
	%	\Box G_{xy}-\dot{\Theta}  \partial_z G_{yy}+\dot{\Theta} \partial_x G_{zy}=0\\
	%	\Box G_{yx}-\dot{\Theta}  \partial_x G_{zx}+\dot{\Theta} \partial_z G_{xx}=0\\
	%	\Box G_{zy}+\dot{\Theta}  \partial_x G_{yy}-\dot{\Theta} \partial_z G_{xy}=0
%	\end{cases}
%\end{equation}
%and similar analogous expressions.\\
The solution of Eq.~(\ref{Green1time}) is generally dependent on $ \vec{r}-\vec{r'}$ and $t-t'$, but also explicitly on $t$, then it is not trivial to find solutions for any $\Theta(t)$. We will consequently limit ourselves to two specific cases of interest, i.e. $\Theta=\alpha_0 t$ and $\Theta=\Theta_0 \sin(\omega_a t)$, as already discussed.
\subsection{Case of constant axion time derivative $\alpha_0$}\label{constdot}
If we consider $\dot{\Theta}=\alpha_0$ as constant in time and uniform in space, we can take, similarly to \cite{PhysRevD.102.123011}, the Fourier Transform of the Green's function and obtain from Eq.~(\ref{Green1time}) the following equation:
\begin{equation}\label{Greenconst}
	\left[ \delta_{ik} (-\mu \omega^2+ k^2)+i \mu \alpha_0 \epsilon_{lik} k_l \right]  \tilde{G}_{ k j}(\vec{k}, \omega)=\mu \delta_{ij}.
\end{equation}
%The calculations in this case are of conceptual interest and can be applied similarly in a physically more interesting case, such as the oscillating axion field.

%We will revisit this topic in the next section.%verify if it is a section

To obtain the Green's function in the vacuum, we perform the calculations with a magnetic permeability $\mu=1+i \tilde{\mu}(\omega)$, where $\tilde{\mu}(\omega)\ll1$ is first order, and evaluate the limit for $\tilde{\mu} \rightarrow 0$, as analogously done in Ref.~\cite{landau1987statistical}.
The usual approach in the literature is with an electrical permittivity $\varepsilon(\omega)=1+i \epsilon(\omega)$, to obtain the quantum mechanical result for the Green's function.
%A magnetic permeability of the form $\mu=\mu_r+ i \mu_I$, where $\mu_r$ and $\mu_I$ are respectively the real and imaginary parts.
However, there exist real media where magnetic viscosity is present (see e.g. Ref.~\cite{WanjunKu1997}):
\begin{equation}
	\mu=\Re{(\mu)}+i \,\epsilon \,\frac{\omega}{\omega_0}.
\end{equation}
This is analogous to the case of electric permittivity, and we have the imaginary part $\Im{(\mu)}>0$ when $\omega>0$. This justifies our choice of $\sgn(\omega)$ in the following.

From Eq.~(\ref{Greenconst}) we get, similarly to Ref.~\cite{PhysRevD.102.123011}, the retarded Green's function as the inverse operator of the expression in square brackets on the LHS of Eq.~(\ref{Greenconst}) in the limit of $\mu=1+i \tilde{\mu} \sgn(\omega) \rightarrow 1+i0 \sgn(\omega)$:
\begin{equation}\label{solutotutto}
	\tilde{G}_{jk}(\vec{k}, \omega)= (\omega^2-|\vec{k}|^2)\tilde{A}(\omega, \vec{k},\vec{\beta}) \delta_{jk}+i\tilde{A}(\omega, \vec{k},\vec{\beta}) \epsilon_{jkl}\beta_l+\frac{1}{\omega^2-|\vec{k}|^2+i 0 \sgn(\omega) }\tilde{A}(\omega, \vec{k},\vec{\beta}) \beta_j \beta_k,
\end{equation}
where $\vec{\beta}$ is a 3-vector with components $\beta_j=\alpha_0 k_j $, whose squared module is $\beta^2$ and
\begin{equation}
		\tilde{A}(\omega, \vec{k},\vec{\beta})=\frac{1}{(\omega^2-|\vec{k}|^2)^2-\beta^2+i 0 \sgn(\omega) } .
\end{equation}
The ordinary electrodynamics limits can be obtained by setting $\alpha_0=0$.
We can now use the Fourier transforms for evaluating the spectral energy density $\rho_{em}(\omega)$ of the electromagnetic field at temperature $T$ and the Casimir force between two parallel conducting plates in such a fixed axionic background.
\subsubsection{Spectral energy density $\rho_{em}(\omega)$ of the electromagnetic field}
To get the zero-point spectral energy density, we calculate the inverse Fourier transform, which we call $\tilde{G}(\vec{r},\omega) $,  of the Green's function (\ref{solutotutto}) in the position space and take the correlation functions in the limit $\vec{r} \rightarrow 0$.
The usual spectral energy density of the electromagnetic field is
\begin{equation}
	\rho(\omega,\vec{r})\, d \omega= \frac{1}{2} \left[2 \langle{\vec{E}^2(\vec{r})}\rangle_{\omega,T}+2\langle{\vec{B}^2(\vec{r})}\rangle_{\omega,T} \right]  \frac{d \omega}{2 \pi},
\end{equation}
where the factors two inside the brackets are inserted because we follow the definition of spectral densities given in  Ref.~\cite{landau1987statistical}:  average spectral densities are defined as integrals in $\omega$ from $-\infty$ to $+\infty$.
We also exploit the distributional relation:
\begin{equation}\label{distro}
	\frac{1}{x \pm i 0}=\mathcal{P}{\frac{1}{x}} \mp i \pi \delta(x).
\end{equation}
where $\mathcal{P}$ is the principal part and $\delta$ is the Dirac delta distribution \cite{landau1987statistical,bagarello2007fisica,stakgold2011green}.
We expand the spectral energy density in a perturbative series:
\begin{equation}
	\rho(\omega,\vec{r})=\rho^{(0)}(\omega,\vec{r})+\rho^{(1)}(\omega,\vec{r})+\rho^{(2)}(\omega,\vec{r})+\text{(higher-order terms).}
\end{equation}
We can develop a perturbative expansion for the Green's function in terms of the perturbative factor $\alpha_0$ :
\begin{equation}
	\tilde{G}_{jk}(\vec{k}, \omega) \sim \tilde{G}^{(0)}_{jk}(\vec{k}, \omega)+\tilde{G}^{(1)}_{jk}(\vec{k}, \omega)+\tilde{G}^{(2)}_{jk}(\vec{k}, \omega),
\end{equation}
where
\begin{subequations}\label{consthetaG}
	\begin{align}
		\tilde{G}^{(0)}_{jk}(\vec{k}, \omega)&=\frac{1}{\omega^2-k^2+i0 \sgn{\omega}}, \\
		\tilde{G}^{(1)}_{jk}(\vec{k}, \omega)&=\frac{i}{(\omega^2-k^2)^2+i0 \sgn{\omega}} \epsilon_{jkl}\beta_l=\frac{i}{(\omega^2-k^2)^2+i0 \sgn{\omega}} \epsilon_{jkl} (\alpha_0 k_l),\\
		\tilde{G}^{(2)}_{jk}(\vec{k}, \omega)&=\frac{\beta^2}{(\omega^2-k^2)^3+i0 \sgn{\omega}}\left[ \delta_{jk}+\frac{\epsilon_{jkl}\beta_l}{\omega^2-k^2} \right]+\frac{1}{(\omega^2-k^2)^3+i0 \sgn{\omega}}\alpha^2_0 k_j k_k.
	\end{align}
\end{subequations}

Analogously to Ref.~\cite{landau1987statistical}, we can obtain the spectral density associated with the zero-order term $\rho^{(0)}(\omega)$ by integrating over the domain of $\vec{k}$ 
\begin{equation}
	\rho^{(0)}(\omega) d \omega=  d \omega \; \omega^2 \left[\frac{1}{2} \omega + \frac{\omega}{e^{\frac{\omega}{T}}-1} \right].
\end{equation}
This is the familiar result from QED: the zero-point energy of the electromagnetic field at a given temperature $T$ is the sum of the vacuum zero-point energy at zero Kelvin temperature and the black-body radiation energy at the given temperature $T$.

The first order term $\rho^{(1)}(\omega)$ is trivially zero, while we have a contribution of  order $\mathcal{O}(g^2_{a \gamma \gamma}) $:
\begin{equation}\label{dothecon}
	\rho^{(2)}(\omega)\; d \omega=  d \omega \; \omega^2 \left[\frac{1}{2} \frac{ \dot{\alpha_0}^2}{8\omega} + \frac{1}{e^{\frac{\omega}{T}}-1}\frac{ \alpha_0^2}{8\omega} \right].
\end{equation}
We notice the frequency dependence of the spectral energy density in Eq.~(\ref{dothecon}), from which we can find the spectral emissivity using Planck's law:
\begin{itemize}
    \item The first term on the RHS is proportional to the frequency and can be physically interpreted as a blue noise in the frequency domain.
    \item The second term is similar to the first one in terms of frequency dependence, but it has an additional Bose-Einstein temperature-dependent weight. For $T \gg \omega$ it behaves as $\simeq T $, so it can be regarded as a flat noise, and is bigger than the zero-temperature term. For $T \ll \omega$, the Bose-Einstein weight would give approximately the Boltzmann weight $e^{-\frac{\omega}{T}}$, so an exponential noise.
\end{itemize}
\subsubsection{Dispersion relations}
Here, we briefly discuss the dispersion relations. The Green's function (\ref{solutotutto}) has poles, to which the dispersion relations are associated :
\begin{equation}\label{equodisperdo}
\omega_{\pm}=\sqrt{|\vec{k}|^2 \pm \alpha_0 |\vec{k}| } \qquad
	\omega_0=|\vec{k}|.
\end{equation}
These dispersion relations can be interpreted physically as follows: consider the equations of motion of the vector potential for a solution propagating along the direction of the $z$-axis:
	\begin{align}
		\begin{split}
			(-\omega^2+k_z^2) \mathcal{A}_x &=\dot{\Theta} k_z  \mathcal{A}_y,\\
			(-\omega^2+k_z^2) \mathcal{A}_y &=-\dot{\Theta} k_z  \mathcal{A}_x,	\\
			(-\omega^2+k_z^2) \mathcal{A}_z &=0.
		\end{split}
	\end{align}
    where $\mathcal{A}_i$ are components of the Fourier transforms of the vector potential $\vec{A}$.
This assumption on the propagation direction does not limit generality, since we treat a system in the vacuum which is isotropic and it is always possible to find an inertial reference frame where an electromagnetic plane wave propagates along an assigned $z$-axis.

It is then easy to demonstrate the dispersion relation $\omega_{\pm}$ of (\ref{equodisperdo}) for real transverse photons. Indeed, if we define the fields $\mathcal{A}_{\pm}=\mathcal{A}_x \pm i \mathcal{A}_y$, we obtain the dispersion relations $\omega_{\pm}$ with $k_x=k_y=0$ . However, for the full expression of $\omega_{\pm}$ in Eq.~(\ref{equodisperdo}) we need to use the relativistic invariance of the magnitude of the wave 4-vector. 

Its physical meaning is as follows: a left-circular polarized wave has a different frequency than a right-circular one with the same $\vec{k}$ and the optical angle rotates. This is due to the optical activity of the vacuum in Axion Electrodynamics as we shall discuss it in Section~(\ref{optical}).

We also notice that a solution for the equation of $\mathcal{A}_z$, one with the dispersion relation $\omega=|\vec{k}|$ and corresponding to longitudinal photons, is simply zero, so that the dispersion relation corresponds to virtual photons.

\subsubsection{Casimir force between parallel plates}
Here we evaluate the axion modifications to the usual Casimir force between two parallel perfectly conducting plates (see Refs.~\cite{https://doi.org/10.48550/arxiv.hep-th/9901011,schwartz2014quantum} for a treatment without an axion background).
Our result is comparable to that obtained by Ref.~\cite{PhysRevD.81.025015}, while improving upon their result.
\begin{figure}
\centering
\begin{tikzpicture}[scale=0.9,>=stealth]
  \draw[fill=yellow!50,draw=black] (-2,0.8) rectangle (2,0.96);
  \draw[fill=yellow!50,draw=black] (-2,-0.8) rectangle (2,-0.96);
  \draw[<->] (2.25,-0.8) -- node[right] {$L$} (2.25,0.8);
  %\node at (0,-1.35) {\small Parallel plates (Casimir)};
\end{tikzpicture}
\caption{Scheme of a system of two metallic parallel plates, for which we measure the Casimir force per unit area. The controlled separation $L$ is experimentally the metrological "lever arm". The z-axis is directed towards the high. We do not use a usual notation of $d$ for the lever arm distance to not confuse it with the $d$ we adopt for denoting the extended number of dimensions, which we need for dimensional regularization in the following.}
    \label{fig:placeholder}
\end{figure}
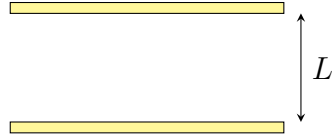

We can evaluate the zero-point energy $u_{em}(L,\alpha_0)$ per unit transverse area by evaluating the following expression:
\begin{equation}\label{uem}
	u_{em}(L,\dot{\Theta})=\frac{1}{2} \sum_{\pm} \sum_n \int \frac{d^2 k}{(2 \pi)^2} \sqrt{|\vec{k}|^2+ \frac{n^2 \pi^2}{L^2} \pm \alpha_0 \sqrt{|\vec{k}|^2+ \frac{n^2 \pi^2}{L^2}  }}.
\end{equation}
This can be expected by a Casimir approach, as given by the dispersion relations.

The expression (\ref{uem}) can be demonstrated by a Green's function approach. We assume the $z$-axis to be the direction normal to the plates.
We need to solve the Green's function equations with the proper boundary conditions. Since we assume to have two perfectly conducting plates, we have boundary conditions for the electric field and the magnetic field at $z=0, L$:
\begin{equation}
B_z=0 \qquad \vec{E}_{\parallel}=0,
\end{equation}
where $\vec{E}_{\parallel}$ is the parallel component of the electric field, corresponding to the Green's function:
\begin{subequations}\label{BCs}
	\begin{align}
g_{ij}(x-x',y-y',z,z')|_{z=0,z=L}&=0	\quad i \neq z, j \neq z,\\
\partial_z g_{zz}(x-x',y-y',z,z')|_{z=0,z=L}&=0.
	\end{align}
\end{subequations}
These conditions, along with the translational invariance along x and y directions, allow us to write the Green's function in the following Fourier expansion for $0<z<L$ and $0<z'<L$:
\begin{subequations}\label{fouseries}
	\begin{align}
	G_{ij}(x,x')&=\int_{-\infty}^{+\infty} \frac{d \omega}{2 \pi}\int_{-\infty}^{\infty} \frac{d^2 \tilde{k}}{(2 \pi)^2} e^{i \tilde{k} \cdot (\tilde{r}-\tilde{r}')} \frac{2}{L}\sum_{n=0}^{\infty}	\sin{\left(\frac{n \pi}{L}z\right)} \sin{\left(\frac{n \pi}{L}z'\right)} \tilde{g}_{ij}(\tilde{k},n) \quad i \neq z, j \neq z,\\
	G_{zz}(x,x')&=\int_{-\infty}^{+\infty} \frac{d \omega}{2 \pi}\int_{-\infty}^{\infty} \frac{d^2 \tilde{k}}{(2 \pi)^2} e^{i \tilde{k} \cdot (\tilde{r}-\tilde{r}')} \frac{2}{L}\sum_{n=0}^{\infty}	\iota(n)\cos{\left(\frac{n \pi}{L}z\right)} \cos{\left(\frac{n \pi}{L}z'\right)} \tilde{g}_{zz}(\tilde{k},n),
	\end{align}
\end{subequations}
where $\tilde{k}=(k_x,k_y)$, $\tilde{r}=(x,y)$ and $\iota(n)=1$ for $n>0$, while $\iota(n=0)=1/2$.\\
The reduced function $\tilde{G}(\tilde{k},n)$ consequently satisfies the equations (\ref{Greenconst}), with solution Eq.~(\ref{solutotutto}) ,where $k_z= \frac{n \pi}{L}$.
We now have two perspectives of looking at Eq.~(\ref{uem}), which will allow us to calculate the Casimir force.

\paragraph{Energy density per transversal area }

Here we adopt the form (\ref{solutotutto}) for the Green's function.
If we exploit the relation (\ref{distro})
and the expression of the Green's function (\ref{solutotutto}) we get
\begin{equation}
	\frac{1}{2}\langle{\vec{E}^2+\vec{B}^2}\rangle_{\omega,k_x,k_y,n} =\pi \frac{\omega}{2} \sum_{\pm} \left[\delta\left(\omega - \sqrt{|k|^2 \pm \alpha_0 |k|}\right)- \delta \left(\omega+ \sqrt{|k|^2 \pm \alpha_0 |k|}\right)      \right],
\end{equation}
where $ |k|^2=k_x^2+k_y^2+ \left(\frac{n \pi}{L}    \right)^2 $.
We can then exploit the properties of the Dirac delta distribution by integrating in the $\frac{\omega}{2 \pi}$ domain from $-\infty$ to  $+\infty$, and then obtain
\begin{equation}
		\frac{1}{2}\langle{\vec{E}^2+\vec{B}^2}\rangle_{k_x,k_y,n}=\frac{1}{2} \sum_{\pm} \sqrt{k_x^2+k_y^2+ \frac{n^2 \pi^2}{L^2} \pm \alpha_0 \sqrt{k_x^2+k_y^2+ \frac{n^2 \pi^2}{L^2}}}.
\end{equation}
We also mention that $	\langle{\vec{A} \cdot \vec{B}}\rangle_{k_x,k_y,n}$ and $	\langle{\vec{E} \cdot \vec{B}}\rangle_{k_x,k_y,n}$  are trivially equal to zero. Consequently, we have $	\langle{T^{00}_{em}}\rangle_{k_x,k_y,n}=	\frac{1}{2}\langle{\vec{E}^2+\vec{B}^2}\rangle_{k_x,k_y,n}$, from which it is easy to get the energy per unit transverse area to be equal to the expression (\ref{uem}).

\paragraph{Casimir force from the reduced Green’s function}

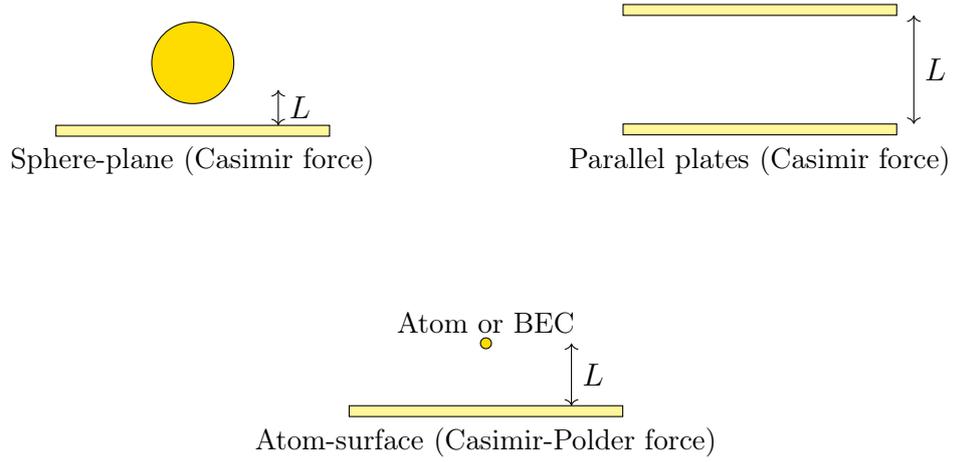
\begin{figure}
    \centering
\begin{center}
% --- Row 1: sphere-plane and parallel plates ---
\begin{tikzpicture}[scale=0.9]
  % Sphere--plane Casimir
  \draw[fill=yellow!50,draw=black] (-2,-0.08) rectangle (2,0.08); % plane
  \draw[fill=yellow!80!orange,draw=black] (0,1) circle (0.6);    % sphere
  \draw[<->] (1.25,0.08) -- node[right] {$L$} (1.25,0.6);        % gap label
  \node at (0,-0.45) {\small Sphere-plane (Casimir force)};
\end{tikzpicture}
\hspace{2cm}
\begin{tikzpicture}[scale=0.9]
  % Parallel plates Casimir
  \draw[fill=yellow!50,draw=black] (-2,0.8) rectangle (2,0.96);   % top plate
  \draw[fill=yellow!50,draw=black] (-2,-0.8) rectangle (2,-0.96); % bottom plate
  \draw[<->] (2.25,-0.8) -- node[right] {$L$} (2.25,0.8);         % separation
  \node at (0,-1.35) {\small Parallel plates (Casimir force)};
\end{tikzpicture}
\vspace{1.5cm}

% --- Row 2: atom-surface CP ---
\begin{tikzpicture}[scale=0.9]
  % Surface
  \draw[fill=yellow!50,draw=black] (-2,-0.08) rectangle (2,0.08);
  % Atom
  \fill[fill=yellow!80!orange,draw=black] (0,1) circle (0.08);
  \node[above] at (0,1) {\small Atom or BEC};
  % Distance arrow
  \draw[<->] (1.25,0.08) -- node[right] {$L$} (1.25,1);
  \node at (0,-0.45) {\small Atom-surface (Casimir-Polder force)};
\end{tikzpicture}
\end{center}
\caption{The three main geometries of the current experiments with Casimir setups. 
       The system with the parallel plates is conceptually clean,overall for theoretical calculations, but it is extremely hard to build experimentally. Plates need to be held  parallel within angles $\theta \ll1 \, \mu$rad and furthermore electrostatic patches are problematic \cite{Decca2005,Decca2007}.
       In the case of the sphere-plane setup, e.g., the  alignment is easy and performed in AFM-style force readout \cite{Mohideen1998,Chen2002}. BEC stands for Bose-Einstein Condensate.}
\label{fig:Casimirs}
\end{figure}

In general, the Casimir force is a physical force that exists between dielectric or metallic surfaces, resulting from quantum fluctuations of the electromagnetic field. 
From a general perspective, it can be treated with different methods, but the essential ingredient is the difference between the normal Maxwell stress components  $T_{zz}$ on the two sides of a boundary, yielding a pressure on them. 
This can be generally done with various geometries of the boundaries and the three main geometries are shown in Fig.~\ref{fig:Casimirs}.

The Casimir force can be derived from equilibrium statistical thermodynamics, as an application of the fluctuation-dissipation theorem, which relates the two-point functions of the electromagnetic field to the imaginary part of retarded Green function, as in Ref.~\cite{landau1987statistical} and before, as well as using other approaches, for instance that in Ref.~\cite{Brevik_Shapiro_Silveirinha_2022}. Alternatively, the effect can be seen as a manifestation of quantum field fluctuations \cite{https://doi.org/10.48550/arxiv.hep-th/9901011}.

We evaluate the expression (\ref{Tzz})  using the equation (\ref{Greenconst}), when $\mu=1$. It is then easy to obtain by means of the expression (\ref{fouseries}) the following expression
\begin{equation}\label{tzo1}
	\langle{T_{zz}}\rangle_{\omega, k_x,k_y}|_{z=z'=0,L}= - 2 i \sum_{n=1}^{+\infty }\frac{n^2 \pi^2}{L^3} \frac{\kappa^2-\frac{n^2 \pi^2}{L^2}}{(\kappa^2-\frac{n^2 \pi^2}{L^2})^2-\alpha^2_0 (k_x^2+k_y^2+\frac{n^2 \pi^2}{L^2})},
\end{equation}
where $\kappa^2=\omega^2-k_x^2-k_y^2$.

This series can be treated by noticing that each $ n$-th term of the series can be written as a sum of two $\pm$ terms, as we have shown in Ref.~\cite{FAVITTA2023169396}, by performing a standard complex frequency rotation $\omega \rightarrow \zeta=i \omega$ , similarly to Refs.~\cite{https://doi.org/10.48550/arxiv.hep-th/9901011}, and integrating over $\zeta/{2 \pi}$. We then obtain
\begin{equation}\label{tzo3}
		\langle{T_{zz}}\rangle_{k_x,k_y}|_{z=z'=0,L}= \sum_{\pm} \sum_{n=1}^{+\infty }\frac{n^2 \pi^2}{L^3} \frac{1}{\sqrt{k_x^2+k_y^2+\frac{n^2 \pi^2}{L^2}\pm\alpha_0 \sqrt{k_x^2+k_y^2+\frac{n^2 \pi^2}{L^2}}}}.
\end{equation}
It is easy to see from Eqs.~(\ref{tzo1}) and (\ref{tzo3}) that,  up to contact terms, we are dealing with divergent expressions. This is because the physical quantity we observe is the discontinuity of $T_{zz}$, not the single values on the two sides of the interface (which is $z=0^+$ in this case).
We also need to calculate $T_{zz}$ at $z=0^{-}$.

We need to solve Eq.~(\ref{Green1time}) with $z<0$ and $z'<0$ (the case with $z>L$ gives the same results for symmetry) with the boundary conditions (\ref{BCs}) at $z=0$ and $g \sim e^{-ikz}$ for $z \rightarrow -\infty$. This can be done similarly to what done in Stakgold's book \cite{stakgold2011green}, by observing that, due to the boundary condition at $z=0$, the sine and cosine Fourier transforms satisfy Eq.~(\ref{Greenconst}).

We can then solve the equations and obtain the following expression:
\begin{equation}\label{tzo}
	\langle{T_{zz}}\rangle_{\omega, k_x,k_y}|_{z=z'=0^-}= - 2 i \int_{0}^{+\infty} k_z^2 \frac{\kappa^2-k_z^2}{(\kappa^2-k_z^2)^2-\alpha_0^2 (k_x^2+k_y^2+k_z^2)} dk_z.
\end{equation}
We similarly manipulate this expression as for the one at $0^+$ and omit contact terms. We obtain the Casimir force per unit area
\begin{equation}
f(L,\alpha_0)=   \int_{-\infty}^{\infty} \frac{dk_x}{2 \pi} \int_{-\infty}^{\infty} \frac{dk_y}{2 \pi}  \frac{1}{L}\sum_{n=1}^{+\infty}  \sum_{\pm} \left[  \frac{\frac{n^2 \pi^2}{L^2} \pm \alpha_0 \frac{n^2 \pi^2}{ L^2 \sqrt{k_x^2+k_y^2+\frac{n^2 \pi^2}{L^2}}} }{\sqrt{|\vec{k}|^2+n^2 \pi^2/L^2\pm \alpha_0 \sqrt{|\vec{k}|^2+ \frac{n^2 \pi^2}{L^2}  }}}\right].
\end{equation}
It is easy to notice that the following formal relation between Casimir pressure and electromagnetic energy density holds:
\begin{equation}\label{dudL}
	f(L,\alpha_0)=-\frac{\partial u_{em}(L,\alpha_0)}{\partial L},
	\end{equation}
confirming our results.
This result can be understood by means of the Thermodynamics' Second Law 
$ dU=T\,dS-p\, dV= -p\, dV $ when the temperature is zero. We observe this since it is in general not valid at temperature $ T$, since entropy can change, but we can calculate the pressure as $p=-\Big(\frac{\partial U}{\partial V}\Big)_{T,S}$.

To evaluate the Casimir force deriving from the expression of the pressure (\ref{uem}), it is convenient to adopt the zeta function regularization method and extend it to a dimension $ d\neq 2$  (see e.g. Milton \cite{https://doi.org/10.48550/arxiv.hep-th/9901011} and Brevik \cite{universe7050133}).

 We employ the Schwinger proper-time representation for the square root \cite{schwartz2014quantum}
\begin{equation}\label{ue}
u_{em}=\frac{1}{2} \sum_{\pm} \sum_n \int \frac{d^d k}{(2 \pi)^d} \int_0^{+\infty} \frac{dt}{t} t^{-\frac{1}{2}} \,e^{-t\left(k^2+n^2 \pi^2/L^2\pm \alpha_0 \sqrt{|\vec{k}|^2+ \frac{n^2 \pi^2}{L^2}  }\right)} \frac{1}{\Gamma(-\frac{1}{2})},
\end{equation}
It is not easy to evaluate it exactly, so we will calculate it perturbatively up to second order in $\alpha_0$.
It is possible to write the Taylor series of the exponential in Eq.~(\ref{ue}) up to the second order in $\alpha_0$ and sum up in the two polarizations explicitly:
\begin{equation}\label{uem}
	u_{em} \simeq \sum_n \int \frac{d^d k}{(2 \pi)^d} \int_0^{+\infty} \frac{dt}{t} t^{-\frac{1}{2}} e^{-t\left(k^2+n^2 \pi^2/L^2 \right)} \left[2+ t^2  \alpha_0^2 \left(k^2+ \frac{n^2 \pi^2}{L^2}\right)\right] \frac{1}{\Gamma(-\frac{1}{2})}.
\end{equation}

As it is usually done in  calculations with integrals, as in Eq.~(\ref{uem}), we carry out the Gaussian integration over k,  adopt the Euler representation of the Gamma function, carry out the sum over $n$ via the definition of the Riemann zeta function \cite{elizalde2012ten}, and we finally obtain
\begin{equation}
	u_{em} \simeq u_0(L,D) +u_{a}(L,D,\alpha_0),
\end{equation}
where
\begin{subequations}
\begin{align}
	&u_0(L,D)=-\frac{1}{2 \sqrt{\pi}} \frac{1}{(4 \pi)^{D/2}} \left( \frac{\pi}{L}\right)^{D+1}  \Gamma \left(-\frac{D+1}{2}\right) \zeta(-D-1) ,\\
&u_{a}(L,D,\alpha_0)=-\frac{\alpha^2_0}{2 \sqrt{\pi}} \frac{1}{(4 \pi)^{D/2}}  \left(\frac{\pi}{L}\right)^{D-1}  \left[ \Gamma \left(\frac{3-D}{2} \right) \zeta(-D-1)+\frac{D}{2} \, \Gamma \left(\frac{1-D}{2} \right) \zeta(-D+1)  \right].
\end{align}
\end{subequations}
The first term $u_0$ is the familiar result of the Casimir energy, while $u_{a}$ is the second-order contribution. For $D=2$ we obtain the Casimir force per unit area
\begin{subequations}
	\begin{align}
		f_0(L)=-\frac{\pi^2}{240} \frac{1}{L^4}  \, \, \, ,\qquad
		f_a(L,\alpha_0)=-\frac{7 \alpha^2_0}{320} \frac{1}{L^2} \, \, \,.
	\end{align}
\end{subequations}
The second-order correction,  going as $\sim \frac{1}{L^2}$, is similar to what was obtained in Ref.~\cite{PhysRevD.81.025015}, although our result differs by a multiplicative factor of $0.63$.

There are some relevant points about the possibilities of physical application of these results:

\begin{itemize}
    \item The approximation $\dot{\Theta} \sim \alpha_0 $ can be used to approximate a coherently oscillating axion field for very small times $t \ll \omega_a^{-1}$, meaning that the spectral theory is a good approximation for frequencies $\omega \gg \omega_a $. This means that it can be adopted when the frequency contributing more significantly to the Casimir force between the plates, which is roughly $\sim L^{-1}$,  is much larger than the axion frequency, so ${L} \ll \omega_a^{-1} $. We have nowadays measurements with percent-level accuracy of the Casimir pressure between parallel plates in the range $d \sim 160$-750~nm \cite{Decca2005,Decca2007}. This means it is a good approximation for the dark matter QCD axion and ALPS with mass $ \lesssim2\, \rm{eV}$.
    
    \item In order the axion correction to be comparable with the usual Casimir pressure in our current Universe with $\Theta_0 \sim 10^{-19}$ \cite{RevModPhys.93.015004}, it needs to be of the order of $L=L_a \sim (\Theta_0 \, \omega_a)^{-1} \sim 10^{14} \, m\left( \frac{10^{-2} \SI{}{eV}}{m_a}\right) $ and this is not viable with possible experiments and the former mass range.
    
    \item It can be of interest for topological insulators, since $b_0 \sim 10^{-1} \, \SI{}{eV}$, corresponding to $\alpha_0 \sim 10^{-1} \,\SI{}{eV}$ with now $L_a \sim \SI{2}{\,\mu m}$, as a percent-level correction.
    It can be of further theoretical interest for applications to vacuum fluctuations in the Early Universe, where we can expect values $\alpha_0 \sim g_{\gamma} m_a \sim 10^{-2} \,m_a$.
    
    %check value
\end{itemize}
The toy model \ref{toy} , which we will develop in the following, is very good for catching the fundamental aspects of the optical properties of the real axion domain walls, since the only ways they could differ are in the expected "smoother" interfaces for the real domain wall. Such effects involve considering the possibility of further multiple reflections and transmissions in the limited region near to interfaces, which are subleading if the reflection coefficients depend on $g_{a \gamma \gamma}$ and then the optical depth is small.  
\subsection{High frequency approximation for the Green's function}
Before working out the case of an oscillating axion field, we can get some quantitative understanding in the following way.  We can substitute $\Theta(t)=\Theta_0 \sin{(\omega_a t)} $ in Eq.~(\ref{greenaqua}), and calculate the Fourier Transform
\begin{equation}\label{Green2time}
	(-\omega^2+|\vec{k}|^2) \tilde{G}_{ij}(\omega,\vec{k}) -i\epsilon_{lik}  k_l \frac{\Theta_0 \omega_a}{2} \left[ \tilde{G}_{ij}(\omega+\omega_a,\vec{k})+\tilde{G}_{ij}(\omega-\omega_a,\vec{k})  \right] =\delta_{ij},
\end{equation}
where we have used the modulation property of the Fourier Transform \cite{Gardiner2004,stakgold2011green}.
The high-frequency approximation consists of assuming $\omega \gg \omega_a$.
In this way, up to the first order in $\omega_a$,  the Green's function satisfies the following equation,
\begin{equation}\label{Green2timeHF}
	(-\omega^2+|\vec{k}|^2) \tilde{G}_{ij}(\omega,\vec{k}) -i\epsilon_{lik} \, k_l \,\Theta_0\, \omega_a \tilde{G}_{kj}(\omega,\vec{k})   =\delta_{ij}.
\end{equation}
It has the same form as the previous equation for (\ref{Greenconst}) and has accordingly the same solutions, with just the substitution $\dot{\Theta} \rightarrow \Theta_0 \omega_a$.
The interpretation of Eq.~(\ref{dothecon}) terms is still valid in this case.
\subsection{Case with an oscillating axion field: Production of real photons}
In order to treat this case properly  we need to work out  Eq.~(\ref{Green1time}) in more detail.\\
We assume that the electromagnetic wave vector is directed along a fixed direction, without losing generality. Actually, the axion field is approximated to be dependent on time but spatially homogeneous, so it is isotropic.

We  perform   a Fourier Transform in space domain and put for simplicity $\vec{k}=|\vec{k}| \hat{e}_z$, since we are in the isotropic vacuum, so we can rotate our $z$ axis in the direction of $\vec{k}$.

%We will solve this coupled differential equation system by a perturbative approach.
We expand the formalism up to second order in $\Theta_0$  and consider $G_{xy}$ and $G_{yy}$, whose equations are coupled only between themselves.
To  order zero, we have
\begin{equation}
	\begin{split}
		G_{yy}(\vec{k},t,t') &= \theta_H(t-t') \; \times \frac{1}{|\vec{k}|}\, \sin\left[|\vec{k}| (t-t')\right] \label{greensol},\\
		G_{xy}(\vec{k},t,t')&=0,
	\end{split}
\end{equation}
according to the null initial conditions we adopt, where the function $\theta_H(t-t')$ is the usual Heaviside function.

We can solve the resulting D'Alembert equation and specify the time dependence of the  axion field in order to obtain as in Ref.~\cite{FAVITTA2023169396} the following perturbative expansion in Fourier space
\begin{equation}
	\tilde{G}_{jk}(\vec{k}, \omega) \sim \tilde{G}^{(0)}_{jk}(\vec{k}, \omega)+\tilde{G}^{(1)}_{jk}(\vec{k}, \omega)+\tilde{G}^{(2)}_{jk}(\vec{k}, \omega),
\end{equation}
where
\begin{subequations}
	\begin{align}
		\tilde{G}^{(0)}_{jk}(\vec{k}, \omega)&=\frac{1}{(\omega^2-k^2)} ,\\
		\tilde{G}^{(1)}_{jk}(\vec{k}, \omega)&=\frac{i}{(\omega^2-k^2)^2}\left[\frac{1}{(\omega-\omega_a)^2-|\vec{k}|^2}+\frac{1}{(\omega+\omega_a)^2-|\vec{k}|^2} \right] \epsilon_{jkl}\beta_l,\\
		\tilde{G}^{(2)}_{jk}(\vec{k}, \omega)&=\frac{\beta^2}{(\omega^2-k^2)}\left[\frac{1}{(\omega-\omega_a)^2-|\vec{k}|^2}+\frac{1}{(\omega+\omega_a)^2-|\vec{k}|^2} \right]\left[ \delta_{jk}+\frac{\epsilon_{jkl}\beta_l}{\omega^2-k^2} \right]+\\+&\frac{1}{(\omega^2-k^2)} \left[\frac{1}{(\omega-\omega_a)^2-|\vec{k}|^2}+\frac{1}{(\omega+\omega_a)^2-|\vec{k}|^2} \right]^2\Theta_0^2 \omega_a^2 k_j k_k.
	\end{align}
\end{subequations}
It is worth mentioning that the results for axion oscillating background are compatible with models of axion echo, treated recently in Ref.~\cite{PhysRevLett.123.131804}.
\begin{figure}[h]
	\begin{center}
		\includegraphics[width=0.7 \textwidth, keepaspectratio=true]{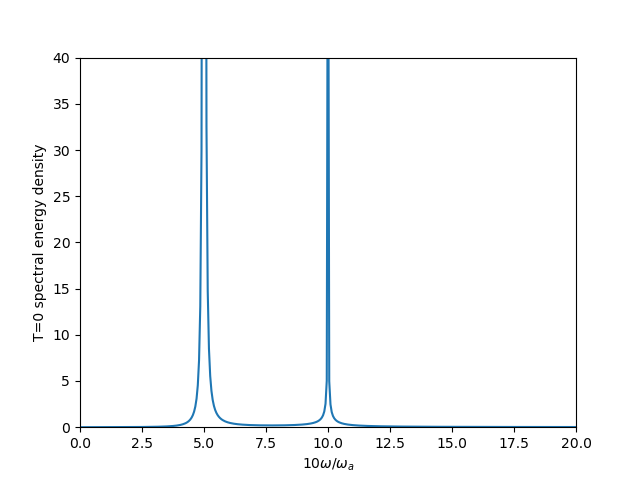}
	\end{center}
	\caption{Plot of the function $f(\omega,\Theta_0,\omega_a )$in Eq.~(\ref{eqf}) with $\Theta_0=1$ for simplicity of visualization.
		 For $\omega >\omega_a$ we have the expected behaviour from Eq.~(\ref{dothecon}), whose infrared cut-off is the peak at $\omega=\omega_a$, associated with the limiting case of the decay of one axion to only one detectable photon. Furthermore, we have a peak at $\omega=\frac{\omega_a}{2}$, a result compatible with the axion echo phenomenon \cite{PhysRevLett.123.131804}, where the physical process behind is the decay of an axion of energy $\omega_a$ to two photons with energy $\frac{\omega_a}{2}$.}
         \label{coherent1}
\end{figure}

It is also possible in this case to evaluate the second-order contribution to energy density:
\begin{equation}
\rho^{(2)}(\omega)\; d \omega=d \omega \; \omega^2 \left[ \frac{1}{2} f(\omega,\Theta_0, \omega_a)+ \frac{1}{e^{\omega/T}-1}  f(\omega,\Theta_0, \omega_a)  \right],
\end{equation}
where
\begin{equation}\label{eqf}
f(\omega,\Theta_0, \omega_a)= \frac{ \Theta_0^2 \omega_a^2 \omega (8 \omega^4+\omega_a^4)}{2(\omega+\omega_a)|\omega-\omega_a| (\omega_a^2-4\omega^2)^2}.
\end{equation}
A numerical plot of the function $f$ is displayed in Figure \ref{coherent1}. From this plot, we observe how the validity of an expression of Casimir force per unit area between two conducting plates analogous to the case of a time-increasing axion field is true for $L^{-1} \gg \omega_a$, as mentioned before. For a bigger $L$, one appreciates the deviations from that case and considers that the behaviour of the modification to energy density going as $1/\omega$ needs to be corrected with a $1/|\omega-\omega_a|$ factor and has a lower cut-off at $\omega=\omega_a$. Furthermore, one needs to take care of the additional contribution of "axion echo" \cite{PhysRevLett.123.131804} at $\omega=\frac{\omega_a}{2}$.

This last contribution has a straightforward physical interpretation: it is associated with the production of virtual photons with frequency $\omega=\frac{\omega_a}{2}$ from the decay of an axion, while the peak at $\omega=\omega_a$ is associated with a decay of one axion to only one detectable photon.
\subsection{A possible way to boost axionic Casimir effect in a current Universe experimental set-up}
In this subsection, we consider a dielectric system containing two interfaces separating media of refractive indices $n_1$ and $n_2$ in a uniform magnetic field $ B_e$. We assume that the media are elastic, so that medium 2 can be "turned back" and glued to the left side of medium 1. This is a ring-formed system that we have treated in Ref.~\cite{PhysRevD.107.043522} and extended in Ref.~\cite{FAVITTA2023169396}.
\begin{figure}[h]
\centering	\includegraphics[scale=0.3]{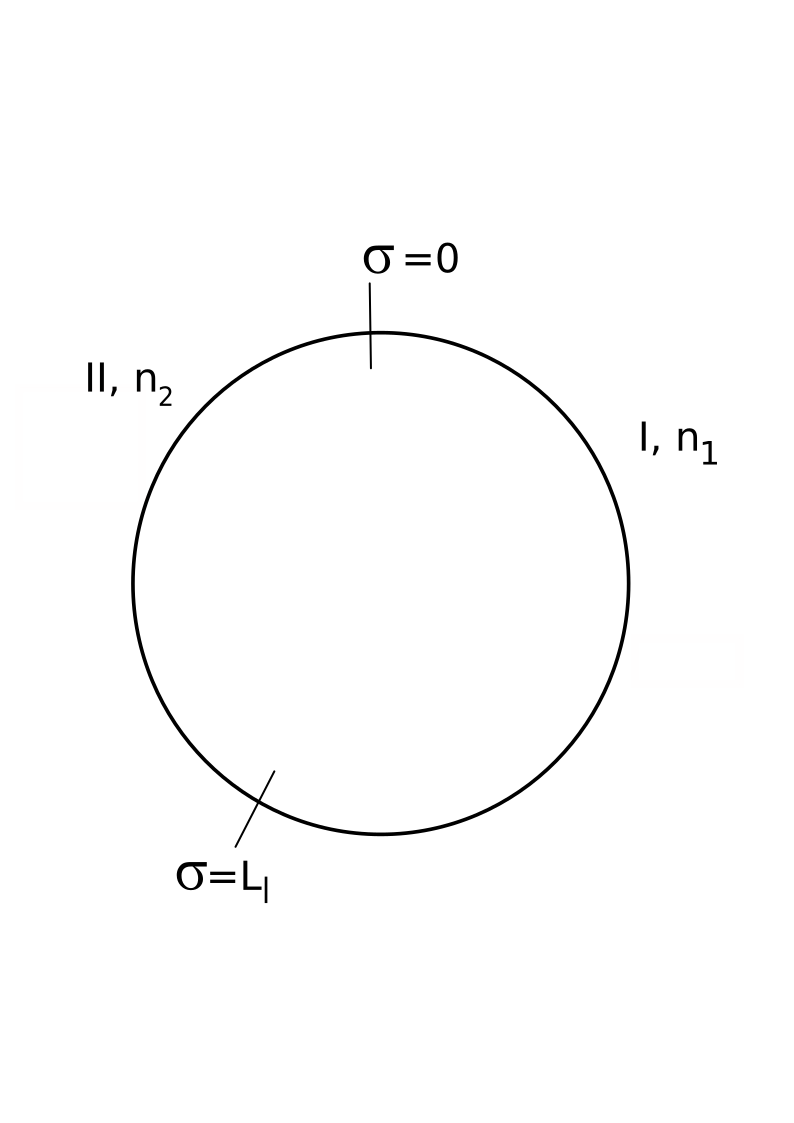}
	\caption{Geometry and notation of the closed string of interest. Figure from Refs.~\cite{PhysRevD.107.043522,FAVITTA2023169396}.}\label{fig:12}
\end{figure}
\\Figure \ref{fig:12} shows the configuration, and here we give the most fundamental results.\\ We take $\sigma$ to be the length coordinate along the string, such that the two junctions are at $\sigma=0$ and $\sigma = L_{1}$, where $L_1$ is the length of region I. The total length of the string is $L=L_1+L_{2}$, where $L_2$ is the length of region II, and the junctions $\sigma=0$ and $\sigma=L$ are coincident.  The string lies in the $xy$ plane, and a strong uniform magnetic field ${\bf B}_e$ is applied in the $z$ direction.

 We evaluate the stationary oscillations of the electromagnetic oscillations in the string. If $E_I(\sigma,t)$ and $E_{II}(\sigma,t)$ are the electric fields in the two regions, we have, using a complex representation, 
\begin{subequations}
	\begin{align}
		E_I(\sigma,t)=\xi_Ie^{in_1\omega \sigma-i\omega t}+\eta_Ie^{-in_1\omega \sigma -i\omega t},\\
		E_{II}(\sigma,t)=\xi_{II}e^{in_2\omega(\sigma-L_1)-i\omega t}+ \eta_{II}e^{ -in_2\omega (\sigma-L_1)-i\omega t},
	\end{align}
\end{subequations}
where $\xi_I, \eta_I,\xi_{II},\eta_{II}$ are constants. Analogously, we obtain the magnetic fields.
One could treat such a system by considering a generic $\omega$, however, as we discussed in Ref.~\cite{PhysRevD.107.043522}, the axion-generated oscillations would be suppressed by a factor $\delta(\omega-\omega_a)$, so the only relevant frequency is $\omega_a$ for our purposes. 

By imposing the appropriate boundary conditions at the junctions,  we can get analytical expressions for particular cases of interest,  namely when $n_2$ becomes large in comparison to $n_1$ so that the ratio $x=\frac{n_1}{n_2}\rightarrow 0$ and when the length $L_I\rightarrow 0$, corresponding to a point defect sitting on an otherwise uniform string. As $\delta_1\rightarrow 0$ in this case, we see  that
\begin{equation}
	\xi_I= \frac{E_0}{2\varepsilon_1}, \quad x\rightarrow 0, L_I \rightarrow 0,
\end{equation}
which is a real quantity.
We also have for this last case:
\begin{align}
	\eta_I=\xi_I,\\
	\eta_{II}=\xi_{II}=0.
\end{align}
We have then an electromagnetic energy density $u_{1def}$ inside the point defect:
\begin{equation}
	u_{1def}=\frac{E_0^2}{4 \epsilon_1}.
\end{equation}
There are interesting similarities and differences with the system of a single dielectric plate treated in Ref.~\cite{Millar_2017}.
Indeed, our system is similar to it, but when the two outward dielectric media are closed in order to form a ring.

We have suggested in Ref.~\cite{FAVITTA2023169396} the possibility that, adopting the same idea of Ref.~\cite{Millar_2017} with a ring of multiple dielectrics, we can boost the stationary oscillations and adopt it for experimental purposes.
\section{Optical properties of domain walls}\label{wavedomain}
The purpose of this subsection is to study the propagation of an electromagnetic wave in the presence of a toy model for the axion domain wall. The toy wall is taken to be a planar sheet in which the axion field increases linearly with the longitudinal coordinate $z$ between $z=0$ and $z=L$ and it is constant and uniform outside. We assume a static wall, so $a=a(z)$ with
\begin{equation}\label{toy}
\Theta(z)=\begin{cases}
		0  \; \; \; \textit{if} \; \; \; z < 0, \\
		\frac{\Theta_0}{L}z \; \; \; \textit{if} \; \; \; 0< z < L,\\
		\Theta_0 \; \; \; \textit{if} \; \; \; z>L.
	\end{cases}
\end{equation}
The model was treated earlier in Ref.~\cite{universe7050133}, and we extended it to use as a toy model for "localised" axion configurations, in particular the axion domain walls in Section~(\ref{axioncosmodw}).
\begin{figure}[h]
	\begin{center}
		\includegraphics[width=0.9 \textwidth, keepaspectratio=true]{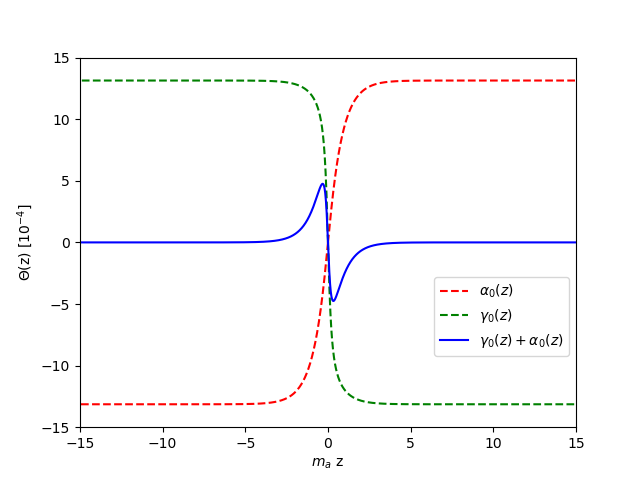}
	\end{center}
	\caption{Explicit graph of the static axion domain wall configuration $\Theta(z)=g_{a \gamma \gamma} a_{\rm phys}(z)$ where $a_{\rm phys}(z)=\gamma_0(z)+\alpha_0(z)$ takes care of respectively pion and $\eta$ contribution. It is reported in the solid blue line, and evaluated as in Ref.~\cite{PhysRevD.32.1560}. We also reported the corresponding values for $\gamma_0(z)$ and $\alpha_0(z)$.}
    \label{axion-domain-wall}
\end{figure}
We will now investigate the effects of the axion on the electromagnetic wave in the system. 
As shown in Ref.~\cite{PhysRevD.32.1560} the 'effective field' in Axion-Modified Maxwell equations for a QCD axion domain wall gets two contributions: one $\alpha_0(z)$ from the phase of Peccei-Quinn field, i.e. the axion field and $\gamma_0(z)$ from the neutral pion field, since the pion has an interaction term to electromagnetic field analogous to Eq.~(\ref{eq0}).
The two contributions are related because of the equilibrium condition for a static wall:
\begin{equation}
	\tan{\gamma_0}=-\xi \tan{\alpha_0},
\end{equation}
where $
\xi=\frac{m_u-m_d}{m_u+m_d}$, where $m_u$ is the mass of the up quark and $m_d$ is the mass of down quark.
In our numerical evaluation here we used the value $\xi=0.3$ as in the original paper and we evaluated $\alpha_0(z)$ by solving numerically the equation:
\begin{equation}\label{eqalpha}
\frac{1}{m_a^2}\frac{d^2 \alpha_0}{dz^2}=\frac{\sin{\alpha_0} \cos{\alpha_0}}{\sqrt{\cos^2{\alpha_0}+ \xi^2 \sin^2{\alpha_0} }},
\end{equation}
We plot one of the possible solutions of Eq.~(\ref{eqalpha}) in Figure \ref{axion-domain-wall}. 
\subsection{Exact calculations for toy model}\label{exactcal-toy}
As above, we consider an axion configuration that is dependent on $z$ only.  At first, we assume $\beta(z) = \partial_z \Theta(z)$   to be an arbitrary function of $z$.
The incident electromagnetic wave is propagating along the $z$-axis and is transversely polarized  (the same configuration was assumed in  Ref.~\cite{PhysRevD.32.1560}).
This configuration is useful because a domain wall  is invariant under Lorentz boosts parallel to the wall surface, so it is always possible to find a reference frame where the wave is incident normally, as also noticed in the same \cite{PhysRevD.32.1560}.

This simplifies the calculations to the first order in the 4-potential because the effective axion charge density is zero.
We then consider the following equation:
\begin{equation}\label{eqHeav}
	\Box \vec{A}=-\nabla \Theta \wedge \frac{\partial \vec{A}}{\partial t},
\end{equation}
and search for an exact solution.
We calculate the Fourier transform of this equation in time $t$ and in the $x$  and $y$ coordinates,  and  get the following expressions for the FT components of the vector potential:
\begin{equation}\label{fracco}
	\begin{split}
		(-\partial_z^2-\kappa^2)\mathcal{A}_x&=i \omega \beta(z) \mathcal{A}_y, \\
		(-\partial_z^2-\kappa^2)\mathcal{A}_y&=-i \omega \beta(z) \mathcal{A}_x, \\
		(-\partial_z^2-\kappa^2)\mathcal{A}_z&=0.
	\end{split}
\end{equation}
When $k_x=k_y=0$, we have $\kappa=\omega$.
Equations (\ref{fracco}) imply that the gauge fields defined as $\mathcal{A}_{\pm}=\mathcal{A}_x \pm i \mathcal{A}_y$  satisfy:
\begin{equation}\label{fracco1}
	\begin{split}
		(-\partial_z^2-\kappa^2)\mathcal{A}_{\pm}&=\pm  \omega \beta(z) \mathcal{A}_{\pm}.
	\end{split}
\end{equation}
From now on, we will assume to deal with the toy model configuration (\ref{toy}).
 We notice that the axion configuration (\ref{toy}) does not have well-defined values of the z derivative at $z=0,L$ since the derivative has a jump in these points. However, this is not problematic for treating Eq.~(\ref{eqHeav}) since vector potential and its first spatial derivative are continuous.

Indeed, in the following calculations, we basically solve Eq.~(\ref{eqHeav}) separately for the regions $z<0$, $z>L$ and $0<z<L$ and then impose continuity of vector potential and its first z-derivative.
This last procedure would also be good in the case of an exact domain wall, as an approximation.

We consider a plane wave propagating along z.
Then, the equations (\ref{fracco1}) can be interpreted as follows.

The gauge fields $\mathcal{A}_{\pm}$ correspond to left/right circular polarizations, then this means that the phase velocities are different for a left circular polarized wave and a right one, and the optical angle rotates inside the slab.

\begin{figure}\centering
	\includegraphics[scale=0.5]{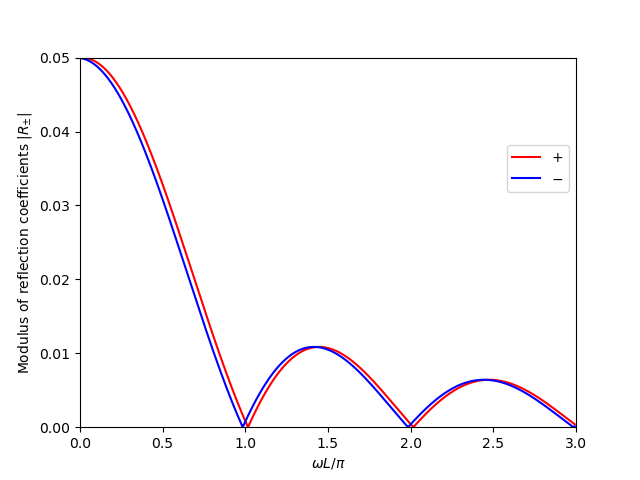}
	\caption{Plot of reflection coefficient modulus for the two circular polarizations.}
	\label{Plotto}
\end{figure}

\begin{figure}\centering
	\includegraphics[scale=0.4]{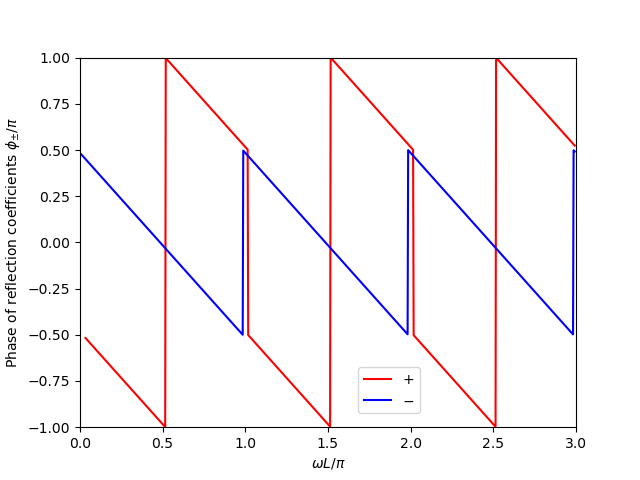}
	\caption{Plot of the reflection coefficient phase for the two circular polarizations.}
	\label{Plotto1}
\end{figure}

We now evaluate the reflected component of an incident orthogonal wave. The continuity conditions at the walls are that the  $A_{\pm}$ fields, as well as their $z$ derivatives, are continuous. Thus, in the leftmost region,
\begin{equation}
	\mathcal{A}_{\pm}(z)_{\text{left}}=A_{\pm}e^{-i \kappa z}+B_{\pm}e^{i \kappa z},
\end{equation}
where the $B$ term is the reflected wave. The wave transmitted outside the slab is
\begin{equation}
	\mathcal{A}_{\pm}(z)_{\text{right}}=C_{\pm}e^{-i \kappa z}.
\end{equation}

\begin{equation}
	A_{\pm}(z)_{\text{in}}=A'_{\pm}e^{-i \alpha_{\pm} z}+B'_{\pm}e^{i \alpha_{\pm} z}.
\end{equation}
where $A'_{\pm}$ is the wave transmitted inside the wall.
To evaluate the reflection coefficient (defined as $R_{\pm}=\frac{B_{\pm}}{A_{\pm}}) $ and the internal transmission coefficient, defined as $T_{\pm}=\frac{A'_{\pm}}{A_{\pm}}$, we need to impose the continuity conditions on the surfaces $z=0$ and $z=L$:
\begin{equation}
	\begin{cases}
		A+B=A'+B' & \text{Continuity of $\mathcal{A}_{\pm} $ at z=0},\\
		\omega A-\omega B=\alpha A'-\alpha B' & \text{Continuity of first derivatives of $\mathcal{A}_{\pm} $ at z=0},\\
		e^{-i \alpha L} A'+e^{i \alpha L} B'=C e^{-i \omega L} & \text{Continuity of $\mathcal{A}_{\pm} $ at z=L}, \\
		-i \alpha e^{-i \alpha L} A'+i \alpha e^{i \alpha L} B'=-i \omega C e^{-i \omega L} & \text{Continuity of first derivatives of $\mathcal{A}_{\pm} $ at z=L}.
	\end{cases}
\end{equation}
From it, we get
\begin{subequations}\label{reflecto}
	\begin{align}
		T_{\pm} &= \frac{2 e^{2 i\alpha_{\pm} L} \omega (\alpha_{\pm} + \omega)}{(-1 + e^{2 i \alpha_{\pm} L}) \alpha^2 + (-1 + e^{2 i \alpha L}) \omega^2 + 2 (1 +e^{2 i \alpha_{\pm} L}) \alpha_{\pm} \omega}, \\
		R &= -\frac{(-1 + e^{2 i \alpha_{\pm} L}) (\alpha^2_{\pm} - \omega^2)}{(-1 + e^{2 i \alpha L}) \alpha^2_{\pm}
			+ (-1 + e^{2 i \alpha_{\pm} L}) \omega^2 + 2 (1 + e^{2 i \alpha L}) \alpha_{\pm} \omega}.
	\end{align}
\end{subequations}
Some fundamental properties of the reflection coefficient are shown in Figs.~(\ref{Plotto}) and (\ref{Plotto1}). Modules and phases are given, where we have taken the unit frequency to be $1/L$ and  $\Theta_0=10^{-3}$ for simplicity.

We notice that there is the most reflection, for both polarizations,  when $\omega=0$ and $\omega=\frac{\Theta_0}{L}$, but there is still reflection at higher frequencies at intervals of $\omega=\frac{\Theta_0}{L}$.
%We  reemphasize that even the simple  domain wall ((\ref{toy})) is useful in order to understand the behavior of  the  axion configuration located to a planar region.
%The relative calculations can be done and are treated in the Appendix \ref{appendixa}.
In the following, we will adopt approximative expressions for the reflection coefficients $R_{\pm}(\kappa)$.
For simplicity we define the useful quantity $m_L=\frac{\Theta_0}{L}$.

The reflection coefficients can be approximated to be $R^2_{+}(k_z) \sim \zeta m_L \, \delta(k_z)$ and $R^2_{-}(k_z) \sim \zeta \, m_L \, \left[\delta(k_z+m_L)+\delta(k_z-m_L)\right]$ where $\zeta$ is a numerical factor of order of unity to fit better the behaviour of the reflection coefficients.
This can be justified by noticing that, although the exact behaviours of the square modula of $R_{\pm}(k_z) $  are complicated, it is clear that a crucial frequency for our system is the same $m_L$, and  by the graphs in Figures~(\ref{Plotto}) and (\ref{Plotto1}) and a qualitative description of them, they can be approximated to be $R^2_{+}(k_z)\sim  \frac{\zeta}{\pi} \Theta_0^2  m_L^2 \frac{\sin^2{(k_z/m_L)}}{k^2_z}$ and  $R^2_{-}(k_z) \sim \frac{\zeta}{\pi} \Theta_0^2  \frac{\sin^2{(k_z/m_L-1)}}{(k_z/m_L-1)^2}$.
Then one exploits the distributional relation:
\begin{equation}
	\frac{\sin^2{(\epsilon t)}}{t^2} \sim \pi\, \epsilon\, \delta(t)  \qquad \text{for} \qquad  \epsilon \rightarrow 0,
\end{equation}
and we can get the former expressions posing $\epsilon=( m_L)^{-1}$.

The adoption of the Dirac delta limit is only for simplicity and takes care of the main contribution coming from the main peak of Figure~(\ref{Plotto}), however, an exact calculation would correspond to taking care of the secondary peaks of the graphs in Figure~(\ref{Plotto})  along with their detailed structure, from which we adopt the numerical factor $\zeta$.
\subsection{Dispersion relation}\label{Casimirdisperdo}
In the previous section we obtained the following dispersion relation for the toy model (\ref{toy}):
\begin{equation}
	\omega^2-k_{z}^2=\mp \omega \beta(z),
\end{equation}
implying
\begin{equation}
	\omega_{\pm}=\sqrt{ \left(k_z^2+ \frac{1}{4}\beta^2(z) \right)} \pm \frac{1}{2}\beta(z).
\end{equation}
This is the same dispersion relation as obtained in  Ref.~\cite{universe7050133}, which is not the general dispersion relation.
This can be understood from our discussion in the previous section, since it applies to just one linear polarization in the reference frame where the wave vector lies along the z direction. 

We obtain the general form by applying an inverse Lorentz boost in x and y directions and obtain the following form, similar to the one in Ref.~\cite{PhysRevD.100.045013}:
\begin{equation}
	\omega=\sqrt{k_x^2+k_y^2+\left(\sqrt{ \left(k_z^2+ \frac{1}{4}\beta^2_z \right)} \pm \frac{1}{2}\beta_z \right)^2   }.
\end{equation}
Having in mind this full expression, we can discuss it physically. It tells us that we have two different dispersion relations between $+$ and $-$ polarization and this lead to a rotation of the optical angle.

\subsection{Optical activity of axion medium and practical application}\label{optical}

It is worth mentioning that the mode splittings in the dispersion relations obtained in Subsections \ref{constdot} and \ref{Casimirdisperdo} are analogous to those in the Faraday effect, where the polarization rotation is proportional to the longitudinal strong magnetic field\cite{prati2003propagation, mansuripur1999faraday}.
Moreover, there is also a strong connection with the Casimir polarization rotation observed in a chiral medium when a strong transverse magnetic field is present
\cite{PhysRevB.99.125403, hoye2020casimir}.

This effect has been treated in much detail for Axion Electrodynamics with only an axion field that holds constant spatial gradients and time derivative, in Refs.~ \cite{RevModPhys.93.015004} and \cite{PhysRevD.41.1231}. %ADJUST
It is indeed easy to notice how our theory with a fixed axion background is equivalent to a modified Electrodynamics with an additional Chern-Simons term:
\begin{equation}\label{carroll}
	\mathcal{L}_a=-\frac{1}{2} p_{\alpha} A_{\beta} \tilde{F}^{\alpha \beta},
\end{equation}
if $p_{\alpha}=g_{a \gamma \gamma}\partial_{\alpha} a(x)$, regardless if axion derivatives are constant and uniform or not.\\
The theory of this modified electrodynamics when $p_{\alpha}$ are constants is very well treated there, where they find that the Axion Electrodynamics medium is characterized by having an optically active vacuum with constant rotation of the optical angle 
\begin{equation}
	\frac{d \Phi}{dt}=\frac{1}{2} g_{a \gamma \gamma}  \sqrt{\frac{\mu}{\varepsilon}}\left(\dot{a}+\frac{\omega}{k^2}  \vec{k} \cdot \nabla a    \right).
\end{equation}
This comes from the dispersion relation:
\begin{equation}
\omega_{\pm}=\frac{|\vec{k}|}{\sqrt{\varepsilon \mu}} \pm \frac{1}{2} g_{a \gamma \gamma}  \sqrt{\frac{\mu}{\varepsilon}}\left(\dot{a}+\frac{\omega}{|\vec{k}|^2}  \vec{k} \cdot \nabla a \right)+ \mathcal{O}(g_{a \gamma \gamma}^2),
\end{equation}
which is compatible with our results. 

We have used this idea in Ref.~\cite{doi:10.1142/S0217751X24500040} for the following system of interest for topological materials.

\subsection{Application to topological material}\label{application}
We consider a usual system with planar symmetry along a z-axis.
We consider the regions $z<L$ and $z>0$ to be perfectly conducting, and the intermediate region $0<z<L$ to be filled with a uniform dielectric with material constants $\varepsilon$ and $\mu$. The effective axion field is also assumed to fill the intermediate region with a constant time derivative and constant gradient along the z-direction and to be zero outside.

\begin{figure}
    \centering
    \includegraphics[width=0.9\linewidth]{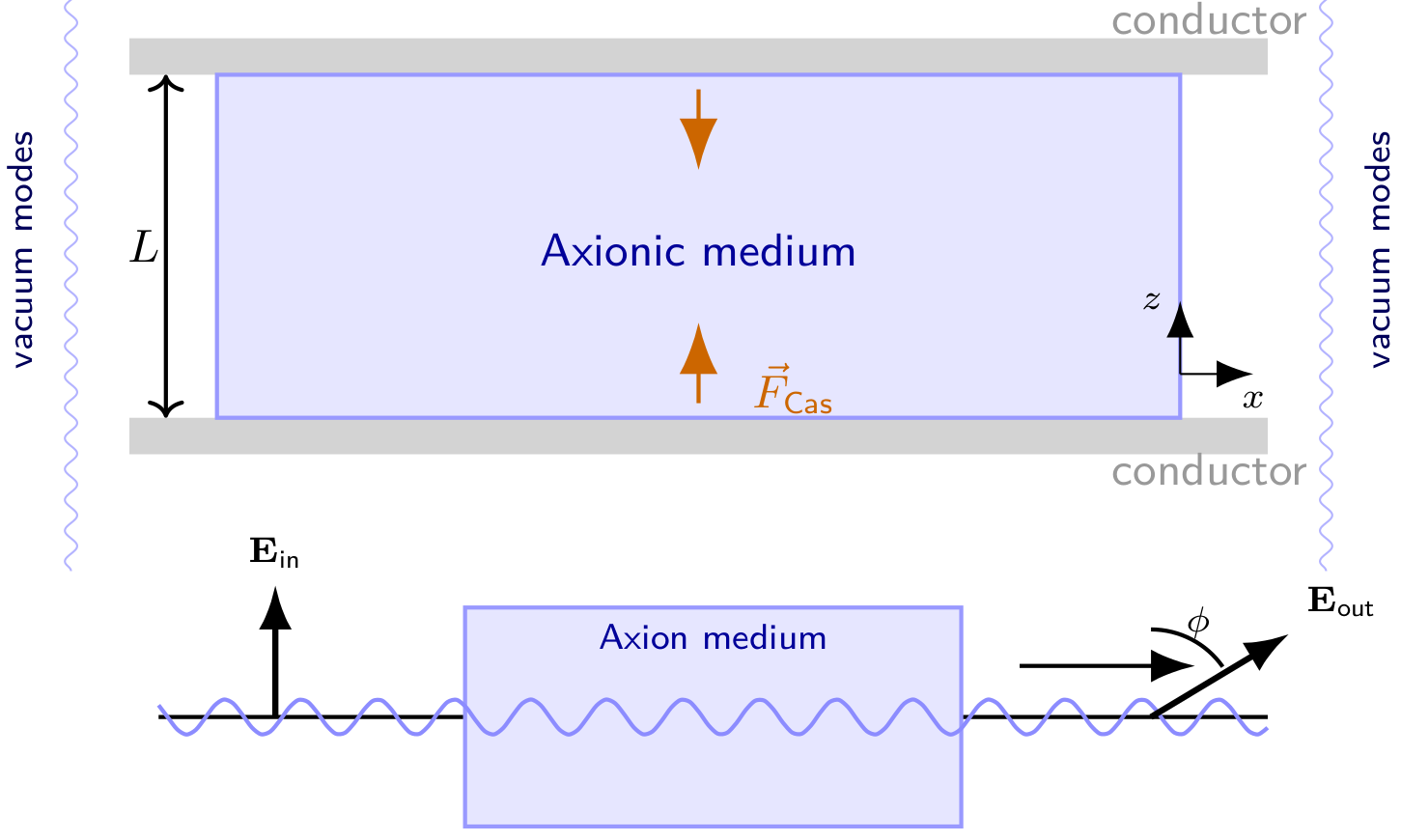}
    \caption{A schematic representation of the Casimir set-up with an axion material in the middle, which we treat here. Since the material is optically active, which means that a propagating plane wave suffers a rotation $\phi$ of the optical angle when passing through it as we show below, we have a further modification to the Casimir force.}
    \label{fig:axiontopo}
\end{figure}

There are no other external fields. We display our system in Fig.~\ref{fig:axiontopo}.

We calculate the Casimir free energy $F$ between the plates per unit area, and begin with the known expression from ordinary electrodynamics at temperature $T$,
\begin{equation}
F= \frac{1}{\pi\beta}{\sum_{m=0}^\infty}^ \prime \int_{n\zeta_m}^\infty \kappa d\kappa \ln (1-e^{-2\kappa L}). \label{energy}
\end{equation}
Here  $\zeta_m=2\pi m/\beta_T$ with $\beta_T =1/T$ is the Matsubara frequency, and $\kappa$ is defined by $\kappa^2=k_\perp^2+n^2\zeta^2$ with $n^2=\varepsilon\mu$. Note that $\kappa$ is defined here in a conventional way, as in Refs.~\cite{https://doi.org/10.48550/arxiv.hep-th/9901011,Brevik2006NJP}). %add brevik 2008

The quantity $\lambda$ defined in Eq.~(\ref{kappa}) is different, although physically related.

We have shown in Ref.~\cite{doi:10.1142/S0217751X24500040}
that the TE and TM modes will rotate between the plates, and we may read this problem as an interaction between harmonic oscillators in the two plates. 
As before, there occurs a slow rotation of the polarization plane, proportional to $z$ and $\chi$,  as the wave propagates through the medium, with a consequent gradual transition of the TM mode into a TE mode, and in the reverse direction.

This is what we treated in Ref.~\cite{doi:10.1142/S0217751X24500040} and we now outline our original results.

We assumed the following form for the fields,
\begin{equation}
{\bf E}({\bf x},t)= {\bf E}(z)e^{i\Phi}, \quad \Phi= {\bf k_\perp \cdot x_\perp}-\omega t,
\end{equation}
 and start from the wave equation for $\bf E$.

 We define $\lambda^2$ as
 \begin{equation}
 \lambda^2= \varepsilon\mu\omega^2-k_\perp^2, \label{kappa}
 \end{equation}
and write out all three component equations,
\begin{equation}
E_x''(z)+\lambda^2 E_x(z)= i\mu k^2\xi E_y(z)-i\mu\alpha k_y E_z(z), \label{modified}
\end{equation}
\begin{equation}
E_y''(z)+\lambda^2 E_y(z) = -i\mu k^2\xi E_x(z) +i\mu \alpha k_x E_z(z),
\end{equation}
\begin{equation}
E_z''(z)+\lambda^2 E_z(z) =i\mu \alpha \omega k_\perp^2\sin kz.
\end{equation}
where we have defined for convenience $\xi=\frac{\alpha \lambda+\beta \omega}{k^2}$.

We solve the equation for $E_x(z)$ as an inhomogeneous differential equation,  observing that the two basic solutions for the homogeneous equation can be chosen as $\psi_1= \sin \lambda z$ and $\psi_2= \cos \lambda z$, with Wronskian $\psi_1 \psi_2'-\psi_2\psi_1'=-\lambda$.  and employ the usual expressions for TE modes. We write the solution $E_x(z)$ as a sum of two terms,
\begin{equation}
E_x(z)= E_x^{(1)}(z)+  E_x^{(2)}(z),
\end{equation}
where $E_x^{(1)}$ and $E_x^{(2)}$ refer respectively to $\psi_1$ and $\psi_2$. Some calculation leads to the expressions
\begin{equation}
E_x^{(1)}(z)= C_1\sin \lambda z -\frac{i}{2}(\mu \omega N)kk_x\xi\left[ \frac{1-\cos (k-\lambda)z}{k-\lambda}+\frac{1-\cos (k+\lambda)z}{k+\lambda}\right]\sin \lambda z,
\end{equation}
\begin{equation}
E_x^{(2)}(z)=   C_2\cos \lambda z   +  \frac{i}{2}(\mu \omega N)kk_x \xi \left[ \frac{\sin(k-\lambda) z}{k-\lambda}-
\frac{\sin(k+\lambda)z}{k+\lambda}\right] \cos \lambda z,
\end{equation}
showing how the axions modify this field component to order $\xi$; $C_1$ and $C_2$ are constants.
  Since the difference between $\lambda$ and $k$ is small, we have replaced $\lambda$ with $k$ in the noncritical nontrigonometric terms. The expressions show that, to  first order, we can make the same replacement in the trigonometric
	terms too. Requiring the total field component $E_x(z)$ to be zero at $z=0$ and $z=L$ we find that $C_1$ is undetermined, while $C_2=0$.  We can thus set $ C_1=N\omega k_y$ to agree with the zeroth-order expression. Altogether,  
\begin{equation}
E_x^{(1)}(z) = N\omega k_y\left[ 1- \frac{i\mu}{4}\frac{k_x}{k_y}\xi (1-\cos 2\lambda z)\right] \sin \lambda z,
\end{equation}
\begin{equation}
E_x^{(2)}(z)= \frac{i}{2}(\mu \omega N)\lambda k_x \xi \left[z-\frac{\sin 2\lambda z}{2\lambda}\right] \cos \lambda z.
\end{equation}
The imaginary terms mean a rotation of the transverse field ${\bf E}_\perp$ in the $xy$ plane. It is of main interest to us the rotation angle proportional to $z$,  similarly to the cases of Faraday effect and chiral electrodynamics. We will therefore focus on this term, and write the full component $E_x$ in the form
\begin{equation}\label{Ex}
E_x(z)= N\omega k_y[\sin \lambda z + i\gamma_x(z)\cos \lambda z].
\end{equation}

However, to evaluate the rotation of the optical angle, we need to consider that, analogously to $E_x$ in Eq.~(\ref{Ex}), we can get the following expression for $E_y$:
 \begin{equation}\label{Ey}
 	E_y(z)= -N\omega k_x[\sin \lambda z - i\gamma_y(z)\cos \lambda z],
 \end{equation}
 where
 \begin{equation}
 	\phi_y(z)= \frac{1}{2}(\mu \lambda z)\frac{k_y}{k_x}\xi.
 \end{equation}
We now observe that we can write the usual fields $E_{\pm}(z)=E_{x}(z) \pm i E_{y}(z)$ , through the equations (\ref{Ex},\ref{Ey}),  as:
\begin{equation}\label{solpm}
E_{\pm}(z)=N \omega (k_y \mp i k_x) \left[ \sin{\lambda z} \mp \frac{1}{2}\mu (\alpha k+\beta \omega )z    \right].	
\end{equation}
To grasp the physical meaning of this expression we can observe that, since we work out the electric and magnetic fields up to the first order in $g_{a \gamma \gamma}$ and for TE mode we have $E_z=0$ at order zero, our results for $E_x$ and $E_y$ is equivalent to get the solution up to the first order of the equations:
\begin{equation}
	E_x''(z)+\lambda^2E_x(z)= i\mu \lambda^2\xi E_y(z), \label{modified1}
\end{equation}
\begin{equation}
	E_y''(z)+\lambda^2E_y(z) = -i\mu \lambda^2\xi E_x(z),
\end{equation}
that can be rewritten in terms of $E_{\pm}$ fields as
\begin{equation}
	E_{\pm}''(z)+[\lambda^2 \mp \mu (\alpha \lambda+\beta \omega) ]E_{\pm}(z)=0,  \label{modified2}
\end{equation}
whose general solution is
\begin{equation}\label{genpm}
	E_{\pm}(z)=A e^{i \sqrt{\lambda^2 \mp \mu (\alpha \lambda+\beta \omega)} z}+B  e^{-i \sqrt{\lambda^2 \mp \mu (\alpha \lambda+\beta \omega)} z}.
\end{equation}
If we employ the boundary conditions $E_x(z=0,L)=E_y(z=0,L)=0$ and our assumption of $\xi \ll 1$ (leading to $\sqrt{\lambda^2 \mp \mu (\alpha \lambda+\beta \omega)}  \sim \lambda \mp\frac{1}{2}  \mu \frac{\alpha \lambda+\beta \omega}{\lambda} $), then we get the same solution (\ref{solpm}).
Now the physical meaning of the solution (\ref{solpm}) is clear thanks to the expression (\ref{genpm}): the phase velocities of left and right circularly-polarised waves are respectively different, so the optical angle rotates from $z=0$ to $z$ of the angle
\begin{equation}
\phi(z)=\frac{1}{2}  \frac{\mu}{n} \frac{\alpha \lambda+\beta \omega}{\lambda}z=\frac{1}{2}  \sqrt{\frac{\mu}{\varepsilon}} \frac{\alpha \lambda+\beta \omega}{\lambda}z.	
\end{equation}
 This rotation of the optical angles consequently results on a gradual transition of the TM mode into a TE mode, and similarly in the reverse direction TE $\rightarrow$ TM.
The value of $\phi$ at $z=L$ is then seen to be
\begin{equation}
\phi(L)= \frac{1}{2}  \sqrt{\frac{\mu}{\varepsilon}}\frac{\alpha \lambda+\beta \omega}{\lambda}L. \label{fi}
\end{equation}
This result is consistent with our previous discussion.

Let now $\phi$ denote the optical rotation angle at $z=L$ and the rotation matrix
\begin{equation}
{\bf A} = \left( \begin{array}{ll}
\cos \phi & \sin \phi \\
-\sin \phi & \cos \phi
\end{array}\right).
\end{equation}
When the wave travels back, the point is whether the rotation occurs in the reverse direction, thus $\phi =0$ in total, or if the rotation continues in the same direction, so that the total $\phi \rightarrow 2\phi$ .

The last case is the only one leading to physical effects of our interest. We therefore need the square of the rotation matrix
\begin{equation}
{\bf A^2} = \left( \begin{array}{ll}
\cos 2\phi & \sin 2\phi \\
-\sin 2\phi & \cos 2\phi
\end{array}\right).
\end{equation}
 %There is then a rotation of the polarization plane proportional to $z$, $\phi(z) = {\cal V}B_0 z$, where $B_0$ is the uniform magnetic field and the material constant $\cal V$ is called the Verdet constant and the reason of the rotation of the optical angle is always the gradual transition of the TM mode into a TE mode, and similarly in the reverse direction.

We now return to the axion problem. We first observe that the logarithmic factor in the energy expression (\ref{energy}) can be written as a trace,
\begin{equation}
2\ln (1-e^{-2\kappa L}) = {\rm Tr} [\ln ({\bf I}-e^{-2\kappa L}{\bf I})],
\end{equation}
where $\bf I$ is the unit matrix in two dimensions. We now replace $\bf I$ with the rotation matrix $\bf A^2$ in the interaction term, containing the exponential term, leading to the effective substitution
 \begin{equation}
 2\ln (1-e^{-2\kappa L}) \rightarrow {\rm Tr}[\ln ({\bf I}-e^{-2\kappa L}{ \bf A}^2)] = \ln[\det ({\bf I}-e^{-2\kappa L}{ \bf A}^2)].
 \end{equation}
where $\phi$ is now the rotation angle $\phi(L)$ of Eq.~(\ref{fi}). The determinant is equal to
\begin{equation}
\det({\bf I}-e^{-2\kappa L}{ \bf A}^2) = 1+e^{-4\kappa L}-2e^{-2\kappa L}\cos 2\phi,
\end{equation}
and we obtain from Eq.~(\ref{energy}) the following expression for the Casimir free energy,
\begin{equation}
F= \frac{1}{2\pi \beta_T}{\sum_{m=0}^\infty}^ \prime \int_{n\zeta_m}^\infty \kappa d\kappa \ln\left[1+e^{-4\kappa L} -2e^{-2\kappa L}\cos\left(\sqrt{\frac{\mu}{\varepsilon}}\frac{\alpha \kappa+\beta \zeta_m}{\kappa}L\right) \right], \label{freenergy}	
\end{equation}
where we have substituted the explicit expression of $\phi$.

We note that with $\beta=0$ the phase $2\phi=\sqrt{\frac{\mu}{\varepsilon}}\alpha L$ is not dependent on $\kappa$ and $\zeta_m$.

The formula combines in a unified fashion the space and the time-varying axion field.
The expression (\ref{freenergy}) is formally the same as for a chiral medium, and has a wide applicability.
For instance, for ideal metal plates in the nonaxion case ($\phi=0$), we have
\begin{equation}
F_{\rm  metal} = \frac{1}{\pi\beta_T}{\sum_{m=0}^\infty}^ \prime \int_{n\zeta_m}^\infty \kappa d\kappa \ln (1-e^{-2\kappa L}),
\end{equation}
whereas  in the repulsive Boyer case ($\phi = 90^o)$,
\begin{equation}
F_{\rm Boyer}= \frac{1}{\pi\beta_T}{\sum_{m=0}^\infty}^ \prime \int_{n\zeta_m}^\infty \kappa d\kappa \ln(1+e^{-2\kappa L}).
\end{equation}
Another known case of considerable interest is the so-called Boyer problem \cite{PhysRevA.9.2078}, where one of the metal plates is replaced by an ideal "magnetic" plate.  This case corresponds to the rotation angle $\phi = 90^o$, and leads actually to a repulsion between the two plates. A further discussion of the Boyer problem can be found, for instance, in Ref.~\cite{hoye18}.

Finally, at zero temperature, the free energy $F$ reduces to the thermodynamic energy $E$.  Making use of the relationship
\begin{equation}
\frac{1}{\beta_T}{\sum_{m=0}^\infty}^ \prime \rightarrow \frac{1}{2\pi}\int_0^\infty d\zeta,
\end{equation}
we then obtain the zero temperature variant of Eq.~(\ref{freenergy}),
\begin{equation}
E_{T=0} = \frac{1}{(2\pi)^2}\int_0^\infty d\zeta \int_{n\zeta}^\infty \kappa d\kappa \ln (1+e^{-4\kappa L}- 2e^{-2\kappa L}\cos 2\phi). \label{nullpunktsenergi}
\end{equation}
As before, $\kappa^2=k_\perp^2+n^2\zeta^2$, but now with $\zeta$ as a continuous variable.

It is noteworthy that for small rotation angles $\phi$, the corrections from axions occur to the order $\phi^2$. We may express this more explicitly by rewriting Eq.~(\ref{freenergy}) as
\begin{equation}
F= F_{\rm metal} -\frac{4}{\pi \beta} {\sum_{m=0}^\infty}^\prime  \int_{n\zeta_m}^\infty  \kappa d\kappa \frac{e^{-2\kappa L}}{(1-e^{-2\kappa L})^2} \phi^2+...\label{freenergy2ndorder}
\end{equation}
From the former results, we have obtained some interesting results for particular cases of interest.
In the case $\beta=0$, we obtain at $T=0$ the following expression
\begin{equation}\label{alphaE0}
	E_{T=0}-E_{T=0,\rm metal}=-\frac{1}{48} \frac{\mu^{1/2}}{\varepsilon^{3/2}}\alpha^2 \frac{1}{L},
\end{equation}
that, if taken with $\mu=\varepsilon=1$, gives a result similar to the analogous in Subsection \ref{time-dependent} and
Ref.~\cite{FAVITTA2023169396}.
 Furthermore, for $\beta=0$ we obtained the high-temperature limit
	\begin{figure} \includegraphics[scale=0.2,width=0.5\paperwidth]{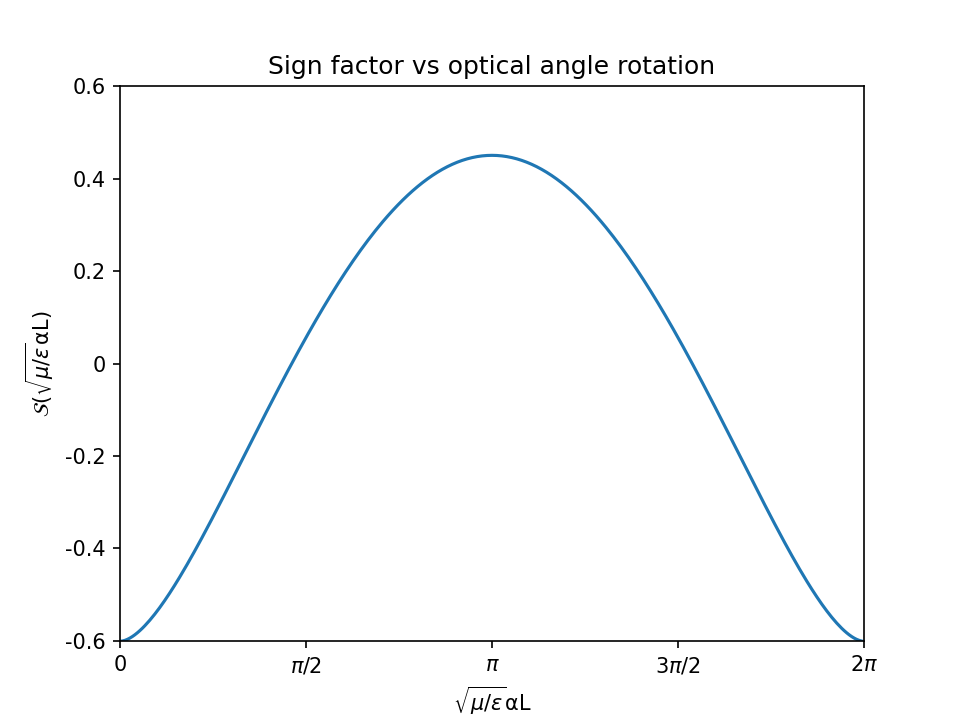}
	\caption{Plot of the sign factor as a function of $2 \phi=\sqrt{\frac{\mu}{\varepsilon}} \alpha L$}\label{fig:1}
\end{figure}
	
\begin{figure}
	\includegraphics[scale=0.2,width=0.5\paperwidth]{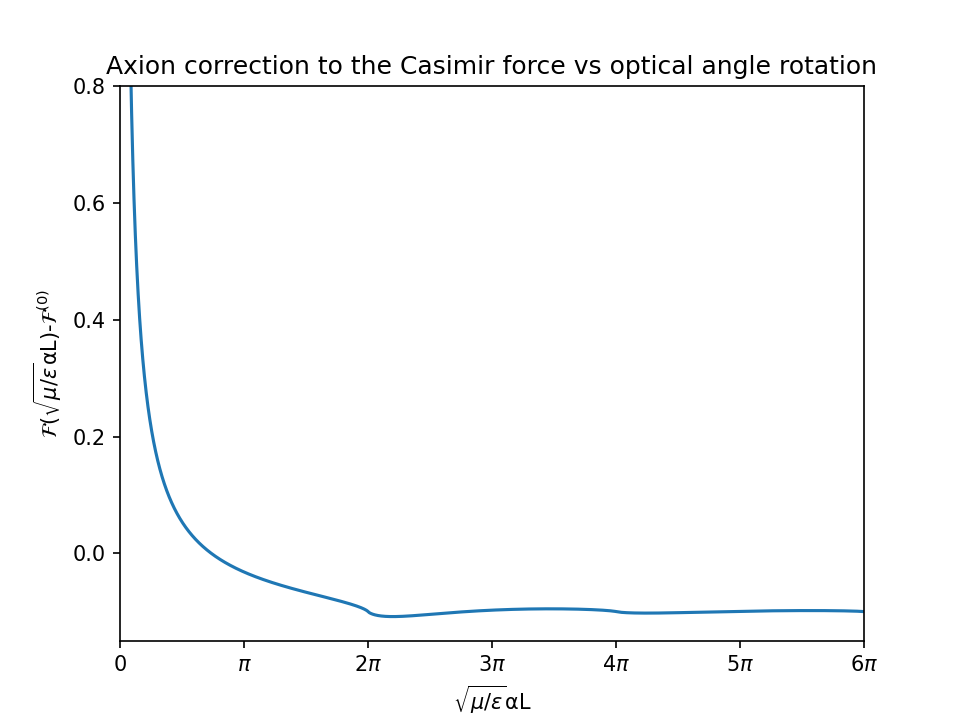}
	\caption{Plot of the ratio $\mathcal{F}\left(\sqrt{\frac{\mu}{\varepsilon}}\alpha L\right)=4 \pi \beta \frac{f^{T \rightarrow +\infty}(L,\alpha)-f^{T \rightarrow +\infty}(L,\alpha=0)}{ (\frac{\mu}{\varepsilon})^{3/2} \alpha^3}$ as a function of $2 \phi=\sqrt{\frac{\mu}{\varepsilon}} \alpha L$. For $\phi \ll \pi/2$ the axion correction is repulsive, so very differently from the case $T=0$ where it is attractive. However, for $2 \phi \sim \pi$ the axion term becomes attractive.  }\label{fig:2}
	\includegraphics[scale=0.2,width=0.5\paperwidth]{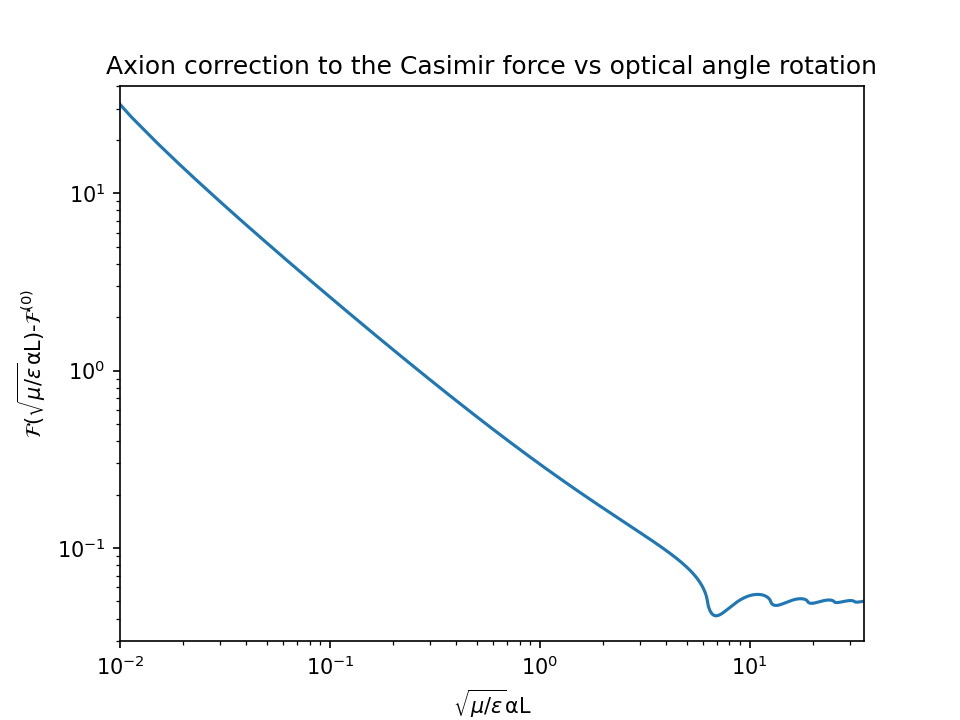}
	\caption{The same plot of Figure (\ref{fig:2}) in a log-log graph and where we have substracted the minimum to have only positive values in the y-axis.It highlights the behaviour of the axion correction to Casimir force for $\phi \ll \pi/2$, that is $\sim 1/L$ differently from the $T=0$ case. Significant deviation from such a behaviour is for $\phi >\pi/2$ as also shown in the same Figure (\ref{fig:2})} \label{fig:3}
\end{figure}
In such a case, as done in usual Casimir calculation, we get this limit by only considering the first term $m=0$ in the series and can evaluate  exactly:
\begin{equation}\label{ftinfinity}
	F^{T \rightarrow +\infty} = \frac{T}{4 \pi L^2} \mathcal{S}\left(\sqrt{\frac{\mu}{\varepsilon}} \alpha L \right),
\end{equation}
where we have defined the function $\mathcal{S}$ as a sign factor for the sake of simplicity, and can be evaluated numerically.
We show its plot in Figure (\ref{fig:1}).
To clarify if such a behaviour is significant for the properties of the Casimir force, if it is repulsive or attractive, we plot in Figure (\ref{fig:2}) the behaviour of the Casimir force, calculated as:
\begin{equation}
	f^{T \rightarrow +\infty}(L)=-\frac{\partial 	F^{T \rightarrow +\infty}}{\partial L}= \frac{2}{L}		F^{T \rightarrow +\infty}-\frac{T}{4 \pi L^2}  \mathcal{S}'\left(\sqrt{\frac{\mu}{\varepsilon}} \alpha L \right),
\end{equation}
and we subtract from it the notorious expression of the Casimir force in the same temperature limit from the usual electrodynamics:
\begin{equation}
	f^{T \rightarrow +\infty}(L,\alpha=0)=-T \frac{\zeta(3)}{8 \pi L^3}.
\end{equation}

We observe how for $\phi \ll \pi/2$ the axion correction goes as $\sim 1/L$ (as shown better in Figure (\ref{fig:3}) ) and it is repulsive, so very different from the case $T=0$ where it goes as $1/L^2$ and it is attractive. However, for $2 \phi \sim \pi$ the axion correction is attractive.
It is worth to notice from Figure (\ref{fig:1}) that the sign factor has its absolute maximum at $\frac{\mu}{\varepsilon} \alpha L=\pi$ and this value corresponds roughly to the threshold between repulsive and attractive regime, as visible in the figures (\ref{fig:2}) and (\ref{fig:3}). This value corresponds to a value of $\alpha$ that is roughly equal to the inverse distance $L^{-1}$ and corresponds to the physical condition of maximum reflection of photons due to the presence of the "wall", analogously to our system in Section \ref{wavedomain} %add section reference
but with no metallic slabs, with an analogous maximum reflectance. The correspondence between the two holds with $\omega \leftrightarrow 1/L$.

Another interesting property of the expression (\ref{ftinfinity}), that is present in the general expression (\ref{freenergy}),  is that, apart of a factor $L^{-2}$, it is periodic in optical rotation angle. This leads to the observable wiggles in the Figures (\ref{fig:2}) and (\ref{fig:3}) at $2 \phi=2 n \pi$, where $n=1,2,...$.
      
We have also discussed the case $\alpha=0$ and we have observed that, while the axion correction is suppressed in the high temperature limit, we have for the limit  $T=0$ the following expression at the second order in $\beta^2$ 
\begin{equation}
	F= F_{\rm metal} -\frac{1}{(2 \pi)^2} \frac{\mu}{\varepsilon} \int_0^{+\infty} d\zeta  \int_{n\zeta}^\infty  \kappa d\kappa \frac{e^{-2\kappa L}}{(1-e^{-2\kappa L})^2} \beta^2 \frac{\zeta^2}{\kappa^2}L^2 .	
\end{equation}
This can be evaluated by a change of variables and using the numerical result of the integral:
\begin{equation}
\iota=\int_0^{+\infty} ds \int_s^{+\infty} dk \,\frac{e^{-2k}}{(1-e^{-2k})^2} \frac{s^2}{k}=0.137078.
\end{equation}
From which, similarly to the case $\beta=0$, we have the attractive term:
\begin{equation}
F-F_{\rm metal}=-\iota \frac{1}{(2 \pi)^2} \frac{\mu^{1/2}}{\varepsilon^{3/2}} \frac{\beta^2}{L},
\end{equation}
whose behaviour with the distance $L$ is the same of Eq.~(\ref{alphaE0}).

%This dispersion relation could be used also to derive the Casimir pressure at $T=0 \,K$ by using the same approach used in Ref.~\cite{PhysRevD.100.045013}.\\
%In the case of interest we need to calculate the Casimir pressure at extremities of the configuration ((\ref{toy})) at zero temperature and then at finite temperatures.
%The first one can be developed straightforwardly: we just need to calculate the Casimir pressure at 0 K:
%\begin{equation}
%	\mathcal{P}=\int \frac{d^3 k}{(2 \pi)^3} \frac{1}{2} \omega=\frac{\Theta_0^4}{16 L^4} \sum_{\pm} \int^{\tilde{\Lambda}_{\pm}} \frac{d \tilde{k}_x d \tilde{k}_y d \tilde{k}_z}{(2 \pi)^3} \sqrt{\tilde{k}^2_x+\tilde{k}^2_y+\left(\sqrt{\tilde{k}^2_z+1} \pm 1 \right)^2}.
%\end{equation}
%Analogously to Ref.~\cite{PhysRevD.100.045013}, it can be evaluated by means of cut-off regularization:
%\begin{equation}\label{nonana}
%	\begin{split}
	%	\sum_{\pm} \int^{\tilde{\Lambda}_{\pm}} \frac{d \tilde{k}_x d \tilde{k}_y d \tilde{k}_z}{(2 \pi)^3} \sqrt{\tilde{k}^2_x+\tilde{k}^2_y+\tilde{k}^2_z \pm 1}=&\\=\sum_{\pm} \frac{2}{\pi} \int_0^{+\infty} \int_{-\infty}^{\infty} dk_r dk_z k_r \sqrt{k_r^2+k_z^2+2-2\sqrt{k_z^2+1}} \; H\left(\tilde{\Lambda}_{\pm}-\sqrt{k_r^2+k_z^2+2-2\sqrt{k_z^2+1}}\right),
%	\end{split}
%\end{equation}
%where H is the Heaviside function.
%The integral  Eq.~(\ref{nonana}) has not  analytical solutions, so we evaluated it numerically and therefore we got after summing on the polarizations the following expression of the attractive pressure:
%%ADJUST
However, the constant axion 4-gradient results are not enough to grasp all the properties of real axion backgrounds, such as an oscillating one and an axion domain wall, which could be of interest for topological materials too. 

These two are characterized by a typical frequency, which are in the case of the oscillating field, it is the axion frequency $\omega_a$, while for the toy model (\ref{toy}) it is $m_L=\frac{\Theta_0}{L}$). 
These results are valid in the limit of axion frequencies and wave numbers much bigger than the typical frequency of our physical system of interest, since it then does not 'appreciate' the space-time variations of the derivatives of the axion field.

It is worth mentioning the deviations we obtained for the two systems: 
\begin{itemize}
    \item The toy model domain  wall (\ref{toy}) is characterized by having a more significant reflection coefficient at  $k_z=m_L=\frac{\Theta_0}{L}$, which means a very steady variation of the polarization plane near the interface. 
    \item When $a(t)=a_0 \sin(\omega_a t)$ we have production of a fainter radiation when there is an input electromagnetic plane wave with frequency $\omega=\frac{\omega_a}{2}$, leading to a different polarization plane rotation. 
    Furthermore, when generating a strong magnetic field, there is production of faint photons with frequency $\omega=\omega_a$.
Those arguments demonstrate that the vacuum of Axion Electrodynamics is achromatic and optically active only in the regime of 'high frequencies', in accordance with Ref.~\cite{PhysRevD.101.123503}%add reference in case
\end{itemize}
\section{Space-dependent axion field}\label{spazio}
We now consider the case of 
an axion field which is constant in time but depends on the longitudinal coordinate $z$, along with a spatial gradient directed along the $z$-axis with gradient $\nabla_z \Theta=\beta(z)$, which is the axion field configuration (\ref{toy}).

We delineate the calculations and the important aspects of the calculation of the discontinuity of the $T_{zz}$ component of the electromagnetic stress-energy tensor at the interface. We developed the full calculations in Ref.~\cite{FAVITTA2023169396}

The discontinuity of the zz-component of the stress-energy tensor can be obtained from adopting the Eq.~(\ref{Tzz}) and calculating the temperature-dependent Casimir force per unit area:
\begin{equation}
	f(T,L)=-T \sum_{\pm} \sum_m^{+ \infty \, '} \int_{\zeta_m}^{+ \infty} \kappa^2 d\kappa \, \frac{|R_{\pm}(\kappa)|^2 e^{-2L\sqrt{q^2 \pm q \beta_z}}}{1-|R_{\pm}(\kappa)|^2 e^{-2 L \sqrt{q^2 \pm q \beta_z}}}.
\end{equation}
Its high temperature limit is when $\delta^{-1} \ll T$, which means that the thickness of the wall $\delta$ is much bigger than the thermal wavelength $\sim T^{-1}$. It can be obtained by taking the first term $m=0$ in the sum and is equal to:
\begin{equation}\label{fo}
	\begin{split}
		&f(T,L) = 2 \int \frac{d^3k}{(2 \pi)^3} \left[R^2_{+}(k_z)+R^2_{-}(k_z)  \right] \frac{k^2_z}{\omega} \frac{1}{e^{\beta \omega}-1} \theta_H(k_z)=\\&2 \int \frac{d^2 k}{(2 \pi)^3} \int_{0}^{+\infty} dk_z \left[R^2_{+}(k_z)+R^2_{-}(k_z)  \right] \frac{k^2_z}{\omega} \frac{1}{e^{\beta \omega}-1} ,
	\end{split}
\end{equation}
after integrating in $k_x$ and $k_y$. This expression is an extension of the analogous kinetical expression in Ref.~\cite{Blasi2023,hassan2025chern}, which is 
\begin{equation}
    P=\frac{2}{(2 \pi)^2} \frac{1}{\beta \gamma} \int^{+\infty}_0 dp \,p^2 \,\mathcal{R}(p) \Bigg[\ln{\Bigg( \frac{f(-v)}{f(v)}  \Bigg)}-2 \beta \,\gamma\, v \,p      \Bigg]
\end{equation}
,since we do not assume a priori the two polarizations to have the same reflection coefficients.

The main difference with a non-high temperature limit comes physically from the possibility for a photon plane wave to be multiply reflected inside the domain wall, as in the system in Ref.~\cite{Ellingsen2007CasimirAI}.

However, it is very hard to develop the calculations with the exact explicit expressions (\ref{reflecto}) of  $R^2_{+}(\kappa) $ of Section~\ref{wavedomain}.%add reference to toy model again

\subsection{Pressure and thermal friction  on the domain wall}\label{dio}
\begin{figure}[h]\label{figuredominio}
	\includegraphics[scale=0.5]{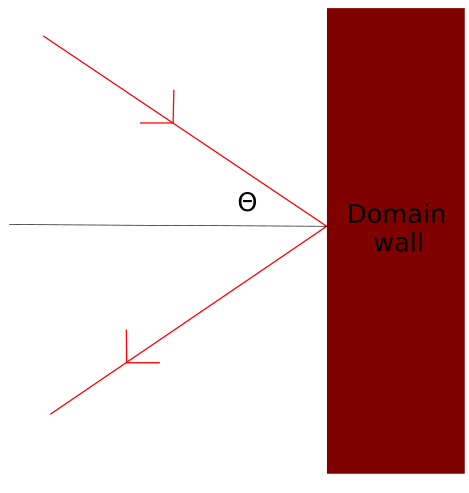}
	\caption{Visual representation of a photon reflected from a planar axion domain wall}
\end{figure}
As mentioned previously, calculating the pressure acting on an axion domain wall is not straightforward.
We can, anyway, get a first physical idea by obtaining the expression (\ref{fo}) and performing calculations similar to those in Refs.~\cite {PhysRevD.32.1560,Blasi2023}, which also highlights its physical meaning and allows us to extend the result to non-static walls.\\
Our results can be first applied to QCD axion domain walls, but also to ALP domain walls \cite{Blasi2023} and topological insulators \cite{PhysRevD.100.045013, nenno2020axion,yan2021majorana}.

This first idea is to calculate kinetically the electromagnetic radiation pressure acting on an axion domain wall at temperature $T$.

If we have a circularly polarized electromagnetic wave incident on the axion domain wall, this is partly reflected as discussed in Section ~(\ref{wavedomain}).
We refer to Figure~(\ref{Plotto1}) for the geometry of the system.

If the incident wave has momentum density that is equal to its energy density $u_{em,\pm}(\omega)$ it is partly reflected with momentum density that is equal to $R^2_{\pm}(k_z) \, u_{em,\pm}(\omega)$ and partly transmitted with momentum density equal to $T^2_{\pm}(k_z) \, u_{em,\pm}(\omega)$, depending on its polarization.

Reflection and transmission coefficients can only depend on $k_z$, since the motion parallel to the domain wall cannot affect the dynamics, since the domain wall is invariant for parallel boosts.
It then experiences a variation of momentum equal to $ \delta t \, \delta A \, \left[2 R^2_{\pm}(k_z) \cos{\theta}^2 \right] u_{em,\pm}(\omega)$ where $\delta A$ is the differential area, $\delta t$ is the differential time and $\theta$ is the incidence angle. This specific expression is a consequence of the relation $R^2_{\pm}(k_z)+T^2_{\pm}(k_z)=1$, from which $1+R^2_{\pm}(k_z)-T^2_{\pm}(k_z)=2R^2_{\pm}(k_z)$, where $T_{\pm}(k_z)$ is the transmission coefficient.
If we divide by $\delta A$ and $\delta t$ we obtain the pressure exerted from that mode with that specific polarization.

Summing up the two polarizations and all frequencies, we get the pressure $\mathcal{P}_L$ acting on the left of the wall, similarly to what is obtained in Ref.~\cite{Blasi2023} and what obtained at the end of the former section
\begin{equation}\label{pressure}
	\begin{split}
	\mathcal{P}_L = \mathcal{P}= 2 \int \frac{d^3k}{(2 \pi)^3} \left[R^2_{+}(k_z)+R^2_{-}(k_z)  \right] \frac{k^2_z}{\omega} \frac{1}{e^{\beta \omega}-1} \theta_{H}(k_z)=\\=2 \int \frac{d^2 k}{(2 \pi)^3} \int_{0}^{+\infty} dk_z \left[R^2_{+}(k_z)+R^2_{-}(k_z)  \right] \frac{k^2_z}{\omega} \frac{1}{e^{\beta \omega}-1} ,
	\end{split}
\end{equation}
which is equal to the pressure $\mathcal{P}_R$ acting on the right of the wall because a static domain wall is left-right symmetric. $\omega=\sqrt{k^2_x+k^2_y+k^2_z}$, and we exploited here the relation $k_z= \omega \cos{\theta}$.

The calculation of $\mathcal{P}$ is not trivial because it is dependent on the details of the axion configuration, from which reflection coefficients depend. We then evaluate it explicitly in the case of the toy model (\ref{toy})
. %%ADJUST.
In this way, the integral (\ref{pressure}) simplifies to
\begin{equation}
	\mathcal{P}_L=2  m_L^3\int \frac{d^2 k}{(2 \pi)^3} \frac{1}{\sqrt{k^2+m_L^2}} \frac{1}{e^{\beta \sqrt{k^2+m_L^2}}-1},
\end{equation}
by using the parity property of the integrand under the change of variable $k_z \rightarrow -k_z$ and the defining property of the Dirac delta $\int_{-\infty}^{\infty} dx f(x) \, \delta(x-x_0)=f(x_0)$.

This integral can be evaluated by noticing that the integrand is only dependent on the variable $k=\sqrt{k_x^2+k_y^2}$, on the physical quantities $m_L$ and $\beta$, so we can use polar coordinates and make the substitution $k_i \rightarrow k_i/m_L$ to obtain an integral dependent only on $\beta m_L$:
\begin{equation}
	\mathcal{P}_L= \frac{m_L^4}{  \pi^2} \int_0^{+\infty} d k' \,  \frac{k'}{\sqrt{k'^2+1}} \frac{1}{e^{\beta m_L \sqrt{k'^2+1}}-1}.
\end{equation}
This integral can be evaluated, as we do in Ref.~\cite{FAVITTA2023169396}, and it can be easily found that
\begin{equation}\label{solo0}
		\mathcal{P}= \frac{m_L^4 }{  \pi^2}  \left[1-\frac{T}{m_L} \ln{(e^{\beta m_L}-1)}\right].
\end{equation}
We get the expressions not agreeing with those mentioned in Ref.~\cite{PhysRevD.32.1560}:
\begin{subequations}\label{Pl0}
	\begin{align}
	\mathcal{P}= \frac{m_L^3}{\pi^2} \,  \, T e^{-\frac{m_L}{T}} \, \, \, \, \, \, \, \, \, \,   \text{for $T \ll m_L$},\label{Pl0a}\\
	\mathcal{P}=  \frac{m_L^3  }{\pi^2}\, \, T \, \ln{\left(\frac{T}{m_L}\right)}  \, \, \, \, \, \, \, \, \, \text{for $T \gg m_L$}.\label{Pl0b}
	\end{align}
\end{subequations}

The reason for this difference is simply that they are two different quantities. This pressure term is not related to thermal friction, but to a planar compression of the quantum vacuum of the electromagnetic field into the wall, which is anyway zero at zero temperature, analogously to fluid dynamics systems treated in Ref.~\cite{landau2013fluid}.%add citazione a Landau fluid //fino a qui

The first limiting case (\ref{Pl0a})  can be found by adopting the asymptotic expression $\ln{(x-1)} \sim \ln{x}-\frac{1}{x}$ for $x \rightarrow +\infty$, where $x$ is in our case $e^{\beta m_L}$.
The second case (\ref{Pl0b}) can be obtained by using  $\ln{(e^x-1)} =\ln{x}+\frac{x}{2}+ \mathcal{O}(x^4)$ for $x \rightarrow 0$ and leaving the more relevant term for $T \gg m_L$, i.e. the logarithm $\ln{x}$.\\
Another relevant aspect of this pressure is that it is the frequency $m_L$, instead of $L^{-1}$, where the reflection coefficients are resonant.

\begin{figure}
    \centering
    \includegraphics[width=0.7\linewidth]{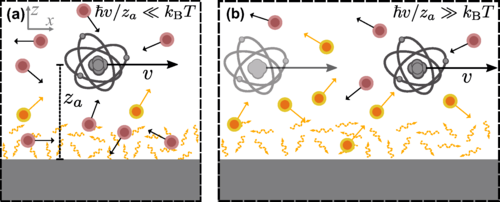}
    \caption{Representation of the thermal friction acting on an atom moving with velocity $v$ in front of a metallic plate in an environment at temperature $T$. It arises from the interaction of moving objects with the thermal fluctuations of the surrounding environment. Image from Ref.~\cite{oelschlaeger2021electromagnetic}.}
    \label{fig:placeholder}
\end{figure}

It is also of interest the case of a domain wall moving at constant velocity $v$, which we fix to be along the direction of the $z$-axis, relatively to the reference frame where the electromagnetic radiation background is an isotropic blackbody one at temperature $T$.
In such a case, expressions for $\mathcal{P}_L$ and $\mathcal{P}_R$ are similar to the ones in the static case $v=0$, but there are two main differences. 

The first is that we do not expect $\mathcal{P}_L$ and $\mathcal{P}_R$ to be equal, since left-right symmetry is broken in such a case; the second concerns the need to account for the Doppler effect.
This is the same logic adopted by the works in the literature treating the thermal friction for the Casimir effect \cite{PhysRevLett.117.100402,PhysRevA.102.050203,oelschlaeger2021electromagnetic,PhysRevA.110.042814,milton2025perspectivesquantumfrictionselfpropulsion}.

%add citazione milton e intravaia           
Consequently, we obtain
\begin{equation}\label{PLR}
	\mathcal{P}_{L,R}= \frac{m_L^4}{  \pi^2} \int_0^{+\infty} d k' \,  \frac{k'}{\sqrt{k'^2+1}} \frac{1}{e^{ \gamma(v)\beta m_L (\sqrt{k'^2+1}\pm v)}-1}.
\end{equation}
where the $+$ is valid for $\mathcal{P}_{L}$ while $-$ for $\mathcal{P}_{R}$.

We observe we have not furtherly transformed the pressure, if not just rewriting the terms of the integrand, since, as already shown in Section~(\ref{axion-domain-wall}), the $T_{zz}$ of a stress-energy tensor does not transform under a Lorentz boost in the z-direction. That observation also suggests us that we also have tangential stresses $T_{xx}=T_{yy}=\gamma^2 T_{zz}$.
All these aspects, along with planar decompression and scalar radiation in domain wall networks, provide strong evidence against the general validity of approximating the domain wall as a rigid body.
Furthermore, the asymmetry of the pressures with a velocity $v$, along with the tangential stresses, is also a signal for an effect of the plasma on contributing to the bending of the wall, along with the axion zero-mode perturbations in the wall. 

The integral~(\ref{PLR}) can be calculated
\begin{equation}
	\mathcal{P}_{L,R}= \frac{1}{  \pi^2} m_L^4 e^{ \mp \gamma(v) \beta m_L v} \left[1-\frac{T}{m_L} \ln{(e^{\beta m_L(1 \pm v)}-1)}\right].
\end{equation}
We highlight, as mentioned before, that the pressures in the limit $\beta m_a \gg 1$ can be evaluated by just substituting the Bose-Einstein distribution with the limiting Boltzmann factor $e^{ -\gamma(v)\beta m_a (\sqrt{k'^2+1} \pm v)}$, which is indeed a good approximation for $\beta m_a \gg 1$,  since it is surely much bigger than 1. This substitution is also true in the ultrarelativistic limit $v \simeq 1$ for the same reason.
We have for such limiting cases ($\beta m_a \gg 1$ and/or $v \simeq 1$):
\begin{equation}
	\mathcal{P}_{L,R} \propto m_L^3 T \sqrt{1-v^2} e^{- \gamma(v)\beta m_L(1 \pm v)}
\end{equation}
The non-relativistic regime $v \ll 1$ can be well approximated by the solution (\ref{solo0})  with $v=0$ and brings us back to the case of just planar decompression, while the difference of the two pressure terms in the case of $v=0$ leads to thermal friction.

Our approach only considers a background which is just the isolated quantum vacuum of electromagnetic field or its thermal state at temperature $T$.
Former approaches have adopted the same approximation. 

A result, coming from WKB approach \cite{PhysRevD.32.1560} with neglecting plasma effects, leads to the following behaviour

 \begin{equation*}
         P=P_L-P_R \sim \Big(\frac{\alpha}{\pi}\Big)^2 \times\begin{cases}
              m_a^3\, T \, e^{-m_a/T} \quad &{T \ll m_a}\\ 
              m_a^2\, T^2 \quad &{T \gg m_a}
             \end{cases}
         \end{equation*} 
and the result from Ref.~\cite{Blasi2023}.

A further improvement was recently done in Ref.~\cite{hassan2025chern}, where they highlight the importance of plasma effects, when interesting to the domain walls in the Early Universe.
By using a Linear response theory, where they keep the field of the axion domain wall as a background entity which perturbs the plasma, they obtain interesting results for the pressure, dependently on the energy scales of the plasma.

In particular, they limit themselves to a plasma of photons and electrons and obtain results for the pressure, which can be summarized as follows\footnote{They also consider the possibility of primordial magnetic field. We will not summarize them for simplicity and being not of direct interest for this work.}\cite{hassan2025chern}:
\begin{equation*}
                P \sim \alpha^2 \times \begin{cases}
                     (\gamma m_a)^2 \,m_D^2 \Big(\frac{\gamma\, m_a}{\Gamma}\Big)^2 v &\qquad \gamma m_a \lesssim \Gamma \\
                    (\gamma\, m_a)^2\, m_D^2 v &\qquad \Gamma \lesssim \gamma m_a \lesssim m_D\\
                 m_a^3 \,T \ln{\Big(\frac{2}{1-v}  \Big)} &\qquad m_D \lesssim \gamma m_a \lesssim T\\
                \end{cases}
            \end{equation*}
where $m_D$ is the Debye mass and $\Gamma$ is the plasma scattering rate, which we present in more details in Section~\ref{DWtherm}.
Our approach with other non-equilibrium QFT comes from taking into account the effect of the planar decompression and the bending of the wall.

\chapter{Non-equilibrium Quantum Field Theory in curved spacetime}\label{noneqQFT}
\section{Introduction}
In this chapter, we review theoretical models and original results regarding the non-equilibrium QFT in a curved spacetime, which can be applied to two interesting cosmological scenarios which are:

\begin{itemize}
    \item Pre-inflationary case for high mass ALPs produced via freeze-in with a late-time inflation model \cite{Blum:2014vsa,Baumholzer_2021,baumann2022cosmology,Ai_2024,ai2024qft,ohare2024cosmology}: the main approximations we can adopt are, in particular, the small gradient approximation, the on-shell limit, and the validity of the non-relativistic limit (we will consider the concrete case of photophilic ALPs). We will constraint the parameter space in  this region considering the resulting dark matter abundancy and comparing with the current dark matter abundancy, up to the limit parameters for which it is also expected to obtain early matter domination and further several experimental constraints need to be taken into account, coming from the obvious modifications to CMB and structure formation, along with the straightforward limits from BBN.

    \item Post-inflationary case for QCD axions and high-mass ALPS ($m_a > 10 \, \mathrm{keV}$ with production of domain walls \cite{PhysRevLett.48.1156,Sikivie2008,Forbes:2000et,Blasi:2023sej,Blasi:2024xvj} (and former production of cosmic strings): we will need the complete BH equation since we expect for our system in the process between production of domain wall network and collapse to switch between condensate, e.g. initial condition with just classical domain walls, and kinetic regime/condensate regime, e.g. emission of highly and mildly relativistic axions, contributing when redshifted to the cold dark matter or to the "hotness" of axion dark matter. 
    
    We are interested to general aspects of the dynamics of the network, although we focus about finding the parameter space where they do not collapse before dominating the energy density of the Universe,which need to be excluded since cosmological observations show us our universe is not inhomogenous as a Universe dominated by a domain wall network and is dominated by a Dark Energy component, which state law is not compatible with domain wall network's one.  

\end{itemize}
We will outline our original results in Chapter~\ref{Schwinger}.
\section{Three toy models}\label{Berges}
We elaborate on three toy models in this section, inspired by two examples from the excellent introduction to non-equilibrium QFT by Jürgen Berges\cite{10.1063/1.1843591}.
\subsection{"Condensate evolution"}\label{Bergescondensate}
Let us consider the following non-linear second-order differential equation
\begin{equation}
    \ddot{y}+y=-\frac{(\varepsilon y)^3}{1-\varepsilon^2 y^2}
\end{equation}
where $\varepsilon$ is a real parameter and the independent variable is $t$, and we denote with a dot its derivative.

\begin{figure}
    %\centering
    \includegraphics[scale=0.6]{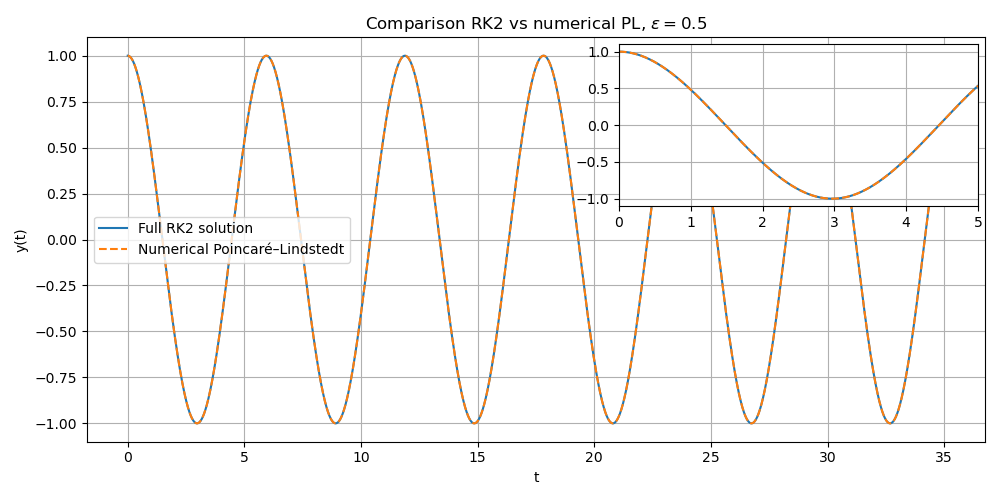}
    \caption{Plot of the numerical solution with a second-order Runge-Kutta (RK2) method and a full numerical Poincaré-Lindstedt (PL) method when $\varepsilon=0.5$.}
    \label{Fig:PLRK}
\end{figure}

This differential equation could be handled numerically, with initial conditions $y(0)=1$ and $\dot{y}(0)=0$, as we fix also for the following, to obtain the 
plot in Figure~\ref{Fig:PLRK}.

However, we could handle the differential equation perturbatively, if we assume $|\varepsilon| \ll 1$.
Indeed, we could then write
\begin{equation}\label{eqpert}
    \ddot{y}+y=-\varepsilon y -(\varepsilon y)^3-(\varepsilon y)^5-(\varepsilon y)^7-...
\end{equation}
coming from the fact that the geometric series satisfies
\begin{equation}
    \sum_{n=1}^{+\infty} (\varepsilon y)^n=\frac{(\varepsilon y)^3}{1-\varepsilon^2 y^2}
\end{equation}
if $|\varepsilon y|<1$. However, this perturbative expansion fails if $|\varepsilon y|\geq1$.%add comment here

Within the former hypothesis, we can solve Eq.~\ref{eqpert} by writing and substituting
\begin{equation}
    y_{\text{pert}}=y_0(t)+ \varepsilon y_1(t)+\varepsilon^2 y_2(t)+\mathcal{O}(\varepsilon^3)
\end{equation}
which gives the solution
\begin{equation}
   y_{\text{pert}}=\frac{1}{2} e^{it} \Big(  
 1- \frac{\varepsilon}{2}t+ \frac{\varepsilon^2}{8}\Big[t^2-it \Big]+...\Big). 
\end{equation}
%add details on validity of perturbative approach and secular terms

The presence of these anharmonic terms to arbitrary orders in the perturbative expansion is reminiscent of when quantum fluctuations can induce self-interactions to higher orders in the field, and we will see that this is concretely analogous for the effective potential and the expansion of the self-energy of the axion field in Chapter \ref{Schwinger}. 
However, an alternative expansion can be adopted, which is analogous to the nPI approach outlined in Chapter \ref{Schwinger} and follows the Poincaré–Lindstedt method \cite{goldstein19801,verhulst2012differential}.

We consider Eq.~\ref{eqpert}, and we divide it in terms of the orders $\varepsilon^0$,$\varepsilon^1$, etc.
\begin{align}
    \text{$\varepsilon^0$ order} &\rightarrow \ddot{y}^{(0)}_{\text{2PI}}+y^{(0)}_{\text{2PI}}=0\\
    \text{$\varepsilon^1$ order} &\rightarrow \ddot{y}^{(2)}_{\text{2PI}}+y^{(2)}_{\text{2PI}}=-\varepsilon y^{(2)}_{\text{2PI}}
\end{align}
from which we obtain 
\begin{equation}
    y_{\text{2PI}}=y^{(0)}_{\text{2PI}}+y^{(2)}_{\text{2PI}}=\frac{1}{2} e^{it}+\frac{1}{2}e^{it \sqrt{1-\varepsilon^2/4} - \varepsilon t/2}  +c.c.
\end{equation}

%add details on difference between perturbative and nPI and the advantages of nPI. After that, add details on energy conservation

These differences between the two approaches reflect on the energy conservation law.
We can multiply in the original equation the term $\dot{y}$ and obtain 
\begin{equation}
    \frac{d}{dt}\Big(\frac{1}{2} \dot{y}^2+y^2 \Big)=-\frac{(\varepsilon y)^3}{1-\varepsilon^2 y^2} \dot{y}
\end{equation}

To leading order, we can write
\begin{equation}
y_{\text{pert}} \approx A_{\text{pert}}(t) e^{i t}, \quad
A_{\text{pert}}(t) \approx \frac{1}{2} \Big( 1 - \frac{\varepsilon}{2} t + \frac{\varepsilon^2}{8} t^2 \Big),
\end{equation}
so the corresponding energy is
\begin{equation}
E_{\text{pert}}(t) = \frac{1}{2} \dot{y}_{\text{pert}}^2 + \frac{1}{2} y_{\text{pert}}^2 
\approx A_{\text{pert}}^2(t) \approx \frac{1}{4} \Big( 1 - \varepsilon t + \frac{3 \varepsilon^2}{16} t^2 \Big).
\end{equation}

We see that $E_{\text{pert}}(t)$ contains secular terms that grow with time, signaling that the naive perturbative expansion is only valid for short times $t \ll 1/\varepsilon$.

Focusing on the nonlinear correction, we can define the slowly varying amplitude
\begin{equation}
A_{\text{2PI}}(t) = \frac{1}{2} e^{-\varepsilon t/2},
\end{equation}
so that the energy becomes
\begin{equation}
E_{\text{2PI}}(t) \approx A_{\text{2PI}}^2(t) = \frac{1}{4} e^{-\varepsilon t}.
\end{equation}

This shows a physically meaningful exponentially decaying energy, with a small frequency shift $\omega_{\text{2PI}} = \sqrt{1 - \varepsilon^2/4} \approx 1 - \varepsilon^2/8$, which is a typical Poincaré–Lindstedt energy shift. 
All these aspects show us that it is for this reason that perturbation theory methods in QFT are affected by infinities, since already our simple model shows us that secular terms can grow dangerously with time. 
\begin{comment}
More reasonably, if we consider theories where   
To summarize
\begin{itemize}
    \item \textbf{Perturbation theory:} Secular terms in $E(t)$, energy grows or decays incorrectly for long times.
    \item \textbf{2PI / resummed solution:} Exponential amplitude decay and small frequency shift, giving a uniform and physically consistent energy evolution.
\end{itemize}

\begin{center}
\textbf{Summary of energy laws:}
\[
E_{\text{pert}}(t) \approx \frac{1}{4} \Big( 1 - \varepsilon t + \frac{3 \varepsilon^2}{16} t^2 \Big), \quad
E_{\text{2PI}}(t) \approx \frac{1}{4} e^{-\varepsilon t}.
\]
\end{center}
\end{comment}
\subsection{"Kinetic evolution"}\label{Bergeskinetic}

Let us consider an integral-differential equation of the form
\begin{equation}
    \dot{y}_{\vec{p}}= \lambda^2\int_{\vec{q} \,\vec{k}} \left[ 
 (1+y_{\vec{p}})\,(1+y_{\vec{q}} )\, y_{\vec{k}} \,y_{\vec{p}-\vec{q}-\vec{p}}\,-y_{\vec{p}} \,y_{\vec{q}} \,(1+y_{\vec{k}})\,(1+y_{\vec{p}-\vec{q}-\vec{k}})\,\right]
\end{equation}
where $\lambda$ is a real parameter that we can associate with quartic self-interaction \footnote{As we will discuss in the following section, this form corresponds to the collisional Boltzmann term for $\phi^4$ interactions}.
We know that the late-time stationary solution is the Bose-Einstein weight
\begin{equation}
    y_{\vec{p}}=\frac{1}{e^{\beta(|\vec{p}|-\mu)}-1}
\end{equation}
%add comment on universality

However, a linear approximation can lead to a "non-self-consistent" approximation, which will not show the required universality.

Indeed, a linearized approximation 

\begin{equation}
    \dot{y}_{\vec{p}}=(1+y_{\vec{p}}) \sigma_{\vec{p}}^0 -y_{\vec{p}} \bar{\sigma}_{\vec{p}}^0
\end{equation}
where 
\begin{equation}
    \sigma_{\vec{p}}^0= \lambda^2\int_{\vec{q} \vec{k}} [\ 1+y_{\vec{q}}(0)]\ \,y_{\vec{k}}(0)\,y_{\vec{p}-\vec{q}-\vec{k}}(0)
\end{equation}
and
\begin{equation}
    \bar{\sigma}_{\vec{p}}^0= \lambda^2\int_{\vec{q} \vec{k}} y_{\vec{q}}(0)\,[\ 1+y_{\vec{k}}(0)]\ \,[\ 1+y_{\vec{p}-\vec{q}-\vec{k}}(0)]\
\end{equation}

The solution we obtain (by imposing $y_{\vec{p}}(0)=y_0(\vec{p})$) is
\begin{equation}
    y_{\vec{p}}(t)=\frac{\sigma_{\vec{p}}^{(0)}}{\gamma_{\vec{p}}^0}+\Bigg[y_0(\vec{p})-1+\frac{\bar{\sigma}_{\vec{p}}^0}{\gamma_{\vec{p}}^0} \Bigg]e^{-\gamma_{\vec{p}}^0 t}
\end{equation}
where $\gamma_{\vec{p}}^0=\sigma_{\vec{p}}^{0}-\bar{\sigma}_{\vec{p}}^0$.
%finish to define everything and put the discussion on this

\section{The model}\label{model}
We consider the following effective low-energy action for the theory \cite{PhysRevD.81.123530,cao2023nonequilibrium,braaten2018axion, Filippini2019}:
\begin{equation}\label{actiontheory}
\begin{aligned}
&\mathcal{S}=\int d^4 x \sqrt{|g|} \left[\frac{1}{2} \partial_{\mu} \Phi \, \partial^{\mu} \Phi-V(\Phi)+ K_{\chi}(x)+g_{\phi \chi} \Phi \, \mathcal{O}_{\chi}(x)  \right],
\end{aligned}
    \end{equation}
where $\Phi$ is the pseudoscalar ALP field, and we consider, in particular, the flat FLRW metric \ref{metric0} as a fixed metric background.

 $K_{\chi}$ and $ \mathcal{O}_{\chi}$ are, respectively, the free Lagrangian term and the interacting pseudoscalar functional of the SM fields that we consider in the theory. This form of interaction Lagrangian is quite general and encompasses the interactions we focus on.
 
We consider in particular an ALP, which are the photophilic ALPs. They couple to photons only, and their characteristic interaction Lagrangian with the SM at UV scales is
    \begin{equation}\label{Liphoto}
    \mathcal{L}_{I}= -\frac{1}{4}\frac{E_W}{N} \frac{g_2^2}{8 \pi^2 f_\phi}  \Phi \, \tilde{W}_b^{\mu \nu} W^b_{\mu \nu}-\frac{1}{4}\frac{E_B}{N} \frac{g_1^2}{8 \pi^2 f_\phi}  \Phi \, \tilde{B}^{\mu \nu} B_{\mu \nu}   
    \end{equation}
    After integrating out the heavy fields at the electroweak symmetry breaking scale, it is easy to obtain that simply
     \begin{equation}\label{Liphoto1}
    \mathcal{L}_{I}= -\frac{1}{4}\frac{E}{N} \frac{e^2}{8 \pi^2 f_\phi} \Phi \, \tilde{F}^{\mu \nu} F_{\mu \nu}   
    \end{equation}
    and $E_W$,$E_B$, and $E$ are quantized and not evolving through RG equations since the couplings with gauge fields in equations~(\ref{Liphoto}) and ~(\ref{Liphoto1}) are topological \cite{choi2024axiontheorymodelbuilding,Benabou2024jlj}.
    The corresponding coupling is obviously, in this case
    $g_{\phi \chi}=g_{\phi \gamma \gamma}= \frac{\alpha_{EM}}{2 \pi f_\phi} C_{\phi \gamma \gamma}$, where here $C_{\phi \gamma \gamma}=\frac{E}{N}$.
   It is worth noticing that, in all the cases, we will consider the interaction between leptons and photons, so we include the QED Lagrangian 
   $\mathcal{L}_{QED}=\sum_f \bar{\Psi}_f (i \slashed{\partial}-m_f-q_f \slashed{A})\Psi_f $ in the $K_{\chi}$ term \cite{schwartz2014quantum}.

 We take $\Phi=\varphi+\phi$ with $\varphi(x)=\angi{\Phi}$ the quantum average field and $\phi$ the fluctuating field (the part with $\angi{\phi}=0$) and extend the work done by Refs. \cite{ai2024qft, Ai_2024,cao2023nonequilibrium,PhysRevD.85.063520,PhysRevD.91.123540,SIKIVIE2017331,Farina:2016tgd,BLUM201430} in both the formalism and the results.
 Namely, the axion part of the Lagrangian density of the action (\ref{actiontheory}) becomes
 \begin{equation}
     \mathcal{L}=\frac{1}{2} \partial_{\mu} \varphi \, \partial^{\mu} \varphi+\frac{1}{2} \partial_{\mu} \phi \, \partial^{\mu} \phi-V(\varphi+\phi) +g_{\phi \chi} (\varphi+\phi) \, \mathcal{O}_{\chi}(x),
 \end{equation}
and in the following, we will also consider the axion potential up to the quadratic order in $\Phi$, so the form of the potential with the fields $\varphi$ and $\phi$ is
\begin{equation}
    V_4(\varphi+\phi)=\frac{m^2_{\phi}}{2} \varphi^2+\frac{m^2_{\phi}}{2} \phi^2+\frac{\lambda}{4!} \varphi^4+\frac{\lambda}{4!} \phi^4+\frac{\lambda}{3!} \phi^3 \varphi+\frac{\lambda}{4}  \varphi^2 \phi^2,
\end{equation}
where the terms linear in $\phi$ have been neglected, as they lead to tadpole diagrams that do not contribute to the system's physics \cite{schwartz2014quantum,ai2024qft}.

\chapter{Keyldish-Schwinger formalism}\label{Schwinger}

\section{The 2PI effective action}\label{2PI}

 To obtain the quantum EoMs as before and give a more quantitative analysis, we extend the 2PI effective action to a case with a curved spacetime.
 In the following, we will adopt at the start the notation $\int d^4 x \sqrt{|g(x)|} =\int d \omega $     for readability\footnote{For mathematical rigour, we do not mean by this that we consider $d\omega$ as an exact form.}.
 
To obtain the explicit expressions for the $ \Gamma_{2PI}$ and the quantum EoMs, we need to emphasise that we work with an initial value problem, and this is relevant to the class of background spacetimes we consider. 

\begin{figure}\label{keldyshcontour}
    \centering
   \begin{tikzpicture}[>=stealth]

  % Time axis
  \draw[->] (-0.5,0) -- (6,0) node[anchor=west] {$\text{Re}(t)$};

  % Contour loop
  \draw[very thick,->,blue] (0,0.2) -- (5,0.2) node[midway, above] {$\mathcal{C}_+$};
  \draw[very thick,dashed,->,red] (5,-0.2) -- (0,-0.2) node[midway, below] {$\mathcal{C}_-$};
  \draw[very thick,gray!60] (5,0.2) -- (5,-0.2);
  \draw[very thick,gray!60] (0,-0.2) -- (0,0.2);

  % Points
  \filldraw (0,0.2) circle (2pt) node[anchor=south east] {$t_0$};
  \filldraw (5,0.2) circle (2pt) node[anchor=south west] {$t_f \rightarrow +\infty$};
  \filldraw (5,-0.2) circle (2pt);
  \filldraw (0,-0.2) circle (2pt);

\end{tikzpicture}
\begin{tikzpicture}[>=stealth]

  % Axes
  \draw[->] (-0.5,0) -- (7,0) node[anchor=west] {$\text{Re}(t)$};
  \draw[->] (0,0.5) -- (0,-3) node[anchor=north] {$\text{Im}(t)$};

  % Contour
  \draw[very thick,->,blue] (0,0) -- (5,0) node[midway, above] {$\mathcal{C}_+$};
  \draw[very thick,->,red] (5,0) -- (0,0) node[midway, below] {$\mathcal{C}_-$};
  \draw[very thick,->,green!60!black] (0,0) -- (0,-2) node[midway,left] {$\mathcal{C}_\beta$};

  % Points
  \filldraw (0,0) circle (2pt) node[anchor=north east] {$t_0$};
  \filldraw (5,0) circle (2pt) node[anchor=north west] {$t_{\text{f}} \rightarrow +\infty $};
  \filldraw (0,-2) circle (2pt) node[anchor=north east] {$t_0 - i\beta$};

\end{tikzpicture}
    \caption{Graphical visualisation of the Keyldish closed time contour. Above, we show the closed time path (CTP) for our initial-value problem and the extension at $t_f \rightarrow +\infty$. Below, we have the same as above, but we also show the Wick rotation procedure \cite{le2000thermal,laine2016basics,schwartz2014quantum} for a QFT at thermal equilibrium at temperature $T$ with inverse temperature $\beta=1/T$.}
    \label{fig:keldysh}
\end{figure}
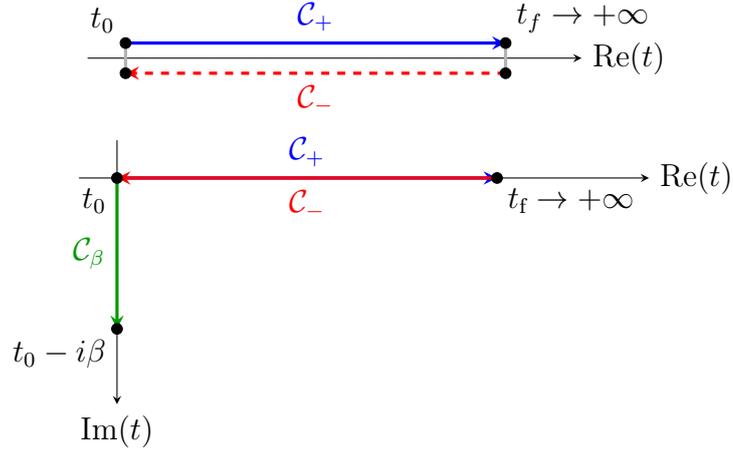

We adopt a Keldysh contour $\mathcal{C}$  for the generating functional that is a closed-time path in the complex plane \cite{Schwinger:1960qe,Keldysh:1964ud,10.1063/1.1843591}.
The closed contour is composed of a forward branch and a backward branch and defines the time ordering operator $\mathcal{T}_{\mathcal{C}}$, as we show in Fig.~ ~(\ref{fig:keldysh}).\\
In such a formalism, to distinguish the forward
and backward branches, we can write the time variable on the forward branch as $t^{+}$ and on the
backward branch as $t^{-}$, but we can adopt an alternative convention where we distinguish the fields on the forward and backward branches, $\Phi^{+}(t, x) = \Phi( t^{+}, x)$ and $\Phi^{-} (t, x) = \Phi( t^{-}, x)$.
In this way, $S_{\mathcal{C}}[\Phi]=S[\Phi^{+},\Phi^{-}]=S[\Phi^{+}]-S[\Phi^{-}]$.
It is then clear that we are limited to background spacetimes where, within the ADM formalism, we can perform a transformation of coordinates for which one coordinate is time-like.%adjust in case

Taking into account both local and 2-point sources, the generating functional can be defined as the path integral \cite{le2000thermal,laine2016basics,PhysRevD.23.2850,bastianelli2005pathintegralscurvedspace,bastianelli2006path,parker2009quantum,bastianelli2017quantum}
\begin{equation}
    \begin{aligned}
        &Z[J,K]=\int \mathcal{D} \Phi \, \exp{ i\Big( S[\Phi]+ \int d \omega \,J(x)\, \Phi(x)+ \frac{1}{2} \int d \omega_1 \int d \omega_2 \, K(x_1,x_2)\, \Phi(x_1)\, \Phi(x_2) \Big)}
    \end{aligned}
    \end{equation}
    We have $\varphi(x)=\angi{\Phi(x)}=\frac{\delta \ln{Z}}{i \sqrt{|g|} \delta J}$ and $\Delta_{\phi}=\angi{\mathcal{T}\Phi(x) \Phi(y)}_c=\frac{\delta^2 \ln{Z}}{i  \sqrt{|g(x)|}  \sqrt{|g(y)|}\delta K(x,y)}-\varphi(x) \varphi(y)$. The 2PI effective action is the Legendre transform of $-i \ln{Z}$:
  \begin{equation}
   \begin{aligned}
       \Gamma_{2PI}[\phi, \Delta_{\phi}]=-i\ln{Z}- \int d \omega\, J(x)\, \varphi(x) - \frac{1}{2} \int d \omega_1 \int d \omega_2\, K(x_1,x_2) \left[\Delta_{\phi}(x_1,x_2)+ \varphi(x_1) \varphi(x_2)\right].
   \end{aligned}
   \end{equation}

   One easily gets the quantum EoMs by varying them 
   \begin{align}\label{2PIactvar}
       \left.\frac{ \delta \Gamma_{2PI}[\phi, \Delta_{\phi}] }{\delta \phi(x)}\right|_{J=0, K=0}=0,
     \quad \quad  \left.\frac{ \delta \Gamma_{2PI}[\phi, \Delta_{\phi}] }{\delta \Delta_{\phi}(x,y)}\right|_{J=0, K=0}=0.
   \end{align}

It can be demonstrated, analogously to what is done in Ref.~\cite{10.1063/1.1843591} for a Minkowski spacetime as a background, that
\begin{equation}\label{lnTr}
   \begin{aligned}
       &\Gamma_{2PI}[\varphi, \Delta_{\phi}, \Delta_{\chi}]=S[\varphi]+ \frac{i}{2} \Tr{\ln{\Delta_{\phi}^{-1}}} +\frac{i}{2} \Tr[G_{\Phi}^{-1}(\varphi) \Delta_{\phi} ]+ \\ &i \sum_{\chi} a_{\chi} \Big[ \Tr{\ln{\Delta_{\chi}^{-1}}}+ i \Tr[G_{\chi}^{-1} \Delta_{\chi} ]\Big]+\Gamma_2(\varphi, \Delta_{\phi}, \Delta_{\chi}),
 \end{aligned}
 \end{equation}
 where $G_{\Phi}^{-1}(\varphi)$ and $G_{\chi}^{-1}$ are the inverse Green's function in position space of respectively the $\Phi $ field and $\chi$ field.
 In particular, we have
 \begin{equation}
    \left. -i \frac{\delta^2 S[\Phi^+,\Phi^-]}{\delta \Phi^a(x_1) \Phi^b(x_2)}\right|_{\varphi}=\delta^{(4)}(x_1-x_2) G_{\Phi}^{ab, \,-1}(\varphi)
 \end{equation}

 and the trace is performed in the position space
 \begin{equation}
     \Tr[G_{\Phi}^{-1}(\varphi) \Delta_{\phi} ]= \sum_{a,b} \int d^4 x \, G_{\Phi}^{ab, -1}(\varphi(x)) \Delta^{ba}_{\phi} (x,x).
 \end{equation}
  We have defined the inverse Green's function and trace with this particular self-consistent choice. Nothing forbids redefining it with factors dependent on $\sqrt{g}$ to see more visibly the covariance of the definition of the trace we took, for example, in the measure $d^4 x$. Our choice is based on the simplicity of the calculations that follow.
  $\Gamma_2$ is proportional to the sum of the 2PI vacuum diagrams, in particular equal to $-i \times \text{sum of 2PI vacumm diagrams}$.

For example, in the case of photons as SM particles
\begin{equation}
\begin{aligned}
   &\Gamma_{2PI}[\varphi, \Delta_{\phi}, \Delta_{\gamma}]=S[\varphi]+\frac{i}{2} \Tr{\ln{\Delta_{\phi}^{-1}}}+\frac{i}{2} \Tr{\ln{\Delta_{\gamma}^{-1}}}+\frac{i}{2} \Tr[G_{\phi}^{-1} \Delta_{\phi} ]\\&+  \frac{i}{2} \Tr[G_{\gamma}^{-1} \Delta_{\gamma} ],
   \end{aligned}
\end{equation}
    where we have explicitly
    \begin{align}
        \Delta_{\gamma \, \mu \nu}=\angi{A_{\mu}(x) A_{\nu}(y)}\\
        G_{\gamma \, \mu \nu}^{ab, -1}=i c^{ab} (\delta_{\mu \nu} \Box -\frac{1}{2} g_{a \gamma \gamma } \varepsilon_{\mu \nu \rho \sigma} \partial^{\rho} \partial^{\sigma} \varphi^{a}).
    \end{align}

 Precisely, we get the following general quantum EoMs for $\varphi(x)=\angi{\Phi}$ and $\Delta_{\phi}(x,y)=\angi{\mathcal{T} \Phi(x) \Phi(y)}$ ,with $\mathcal{T}$ the time ordering operator, from equations ~\ref{2PIactvar} and ~\ref{lnTr}
 \begin{subequations}\label{quantumEoMs}
   \begin{equation}\label{quantumEoMav}
   \frac{1}{\sqrt{|g|}} D^{\mu} \partial_{\mu}\varphi +\frac{ \partial V_{\text{eff}}(\varphi, \Delta_{\phi}, \Delta_{\chi})}{\partial \varphi} - \frac{\delta \Gamma_2}{\delta \phi^+} \Big|_{\phi^+=\phi^-=\varphi} + g_{\Phi \chi} \Delta_{\mathcal{O}_{\chi}} = 0
   \end{equation}
   \begin{equation}\label{quantumEoM2p}
   \begin{aligned}
      -\Bigg( \frac{1}{\sqrt{|g|}}D^{\mu} \partial_{\mu}+\tilde{m}_{\phi}^2\Bigg)\Delta^{ab}(x_1,x_2) - c \int d^4 x_3 \, \Pi^{ac}_{\phi}(x_1,x_3) \, \Delta^{cb}(x_3,x_2) 
       = i c^{ab} \delta(x_1 - x_2) 
   \end{aligned}
    \end{equation}
\end{subequations}
where $ \tilde{m}_{\phi}^2= m_{\phi}^2+\lambda_{\phi}\varphi^2/2+(\text{higher order terms})=\frac{\partial V_{\text{eff}}(\varphi, \Delta_{\phi},\Delta_{\chi})}{\partial \Delta^{ab}_{\phi}}, $and $V_{\text{eff}}$ is the effective potential taking care of quantum and thermal correction from $\phi$ field and the environment, for example correction to thermal mass of $\varphi$ from quartic interactions with $\phi$ giving a term proportional to $\Delta_{\phi}^{++}(x,x)$ as we see in the following.

We have also defined for simplicity the operator $\Box=\frac{1}{\sqrt{|g|}} D^{\mu} \partial_{\mu}$.

The sum over $c=+,-$ is implicit and
\begin{equation}
    \Pi^{ab}_{\phi}(x,y)=-2(ab) \frac{\delta \Gamma_2[\varphi,\Delta_{\phi},\Delta_{\chi}]}{\delta \Delta^{(ab)}_{\phi}(x,y)}
\end{equation}

We can define the advanced and retarded self-energies analogously
\begin{equation}
    \begin{aligned}
         &\Pi^{r}_{\phi}(x,y)= \Pi^{++}_{\phi}(x,y)- \Pi^{+-}_{\phi}(x,y)\\
          &\Pi^{a}_{\phi}(x,y)= \Pi^{++}_{\phi}(x,y)- \Pi^{-+}_{\phi}(x,y),
    \end{aligned} 
\end{equation}
and the analogous for the advanced and retarded propagators.

The Eq.~\ref{quantumEoM2p} can be notoriously treated with the usual treatment with convolution product and Wigner transforms \cite{Calzetta:1986cq,PhysRevD.73.025005,Drewes:2012qw,ai2024qft} to obtain the Kadanoff-Baym (KB) equation and, from this, the Boltzmann equations in Minkowski spacetime. Here, we present an analogous procedure that directly starts with a fixed metric tensor $g$, inspired by, and also extending, Ref.~\cite{HABIB1989335}.

If we introduce the convolution product 
\begin{equation}
 \Pi^r_{\phi} \odot   \Delta^{>}_{\phi}= \int d^4 x_3  \Pi^r_{\phi}(x_1,x_3)  \Delta^{>}_{\phi}(x_3,x_2),
\end{equation}

the EoMs for $\Delta^{><}_{\phi}$ is of the form
\begin{equation}
    (-\Box+ \tilde{m}_{\phi}^2\,) \,\Delta^{><}_{\phi}- \Pi^{\mathcal{H}}_{\phi} \odot \Delta^{><}_{\phi}- i \,\Pi^{><}_{\phi} \odot \Delta^{\mathcal{H}}=\mathcal{C}_{\phi},
\end{equation}

where
\begin{equation}
    \mathcal{C}_{\phi}=\left(\Pi^{>}_{\phi} \odot \Delta^{<}-\Pi^{<}_{\phi} \odot \Delta^{>}     \right).
\end{equation}

This is the usual KB equation, which can be furtherly manipulated by defining the Wigner transform as

\begin{equation}
    \bar{\Delta}_{\phi}(k,x)=\int d^4 r \, e^{i k \cdot r} \Delta_{\phi}\left(x+\frac{r}{2}, x-\frac{r}{2}\right),
\end{equation}
and using the following properties of the Wigner transform and convolutions
\begin{equation}
\begin{aligned}
    \int d^4(X_1-X_2) e^{i k \cdot(x_1-x_2)} \int d^4 x_3 A(x_1,x_3)  A(x_3,x_2)=e^{-i \diamond} \left\{\bar{A}(k,x)  \right\}\left\{\bar{B}(k,x)  \right\},
    \end{aligned}
\end{equation}
where the diamond operator $\diamond$ is defined as
\begin{equation}
\begin{aligned}
    &\diamond \left\{\bar{A}(k,x)  \right\}\left\{\bar{B}(k,x)  \right\}=\frac{1}{2} \left(\frac{\partial \bar{A}(k,x) }{\partial x^{\mu}} \frac{\partial \bar{B}(k,x) }{\partial k_{\mu}}- \frac{\partial \bar{B}(k,x) }{\partial x^{\mu}} \frac{\partial \bar{A}(k,x) }{\partial k_{\mu}}   \right).
    \end{aligned}
\end{equation}
We obtain in analogy with Ref.~\cite{ai2024qft}

\begin{equation}\label{KB}
\begin{aligned}
    &\left(k^2-\frac{1}{4} \partial_x^2-\Gamma^{\mu}_{\nu \rho} (k^{\nu}+i \partial_x^{\nu}) (k^{\rho}+i \partial_x^{\rho}) \partial^{(k)}_{\mu} +i k \cdot \partial_x- \tilde{m}_{\phi}^2 e^{-\frac{i}{2} \overleftarrow{\partial}_x \cdot \partial_k} \right) \bar{\Delta}^{><}_{\phi}-e^{-i \diamond} \left\{\bar{\Pi}^{\mathcal{H}}_{\phi}  \right\}\left\{\bar{\Delta}^{><}_{\phi}      \  \right\}\\ &-i e^{-i \diamond} \left\{\bar{\Pi}^{><}_{\phi}  \right\}\left\{\bar{\Delta}^{\mathcal{H}}_{\phi} \right\}=\bar{C}_{\phi} ,
    \end{aligned}
\end{equation}

where \begin{equation}
    \bar{C}_{\phi} =\frac{1}{2}  e^{-i \diamond} \left( \left\{\bar{\Pi}^{>}_{\phi}  \right\}\left\{\bar{\Delta}^{<}_{\phi} \right\}-\left\{\bar{\Pi}^{<}_{\phi}  \right\}\left\{\bar{\Delta}^{>}_{\phi} \right\}   \right).
\end{equation}

The terms involving convolutions and the left partial derivative $\overleftarrow{\partial}_x$ take care of memory effects and so non-Markovian effects in the system.

.%add and adjust everything

\section{Axion Freeze-in}\label{axionfreezein}
\subsection{Boltzmann equation and non-relativistic Gross-Pitaevskii equation}\label{boltzoptical}
The equation \ref{KB} can be, in general, treated similarly to \cite{PhysRevD.104.123504}, with an infinite gradient expansion.
If we stop at the first order, accordingly to an assumption of neglecting non-Markovian processes, we get

\begin{equation}
\begin{aligned}
    &\left[k_{\mu} \partial^{\mu}_{(x)}-\Gamma^{\mu}_{\nu \rho} k^{\nu} k^{\rho} \partial^{(k)}_{\mu}+\frac{1}{2} \left(\partial_{\mu}^{(x)} M^2_{\phi} \right)\partial_{(k)}^{\mu}    \right] \bar{\Delta}^{><}_{\phi}=\left( \left\{\bar{\Pi}^{>}_{\phi}  \right\}\left\{\bar{\Delta}^{<}_{\phi} \right\}-\left\{\bar{\Pi}^{<}_{\phi}  \right\}\left\{\bar{\Delta}^{>}_{\phi} \right\}   \right).
    \end{aligned}
\end{equation}
and if we further assume the on-shell limit, also called the quasi-particle approximation, 

\begin{equation}
    \begin{aligned}
       &\bar{\Delta}_{\phi}^{<}(k,x)=2 \pi \delta(k^2-\tilde{M}_{\phi}^2) \sign{(k_0)} f_{\phi}(k,x),\\
        &\bar{\Delta}_{\chi}^{>}(k,x)=2 \pi \delta(k^2-\tilde{M}_{\phi}^2) \sign{(k_0)} (1 + f_{\phi}(k,x)),  
    \end{aligned}
\end{equation}
we obtain, assuming the validity of the perturbation theory

\begin{equation}
   \frac{1}{k_0}\left[ k_{\mu} \partial_{\mu}^{(x)}-\Gamma_{\alpha \beta}^{\mu} k^{\alpha}   k^{\beta} \partial_{\mu}^{(k)} + \frac{1}{2} \left( \partial_{\mu}^{(x)}  M^2_{\phi} \right) \partial^{\mu}_{(k)}          \right]f_{\phi}(k,x)=\mathcal{C}[f_{\phi},f_i],
\end{equation}

where $\mathcal{C}[f_{\phi},f_i]$ is the usual Boltzmann collisional term
\begin{equation}
\begin{aligned}
    &\mathcal{C}[f_{\phi},f_i]= \sum \left[ \frac{1}{2k_0} \int \Pi_i \frac{d^3 \vec{k}_i}{(2\pi)^3 2 k_i^0} (2\pi)^4  \delta^{(4)}(k+k_{A_1}+...-k_{B_1}-...)  |\mathcal{M}_{\phi A_1 ... \rightarrow B_1 ...}  |^2 \right. \qquad \qquad \qquad \qquad \qquad \qquad \qquad \qquad \qquad \qquad \qquad \qquad\\ &\left.  \times \left\{ (1+ f_{\phi})(1\pm f_{A_1})...f_{B_1}...-f_{\phi}f_{A_1}...(1\pm f_{B_1}) \right\}               \right].
    \end{aligned}
\end{equation}
This is obtained by using the usual relations
\begin{equation}
    \begin{aligned}
        &\delta(k^2-M^2_i)\, \sgn{(k_0)}=\frac{\delta(k_0+E_k^i)+\delta(k_0-E_k^i)}{2k_0},\\
        &f_i(-k,x)=-\left(1\pm f_i(k,x)\right).
    \end{aligned}
\end{equation}

We will assume the SM particles in our models to be in thermal equilibrium, which means that we take all the 2-point functions $\Delta_{\chi} = \Delta^{(0)}_{\chi}(T)$ to be the unperturbed thermal ones.
We also adopt the shell limit for the $\bar{\Delta}_{\chi}$ limit, following an analogous approach.

\begin{equation}
   \begin{aligned}
       &\bar{\Delta}_{\chi}^{<}(k,x)=2 \pi \delta(k^2-\tilde{M}_{\chi}^2) \sign{(k_0)} f_{\chi}^{eq}(k_0),\\
        &\bar{\Delta}_{\chi}^{>}(k,x)=2 \pi \delta(k^2-\tilde{M}_{\chi}^2) \sign{(k_0)} (1 \pm f_{\chi}^{eq}(k_0)),
   \end{aligned} 
\end{equation}
We observe that, based on the adiabatic expansion, we have treated the frequencies as if they were on Minkowski.

%For the cases of photophilic and photophobic scenarios, this means that 

%while for the QCD axion we have that $\Delta_{\tilde{G}G}=\braket{-\frac{1}{4} \tilde{G}_{\mu \nu}^a G^{\mu \nu a}}=\frac{1}{2} m_{a \, QCD}^2(T) \Phi$

For the condensate part, we follow a different approach, which is inspired, as mentioned before, by Refs ~\cite{ai2024qft, PhysRevD.85.063520,SIKIVIE2017331,HABIB1989335} and is necessary based on the preceding considerations.

We consider equation~(\ref{quantumEoMav}) and treat it in the particular case of our FLRW metric
%slightly show the former picture on the topic with the 2PI formalism as in Ref.~\cite{ai2024qft}.
\begin{equation}
\begin{aligned}
    \left(\partial_t^2+3H \partial_t -\frac{\nabla^2}{a^2}\right)\,\varphi+V'_{\text{eff}}(\varphi,\Delta_{\phi},T)+      \int d^4 x'\,  \Pi^r_{\varphi}(x,x') \varphi(x')  =0
    \end{aligned}
\end{equation}
This equation can also be rewritten in a similar expression to the ones of Refs~\cite{ai2024qft, PhysRevD.85.063520,SIKIVIE2017331,HABIB1989335} for the case of a quartic potential
\begin{equation}
\begin{aligned}
   & \left(\partial_t^2+3H \,\partial_t -\frac{\nabla^2}{a^2}+m_{\phi}^2+\frac{\lambda_{\phi}}{6} \Delta_{\phi}^{++}(x,x) \right)\varphi(x)\\&+ g_{\phi \chi} \Delta_{O_{\chi}}(x)+     \int d^4 x'\,  \Pi^r_{\varphi}(x,x') \varphi(x') + \int d^4 x' \,V^r_2(x,x') \varphi^2(x') +\int d^4 x' \,V^r_3(x,x') \varphi^3(x')=0 \qquad \qquad
    \end{aligned}
\end{equation}
where in the particular case of quartic potential $V^r_3(x-x')=\frac{\lambda_{\phi}}{6} \delta^{(4)}(x-x') $ and $V^r_3(x-x')$ takes care of interactions with $\phi$ .
We notice that the contribution from interaction processes with one quantum of $\varphi$ depends on 
\begin{itemize}
    \item The $\Delta_{O_{\chi}}(x)$ contribution, which has a "more deterministic nature" and is a classical-like term, which we would have obtained from the classical equation of motion (not as an average value) and depends on the surrounding plasma;
    \item The contribution from the retarded self-energy $\Pi^r_{\varphi}(x,x')$, so from the rest $\Gamma_2$, which comes from the interactions between the axions and the background plasma, giving the collision terms, but also holds noisy terms and non-Markovianity, as already noticed in Ref.~\cite{proukakis2023unifieddescriptioncorpuscularfuzzy} for non-relativistic case.
    \item The contribution from self-interactions $V^r_2$ and $V^r_2$, which can be non-perturbative, but also in the perturbative regime, can lead to both condensate and kinetic regimes.
    
\end{itemize}

We will assume in the following of this section, in leading order and justified by the fact that we are interested in systems in a radiation-dominated era, that the average $\Delta_{O_{\chi}}(x)$ is not significantly affected by the backreaction of the axion, so we have no creation of SM condensates. We then take $\Delta_{O_{\chi}}(x)=0$ by statistical isotropy of the Universe and because it is a pseudoscalar. 

For the case of freeze-in, we are interested in the evolution of photophilic ALPs for low reheating temperature as $T_{\text{RH}}=10-100 \mathrm{MeV}$, so after the QCD phase transition, or any $SU(3)$ dark sector transition giving them mass. 

As we discuss in the following, they are furthermore in a non-relativistic regime. If the decay rate $\Gamma_\Phi$ is high enough compared to the Hubble parameter $H$, they can significantly decay into photons and contribute to $\Delta N_{\text{eff}}$.

Consequently, we assume perturbation theory for both self-interactions and interactions with the SM plasma and neglect Bogoliubov particle production (since in such regimes $\frac{\dot{\Omega_k}}{\Omega_k} \sim H/m_{\Phi}$ for axions and $\frac{\dot{\Omega_k}}{\Omega_k} \sim H/T$ for the hot plasma particles are much smaller than $1$).
We can then adopt the generalized optical theorem and polology properties of Green's function \cite{schwartz2014quantum}, and from the hypothesis of negligible Bogoliubov particle production and perturbation theory, 
we can substitute the field $\varphi(x)=\int \frac{d^3{\vec{k}}}{(2 \pi)^3} \frac{1}{\sqrt{\omega_k}}\left(\Delta_{\vec{k}} e^{-i k \cdot x}+  \Delta^*_{\vec{k}}  e^{i k \cdot x}   \right)
$ and the analogous mode expansion for the other quantum fields.

All these hypotheses, disregarding the non-relativistic limit and neglecting thermal corrections to the mass, bring us to 

\begin{equation}\label{basicond}
    \dot{\rho}_{\varphi}+3H \rho_{\varphi}+H \,\sum_i p_{i \,\varphi}= -2 \gamma M_{\phi}(T)\, n_{\varphi}-2 \sigma {n}_{\varphi}^2
\end{equation}

where $M_{\varphi}^2(T)=m_{\phi}^2+\frac{\lambda_{\phi}}{2} \Delta_{\phi}^{++}(x,x)$ and explicitly
\begin{equation}
    \rho_{\varphi}=\tfrac12 \dot\varphi^2 + \tfrac{1}{2a^2} (\nabla \varphi)^2 + V(\varphi)
\end{equation}
and 
\begin{equation}
    p_{i} = \tfrac1{a^2} (\partial_i \varphi)^2 \,  - \left[\tfrac12 \dot\varphi^2+\tfrac{1}{2a^2}(\nabla \varphi)^2+   V(\varphi) \right],
\end{equation}

Furthermore, we obtain from using a similar procedure to Chapter 24.3 of Ref.~\cite{schwartz2014quantum} the following collision terms, as the ones obtained in Ref.~\cite{ai2024qft} for the non-relativistic limit of $\varphi$
\begin{equation}
\begin{aligned}
    &-2 \gamma=\sum_{\text{proc}} \left[\frac{1}{2 M_{\phi}} \int \Pi_i \frac{d^3 \vec{k}_i}{(2\pi^3) 2k_i^0} (2 \pi)^4 \times \delta^{(4)}(\bar{k}+k_{A1}+...-k_{B1}-...) |\mathcal{M}_{\varphi A_1... \rightarrow B_1... }|^2  \qquad \qquad \qquad \qquad \qquad \qquad \qquad \right.\\ &\left. \times \left\{(1 \pm f_{A1})...f_{B1}...-f_{A1}...(1 \pm f_{B1})...        \right\}  \right]\qquad \qquad \qquad \qquad,
    \end{aligned}
\end{equation}
and 
\begin{equation}
\begin{aligned}
   & -2 \frac{\sigma}{M_{\phi}}=\sum_{\text{proc}} \left[\frac{1}{(2 M_{\phi})^2} \int \Pi_i \frac{d^3 \vec{k}_i}{(2\pi^3) 2k_i^0} (2 \pi)^4 \delta^{(4)}(2\bar{k}+k_{A1}+...-k_{B1}-...)\qquad \qquad \qquad \qquad \qquad \qquad \qquad \qquad \qquad \right. \\ &\left. \times \left\{(1 \pm f_{A1})...f_{B1}...-f_{A1}...(1 \pm f_{B1})...        \right\}   |\mathcal{M}_{(2\varphi) A_1... \rightarrow B_1... }|^2 \right].\qquad \qquad \qquad \qquad \qquad \qquad
    \end{aligned}
\end{equation}
%and identified $\Delta_k \Delta_k^* \rightarrow f_k(t)$
where the sum is over all processes, with SM particles and $\phi$ quanta and
%we have used $ f_{\vec{k}}(t)  f^*_{\vec{k}}(t)=\frac{1}{2\Omega_k(t)}$ and then
we have identified $\Delta^*_{\vec{k}}\Delta_{\vec{k}} \rightarrow \mathcal{N}_{\varphi}(\vec{k})$ the number operator of mode $\vec{k}$.
We obtained it in a similar fashion to Refs.\cite{proukakis2023unifieddescriptioncorpuscularfuzzy,Proukakis:2024pua,Domcke:2025lzg}, which produced similar partial results in non-relativistic and neutrino scenarios.

To implement the non-relativistic limit and neglect the thermal correction to the axion mass, we can adopt the same procedure as Refs.\cite{Ai_2024, ai2024qft}. 

We take the following form of the field $\varphi$
\begin{equation}
    \varphi(t)=A(t) \cos{\left[ \int^t M_{\phi}(t') dt'    \right]} \sim A(t) \cos{(M_{\phi}(t) t)}
\end{equation}
The approximative EoMs for $A(t)$ is then

\begin{equation}\label{EoMA(t)0}
    \frac{dA}{dt}+\left(\gamma+\frac{3}{2} H +\frac{1}{2 M_{\phi}(t)} \frac{d M_{\phi} }{dt}               \right)A(t)+\frac{\sigma}{2} A^3(t)=0,
\end{equation}

where 
\begin{equation}
    \gamma= -\frac{Im \left[\tilde{\pi^r}_{\varphi}(M_{\phi})    \right]}{2 M_{\phi}}, \qquad \sigma= -\frac{Im \left[\tilde{v^r}_{\varphi}(2M_{\phi})    \right]}{24 M_{\phi}}, \qquad 
\end{equation}
and the tilde denotes the Fourier transform:

\begin{equation}
    \tilde{\pi^r}_{\varphi}(\omega)=\int_{-\infty}^{+\infty} dt' e^{i \omega (t-t')} \pi^r_{\varphi}(t,t').
\end{equation}
where $\pi^r_{\varphi}$ and $v^r_{\varphi}$ are the retarded self-energy and potential.

We can rewrite Eq.~(\ref{EoMA(t)0}) in the following way (assuming both $\frac{\dot{A}}{A} \ll M_{\phi}$ and $\frac{\dot{M_{\phi}}}{M_{\phi}} \ll M_{\phi}$:
\begin{equation}\label{boltzmanncondo}
     \dot{n}_{\varphi}+3 H n_{\varphi}=2 \gamma n_{\varphi}+\frac{2 \sigma}{M_{\phi}} n^2_{\varphi}.
\end{equation}
that is the same form of Ref.~\cite{ai2024qft}.

 \subsection{\texorpdfstring{ Axion freeze-in: numerical solutions and results}{Axion freeze-in: numerical solutions and results}}.
As mentioned before, we firstly specialize in an axion freeze-in scenario where we consider an inflation scenario with low reheating temperature (reference values we take are $T_{\text{RH}} =5 \,\, \mathrm{MeV}$, $T_{\text{RH}} =10 \,\, \mathrm{MeV}$ and over to $T_{\text{RH}} =100 \,\, \mathrm{MeV}$ that are compatible with the Big Bang Nucleosynthesis \cite{baumann2022cosmology,Marsh:2024ury,ohare2024cosmology}). There is no contribution from topological defects since we deal with a preinflationary scenario, so we consider only a standard misalignment angle $\varphi=\varphi(t)$, as mentioned before.

In the approximations taken in the former subsection, the quantum equations become of the form
\begin{equation}\label{ntequaz}
    \begin{aligned}
        &\dot{n}_{\phi}+3 H n_{\phi}=\mathcal{C}[f_{\phi},f_{\chi}] \\
          &\dot{n}_{\varphi}+3 H n_{\varphi}= \mathcal{C}_{1}[f_{\chi}]  n_{\varphi}+ \mathcal{C}_{2}[f_{\chi}] n^2_{\varphi}
    \end{aligned}
    \end{equation}
   where $\mathcal{C}$, $\mathcal{C}_1$ and $\mathcal{C}_2$ are collision terms as before and it is relevant to underline that $\mathcal{C}_1$ and $\mathcal{C}_2$ are collision terms with reduced phase space since we have obtained those in the non-relativistic limit for the axion, where the phase space function $f_{\varphi}$ is of the form $n_{\varphi} \delta^{(3)}(\vec{k}_{\varphi})$ .
   
    The relevant aspect we point out here is the significant difference in our treatment from former references \cite{Blum:2014vsa,Baumholzer_2021,LanghoffPhysRevLett.129.241101}, since here we obtain that, apart from the irreducible axion misalignment contribution, even the axion misalignment condensate can freeze-in and contribute to DM and has a different form for the collision term in comparison to pure Boltzmann approximation, even assuming a tiny initial angle so that we can improve the estimation of the axion abundances coming from both freeze-in and misalignment.
    This aspect is particularly relevant for the freeze-in production arising from the Primakoff process and lepton-antilepton annihilation, as follows. 
    
    Given the form of the collision terms in Eq.~\ref{ntequaz}, the contribution of the average field is significant, in comparison with a pure Boltzmann equation, if the scattering amplitude does not vanish for small axion momenta $\vec{k}_{\varphi}$.
    This is true for those processes in the vacuum; however, we deal with axions in the primordial plasma and plasma effects arise, leading to non-vanishing limits when we take care of it by considering a plasma mass $m_{\gamma}$.

    However, as we will see, there is a further difference arising from the distinction between the condensate collision term and the standard Boltzmann collision term. This more precise evaluation leads to a total effect of reduced Dark Matter production, as the Boltzmann collision term overestimates production from low-energy modes.

We assume the axions to be non-relativistic at late times. This assumption is helpful since we then have $m_{\Phi} n _ {\Phi}=m_{\Phi}(n_{\phi}+n_{\varphi}) = \rho _{\Phi} $ and can be justified in the following way for the parameter space of our interest, as similarly done for the photophilic and photophobic cases in Ref.\cite{LanghoffPhysRevLett.129.241101}. Since we are dealing with axions produced via freeze-in, we expect an axion energy of approximately the same order as the reheating temperature that will redshift proportionally to the primordial plasma temperature. For the case of an axion with mass above a minimal reheating temperature as $T_{RH}=10 \, \mathrm{MeV}$, the non-relativistic limit is trivial; for masses below this, it comes from noticing that we are considering masses up to $\sim 5 \times 10^4 \, \mathrm{eV}$.

%$ with $\rho_i$ the energy density of the i species and $m_i$ the mass.
\subsection{Some important aspects on collision terms and their numerical treatment}

    We have solved these equations numerically with  MicrOmegas 6.1.15 \cite{alguero2024micromegas} by solving the resulting equations for the comoving number densities $Y_i(T) =n_i(T)/s(T)$ where $n_i$ is the number density of the i species, $T$ is the temperature of the primordial plasma, related to the cosmological time $t$ by 
\begin{equation}\label{Tvst}
    dt=-\frac{dT}{\bar{H}(T)T},
\end{equation}

$s(T)=\frac{2 \pi^2}{45} T^3 h_{\text{eff}}(T)$ is the entropy density of the primordial plasma, $h_{eff} $ the effective entropy degrees of freedom and 
\begin{equation}
    \bar{H}(T)=\frac{H(T)}{1+\frac{1}{3} \frac{d \ln(h_{\text{eff}}(T))}{d \ln T}},
\end{equation}
with the relation \ref{Tvst} coming from the standard assumption of the entropy conservation law 
\begin{equation}
    \frac{ds}{dt}=-3Hs,
\end{equation}
and substituting the expression for the entropy density $s$ in terms of temperature $T$ and $h_{\text{eff}}$.

This comes from the fact that we have, in general, for the comoving entropy density the following relation coming from the second law of Thermodynamics
\begin{equation}
    s=\frac{a^3}{T} \Big(\rho+p-\sum_i \mu_i n_i          \Big)= a^3 \Big( g_s T^3-\sum_i \mu_i g_i T^2 \Big),
\end{equation}
where $\rho$ and $p$ are the energy density and pressure of the plasma, $n_i$ the number density of the ith species in the plasma with chemical potential $\mu_i$.

In the usual case, the above $s$ remains constant during adiabatic expansion when there is no entropy ejection/injection from or into the plasma. However, the production of ALPs (and later their possible decay
back into the photons) leads to non-conservation of this entropy, reflected as a deviation in the temperature’s usual evolution as $T \sim 1/a$ and deviations from the value of the effective number of neutrino species ${N}_{\mathrm{eff}}$ obtained in SM \cite{Sikivie2008,ohare2024cosmology,baumann2022cosmology}. We can compute this change using an equation for the temperature change, as described in Ref.~\cite{jain2024new}. 

The idea is that the only significant contribution to the change in the plasma temperature would come from ALP decays to photons when $\Gamma_{\text{decay}} \gg H$ before recombination. 
This causes a heating of the plasma, and we have

\begin{equation}
    ds=-\frac{a^3}{T} d\rho_{\Phi},
\end{equation}
where we neglect the pressure term since ALP is already in a non-relativistic regime when decaying. Using the relation between the total temperature $T=\bar{T}(1+\delta)$, where $\bar{T}$ is the unperturbed plasma temperature, and we define the useful quantity $\delta$, and assuming $\delta \ll 1$, one obtains

\begin{equation}\label{eqdeltaNeff}
   \frac{d \delta}{d \ln x} =-\frac{15}{2 \pi^2} \frac{x^4}{m_{\Phi}^4}  \frac{d \rho_{\Phi}}{d \ln x}
\end{equation}
where $x=m_{\Phi}/\bar{T}$.

We can then estimate $\Delta {N}_{\mathrm{eff}}$ from the definition of ${N}_{\mathrm{eff}}$ as
\begin{equation}
    \rho_{r}=\rho_{\gamma}+\rho_{\nu}=\frac{\pi^2}{15} \Big[ 1+\frac{7}{8}      \Big(  \frac{4}{11}\Big)^{4/3} {N}_{\mathrm{eff}} \Big]
\end{equation}
where $\rho_{\gamma}=2 \frac{\pi^2}{30} T^4$ is the photon energy density, $\rho_{\nu}=3 \times 2 \times \frac{7}{8} \times \frac{\pi^2}{30} T^4_{\nu}$ the relativistic neutrinos energy density and $T_{\nu}$ is the neutrino temperature that is different from plasma temperature after decoupling.
Then
\begin{equation}
    \Delta {N}_{\mathrm{eff}}={N}_{\mathrm{eff}}-\bar{N}_{\text{eff}}=\bar{N}_{\text{eff}} \Bigg[\frac{1}{(1+\delta)^4} -1     \Bigg] \sim -12.18\, \delta
\end{equation}
where $\bar{N}_{\text{eff}} \simeq 3.044$ \cite{Akita_2020,Froustey_2020,Bennett_2021,PhysRevD.108.L121301,Drewes_2024} for the usual cosmology with just SM particles, without any production or loss of entropy.
The best experimental estimation of ${N}_{\mathrm{eff}}$ comes from the Planck Collaboration, in particular the last Planck+BAO 2018 measures,  which obtained in the framework of fitting with the $\Lambda$ the value ${N}_{\mathrm{eff}} =2.99  \pm0.17$ at $68 \%\, \, \text{CL}$ \cite{refId0}.

Furthermore, this estimation is in agreement with the value inferred from the abundances of primordial SM particles and elements within the framework of BBN, $N_{\mathrm{eff}}=2.89 \pm 0.28$. Both the two values are compatible with $\bar{N}_{\text{eff}} \simeq 3.044$ from SM, but it also mean from these estimations that a value of $ \Delta {N}_{\mathrm{eff}}>0.5$ would have been spotted already as a $2 \sigma$ discrepancy with the theory \cite{Sikivie2008,ohare2024cosmology}.

%ALERT need to integrate it with definition of thermodynamical quantities and cosmological ones

Having this in mind, we can use the equations (\ref{ntequaz}) to obtain the density parameter $\Omega_\Phi h^2$ as
\begin{equation}
 \Omega_\Phi h^2=\frac{m_\phi Y_\Phi^0 s_0 h^2}{\rho_c}   
\end{equation}
where $\rho_c$ is the critical density of the Universe, $s_0 \simeq 2.9 \times 10^9 \mathrm{m}^{-3}$ is the entropy density at the current time, and $Y_\Phi^0$ is the total axion abundance calculated at the final temperature $T_0$ of today.

Furthermore, we use the former results from solving the equations (\ref{ntequaz}) and (\ref{eqdeltaNeff}) to get the $\Delta {N}_{\mathrm{eff}}$, helpful in constraining the axion parameters along with the requirement of $\Omega_\Phi h^2<0.12=\Omega_{\text{DM}} h^2$.

We notice how the comoving number density is related to the fraction $\xi$ adopted in Ref.~\cite{jain2024new} by
\begin{equation}
   \xi(T)=Y(T) \times \frac{m \, s_0}{\rho_c \, \Omega_{DM} h^2}  .
\end{equation}
It has a simple physical meaning: the fraction with which our axion contributes to cold dark matter, if it were redshifted from temperature $T$ to $T_0$ with a cold dark matter behaviour $\rho \propto a^{-3}$.%add anything?

The appendix provides additional details on the numerical methods used to solve such equations. In the following, we discuss the collision terms, specifically the relevant processes, with the 2PI and Feynman diagrams,  and some limiting analytical expressions, as well as the results and bounds for each model. %The appendix also shows details on the 2PI and Feynman diagrams.

We do not assume Maxwell-Boltzmann statistics for the axions.

\subsection{Photophilic ALPs}
As previously discussed in the literature \cite{LanghoffPhysRevLett.129.241101,Blum:2014vsa,Baumholzer_2021,ai2024qft}, the relevant tree-level processes for particles and condensates in the photophilic case are the Primakoff process, lepton-antilepton annihilation (with the main contributions coming from electrons and muons from our analysis, analogously as obtained in Ref.\cite{ai2024qft}), and axion decay to two photons. We obtain compatible results for the collisional terms with the former works and extend their analysis to processes with particles and condensates.
We have considered the relevant $\Gamma_2$ and Feynman diagrams as shown in the figures (\ref{Fig:Fey})-(\ref{Fig:Fey9}), taking care of the interaction term
\begin{equation}
    \mathcal{L}_{int}=-\frac{1}{4} g_{\phi \gamma \gamma} (\varphi+\phi) \tilde{F}_{\mu \nu} F^{\mu \nu},
\end{equation}
and the QED Lagrangian.
\begin{figure}[H]
\centering

% Resize to fit column width
\resizebox{0.9\linewidth}{!}{
\begin{tikzpicture}
\begin{feynman}

  % === FIRST ROW ===
  \vertex (start) {$\Gamma_2 \sim -i \, \Bigg($};
  \vertex [right=0.9cm of start] (a) {$\bigotimes$};
  \vertex [dot, right=0.9cm of a] (b) {};
  \vertex [dot, right=0.9cm of b] (c) {};
  \vertex [right=0.9cm of c] (d) {$\bigotimes$};
  \vertex [right=0.3cm of d] (plus1) {+};

  \vertex [dot, right=0.3cm of plus1] (b1) {};
  \vertex [dot, right=0.9cm of b1] (c1) {};
  \vertex [right=0.3cm of c1] (plus2) {+};

  % === SECOND ROW ===
  \vertex [right=0.3cm of plus2] (a2) {$\bigotimes$};
  \vertex [dot, right=0.8cm of a2] (b2) {};
  \vertex [dot, right=0.8cm of b2] (c2) {};
  %\vertex [right=0.3cm of c2] (plus31) {+};
  \vertex [right=0.8cm of c2] (d2) {$\bigotimes$};
  \vertex [right=0.5cm of d2] (plus3) {+};

  % Add more rows if needed:
  % \vertex [below=0.8cm of a2] (...) ...

			%	\vertex [below left=of c] (b) {\(\bar{f}\)};

\diagram* {(a) -- [scalar] (b),
(b) -- [scalar] (c),
				(b) -- [scalar, half left] (c),
				(b) -- [scalar, half right] (c),
				(c) -- [scalar] (d),

				(b1) -- [scalar] (c1),
				(b1) -- [scalar, half left] (c1),
				(b1) -- [scalar, half right] (c1),
				
				(a2) -- [scalar] (b2),
				(b2) -- [scalar] (c2),
				(b2) -- [boson, half left] (c2),
				(b2) -- [boson, half right] (c2),
				(c2) -- [scalar] (d2),

				%	(f4) --[scalar] (d), 
				%	(f4) --[boson] (e),
				
			}; 
			
			%\vertex [right=0.5 cm of e2](01) {$\Gamma_2 \sim -i \,\,\, \Bigg( \Bigg.$};
			
			% Axion to start
			\vertex [right=0.5 cm of plus3] (a31) {$\bigotimes$};
			\vertex [dot, right=1.2cm of a31] (b31) {};
			\vertex [dot, above right=0.5cm of b31] (b311) {};
			\vertex [dot, right=0.5cm of b311] (b321) {};
			\vertex [dot, right=1cm of b31] (c31) {};
			\vertex [right=1.2cm of c31] (d31) {$\bigotimes$};
			\vertex [right=0.5 cm of d31] (e31) {$\Bigg. \Bigg)$};
			
			% Diagram
\diagram* {
				(a31) -- [scalar] (b31),
				(b31) -- [scalar] (c31),
				(b31) -- [boson, half left] (b311),
				(b311) -- [fermion, half left, shorten <=0.1cm] (b321),
				(b311) -- [anti fermion, half right, shorten <=0.1cm] (b321),
				(b321) -- [boson, half left] (c31),
				(b31) -- [boson, half right] (c31),
				(c31) -- [scalar] (d31),
			};
		\end{feynman}
	\end{tikzpicture}
    }
	\caption{$\Gamma_2$ for the photophilic case with the relevant 2PI diagrams. The $\bigotimes$ denote the condensate $\varphi$.}
    \label{Fig:Fey}
\end{figure}
\begin{figure}[H]
\centering

    % Diagram 1: Primakoff process
    \begin{minipage}{0.4\linewidth}  % Adjust width to fit two columns
    \centering
    \begin{tikzpicture}[scale=0.6]  
        \begin{feynman}
            \vertex (a1) at (-2, 2) {$\gamma$};  
            \vertex (a2) at (-2, -2) {$f$};    
            \vertex (b1) at (2, 2) {$\bigotimes$};   
            \vertex (b2) at (2, -2) {$f$};  
            \vertex [dot](c1) at (0, 1) {};  
            \vertex [dot](c2) at (0, -1) {}; 
            \diagram*{
                (a1) -- [photon] (c1),      
                (a2) -- [fermion] (c2),     
                (c1) -- [scalar] (b1),     
                (c2) -- [fermion] (b2),      
                (c1) -- [photon, edge label'=$\gamma$] (c2), 
            };
        \end{feynman}
    \end{tikzpicture}
    \caption{Primakoff process producing one axion condensate}
    \label{Fig:Fey1}
    \end{minipage}
    \hfill  % Horizontal space between minipages
    % Diagram 2: Fermion-antifermion annihilation
    \begin{minipage}{0.4\linewidth}
    \centering
    \begin{tikzpicture}[scale=0.6]
        \begin{feynman}
            \vertex (a1) at (-2, 2) {$\bar{f}$};  
            \vertex (a2) at (-2, -2) {$f$}; 
            \vertex (b1) at (2, 2) {$\gamma$};  
            \vertex (b2) at (2, -2) {$\bigotimes$}; 
            \vertex [dot](c1) at (0, 1) {};  
            \vertex [dot](c2) at (0, -1) {}; 
            \diagram*{
                (c1) -- [fermion] (a1),       
                (a2) -- [fermion] (c2),       
                (c1) -- [photon] (b1),        
                (c2) -- [scalar] (b2),        
                (c1) -- [photon, edge label'=$\gamma$] (c2),  
            };
        \end{feynman}
    \end{tikzpicture}
    \caption{Fermion-antifermion annihilation mediated by one photon propagator}
    \label{Fig:Fey2}
    \end{minipage}

    \vspace{1em}  % Vertical space between rows of minipages
    
    % Diagram 3: Scalar particle decay into photons
    \begin{minipage}{0.4\linewidth}
    \centering
    \begin{tikzpicture}[scale=0.6]
        \begin{feynman}
            \vertex (a1) at (0, 2) {$\bigotimes$};  
            \vertex (b1) at (2, 3) {$\gamma$};  
            \vertex (b2) at (2, 1) {$\gamma$};  
            \vertex [dot] (c) at (1, 2) {}; 
            \diagram*{
                (a1) -- [scalar] (c),       
                (c) -- [photon] (b1),       
                (c) -- [photon] (b2),       
            };
        \end{feynman}
    \end{tikzpicture}
    \caption{Axion condensate decay to two photons}
    \label{Fig:Fey3}
    \end{minipage}
  \hspace{2.5 cm} % Horizontal space between minipages
     \begin{minipage}{0.4\linewidth}
    \centering
    \begin{tikzpicture}[scale=0.6]
        \begin{feynman}
            \vertex (a1) at (0, 2) {$\phi$};  
            \vertex (b1) at (2, 3) {$\gamma$};  
            \vertex (b2) at (2, 1) {$\gamma$};  
            \vertex [dot] (c) at (1, 2) {}; 
            \diagram*{
                (a1) -- [scalar] (c),       
                (c) -- [photon] (b1),       
                (c) -- [photon] (b2),       
            };
        \end{feynman}
    \end{tikzpicture}
    \caption{Axion particle decay to two photons}
    \label{Fig:Fey4}
    \end{minipage}

   \vspace{1em}  
    % Diagram 4: phi-phi scattering
    \begin{minipage}{0.4\linewidth}
    \centering
    \begin{tikzpicture}[scale=0.5]
        \begin{feynman}
            \vertex (a1) at (-2, 2) {$\bigotimes$};  % Incoming axion
            \vertex (a2) at (-2, -2) {$\bigotimes$}; % Incoming axion
            \vertex (b1) at (2, 2) {$\bigotimes$};   % Outgoing axion
            \vertex (b2) at (2, -2) {$\bigotimes$};  % Outgoing axion
            \vertex [dot](c1) at (0, 0) {};    % Interaction vertex
            \diagram*{
                (a1) -- [scalar] (c1),        % Incoming axion to vertex
                (a2) -- [scalar] (c1),        % Incoming axion to vertex
                (c1) -- [scalar] (b1),        % Outgoing axion from vertex
                (c1) -- [scalar] (b2),        % Outgoing axion from vertex
            };
        \end{feynman}
    \end{tikzpicture}
    \caption{$\varphi \varphi \to \varphi \varphi$ scattering}
    \label{Fig:Fey5}
    \end{minipage}
\hfill
     % Diagram 4: phi-phi scattering
    \begin{minipage}{0.4\linewidth}
    \centering
    \begin{tikzpicture}[scale=0.6]
        \begin{feynman}
            \vertex (a1) at (-2, 2) {$\phi$};  % Incoming axion
            \vertex (a2) at (-2, -2) {$\phi$}; % Incoming axion
            \vertex (b1) at (2, 2) {$\phi$};   % Outgoing axion
            \vertex (b2) at (2, -2) {$\phi$};  % Outgoing axion
            \vertex [dot](c1) at (0, 0) {};    % Interaction vertex
            \diagram*{
                (a1) -- [scalar] (c1),        % Incoming axion to vertex
                (a2) -- [scalar] (c1),        % Incoming axion to vertex
                (c1) -- [scalar] (b1),        % Outgoing axion from vertex
                (c1) -- [scalar] (b2),        % Outgoing axion from vertex
            };
        \end{feynman}
    \end{tikzpicture}
    \caption{$\phi \phi \to \phi \phi$ scattering}
    \label{Fig:Fey6}
    \end{minipage}
\end{figure}

\begin{figure}[H]
\centering

    % Diagram 1: Primakoff process
    \begin{minipage}{0.4\linewidth}  % Adjust width to fit two columns
    \centering
    \begin{tikzpicture}[scale=0.6]  
        \begin{feynman}
            \vertex (a1) at (-2, 2) {$\gamma$};  
            \vertex (a2) at (-2, -2) {$f$};    
            \vertex (b1) at (2, 2) {$\phi$};   
            \vertex (b2) at (2, -2) {$f$};  
            \vertex [dot](c1) at (0, 1) {};  
            \vertex [dot](c2) at (0, -1) {}; 
            \diagram*{
                (a1) -- [photon] (c1),      
                (a2) -- [fermion] (c2),     
                (c1) -- [scalar] (b1),     
                (c2) -- [fermion] (b2),      
                (c1) -- [photon, edge label'=$\gamma$] (c2), 
            };
        \end{feynman}
    \end{tikzpicture}
    \caption{Primakoff process producing one axion particle}
    \label{Fig:Fey7}
    \end{minipage}
    \hfill  % Horizontal space between minipages
    % Diagram 2: Fermion-antifermion annihilation
    \begin{minipage}{0.4\linewidth}
    \centering
    \begin{tikzpicture}[scale=0.6]
        \begin{feynman}
            \vertex (a1) at (-2, 2) {$\bar{f}$};  
            \vertex (a2) at (-2, -2) {$f$}; 
            \vertex (b1) at (2, 2) {$\gamma$};  
            \vertex (b2) at (2, -2) {$\phi$}; 
            \vertex [dot](c1) at (0, 1) {};  
            \vertex [dot](c2) at (0, -1) {}; 
            \diagram*{
                (c1) -- [fermion] (a1),       
                (a2) -- [fermion] (c2),       
                (c1) -- [photon] (b1),        
                (c2) -- [scalar] (b2),        
                (c1) -- [photon, edge label'=$\gamma$] (c2),  
            };
        \end{feynman}
        
    \end{tikzpicture}
    \caption{Fermion-antifermion annihilation mediated by one photon propagator}
    \label{Fig:Fey8}
    \end{minipage}

    %\hspace{0.2 cm} % Horizontal space between minipages
    \vspace{1em}  
    % Diagram 4: phi-phi scattering
    \begin{minipage}{0.4\linewidth}
    \centering
    \begin{tikzpicture}[scale=0.5]
        \begin{feynman}
            \vertex (a1) at (-2, 2) {$\bigotimes$};  % Incoming axion
            \vertex (a2) at (-2, -2) {$\bigotimes$}; % Incoming axion
            \vertex (b1) at (2, 2) {$\varphi$};   % Outgoing axion
            \vertex (b2) at (2, -2) {$\varphi$};  % Outgoing axion
            \vertex [dot](c1) at (0, 0) {};    % Interaction vertex
            \diagram*{
                (a1) -- [scalar] (c1),        % Incoming axion to vertex
                (a2) -- [scalar] (c1),        % Incoming axion to vertex
                (c1) -- [scalar] (b1),        % Outgoing axion from vertex
                (c1) -- [scalar] (b2),        % Outgoing axion from vertex
            };
        \end{feynman}
    \end{tikzpicture}
    \caption{$\varphi \varphi \to \phi \phi$ scattering diagram}
    \label{Fig:Fey9}
    \end{minipage} 
\hfill
     % Diagram 4: phi-phi scattering
    \begin{minipage}{0.4\linewidth}
    \centering
    \begin{tikzpicture}[scale=0.6]
        \begin{feynman}
            \vertex (a1) at (-2, 2) {$\bigotimes$};  % Incoming axion
            \vertex (a2) at (-2, -2) {$\phi$}; % Incoming axion
            \vertex (b1) at (2, 2) {$\phi$};   % Outgoing axion
            \vertex (b2) at (2, -2) {$\bigotimes$};  % Outgoing axion
            \vertex [dot](c1) at (0, 0) {};    % Interaction vertex
            \diagram*{
                (a1) -- [scalar] (c1),        % Incoming axion to vertex
                (a2) -- [scalar] (c1),        % Incoming axion to vertex
                (c1) -- [scalar] (b1),        % Outgoing axion from vertex
                (c1) -- [scalar] (b2),        % Outgoing axion from vertex
            };
        \end{feynman}
    \end{tikzpicture}
    \caption{$\varphi \phi \to \varphi \phi$ scattering}
    \end{minipage}

\end{figure}

The usual collision operators for such processes, when considering a thermal plasma mass $m_{\gamma}$, are the following, as in Ref.~\cite{jain2024new}
\begin{equation}
    \begin{aligned}
        &\mathcal{C}_{\rm Prim}(\vec{k})=\int \frac{d^3 \vec{p}\,d^3 \vec{q}\,d^3 \vec{l}}{(2\pi)^5}  \frac{\delta^{3}( \vec{l}+\vec{q}-\vec{p}-\vec{k}) \, \delta(     \omega^{\gamma}_{\vec{l}}+\omega^{e}_{\vec{q}}- \omega^{\phi}_{\vec{k}} - \omega^{e}_{\vec{p}}) }{(2 \omega^{\gamma}_{\vec{l}})(2 \omega^{e}_{\vec{q}})(2 \omega^{\phi}_{\vec{k}})(2  \omega^{e}_{\vec{p}})  }\\
        &\times \frac{4 \pi \alpha_{em} g_{\phi \gamma \gamma}^2 }{(t-m_{\gamma}^2)^2} \left[-2 m_e^2 m_{\gamma}^2-2t^2(s-m_{\phi}^2)-t^3 \right.\\ &\left.-t\left(m_{\phi}^4+2(s-m_e^2)^2-2m_{\phi}^2(s+m_e^2)\right)             \right]  \\
        &\times \left[(1-f^e_{\vec{p}})f^{\gamma}_{\vec{l}} f^e_{\vec{q}} -f^e_{\vec{p}}(1+f^{\gamma}_{\vec{l}})(1-f^e_{\vec{q}})     \right],
    \end{aligned}
\end{equation}
for the Primakoff process with the electron
and
\begin{equation}
    \begin{aligned}
        &\mathcal{C}_{\rm ann}(\vec{k})=\int \frac{d^3 \vec{p}\, d^3 \vec{q}\, d^3 \vec{l}}{(2\pi)^5} \frac{ \delta^{3}( -\vec{l}+\vec{q}+\vec{p}-\vec{k}) \delta(     -\omega^{\gamma}_{\vec{l}}+\omega^{e}_{\vec{q}}- \omega^{\phi}_{\vec{k}} - \omega^{e}_{\vec{p}}) }{(2 \omega^{\gamma}_{\vec{l}})(2 \omega^{e}_{\vec{q}})(2 \omega^{\phi}_{\vec{k}})(2  \omega^{e}_{\vec{p}})  }\\
        &\times \frac{4 \pi \alpha_{em} g_{\phi \gamma \gamma}^2 }{(p+q)^4} \left[8 (p \cdot q +2m_e^2)(l \cdot (p+q))^2- 8(p+q)^2(l \cdot q) (l \cdot p)       \right]  \\
        &\times \left[(1+f^{\phi}_{\vec{k}}) (1+f^{\gamma}_{\vec{l}}) f^e_{\vec{p}} f^e_{\vec{q}} -f^{\phi}_{\vec{k}}f^{\gamma}_{\vec{l}}(1-f^e_{\vec{p}}) (1-f^e_{\vec{q}})     \right],
    \end{aligned}
\end{equation}
for electron-positron annihilation
and 
\begin{equation}
\begin{aligned}
  &\mathcal{C}_{\text{decay}}(\vec{k})=  \int  \frac{d^3 \vec{q} \,d^3 \vec{l}}{(2\pi)^2} \frac{\delta^{3}( -\vec{l}+\vec{q}-\vec{k}) \delta(     \omega^{\gamma}_{\vec{l}}+\omega^{\gamma}_{\vec{q}}- \omega^{\phi}_{\vec{k}}) }{(2 \omega^{\gamma}_{\vec{l}})(2 \omega^{\gamma}_{\vec{q}})(2 \omega^{\phi}_{\vec{k}})  } \left((l \cdot q)^2-m_{\gamma}^4         \right) \\ &\times \left[(f^{\gamma}_{\vec{l}}) f^{\gamma}_{\vec{q}} (1+f^{\phi}_{\vec{k}})  -f^{\phi}_{\vec{k}}(1+f^{\gamma}_{\vec{l}})(1+f^{\gamma}_{\vec{q}})     \right],
  \end{aligned}
\end{equation}
for axion decay to two photons.
The variables $s$ and $t$ are the usual Mandelstam variables \cite{schwartz2014quantum}, while the ones for $\varphi$ are of the same form but with the reduced phase space, as discussed before.
 It is easy to notice how the most significant contribution from fermion-antifermion annihilation comes from electrons (mostly) and muons, the lightest leptons.

 Some preliminary rough estimates can be performed, considering some limiting expressions and adopting the usual logic of current literature on distinguishing between the Boltzmann-suppressed limit, which is valid at the moment of $T \lesssim m_a$,  and the standard freeze-in regime, as in recent works~\cite{jain2024new,Arias:2025nub}.

 The main differences arising from taking into account the plasma effect for the scattering matrices $\mathcal{M}$ of the processes we are interested in, in comparison with the vacuum case, can be summarized as follows. For the Primakoff process, the structure of the scattering matrix is $\mathcal{M}_{\rm{prim}} \sim g_{a\gamma\gamma}\, \frac{1}{q^2}\, 
(\mathbf{k}\times\boldsymbol{\epsilon}_\gamma)\cdot \mathbf{q},
$ where $\mathbf{q}$ is the momentum transfer, $k$ the axion momentum and $\boldsymbol{\epsilon}_\gamma$ is the photon polarization. For the vacuum case, we expect it to behave $\mathcal{M}_{\rm{prim}} \sim k^2$ for small axion momenta, while with the plasma, the photons acquire an effective mass
\begin{equation}
m_\gamma = \omega_{\text{p}}, \qquad
\omega_{\text{p}}^2 = \frac{4\pi \alpha n_e}{m_e},
\end{equation}
with just a plasma of electrons here, for simplicity.
The photon dispersion relation becomes
\begin{equation}
\omega^2 = |\mathbf{k}|^2 + m_\gamma^2.
\end{equation}
The presence of a plasma mass modifies both the kinematics and
the small-$k$ scaling of the amplitudes also for fermion-antifermion annihilation and axion decay, allowing for contributions from small $k$, overall condensate contributions.

This significantly changes the threshold for which production is Boltzmann-suppressed.

\begin{figure}
    \centering
    \includegraphics[width=1.0\linewidth]{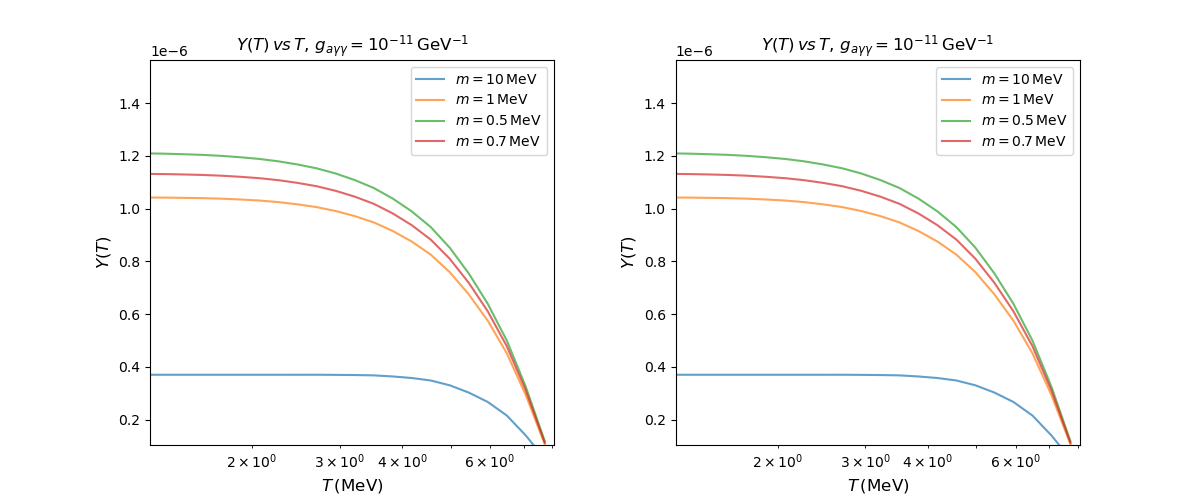}
    \caption{Plot with comparisons between the two methods with $T_{RH}=10 \, \mathrm{MeV}$ }
    \label{1e-11GeV-1}
\end{figure}

We show in the figures ~\ref{1e-11GeV-1} a plot of the values of the fraction $Y$ as a function of $T$ with $g_{a \gamma \gamma}=10^{-11} \mathrm{GeV}$ and $T_{RH}=10 \mathrm{MeV}$ and we show for each value of the mass two different results, one coming from the usual Boltzmann equation of $\Phi$ and neglecting self-interactions on the left side and the other from the coupled equations (but leaving the average field in a Boltzmann kinetic regime) and considering self-interactions on the right. In both cases, we account for plasma effects through thermal mass and contributions from electrons and muons. 
We observe that the results remain the same, provided self-interactions are significantly minor than photophilic processes but not negligible.

\begin{figure}
    \centering
    \includegraphics[width=0.8\linewidth]{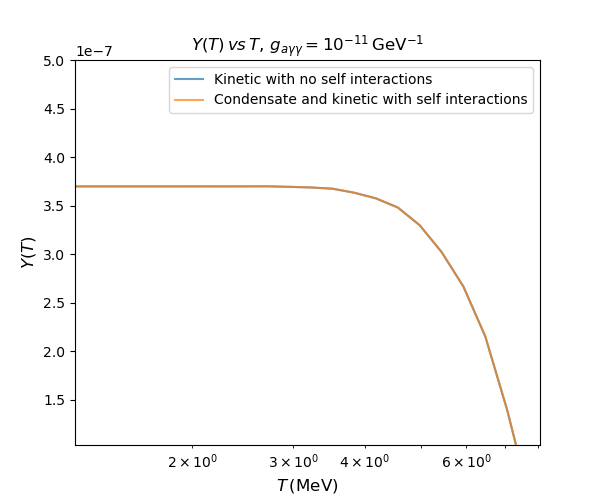}
    \caption{Plot with comparisons between the two methods with $T_{RH}=10 \, \mathrm{MeV}$ and $m_{\phi}=0.7 \, \mathrm{MeV}$ }
    \label{kinvskinstrano}
\end{figure}

We add a further check in the plot~\ref{kinvskinstrano}.

We observe that varying the masses alone results in a production enhancement for $m \sim T_{\text{RH}}$, leading to constraints with the lowest coupling constants at roughly those masses, as expected for a freeze-in mechanism.

\begin{figure}
    \centering
    \includegraphics[width=0.7\linewidth]{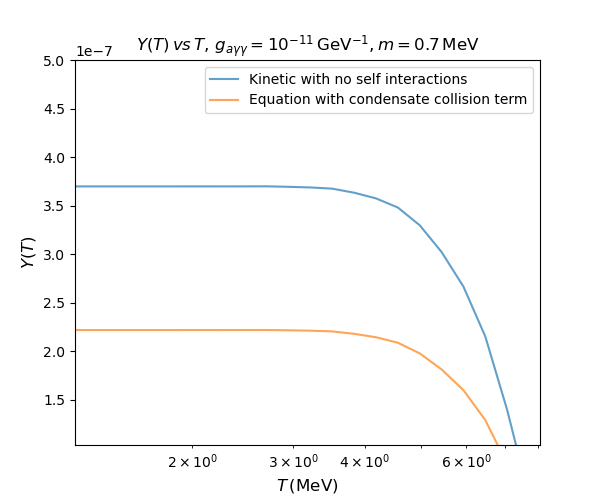}
    \caption{Comparison between usual Boltzmann method and ours for $T_{RH}=10 \, \mathrm{MeV}$  }
    \label{usualvsus}
\end{figure}

We further analyse the plot \ref{usualvsus}, comparing the standard Boltzmann kinetic equation for $\Phi$ with self-interactions to our set of equations \ref{ntequaz}. We see that the usual Boltzmann collision term overestimates the contribution from condensate low-energy modes.

\begin{figure}[h]
    \centering
    \includegraphics[scale=0.5]{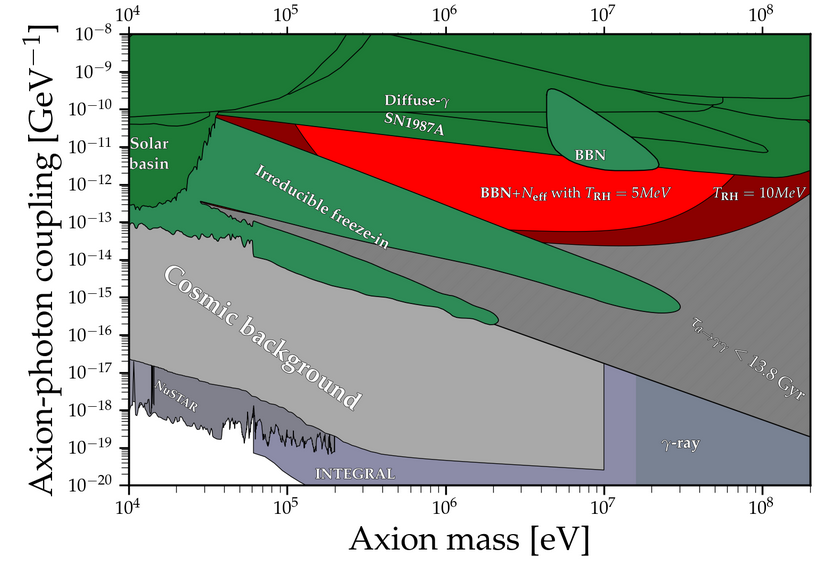}
    \caption{Plot of resulting constraints for the photophilic model with $T_{\text{RH}}=5-10 \,\mathrm{MeV}$. We show the regions in red that do not overlap with the former constraints from experiments, astrophysical observations, and especially the corresponding BBN ones.  Additionally, we observe a significant extension of the constraint region due to DM underproduction from the condensate part, and, mostly on the right of the parameter space, the muon contribution. Image adapted from the code in Ref.~\cite{axionLimits} .}
    \label{fig:magayyplot}
\end{figure}

Other significant differences are visible in the following graph \ref{fig:magayyplot} of our resulting anti-constraints, since we claim they do decay slower than the usual estimation of $\tau_{a \rightarrow \gamma \gamma}$ since it worked if the Boltzmann kinetic limit were valid at all the energies and were not a misalingment angle (even if very small), ranging in the two regions between respectively $m_a\sim 10^5-10^7 \,\mathrm{eV}$ with $g_{a \gamma \gamma}\sim10^{-10}- 10^{-13}\, \mathrm{GeV}^{-1}$  and $m_a\sim 10^7-10^9$ with $g_{a \gamma \gamma}\sim 10^{-11}-10^{-13}\, \mathrm{GeV}^{-1}$

The new region on the left primarily results from considering self-interactions and interactions between $\varphi$ and $\phi$ quanta. This is because $m_a g_{a\gamma\gamma}$ is higher in this region compared to the former BBN constraint region, implying a higher $m_a/f_a$.

In contrast, the right one mainly contributes from the condensate misalignment angle (collision terms for both $\varphi$ and $\phi$ are Boltzmann-suppressed for $m_a/T \gg 1$, but with the same behaviour)  and muon contribution for masses of the order or higher than muon mass.
A further important aspect of our analysis is that we limited ourselves to tree-level processes.
However, for energies of the order of the muon mass, we expect contributions from electron and muon loops.
We have considered them in our numerical analysis, by just taking into account the running of the fine-structure constant $\alpha_{\text{em}}(E)$ and using the code alphaQED \cite{JEGERLEHNER2008135,PhysRevD.97.114025,Jegerlehner:2019lxt}.

\section{Analytical model for DW networks}\label{DWtherm}

Here, we consider a post-inflationary scenario in which the axion model may lead to a domain wall problem. We outline our theoretical approach and then show some preliminary results for photophilic axions in a region of the parameter space $(m_a,g_{a \gamma \gamma}) $, in particular near to $m_a=10 \,\rm{MeV}$ and $g_{a \gamma}=10^{-11}\, \rm{GeV}$,   for which we can expect a relevance for the domain walls of both self-interactions (depending on $\tfrac{m_a^2}{f_a^2}$) and axion-photon interaction (depending on $g_{a \gamma \gamma}^2 $).

\subsection{On the VOS model of axion domain walls}

The approximations and procedures to obtain an approximating equation for the average energy density and the root-mean-square (rms) velocity for a DW network, as already sketched in Refs.\cite{Hindmarsh:1996xv, Oliveira:2004he, PhysRevD.93.043534} and in Section~(\ref{firstVOS}),
use precise approximations necessary to obtain the literature form of Ref.~\cite{PhysRevD.93.043534} for the VOS model, as already mentioned in the same Section~\ref{firstVOS}.

An important one is to assume $\angi{v^4}=v_{\rm rm}^2$ and a lot of hypothesis to obtain the VOS model without phenomenological energy loss are compatible with assuming the whole system as an ensemble of isolated domain walls that do not intersect with each other or bend with the Hubble horizon.

\subsection{Some useful orders of magnitude}

A system of a network of axion domain walls inside the SM primordial plasma is, in general, characterized by numerous processes; it is then essential to distinguish its relevant energy and length scales and study in each regime what is more relevant and what is not, or if there are regimes where all processes are essential.

Starting from the domain walls in the FLRW metric, we have the length scales
\begin{itemize}
    \item The width $\delta \sim (\gamma m_a)^{-1}$ of the DW, without any compression acting on it.
    \item The curvature radii $R_1$ and $R_2$ of the worldsheet of wall.
    \item Mean free path related to the axion self-interactions (mean free path for a center-of-mass collision between two DWs, correlation length for self-production of axions from %add here
    , length involved when two DWs collide violently)
    \item Hubble radius $r_H \sim H^{-1}$
\end{itemize}

Furthermore, we have the SM primordial plasma, which adds effects as the thermal friction and the compression on the wall, and it is also characterized by the plasma lengths.

In particular, the relevant plasma length scales—each associated with a characteristic energy scale—are the following, listed from the largest to the smallest \cite{le2000thermal,Hassan:2024nbl}:
\begin{itemize}
    \item Mean free path: 
    \[
        \tau_{\rm MFP} \sim (\alpha_{\rm em}^2\, T)^{-1} \coloneqq \Gamma^{-1},
    \]
    where $\Gamma$ represents the typical scattering rate between charged particles in the plasma. It characterizes the distance a particle travels, on average, before undergoing a significant collision.
    
    \item Debye length:
    \[
        l_D = m_D^{-1} \sim (\sqrt{\alpha_{\rm em}}\,T)^{-1}.
    \]
    This length quantifies the range of electrostatic screening in the plasma, arising from collective charge rearrangements. It governs how electric fields are exponentially suppressed beyond $l_D$, which is why it is often referred to as the \emph{electric scale}.
    
    \item Thermal wavelength:
    \[
        l_T \sim T^{-1}.
    \]
    This sets the microscopic length scale corresponding to the typical de Broglie wavelength of thermal excitations. It determines when quantum effects become important and serves as the shortest characteristic scale in a relativistic plasma.
\end{itemize}

\begin{table}[h!]
\centering
\begin{tabular}{|c|c|c|}
\hline
\multicolumn{3}{|c|}{\textbf{Table of main physical quantities}} \\ 
\hline
\textbf{Quantities} & \textbf{QCD axion} & \textbf{High mass ALPs} \\ 
\hline
Hubble parameter $H$ & $\simeq10^{-16}-10^{-11}\, \mathrm{eV}$  & $\simeq10^{-11}-10\, \mathrm{eV}$ \\
\hline
Hubble radius $r_H$ &  $\simeq10^4-10^9 \,\rm{m}$   & $\simeq10^{-11}-10^1 \,\rm{m}$ \\
\hline
Axion mass $m_a$ & $\simeq 10^{-6} -10^{-2} \,\rm{eV}$ & $\simeq 10^6 - 10^8  \,\rm{eV}$ \\
\hline
Axion Compton wavelength $m_a^{-1}$ & $\simeq 10^{-4} -1\, \rm m $ & $10^{-14}-10^{-12}\, \rm m$ \\
\hline
Curvature radii  & $ (\gamma m_a)^{-1} \lesssim R \lesssim r_H$ & $r_D \lesssim R \lesssim  r_H$ \\
\hline
\end{tabular}
\caption{Useful orders of magnitude for a generic axion domain wall network}
\end{table}

\subsection{Analytical model from nonequilibrium QFT}

Now comes the most crucial ingredient of our whole work.

We have mentioned in Section~\ref{boltzoptical} that we can obtain the usual Boltzmann collision terms by adopting the generalized optical theorem and, after assuming a plane wave expansion where we quantize the Fourier coefficients, the politopy properties of Green's functions.
The adoption of the plane wave expansion is not convenient with the non-linear dynamics of axion domain walls, since it is a proper expansion for a free particle with a squared potential. 

We can consider, with the instanton potential, at first glance, the planar domain wall solutions, moving rigidly with velocity $\vec{v}$ of the center-of-mass.
We follow then an idea similar to moduli quantization, where we write the classical field as a superposition of "more convenient" classical solutions of planar domain walls labeled not by $\vec{k}$, but $\vec{v}$.

Following the new procedure , we obtain a Boltzmann collision term which has an analogous structure to the usual one, since the combinatorial aspects and Feynman diagrams are not affected, but the structure of the scattering matrix $\mathcal{M}$ is different since different Feynman rules are valid.

Using such results, we obtain the analytical model for an axion domain wall network by assuming a statistical isotropy for the velocity, as we do for VOS models, and obtain the following Fokker-Planck equations

\begin{flalign*}
\begin{cases}
\dot{\rho} &= -H \rho (1+3v^2) -\frac{1}{l_f} \rho- \frac{c_w \rho}{L} v+P_{\rm lor}(T,v,\rho), \\
\dot{v} &= (1-v^2)\left[\frac{k_w(v)}{L} - 3Hv -\frac{\gamma v}{m_a \,l_f}+A_{\rm drg}(T,v,\rho)\right]
\end{cases}
\end{flalign*}
where we obtain the terms which can be associated to the usual drag force and energy-loss rate terms in Fokker-Planck equations

    \begin{equation*}
        A_{\rm drag}(T,v,\rho)=\sqrt{\int \frac{d^3 \vec{p}}{(2\pi)^3} \Big(\frac{\vec{p}}{m_a}\Big)^2 \,\mathcal{C}[f_{\vec{p}}] }
    \end{equation*} and 
    \begin{equation*}
        P_{\rm lor}(T,v,\rho)=\int \frac{d^3 \vec{p}}{(2\pi)^3} E_{\vec{p}} \,\mathcal{C}[f_{\vec{p}}]
    \end{equation*}

which includes all the phenomena we have mentioned and the additional contribution coming from the interaction with the $\phi$ field.

\begin{figure}
    \centering
    \includegraphics[width=0.7\linewidth]{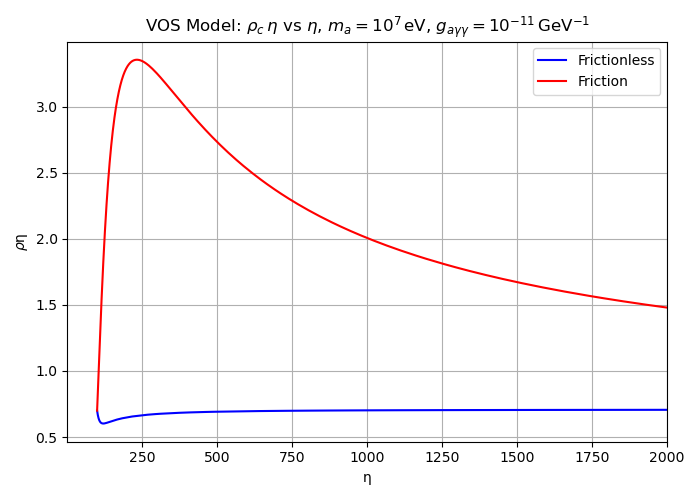}
   % \captionof{figure}{$\xi_c \eta$ vs $\eta$}

    \includegraphics[width=0.7\linewidth]{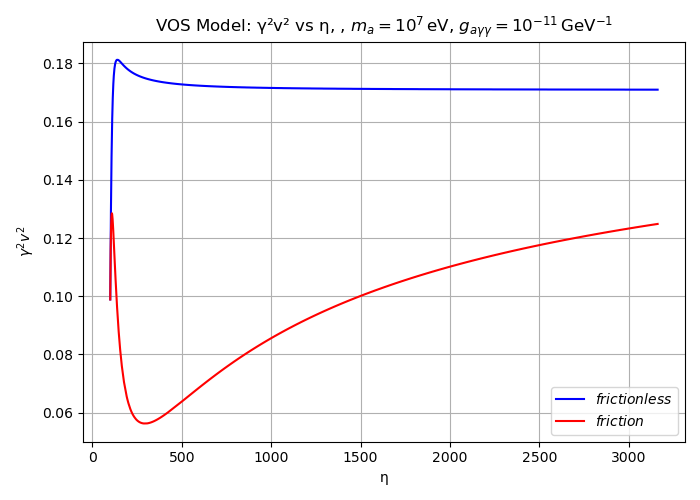}
   % \captionof{figure}{$v$ vs $\eta$}
   \label{anal}
   \caption{The two plots show the evolution with the conformal time $\eta$ of two significant quantities for domain wall network evolution.}
  \end{figure}

We show our preliminar results for a case of a photophilic axion in the plots of Fig.~(\ref{anal}).
The results we show already improve with the final considerations of Ref.~\cite{hassan2025chern} for which a significant or dominant contribution from thermal friction to the dynamics of a DW network was expected to be for $m_a \gtrsim 10^8\,\, \rm eV$, while our preliminary result shows dominance for a smaller mass and a reasonable $g_{a \gamma \gamma}$ coupling constant.

This is due to the additional effects we consider, such as planar decompression, thermal bending, and further decoherence effects.

 The plots show above the energy density, which is reported with the product $\rho \eta$, and below is the rms velocity, which is plotted as $\gamma^2 v^2$.
The reasons for these choices of derived quantities are the same in the literature and related to convenience, since such new quantities are constants when we have the scaling regime, so an asymptote is a signal of a scaling regime.

The plasma effects are taken care of by our collisional terms, where our Feynman diagrams are the same as the ones adopted for axion freeze-in, and we interpret the effects depending on the plasma length scales in terms of resonant production from the Primakoff process and fermion-antifermion annihilation (in particular for electron and muons, similarly to what was already done for axion freeze-in).

%birefringence 

%\cite{Ganoulis:1986rd,PhysRevD.32.1560,PhysRevD.41.1231,Harari:1992ea,FAVITTA2023169396}

%Numerical results from analytical models (preliminary)\\
   %\centering

\begin{comment}
A more general form can be obtained by just keeping the term $\ddot{\Phi}$ and it is similar to the approach in Ref.~\cite{Briaud2024} from which  we use the same general cutoff function $W\Big(\frac{k}{\epsilon R H}\Big)$.
We obtain
\begin{equation}
    \ddot{\bar{\Phi}}+3H \dot{\bar{\Phi}}+V'({\bar{\Phi}})=\hat{g}(\vec{x},t)
\end{equation}
where 
\begin{equation}
    \hat{g}(\vec{x},t)=\frac{1}{3}\epsilon \,R(t)\, H \int \frac{ d^3\vec{k}}{(2 \pi)^{3/2}} \Bigg\{\Bigg[\frac{d^2}{dt^2}+3H\frac{d}{dt}\Bigg]W(k-\epsilon R(t) H)\Bigg\} \left[ d_{\vec{k}} f_{\vec{k}}(t)e^{-i \,\vec{k} \cdot \vec{x}}  +d^{\dagger}_{\vec{k}} f^*_{\vec{k}}(t)e^{i \,\vec{k} \cdot \vec{x}}      \right].
\end{equation}

This general result for the noise can be put in a more interesting perspective in several ways.
If we use the Heaviside cutoff function and we notice we get an expression depending on a term with the dirac delta and a term with the derivative of dirac delta, which can be interpreted as a distributional limit of an Ornstein-Uhlenbeck noise

%arrived here

This is a relevant result to keep in mind, since we will find a more general form coming from the quantum EoM of the average field $\angi{\Phi}=\varphi$ from the 2PI effective approach in FLRW metric 

\end{comment}

%\subsection{Axion string-wall networks}

%\include{./AEDlab}
%\include{./diehaloscope}
%\include{./propagation}
%\include{./electro}
%\include{renormostringa}
%\include{lechneremio}
%\include{mioloop}

\chapter{Conclusion and discussion}
The axion possesses curious and distinctive properties, and research on its properties, along with its detection, could provide clues to solving the Strong CP problem in Quantum Chromodynamics (QCD) and understanding the origin of dark matter and the universe's early history.\\
In this thesis, we have explored several interconnected aspects of axion physics, with particular attention to Axion Cosmology and Axion Electrodynamics. These two areas have been investigated through analytical and field-theoretical approaches, contributing with new insights and results developed throughout the three years of this doctoral work.

After reviewing the theoretical foundations of Quantum Field Theory in curved spacetimes and the essential elements related to Cosmology and the concept of particle, we discussed the origin and motivation for the introduction of the QCD axion as a solution to the Strong CP problem in Quantum Chromodynamics. In this context, we presented the Peccei-Quinn mechanism, the original PQWW model, and the more recent invisible QCD axion models. We also analyzed two main classes of ultraviolet completions of axion theories, namely the field-theoretic and extra-dimensional models, in connection with the axion quality problem and as a reason for the possible emergence of additional axion-like particles, along with the QCD axion.

A significant part of this thesis focused on the theoretical and phenomenological aspects of Axion Electrodynamics, where we examined how a classical axion field modifies Maxwell’s equations and the corresponding physical observables. These modifications have important implications for axion detection experiments based on strong magnetic fields, as well as for theoretical and cosmological questions such as energy-momentum conservation and thermal friction.

We investigated several aspects that remain less explored in the literature and were objects of our works~\cite{PhysRevD.107.043522,FAVITTA2023169396,doi:10.1142/S0217751X24500040}, including:
\begin{itemize}
    \item The derivation and interpretation of the modified Maxwell equations in the presence of an axion field;
    \item The analysis of energy-momentum conservation and its physical meaning;
    \item Theoretical models for detection schemes and topological materials;
    \item The computation of Green’s functions in the axion-photon system;
    \item The impact of both time-dependent and spatially varying axion backgrounds on Casimir forces and zero-point energies;
    \item The role of thermal friction and plasma compression effects on axion domain walls;
    \item The study of optical properties and dispersion relations in axion media.
\end{itemize}
Those aspects will be furtherly discussed in a work in preparation, such as Ref.~\cite{Campello2025}.

In the second part of this work, we introduced techniques from non-equilibrium Quantum Field Theory, specifically the path integral formalism and the 2PI  effective action to describe the self-interacting axion field and its couplings to Standard Model and dark sector fields. This approach allowed us to go beyond the limitations of conventional perturbative or mean-field approximations and to properly account theoretically for non-linear and out-of-equilibrium effects in axion dynamics.

These theoretical developments are then skectched to understand how they are applied to two cosmological contexts:
\begin{enumerate}
    \item Pre-inflationary scenario for high-mass photophilic ALPs: we analyzed the cosmological constraints on the parameter space $(m_a, g_{a\gamma\gamma})$, focusing on the contribution to $\Delta N_{\mathrm{eff}}$ from axion freeze-in production mechanisms.
    \item Post-inflationary scenario for the QCD axion and high-mass photophilic ALPs: we addressed the domain wall problem and investigated the dynamics of axion topological defect networks through both analytical models based on the Velocity-One-Scale framework and an extended non-equilibrium QFT approach inspired by moduli-space quantization.
\end{enumerate}
They will be further developed in next works in preparation, such as Ref.~\cite{Favitta2025-AxionBounds-prep,Favitta2025-AxionPoS-prep}.
In both scenarios, we studied the frictional effects experienced by photophilic ALPs due to interactions with the primordial plasma, identifying the dominant contributions from electrons and muons. We verified the consistency of our approach with existing models and constraints and presented preliminary results that highlight the potential of this framework for future studies.

The methods and results presented in this thesis open several promising directions for further research. On the theoretical side, extensions of the non-equilibrium formalism could improve our understanding of axion dynamics in more complex environments, including strong-gravity or magnetized systems. On the phenomenological side, the study of Casimir forces and optical effects in effective axion backgrounds may provide new experimental ways for detecting axion-induced signatures at laboratory and astrophysical scales, or be useful for practical application with topological materials. 

Overall, this work contributes to strengthening the theoretical foundations and phenomenological relevance of axion physics, bridging between cosmology, quantum field theory, and experimental observables, and supporting the ongoing research for one of the most compelling dark matter candidates.

We also found a theoretical formalism which allows us to better deal with the average evolution of axion topological defects, both for self-interactions and interactions with plasma and dark sectors.
%\item  It can help to link QFT with stochastic processes 
It can help us solve the tension on the numerical simulations of axion topological defects
  % \item  Thermal friction is a phenomenon that can be relevant for ALP networks in the early dynamics and can be relevant for the stability of spherical domain walls, even in the later dynamics of QCD axion networks.
and put future constraints for interesting ALP models and QCD axion models.
    %  \item <3->Axion field backgrounds, which are both time and space-dependent, are of interest for compact objects. 

%\begin{frame}[allowframebreaks]
%\frametitle{Some references}
%\bibliographystyle{mla}
	%Call the file
%\bibliography{phd_favitta}
 %\printbibliography
% \end{frame}

%adj until here
%We have discussed older experimental schemes to detect the axion, such as the Haloscope model and its 'updated version', the Dielectric Haloscope, as well as new methods.
%We then extended and improved the axion-to-photon conversion processes and introduced possible detection schemes for the axion.

%We also investigated the modifications to the Casimir force for an axion field with constant derivatives, providing a comprehensive picture (also at finite temperature T) that is valid in certain approximations for the time-oscillating axion field.\\
%We have calculated the temperature-dependent Casimir effect for a 'toy model' of an axion domain wall and found some interesting properties that could be observed in a real domain wall, resulting in a collapsing radiation pressure that could lead to instability for axion domain walls in every cosmological scenario.

%However, the actual treatment of the axion domain wall requires a computational approach to the problem because the effective functional form of the Axion domain wall, as given by Huang and Sikivie \cite{PhysRevD.32.1560}, is not known analytically but only numerically. This can serve as the basis for future work.

\appendix
\counterwithin{figure}{section}
%\chapter{Some basics of Cosmology}\label{Appendixcosmo}

%\chapter{Some basics of Quantum field theory in curved spacetimes and fixed backgrounds}
%\chapter{Boltzmann transport equation in FLRW metric and Thermodynamics in expanding universe}\label{Appendixboltz}

%The general form for 

\chapter{Numerical details of Boltzmann equations}\label{Appendixnum}

If we are interested in the evolution of the number density of a particle species $a$ in  the Early Universe, assuming the metric to be a FLRW one and so without significant backreaction from $a$, we can describe it in the collisional approximation with a Boltzmann equation, which is in a full general form
\begin{equation}\label{botzgenmicromega}
    \dot{n}_{a}+3 H n_{a}= \sum_{A,B} (\, \xi_B-\xi_A ) \, \mathcal{N}(A \rightarrow B),
\end{equation}

with $A$ and $B$ denoting respectively the generic initial and final states containing $\xi_{A,B}$ particles of type $a$ and $\mathcal{N}(A \rightarrow B)$ is the integrated collision term for the reaction $A \rightarrow B$, with the physical meaning of being the number of $A \rightarrow B$ reactions happening in the thermal bath per unit space-time volume.

The explicit form of the integrated collision term is the following in the Boltzmann limit
\begin{equation}
 \mathcal{N}(A \rightarrow B)= \int \underset{i \,\in A}{\Huge \Pi} \Bigg(\frac{d^3 p_i}{(2 \pi)^3 2 E_i} f_i    \Bigg) \underset{j \,\in B}{\Huge \Pi}\Bigg(\frac{d^3 p_j}{(2 \pi)^3 2 E_j} (1 \mp f_j)    \Bigg) (2 \pi)^4 \, \delta^4\Bigg(\underset{i \,\in A}{\sum} P_i -\underset{j \,\in B}{\sum} P_j \Bigg) C |\mathcal{M}|^2,
\end{equation}
where $f_i$ are the distribution functions of particles $i$, $P_i $ their 4-momenta involved, $C$ a combinatorial factor\footnote{For example, in the case of $2 \rightarrow 2 $ reactions, it is equal to $1/2$ for identical incoming particles and one otherwise} and $|\mathcal{M}|^2$ the squared Feynman amplitude summed over initial and final polarisations.

%introduce here useful thermodynamical quantities 

%put assumptions here

If we consider a scenario with only freeze-in, the dark matter particle $a$ is characterised by very small, then "feeble", couplings with the visible sector and a negligible initial abundance $n_a \ll 1$, so one can assume $\xi_A=0$ in Eq.~\ref{botzgenmicromega}, consider processes with just production of DM.
This assumption of a stable DM can constitute a good approximation as long as $H(T) \gg n \angi{\sigma v}$
However, we will also deal with the axion cases of our interest, also with processes with $a \rightarrow bath$.

Using the time-temperature relation   for Eq.~\ref{botzgenmicromega}, we can obtain the dark matter yield $Y_{a}^0$ at the present temperature $T_0$ by integrating the collision term from the temperature at which DM production starts, which we take to be the reheating temperature $T_R$, to $T_0$

\begin{equation}
    Y_{\phi}(T=T_0)=Y_{\phi}^0= \int_{T_0}^{T_R} \frac{dT}{T \bar{H}(T) s(T)} \Bigg(\mathcal{N}(bath \rightarrow \phi+X)+\mathcal{N}(\phi \rightarrow bath)  \Bigg),
\end{equation}%adjust equation

%talk about 1->2 and 2->1 cases, details and so on 

We are interested in processes of the kind $1,2 \rightarrow$, for which the collision term is of the form
\begin{equation}\label{12abcollterm}
\begin{aligned}
    \mathcal{N}(1,2 \rightarrow a,b )=C_{12} \int \Bigg(\frac{d^3 p_1}{(2 \pi)^3 2 E_1}    \Bigg) \Bigg(\frac{d^3 p_2}{(2 \pi)^3 2 E_2}    \Bigg) \Bigg(\frac{d^3 p_a}{(2 \pi)^3 2 E_a}    \Bigg)\Bigg(\frac{d^3 p_b}{(2 \pi)^3 2 E_b}    \Bigg) \\ \times (2 \pi)^4\, \delta^4(P_1+P_2-P_a-P_b)\, |\mathcal{M}|^2\, f_1\, f_2 \,(1 \mp f_a) \,(1 \mp f_b) 
    \end{aligned}
\end{equation}
where $f_i$ are the phase-space distributions of the particles, and $|\mathcal{M}|^2$ is the Feynman scattering amplitude.
If one can approximate $(1 \mp f_a) \,(1 \mp f_b) \sim 1$, the expression \ref{12abcollterm} can be reduced to
\begin{equation}\label{12abcolltermapprox}
   \mathcal{N}(1,2 \rightarrow a,b )= \frac{T \, g_1 \, g_2 |\eta_1 \eta_2|}{8 \pi^4} C_{12} \int ds \,(p_{1,2}^{CM})^2 \sqrt{s} \, \sigma(s) \tilde{K}_1(\sqrt{s}/T, x_1,x_2, 0, \eta_1,\eta_2) .
\end{equation}

Analogously to the Refs.~\cite{BELANGER2018173,alguero2024micromegas}, we can approximate the integral \ref{12abcollterm}  by adopting the approximative expression of Eq.~\ref{12abcolltermapprox}   and introducing in the integrand the correction factor 
\begin{equation}
    K\Big(\sqrt{s}/T,\, x_1,\,x_2, \eta_2)=\frac{ \tilde{K}_1\Big(\sqrt{s}/T,\, x_1,\,x_2,0,0,  \eta_2\Big)}{\tilde{K}_1\Big(\sqrt{s}/T,\, x_1,\,x_2, 0,0,0\Big)}
\end{equation} %add something on this and 1->2 processes
where 
$x_i= \frac{m_i}{T}$,which accounts for the initial-state Pauli/Bose effects and $\tilde{K}$ and $K$ are defined as in Ref.~\cite{BELANGER2018173}.

If the integral over $d^3 p_a$ collapses (as $\varphi$):

\begin{equation}
\frac{d^3 p_{\varphi}}{2 E_\varphi} f_\varphi(\mathbf{p}_\varphi) \to \frac{n_\varphi}{2 m_\varphi},
\end{equation}
the collision term reduces to

\begin{equation}
\mathcal{N} = \frac{n_{\varphi}}{2 m_{\phi}} \int \frac{d^3 p_1}{2 E_1} \frac{d^3 p_2}{2 E_2} \frac{d^3 p_b}{2 E_b} \, (2\pi)^4 \delta^4(P_1+P_2-P_b-P_a) \, |\mathcal{M}|^2|_{\vec{p}_{\varphi}=0} \,f_1 f_2 (1 \mp f_b),
\end{equation}
as already seen in the former section.

Using the center-of-mass energy $s = (P_1 + P_2)^2$, the phase space reduces effectively to a 2$\to$1 process:

\begin{equation}
\mathcal{N}_{2\to2}^{\varphi} 
= \frac{n_{\varphi}}{2 m_{\phi}} \, \frac{T \, g_1 g_2 |\eta_1 \eta_2|}{8 \pi^4} \int ds \, (p_{1,2}^{\rm CM})^2 \, \frac{|\mathbf{p}_b|}{\sqrt{s}} \, \sigma(s) \, \tilde{K}_1^{(b)}\Big(\frac{\sqrt{s}}{T}, x_1, x_2, x_b, \eta_1, \eta_2 \Big),
\end{equation}

with

\begin{align}
|\mathbf{p}_b|&= \frac{\sqrt{(s-(m_a+m_b)^2)(s-(m_a-m_b)^2)}}{2 \sqrt{s}} ,
\end{align}  

and

\begin{equation}
K_1^{(b)}\Big(\sqrt{s}/T, x_1, x_2, \eta_1, \eta_2\Big) = \frac{\tilde{K}_1^{(b)}(\eta_1, \eta_2)}{\tilde{K}_1^{(b)}(0,0)},
\end{equation}
which accounts for the initial-state Pauli/Bose effects in the same spirit.

Other collision terms can be obtained with the same logic.
All these collision terms are evaluated numerically with MicrOmegas 6.1.15, using the self-checking adaptive algorithm for Simpson integration adopted in the code, precisely using the closed Newton–Cotes quadrature formulas with 3, 5, and 9 nodes \cite{alguero2024micromegas}.

\begin{acknowledgements} 	
	I would like to express my immense gratitude to my tutor, Roberto Passante, and my co-tutor, Lucia Rizzuto, who guided and supported me throughout my PhD activity.\\ 
	I want to thank the University of Palermo for allowing me to pursue a PhD program and the COST Action Cosmic WISPers for the research opportunities they have provided.
	\\
	Special thanks to Miguel Escudero, Ben Safdi, Amelia Drew, Viatcheslav Mukhanov, Pierre Sikivie, Hyungjin Kim, Francesca Calore, Giuseppe Lucente, Alessandro Lella, Giovanni Pierobon, Elisa Todarello,  Mario Reig, Ken'ichi Saikawa, Javier Redondo, Elisa Ferreira, Junu Jeong, and Caterina Braggio for their valuable suggestions and discussions.
    
I would be remiss in not mentioning "my son" Sirio, Nicola, Giulia, Alessia, Andrea, Giuseppe, and Roberta. Their moral support and belief in me have kept my spirits and motivation together during this process and have been fundamental for completing my PhD.
Many thanks to five colleagues whose physics knowledge, expertise, ability in professional relationships and humanity I greatly appreciate: Luca Cammarata, Tommaso Fazio,  Yeray Garcia del Castillo, Thong Nguyen, ,Edoardo Alaimo.

Furthermore, I would like to thank my family and a few colleagues and professors whose physics knowledge and expertise helped me through this process: Grazia Maria Cottone, Roberto Grimaudo, Luca Innocenti, Vincenzo Intravaia, Salvatore Lorenzo, and Davide Valenti.
   
   % \begin{figure}
    %    \centering
     %   \includegraphics[width=0.1\linewidth]{Wantothankme.png}
      %  \caption{Caption}
       % \label{fig:enter-label}
    %\end{figure}
	%No thanks to many non-serious and unprofessional people who have been obstacles or troublesome during my PhD.
  
\end{acknowledgements}

\bibliographystyle{unsrt}
\bibliography{phd_favitta}
\end{document}